\documentclass[12pt, headings=standardclasses]{scrartcl} 
\usepackage{listings, xcolor}
\usepackage{amsmath}
\usepackage{amssymb}
\usepackage{graphicx}
\usepackage{caption}
\usepackage{subcaption}
\usepackage[T1]{fontenc}
\usepackage[latin1]{inputenc}
\usepackage[english]{babel}
\usepackage[latin1]{inputenc}
\usepackage{lmodern}
\usepackage{bbm}
\usepackage{float} 
\usepackage{epstopdf}
\usepackage{placeins}
\usepackage[normalem]{ulem}

\usepackage{rotating}
\usepackage{pdflscape}
\usepackage{afterpage}

\usepackage{pifont}
\newcommand{\cmark}{\ding{51}}%
\newcommand{\xmark}{\ding{55}}%

\usepackage{threeparttable}
\usepackage{setspace}
\usepackage[round]{natbib}
\usepackage{babelbib}
\usepackage{url}
\usepackage{rotating}
\usepackage{dcolumn}
\newcolumntype{d}[1]{D{.}{.}{#1}}
\usepackage{graphicx}
\usepackage{subcaption}
\usepackage{pgffor}
\pgfkeys{/captions/.is family, /captions,
	1/.initial = {glm1},
	2/.initial = {glm2},
	3/.initial = {xgb},
	4/.initial = {nnet},
	5/.initial = {grf}
}
\pgfkeys{/captions2/.is family, /captions2,
	1/.initial = {Logit},
	2/.initial = {Logit Int},
	3/.initial = {XGBoost},
	4/.initial = {Neural Net},
	5/.initial = {Rand Forest}
}

\usepackage{booktabs}
\usepackage{tabularx}
\usepackage{indentfirst}
\usepackage{multirow}
\usepackage{framed}
\usepackage{enumitem}
\usepackage{lscape}
\usepackage{longtable}
\usepackage{tabulary}  
\usepackage{rotating}

\usepackage{chngcntr}
\counterwithin{figure}{section}
\counterwithin{table}{section}
\counterwithin{equation}{section}
\usepackage{setspace}

\usepackage{colortbl}

\usepackage[left=2.54cm,right=2.54cm,top=2.54cm,bottom=2.54cm,marginparwidth=1.75cm]{geometry}
\usepackage{scrlayer-scrpage}
\usepackage{ragged2e}
\usepackage[overload]{empheq}
\ofoot[\pagemark]{\pagemark}

\usepackage{tikz}
\usetikzlibrary{shapes,decorations,arrows,calc,arrows.meta,fit,positioning}
\tikzset{
	-Latex,auto,node distance =1 cm and 1 cm,semithick,
	state/.style ={ellipse, draw, minimum width = 0.7 cm},
	point/.style = {circle, draw, inner sep=0.04cm,fill,node contents={}},
	bidirected/.style={Latex-Latex,dashed},
	el/.style = {inner sep=2pt, align=left, sloped}
}
\tikzset{
	vertex/.style = {
		circle,
		fill            = black,
		outer sep = 2pt,
		inner sep = 1pt,
	}
}

\tikzstyle{line} = [draw, -latex']
\usetikzlibrary{arrows,decorations.markings,patterns,calc}
\usetikzlibrary{shadows}
\tikzset{shadow scale=1, shadow xshift=-.5ex, shadow yshift=-.5ex,
	opacity=.5, fill=black!50, every shadow}

\color{black}
\usepackage[titletoc,title]{appendix}

\usepackage[hidelinks,breaklinks=true]{hyperref}

\newcommand\ind{\perp\!\!\!\perp}

\newtheorem{theorem}{Theorem}[section]
\newtheorem{ass}{Assumption}[section]
\newtheorem{corr}{Corollary}[section]
\newtheorem{lem}{Lemma}[section]

\newtheorem{prop}{Proposition}[section]

\usepackage{dcolumn}
\newcolumntype{d}[1]{D{.}{.}{#1} }
\newcolumntype{Y}{>{\centering\arraybackslash}X}
\newcolumntype{Z}{>{\flushleft\arraybackslash}X}
\newcommand{\titleinfo}{Treatment Evaluation at the Intensive and Extensive Margins}
\title{\titleinfo}
\def\authora{Phillip Heiler}
\def\authorb{Asbj\o rn Kaufmann}
\def\authorc{Bezirgen Veliyev}
\def\aff{Aarhus University}
\def\emaila{\href{mailto:pheiler@econ.au.dk}{pheiler@econ.au.dk}}

\date{} 

\begin{document}
	\begin{titlepage}
		\title{\titleinfo \thanks{ \scriptsize We would like to thank the Editor Yanqin Fan, Associate Editor, two anonymous referees, Isaiah Andrews, Andres Aradillas-Lopez, Kevin Chen, Juan Carlos Escanciano, Patrik Guggenberger, Sukjin Han, Mark Henry, Keisuke Hirano, Edward Kennedy, Giovanni Mellace, Peter Schochet, Vira Semenova, Neil Shephard, Mikkel S\o lvsten, Elie Tamer, Davide Viviano, Chris Walker, as well as the participants of the Aarhus, Harvard, and PennState Econometrics Seminars, IAAE 2025 Conference, and 2025 World Congress of the Econometric Society for their comments and suggestions that helped to improve the work. All remaining errors are ours. Phillip Heiler and Bezirgen Veliyev thank the Danish National Research Foundation (DNRF186) and the Independent Research Fund Denmark for support (DFF 3099-00077B)}}
		\author{\authora\thanks{{\scriptsize Aarhus University. Department of Economics and Business Economics, Aarhus Center for Econometrics (ACE), TrygFonden's Centre for Child Research, Universitetsbyen 51, 8000 Aarhus C, Denmark, email: \emaila. Part of this research was conducted during main affiliation with Harvard University, Department of Economics. 02138 Cambridge MA, United States.}} \quad 
      \authorb\thanks{{\scriptsize \aff. Department of Economics and Business Economics, Universitetsbyen 51, 8000 Aarhus C, Denmark.}} \quad 
      \authorc\thanks{{\scriptsize \aff. Department of Economics and Business Economics, Aarhus Center for Econometrics (ACE), Universitetsbyen 51, 8000 Aarhus C, Denmark.}}\\ 
      }

		\date{} 
		\maketitle
		\thispagestyle{empty}
  
		\begin{abstract} \singlespacing	\small
            \vspace{-4em}
			This paper provides a solution to the evaluation of treatment effects in selective samples when neither instruments nor parametric assumptions are available. We provide sharp bounds for average treatment effects under a conditional monotonicity assumption for all principal strata, i.e.~units characterizing the complete intensive and extensive margins. Most importantly, we allow for a large share of units whose selection is indifferent to treatment, e.g.~due to non-compliance. 
            The existence of such a population is crucially tied to the regularity of sharp population bounds and thus conventional asymptotic inference for methods such as Lee bounds can be misleading. This problem can be solved using smoothed outer identification regions for inference. We provide semiparametrically efficient debiased machine learning estimators for both regular and smooth bounds that can accommodate high-dimensional covariates and flexible functional forms. Our study of active labor market policy reveals the empirical prevalence of the aforementioned indifference population and supports results from previous impact analyses under much weaker assumptions.  
		\end{abstract}
        \vspace{1em}
  
		\noindent \textbf{Keywords:} Double/debiased machine learning; Lee bounds; Partial identification; Principal strata; Sample selection \\
		\textbf{JEL classification:} C13, C14, C21
	\end{titlepage}
	
\setcounter{page}{1} 
	\newpage

\newpage

\normalsize 
\section{Introduction}
This paper deals with the evaluation of causal effects of a binary treatment when the outcome is only selectively observed and no instruments are available. Such missing outcome data or sample selection is ubiquitous and a threat to internal validity \citep{heckman1974shadow,heckman1979sample}. Typical examples include outcome data such as wages that are only observed if units are employed \citep{lee2009training}, missing survey responses \citep{bernhardt2024howdoes}, or attrition \citep{zhang2003estimation}. In the context of impact evaluation, a treatment can often change the sample selection status for some units. For example, if the treatment is an active labor market policy such as job training, it likely affects the probability of employment for some participants, e.g.~due to accumulation of human capital (positive) or opportunity costs to job search (negative). For some units, however, the policy might be irrelevant with regard to employment or sample selection, but still relevant in terms of their earnings or other outcomes. Hence, there are potentially heterogeneous units at the intensive and extensive margins. These units can be classified within principal strata defined by their potential selection status \citep{frangakis2002principal}. 

We consider set identification, estimation, and inference for all principal strata, encompassing the complete intensive and extensive margins, under a weak conditional monotonicity assumption. Weak monotonicity allows for the simultaneous presence of always-takers, compliers, defiers, and never-takers.\footnote{In contrast to the instrumental variables literature, principal strata here refer to potential selection statuses when treatment is exogenously set to zero and one respectively, see Section \ref{sec_model}.}

The main challenge is the presence of units whose selection probabilities are unaffected by the treatment. The existence of such units is most apparent when the treatment is subject to additional non-compliance after assignment and intent-to-treat effects are analyzed. In this case, units who have information or preferences that lead them to not participate after treatment assignment will likely also not change their selection behavior, e.g.~selection into employment, as a consequence. Further, there are policies that do not boost or hinder employability for some units, e.g.~due to a lack of signal for their potential employers in the short run, but increase human capital boosting productivity and earnings. Similarly, in field experiments where responses cannot be enforced, an intervention might only affect response probabilities to follow-up surveys for some units.

From a discrete choice perspective, our setup admits units who are indifferent to the treatment at any conceivable level of their private information.
From a model selection perspective, unaffected units are equivalent to a sparsity constraint in the selection equation. Modern model selection methods such as $\ell_1$-regularization, subset selection, deep neural networks, boosted trees, and random forests are able to recover such potentially sparse structures in the selection equation. 

\begin{figure}[!h] \centering\caption{Job Corps Relative Sample Selection Probabilities} 
		\includegraphics[width = 0.67\textwidth, trim = 0 120 0 100, clip]{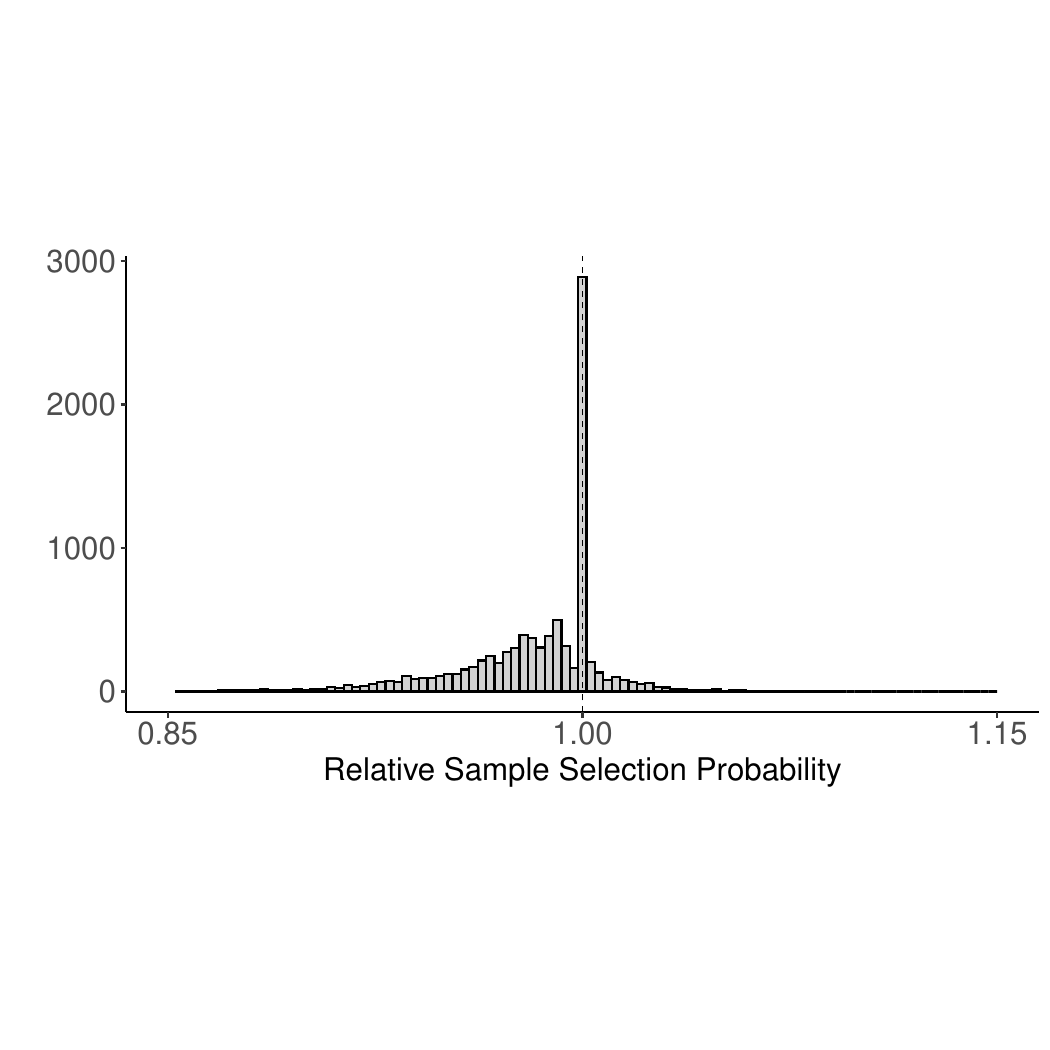}
		\label{fig_p0x_INTRO1}
        \footnotesize \begin{justify}
	Histogram of the estimated relative conditional employment selection probabilities under treatment and control. Time period $t=180$, method XGBoost, $n = 9415$. There are 2616 units at the exact numerical value $1.00$ ($P(\mathcal{X}^0) = 28.61\%$). For more details on data and nuisance model consider Section \ref{sec_empirical1} and Appendix \ref{app_JC}.
	\end{justify}
\end{figure}

Figure \ref{fig_p0x_INTRO1} contains an example from the job training program analyzed in this paper (Job Corps, JC). It provides the histogram of the estimated relative selection probabilities with and without assignment to training using boosted trees. A value of $1.00$ indicates that assignment does not predict employment status for a given unit.
In this case, a point mass of 2616 out of 9415 units occurs at an exact numerical value of one. This pattern is striking and persistent across evaluation periods and estimation methods, see Section \ref{sec_model} and Section \ref{sec_empirical1}.

\begin{figure}[!h] 
	\centering \caption{Job Corps Test Statistics and Rejections of No Point Mass  $H_0: P(\mathcal{X}^0) = 0$} \label{fig_Test_INTRO1}
        \includegraphics[width=\textwidth, clip, trim = 0 100 0 90]{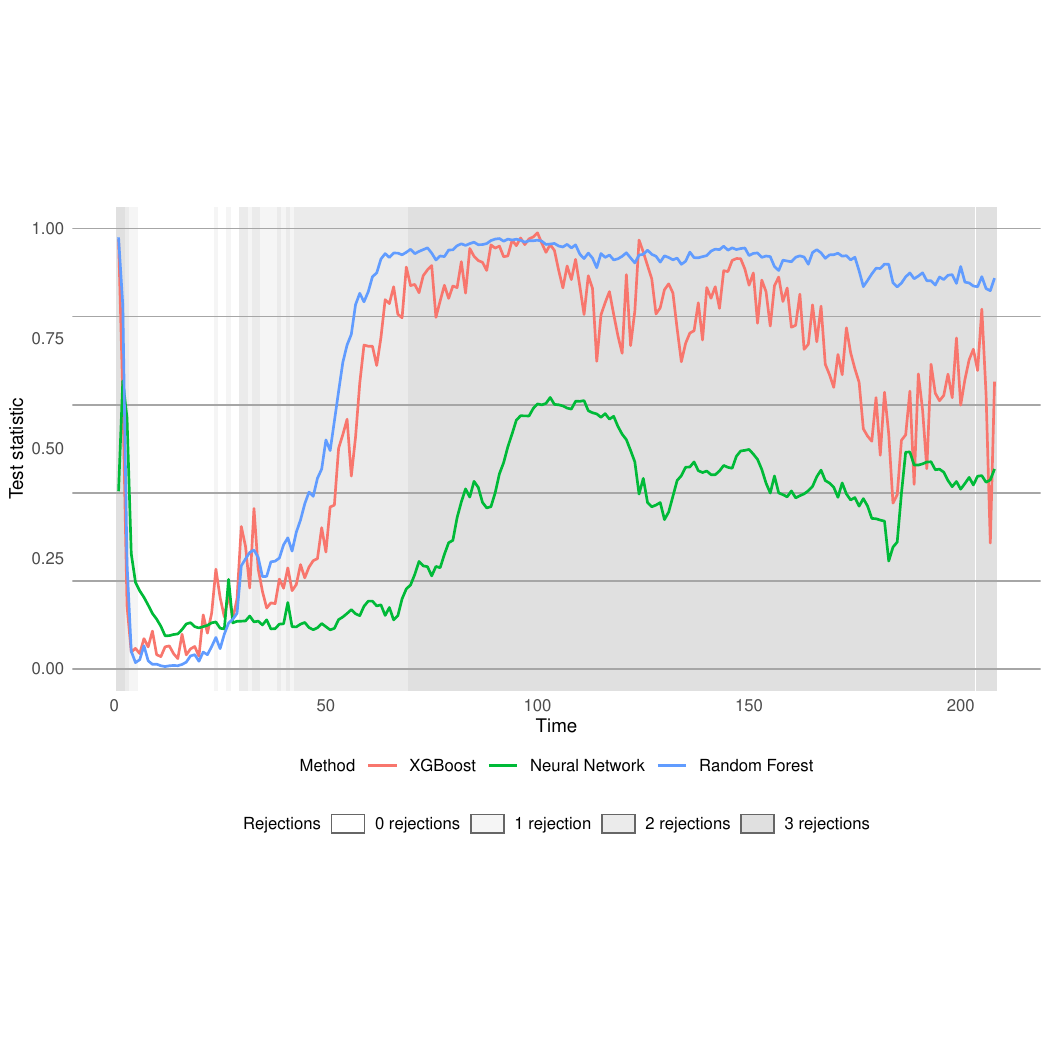}
	\footnotesize
 \begin{justify} This plot contains the value of the test statistic $T_n(\kappa_n)$ for three different nuisance function estimators and the number of rejections for all time periods $t=1,\dots,208$. The total number of rejections is 502/624. Local neighborhood $\kappa_n = 0.0111$ and critical value $c_n = 0.1889$ were chosen via $\kappa_n = n^{-1/4}\log^{-1}(n)$ and $M_n = 2 \log\log n$, see Section \ref{sec_Test-pointMass1} for more details. 
  \end{justify}
\end{figure}

We also provide a diagnostic test for the absence of point mass. Figure \ref{fig_Test_INTRO1} contains the test statistic and rejections using various selection probability estimation methods in JC for all post-treatment weeks. Overall, we reject the null in 502/624 cases supporting the empirical relevance of a method that provides valid inference for such DGPs. In this study, there is significant non-compliance. Thus a large presence of estimated unaffected units seems plausible. 

Mixture distributions for relative conditional selection probabilities with point mass at $1.00$ as in Figure \ref{fig_p0x_INTRO1} are harmless from an identification perspective. However, they directly translate into a fundamental problem for estimation and inference. In particular, sharp population bounds are non-smooth functionals for which no uniformly unbiased estimator exists, i.e.~the problem is irregular by construction \citep{hirano2012impossibility}. More concretely, we show that no point mass is in fact necessary and sufficient for population bounds to be pathwise differentiable in the sense of \cite{bickel1993efficient}. This means that standard errors and statistical inference obtained from methods such as Lee bounds with ``naive'' asymptotics \citep{lee2009training} are invalid in the presence of unaffected units. 

We propose a solution via repeated smoothing steps that effectively act as covers for the sharp identified sets. The smooth outer identification region can converge to the sharp identified set. For both regular and smoothed irregular bounds, we provide efficient influence functions and corresponding estimators using debiased machine learning. Under simple high-level conditions, the estimators are root-$n$ asymptotically normal and reach their respective semiparametric efficiency bounds. 
Moreover, the smooth outer bounds yield valid, but possibly conservative confidence regions for the true parameter of interest as the direction of any bias is known. Smoothing parameters can be of any rate that avoids re-introducing the irregularity asymptotically. Moreover, any specific smoothing choice can be equipped with an ex-post economic interpretation with regard to the size of intensive and extensive margins and/or principal strata underlying the bounds. This further guides the choice of appropriate smoothing levels in finite samples.
Monte Carlo simulations suggest that, in irregular designs, inference using the smooth methods compares favorably to the alternatives from the literature that either ignore the irregularity of the problem or rely on trimming \citep{heiler2024heterogeneous} or moment selection/shifting methods \citep{andrews2010inference,semenova2025generalized}. 



We also explore the role of the propensity score for efficiency, extending the results in \cite{hahn1998role} to sample selection without exclusion. The efficiency bounds for the principal-stratum treatment effect bounds do not depend on the knowledge of the propensity score. 
We quantify the efficiency gap when using the moment functions under knowledge of the propensity score as suggested by \cite{heiler2024heterogeneous} and \cite{semenova2025generalized}. Using the latter is generally inefficient except in knife-edge cases. However, there is a trade-off from a practical perspective: Analogously to inverse probability weighting (IPW) estimation of average treatment effects, these estimators do not require conditional (truncated) outcome means as additional nuisance inputs. Thus, depending on the difficulty in estimating the latter, using these simpler moment functions might be preferable if propensity scores are known. All aforementioned properties regarding (ir)regularity and smoothing also directly apply to these inefficient moment functions and their smoothed counterparts.  


The empirical study is a comprehensive re-evaluation of the National Job Corps Study \citep{burghardt1999national,schochet2008does}. We demonstrate the empirical relevance of flexible estimation via machine learning methods, in particular with regard to sparsity in the selection equation. Aggregate results of the new semiparametrically efficient smoothing bounds with optimized learners rule out moderate to large negative impact of Job Corps on hourly earnings at the end of the evaluation period. At the intensive margin, results over various time periods are surprisingly close to the more restrictive impact analysis by \cite{lee2009training} that relies on a stronger monotonicity assumption rejected by the data \citep{semenova2025generalized}. At the extensive margin, results are uncertain due to the small aggregate employment effects of the Job Corps intervention.

The paper is organized as follows. Section \ref{sec_literature} discusses the relevant literature. Section \ref{sec_model} introduces the nonparametric sample selection model and provides an in-depth discussion of conditional monotonicity. Section \ref{sec_identification} presents the sharp and smooth outer identification regions for all principal strata. Section \ref{sec_regularity} provides the results on pathwise differentiability. Section \ref{sec_estimationinference} contains the assumptions for debiased machine learning estimation and asymptotic inference as well as the efficiency gap under known propensity scores.
Section \ref{sec_smoothingALL} discusses smoothing parameter choice.
Section \ref{sec_Test-pointMass1} provides the theory for the no point mass test. 
Section \ref{sec_empirical1} presents the empirical study. Section \ref{sec_S25} contains additional comparisons to \cite{semenova2025generalized}. Section \ref{sec_conclusion} concludes. All additional results, extensions and supplementary derivations are in the Appendix.

\section{Literature} \label{sec_literature}
Bounds on causal effects for the intensive margin under monotonicity have first been introduced by \cite{zhang2003estimation}, see also \cite{lee2009training} for an extension to a form of conditional monotonicity that excludes units whose selection behavior is unaffected by the treatment. Such ``Lee'' or ``Zhang-Rubin-Lee'' bounds \citep{andersen2023guide} are defined by conditional expectations trimmed at conditional quantiles. \cite{imai2008sharp} shows the sharpness of such trimming bounds.
\cite{huber2015sharp} derive bounds for compliers under strong monotonicity. We extend their ideas allowing for compliers and defiers simultaneously. 
\cite{honore2020selection,honore2022sample} discuss bounds in sample selection models without exclusion restrictions under additional parametric assumptions. Our bounds do not require any parametric structure. 
\cite{bartalotti2021identifying} provide bounds with and without monotonicity within a marginal treatment effect (MTE) framework. They focus on strong monotonicity and do not provide inference theory. \cite{heiler2024heterogeneous} considers heterogeneous intensive margin bounds under a strong margin assumption effectively ruling out subgroups for which selection probabilities do not depend on treatment. Our new moment functions can be used in conjunction with this approach to provide heterogeneous effect bounds for any margin. 
\cite{semenova2025generalized} introduces orthogonal moments for estimation of intensive margin bounds in high-dimensional setups. We provide a more thorough discussion on the relationship with \cite{semenova2025generalized} in Section \ref{sec_S25}. 
\cite{okamoto2025robustify} discusses sensitivity analysis of intensive margin and MTE bounds under partial violation of strong monotonicity. His stochastic monotonicity assumption is distinct from our approach as it assumes strong monotonicity for a limited share of the population instead of conditional monotonicity. The use of intensive margin bounds has also been advocated by \cite{chen2023logs} when encountering dependent variables with (many) zeroes. Other than \cite{semenova2025generalized}, all these papers consider either strong monotonicity or the restrictive version of conditional monotonicity \citep{lee2009training}.  We explicitly want to allow for units whose selection is unaffected by treatment, i.e.~mixture distributions for relative conditional selection probabilities with point mass at one.\footnote{Monotonicity also plays a crucial role for instrumental variable (IV) based methods \citep{imbens1994identification,angrist1995two,abadie2003semiparametric,heckman2007econometric2,sloczynski2020should,heiler2022efficient}. 
It rationalizes the same choices as a structural model for treatment selection that is additively separable in observables and unobservables \citep{vytlacil2002independence} as in our setup. 
The key distinction to the IV setup is that we are interested in the causal effect of the treatment accounting for endogenous selection. Thus, our treatment takes on the role of the instrument while the selection is fully endogenous as the treatment in the IV case. 
The weak monotonicity assumption posited in this paper allows for the presence of all latent subtypes in the population but limits their presence within ex-ante unknown partitions of the covariate space. The IV framework does not allow for nonparametric identification in any population without switchers (no first stage). For identification and inference of complier effects at the intensive margin under (weak) monotonicity in both treatment and selection response consider \cite{dong2026sharpboundsinferencesample}.}

Such mixtures imply that bounds are no longer smooth in the underlying data distribution and thus there exists no estimator sequence that is locally asymptotically unbiased in the sense of \cite{hirano2012impossibility}. 
We construct an outer identification region that circumvents the non-regularity of the underlying functional of interest. Our identified set has non-empty interior and a zero duality gap \citep{kaido2014asymptotically}. Thus, its semiparametric efficiency bound can be characterized and root-$n$ consistent regular estimators of the support function exist in a uniform sense. As a by-product of the strict convexity of the smooth identified set, confidence intervals for the effects can be made tighter using modified critical values \citep{imbens2004confidence,stoye2020simple}. 

Our estimands are ratios of weighted conditional expectations normalized by their share. For the intensive margin, estimating the share is 
analogous to estimating 
the optimal welfare function. \cite{luedtke2016statistical} show that the latter is pathwise differentiable if and only if units which are indifferent between treatments have an \textit{exceptional} law with zero conditional variance under both regimes. We extend their result to a class of ratio estimands and show that, for all bounds, the point mass phenomenon is both necessary and sufficient for irregularity, i.e.~there are no \textit{exceptional} laws as the latter would imply a violation of the overlap assumption.

Smoothing non-differentiable functionals has a long-standing tradition in econometrics, see e.g.~\cite{Horowitz1992smoothedmaxscore}. \cite{levis2024covariateassisted}  suggest smoothing for obtaining regular estimators of \cite{balke1997bounds} bounds. They propose a worst-case bias correction arising from the particular approximation. Unlike their approach, our smoothing always widens the identified set and thus one can navigate a tightness-inference trade-off without the need for a worst-case bias component. \cite{whitehouse2025inferenceoptimalpolicyvalues} also suggest smoothing in the context of optimal welfare. Again, translating their approach to our setup would not widen the identified set. Thus, confidence intervals might be overly tight in finite samples. However, \cite{whitehouse2025inferenceoptimalpolicyvalues} achieve learning guarantees under weaker assumptions on the underlying nuisances.

In other contexts, \cite{pakel2023bounds} and \cite{lee2026bounding} note that there can be a trade-off between tightness and precision for inference. They argue that confidence sets obtained from outer bounds can well be tighter compared to using sharp bounds. Our case is more accurately characterized as a tightness-validity trade-off. In particular, under irregularity, naive confidence intervals would rely on wrong first-order approximations while smoothing based inference remains valid but conservative. 

We characterize the efficient influence functions and semiparametric efficiency bounds with and without knowledge of the propensity score. Our results generalize \cite{hahn1998role} to sample selection under weak conditional monotonicity and either (i) regular sharp bounds or (ii) regular smooth bounds that cover the irregular sharp bounds. We also contribute and make use of the expanding literature on the use of debiased machine learning methods for estimation of causal effects in economics \citep{chernozhukov2018double,chernozhukov2022locally} and its combination with partial identification \citep{heiler2024heterogeneous,semenova2023set,semenova2025generalized}. More specifically, we show that the particularities of the relative selection probabilities and the rate at which they can be learned from data crucially interact with the (ir)regularity of the parameters of interest. 

\section{Model and Monotonicity} \label{sec_model}
Assume we observe iid data $W = (YS,S,D,X)'$ where $S \in \{0,1\}$ is a selection variable indicating whether an outcome is observed or not. $Y \in \mathcal{Y}$ is the partially unobserved outcome of interest. $D \in \{0,1\}$ is a binary treatment of interest. $X \in \mathcal{X}$ is a vector of predetermined covariates. We are interested in the causal effect of the treatment on the outcome defined in terms of potential outcomes and potential selection indicators. Outcome and selection variables are defined as \begin{align}
    Y = DY(1) + (1-D)Y(0), \\
    S = DS(1) + (1-D)S(0).
\end{align} \vspace{-24pt}
		\begin{figure}[!h] \centering \caption{Nonparametric Sample Selection Model} \label{fig_dagSS1}
			\begin{tikzpicture}[scale=1]
				\tikzset{line width=1.5pt}

				\node[ellipse,draw,line width = 1.2pt,dashed, drop shadow, fill = white] (-1) at (4.8,0) {\scriptsize$U$};
				\node[ellipse,draw,line width = 1.2pt, drop shadow, fill = white] (0) at (2.4,1.2) {\scriptsize$X$};
				\node[ellipse,draw,line width = 1.2pt, drop shadow, fill = white] (1) at (0,0) {\scriptsize$D$};
				\node[ellipse,draw,line width = 1.2pt, drop shadow, fill = white] (2)  at (3.6,-1.2) {\scriptsize$S$};
				\node[ellipse,draw,line width = 1.2pt, drop shadow, fill = lightgray] (3) at (1.2,-1.2) {\scriptsize$Y$};
				\node[ellipse,draw,line width = 1.2pt, drop shadow, fill = white] (4) at (2.4,-2.4) {\scriptsize$Y\times S$};
				
				\path (1) edge (3);
				\path (2) edge (4);
				\path (3) edge (4);

				\path (0) edge[out=west,in=north west] (1);	
				\path (0) edge (2);	
				\path (0) edge (3);	
				
				\path[<->] (0) edge[out=east,in=north west, dashed]  (-1);	
				
				\path (-1) edge[dashed] (2);	
				\path (-1) edge[dashed] (3);	
				\path (1) edge (2);	
				
			\end{tikzpicture}\footnotesize \begin{justify}
	A graphical representation of a nonparametric sample selection model. Nodes denote variables and edges are structural relationships. A missing arrow from one node to another is an exclusion restriction. Unobserved independent components at each node are omitted. Unobserved variables and their edges are dashed. Gray nodes are partially unobserved.
	\end{justify} \vspace{-12pt}
		\end{figure}
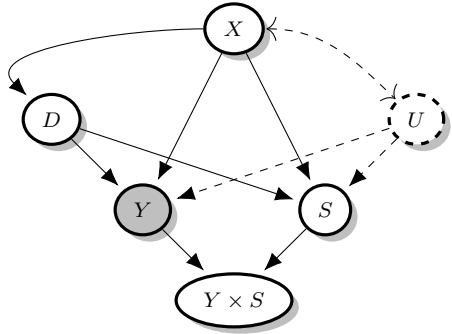

Figure \ref{fig_dagSS1} depicts a prototypical sample selection model with the relevant exclusion restrictions. Such a model implies the following conditional independence assumption for the treatment which we maintain throughout:

\begin{ass}[Conditional Independence] \label{ass_CIA}
    $$D \ind Y(d),S(d)~|~X=x \textit{ for all } x \in \mathcal{X} \textit{ and } d=0,1,$$
where $\ind$ denotes statistical independence. 
\end{ass}

Under Assumption \ref{ass_CIA}, treatment is exogenous conditional on $X$. However, selection is left essentially unrestricted, i.e.~it can be fully endogenous or exogenous. 
In particular, there are no exclusion restrictions available for $S$, i.e.~we cannot use instruments or similar to account for endogenous selection. 
The potential selection indicators define principal strata which represent how the treatment affects selection behavior. Table \ref{tab_principalStrata1} contains all strata using the nomenclature from the IV literature as well as intensive and extensive margin.

\begin{table}[!h]
    \centering 
    \caption{Principal Strata and Margins}
    \label{tab_principalStrata1}
\begin{tabular}{llccl} \hline \hline 
Type &  & $S(0)$ & $S(1)$ & Margin \\ \hline  \\[-1.8ex]
(AT) & Always-takers & 1 & 1 & Intensive margin \\
(C)  & Compliers     & 0 & 1 & Extensive margin (entry) \\
(D)  & Defiers       & 1 & 0 & Extensive margin (exit) \\
(NT) & Never-takers  & 0 & 0 & None \\ [0.5ex]
\hline \hline 
\end{tabular}
\end{table}

To proceed, we postulate the existence of an unknown but identified partitioning of the covariate space $\mathcal{X}$ on which a \textit{weak} or \textit{conditional monotonicity} assumption applies to these principal strata.\footnote{Note that this definition differs slightly from the conditional monotonicity assumption 2(b) of \cite{semenova2025generalized} which does not yield a unique partitioning.}

\begin{ass}[Weak/Conditional Monotonicity]\label{ass_monotone} 
For each $x \in \mathcal{X},$ we have that $
P( S(1) \geq S(0) | X=x)=1 \quad \mbox{or} \quad 
    P( S(1) \leq S(0) |  X=x)=1.
$ \label{ass_weakMon1}
\end{ass}
Assumption \ref{ass_weakMon1} yields a distinct partitioning of $\mathcal{X} = {\mathcal{X}}^+ \cup {\mathcal{X}}^- \cup {\mathcal{X}}^0$ where \begin{align*}
    \mathcal{X}^0 &=\left\{ x \in \mathcal{X}: P(S(1)=S(0)|X=x)=1 \right\}, \\
    \mathcal{X}^+ &= \{ x \in \mathcal{X}: P(S(1) \geq S(0)|X=x)=1, \quad P(S(1)>S(0)|X=x)>0 \}, \\
    \mathcal{X}^- &=\{  x \in \mathcal{X}:   P(S(1)\leq S(0)|X=x)=1, \quad P(S(1)<S(0)|X=x)>0 \}.
\end{align*}

The presence of a potentially non-empty $\mathcal{X}^0$, i.e.~$P(\mathcal{X}^0) > 0$, will be crucial in what follows. 
Substantially, weak monotonicity allows for the presence of all principal strata including defiers. However, it is a partial restriction of types within partitions. Table \ref{tab_partitions1} contains the mixture of types in the different partitions. 

	\begin{table}[!h]
		\centering \footnotesize
		\caption{Partitions, Observed Strata, and Admitted Latent Types}
        \label{tab_partitions1}
        \vspace{6pt}
		\begin{minipage}{0.3\linewidth}
			\centering
			\begin{subtable}{\linewidth}
				\centering
				\begin{tabular}{l|cc}
					$S~\backslash~ D$ & 0 & 1  \\[0.5ex] \hline \\[-1.5ex]
					0 & NT/C & NT \\[1ex]
					1 & AT & AT/C \\[0.5ex] \hline 
				\end{tabular}
				\caption{$x\in \mathcal{X}^+$}
			\end{subtable}
		\end{minipage}
		\begin{minipage}{0.3\linewidth}
			\centering
			\begin{subtable}{\linewidth}
				\centering
				\begin{tabular}{l|cc}
					$S~\backslash~ D$ & 0 & 1  \\[0.5ex] \hline \\[-1.5ex]
					0 & ~~NT~~ & ~~NT~~ \\[1ex]
					1 & AT & AT \\[0.5ex] \hline 
				\end{tabular}
				\caption{$x\in \mathcal{X}^0$}
			\end{subtable}
		\end{minipage}%
		\begin{minipage}{0.3\linewidth}
			\centering
			\begin{subtable}{\linewidth}
				\centering
				\begin{tabular}{l|cc}
					$S~\backslash~ D$ & 0 & 1  \\[0.5ex] \hline \\[-1.5ex]
					0 & NT & NT/D \\[1ex]
					1 & AT/D & AT \\[0.5ex] \hline 
				\end{tabular}
				\caption{$x\in \mathcal{X}^-$}
			\end{subtable}
		\end{minipage}
	\end{table}

For example, in partition $\mathcal{X}^0$, the population consists only of always-takers and never-takers while on $\mathcal{X}^+$ and $\mathcal{X}^-$ only defiers or compliers are ruled out, respectively. This assumption seems most relevant if we have some discrete covariates and/or continuous variable that affect selection behavior discontinuously e.g.~by threshold effects.

\paragraph{\textbf{Remark 1 (Single Index Selection Model):}} \label{sec_singleIndex1} 
\textit{Assumption \ref{ass_weakMon1} is also consistent with a single-index structure in the selection equation that allows for sparsity with respect to the treatment on some parts of the covariate space. 
Consider the following semiparametric selection model with a selection that is additively separable in observables and unobservables
\begin{align}
    D &= \mathbbm{1}(f(X,U_D) \geq 0), \notag \\
    Y &= \mu(X,D,U_Y),  \notag  \\
    S &= \mathbbm{1}(g(D,X) + U_S \geq 0),
\end{align}
where $U_D \ind U_Y,U_S$ conditional on $X=x$ implies Assumption \ref{ass_CIA}.  
Weak monotonicity as in Assumption \ref{ass_monotone} here is implied by separability in the selection equation and an unrestricted $g(D,X)$. This can generate a mixed index structure \begin{align}
    g(d,x) = \begin{cases} g_+(d,x) &\textit{ if } x \in \mathcal{X}^+,\\
    g_-(d,x) &\textit{ if } x \in \mathcal{X}^-,\\
    g_0(x) &\textit{ if } x \in \mathcal{X}^0.\end{cases}
\end{align}
where $g_+(1,x) > g_+(0,x)$ and $g_-(0,x) > g_-(1,x)$. The fact that $g_0(x)$ does not depend on $d$ can be seen as a sparsity assumption within a partition of the covariate space. For example, consider the case of a simple parametric index in the selection equation with $X = (X_1,X_2)'$ where $X_1$ is continuous and $X_2$ discrete with support points $\{-1,0,1\}$
\begin{align}
    S = \mathbbm{1}\big(\gamma_0 + \gamma_1X_1 + \gamma_2D\mathbbm{1}(X_2 = -1) + \gamma_3D\mathbbm{1}(X_2 = 1) + U_S \geq 0\big).
\end{align}
Here $X_2$ can separate the monotonicity types. In particular, $\gamma_2 < 0 < \gamma_3$ or $\gamma_2 > 0 > \gamma_3$ imply conditional monotonicity. If the signs of $\gamma_2$ and $\gamma_3$ are identical, then either $\mathcal{X}^+ = \{ \emptyset\} $ or $\mathcal{X}^- = \{ \emptyset\}$, implying strong monotonicity $P(S(1) \geq S(0)) = 1$. }

Whenever $P(\mathcal{X}^0) > 0$, conditional monotonicity and independence generate a \textit{mixture} distribution for the potential or relative conditional selection probabilities
\begin{align}
  p_0(x) = \frac{P(S=1|D=0,X=x)}{P(S=1|D=1,X=x)} = \frac{P(S(0)=1|X=x)}{P(S(1)=1|X=x)}.  
\end{align}
In particular, it is straightforward to show that $x \in \mathcal{X}^+$ and $x \in \mathcal{X}^-$ if and only if $p_0(x) < 1$ and $p_0(x) > 1$, respectively, while there is potentially a point mass for which $p_0(x) = 1$ achieved if and only if $P(\mathcal{X}^0) > 0$. Thus, conditional monotonicity allows for mixture distributions which are not necessarily smooth in $p_0(x)$. We seek to conduct inference on causal effects for all principal strata under such mixtures.

\begin{figure}[!h]
	\centering
	\caption{Example Distributions for $\hat{p}_0(x)$ at $t = 208$}
	\label{fig:hist_alL^208} 
\begin{subfigure}{0.45\textwidth}
\caption{\footnotesize Logit}
    \begin{flushright} 
\includegraphics[width=0.66667\textwidth, trim =0 10 0 0 , clip]{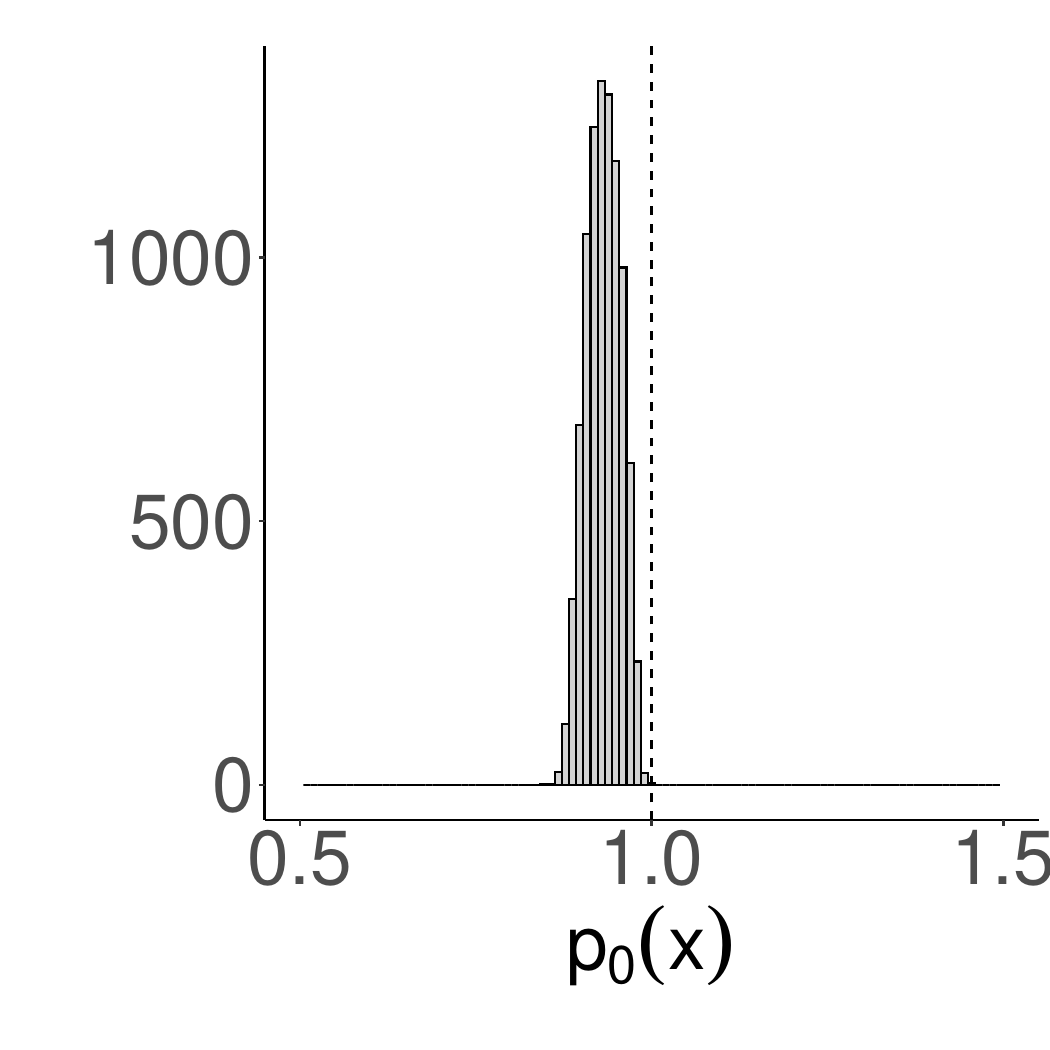} \end{flushright}
\end{subfigure}
\begin{subfigure}{0.45\textwidth}
\caption{ \footnotesize Logit Interacted}
    \begin{flushleft} \includegraphics[width=0.66667\textwidth, clip]{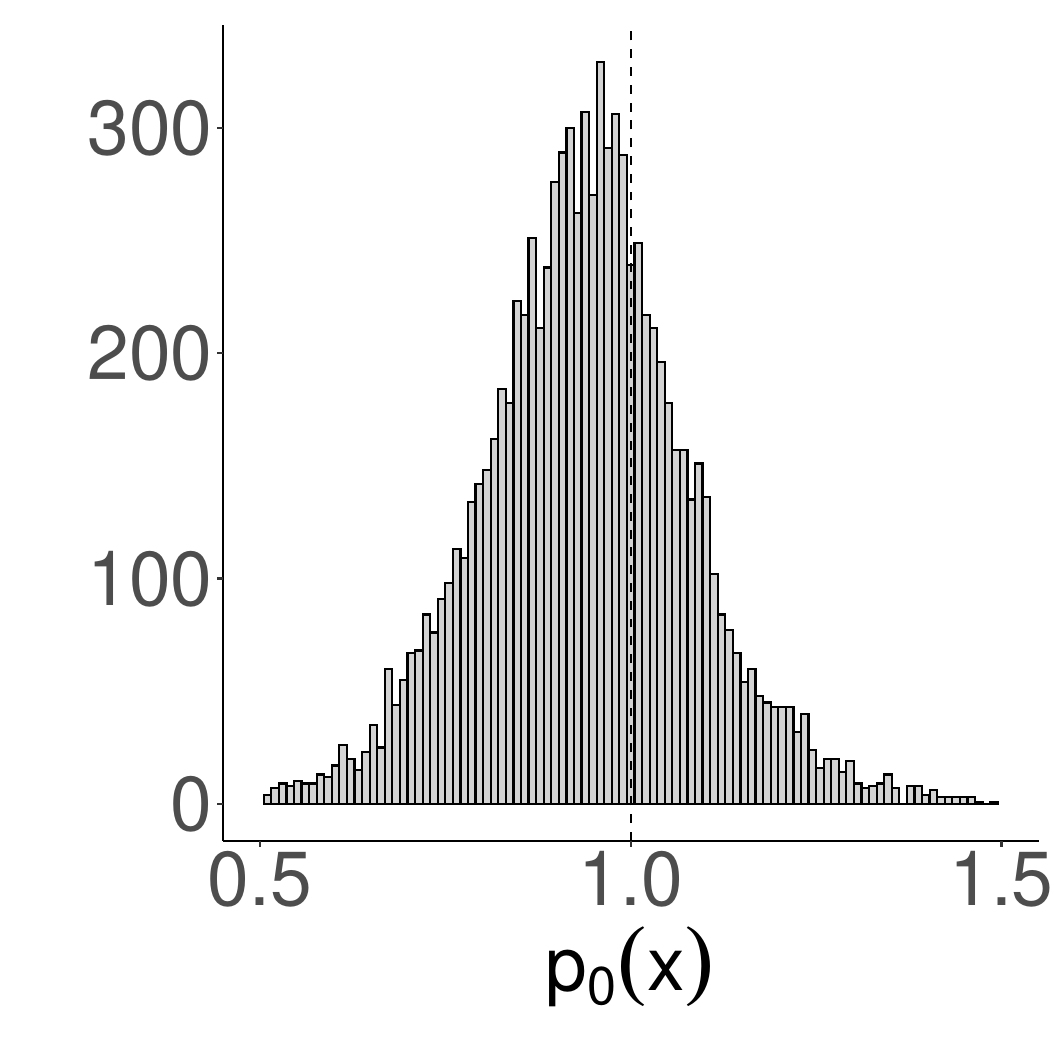} \end{flushleft}
\end{subfigure} \\
\begin{subfigure}{0.3\textwidth}
\caption{\footnotesize XGBoost}
    \begin{flushright} \includegraphics[width=\textwidth, clip]{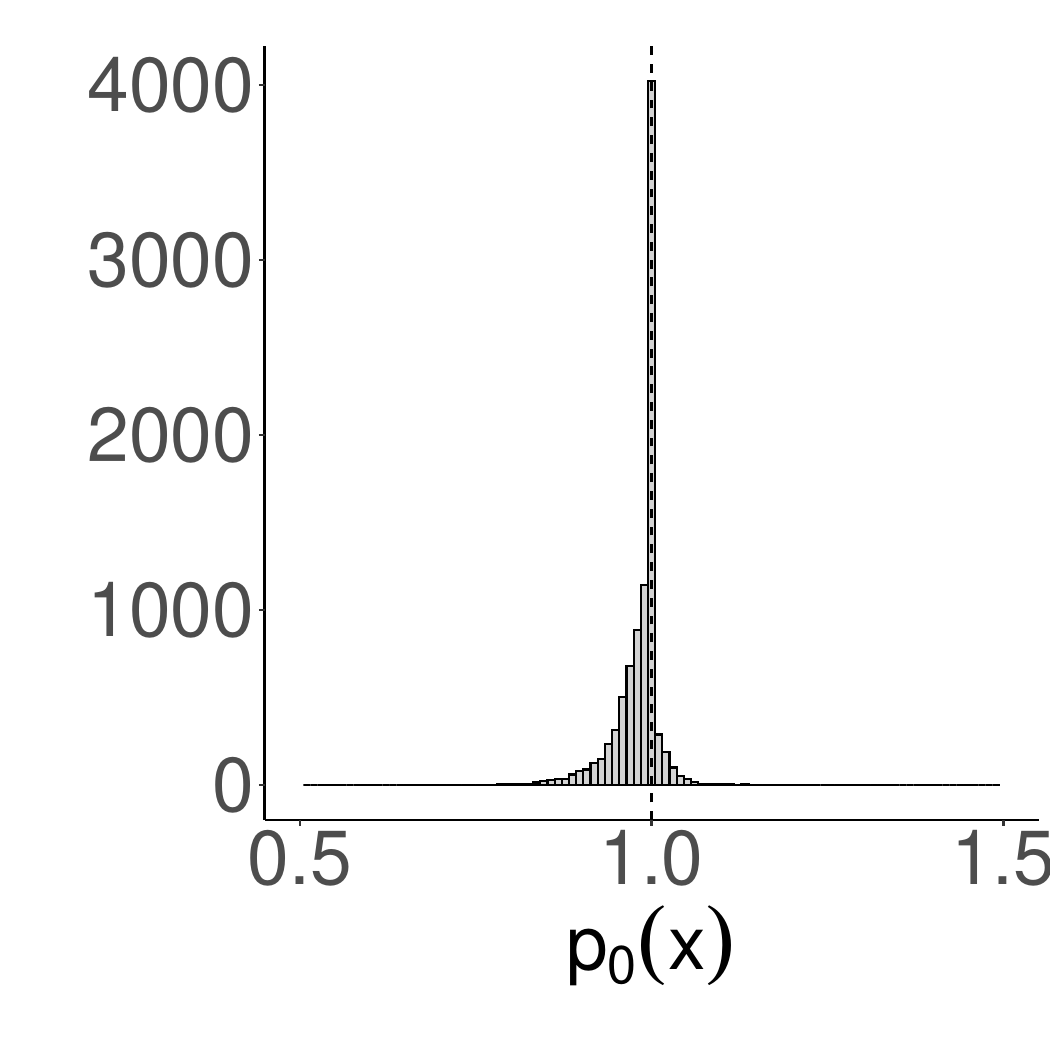} \end{flushright}
\end{subfigure}
\begin{subfigure}{0.3\textwidth}
\caption{\footnotesize Neural Network}
    \begin{flushleft} \includegraphics[width=\textwidth, clip]{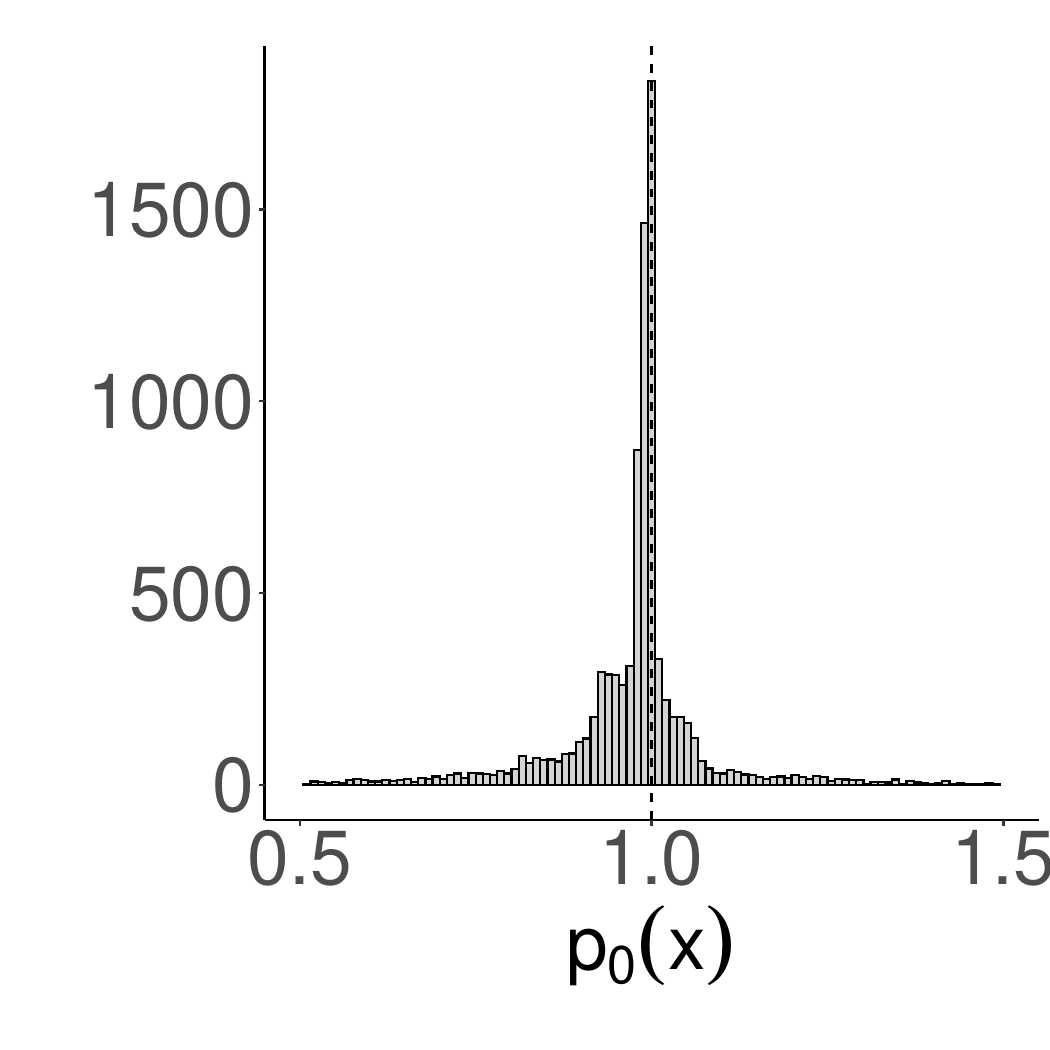} \end{flushleft}
\end{subfigure}
\begin{subfigure}{0.3\textwidth}
\caption{\footnotesize Random Forest}
    \begin{flushleft} \includegraphics[width=\textwidth, clip]{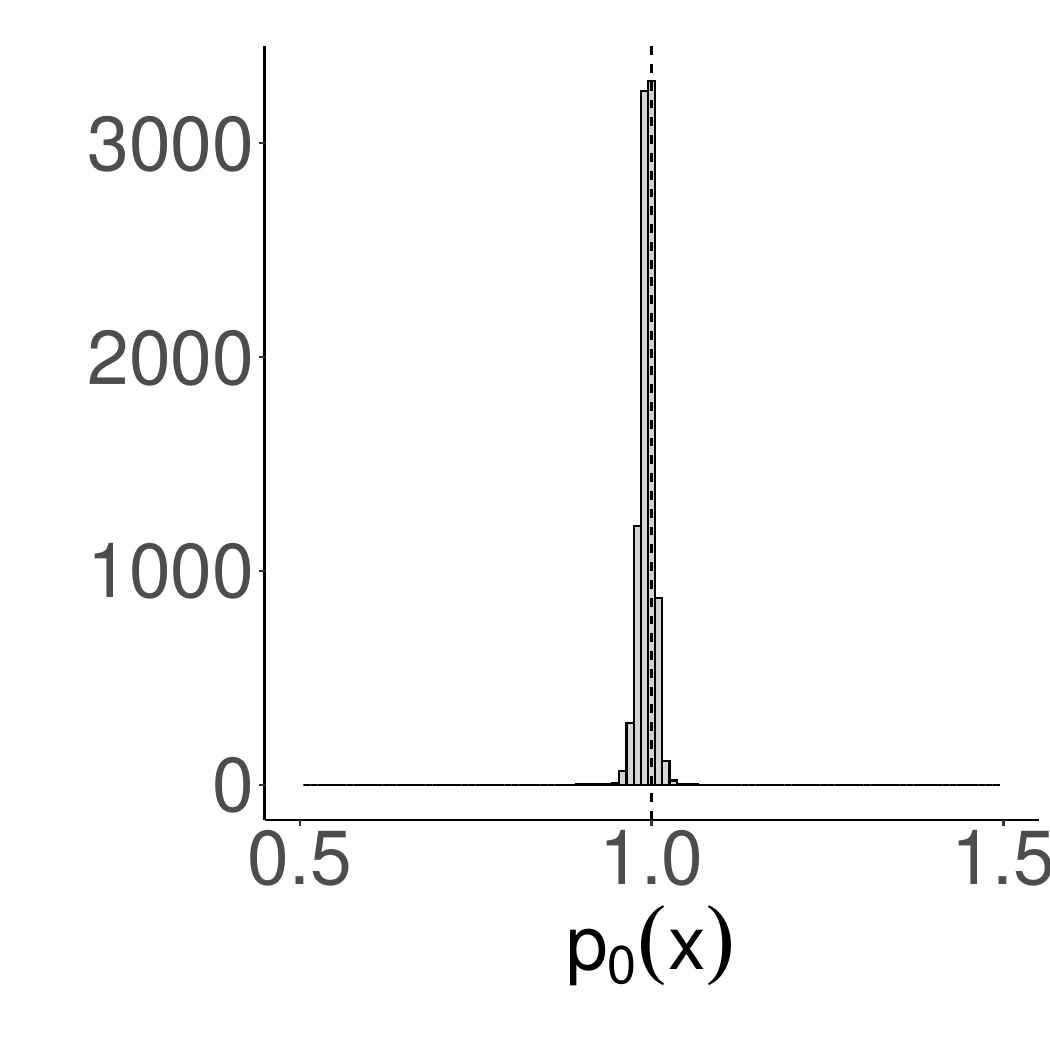} \end{flushleft}
\end{subfigure}
	\begin{justify} \footnotesize
		The figure contains histograms for the estimated $p_0(x)$ using five different estimation methods on the research sample at week $t=208$. (a) Logit is a logistic regression with linear additive index. (b) Logit Interacted is a fully treatment interacted logistic model. (c) XGBoost uses gradient-boosted trees with cross-validated boosting steps using \verb|xgboost|. (d) Neural Network is a batch-trained, fully connected artificial feed-forward neural network with three ReLU hidden layers, drop-out regularization, and a final sigmoid layer using \verb|keras3|. (e) Random Forest is an honest probability forest using \verb|grf| with default parameters. 
	\end{justify}
\end{figure}

Modern model selection methods such as $\ell_1$-regularization, subset selection, neural networks, boosting, and random forests are able to recover and produce potentially sparse structures in the selection equation. We present an example from relative employment probabilities and a job training treatment in Figure \ref{fig:hist_alL^208}. It contains the predicted relative selection probabilities from Section \ref{sec_empirical1} using five different estimation methods: (a) logistic regression, (b) logistic regression with fully interacted treatment, (c) gradient-boosted trees (XGBoost), (d) neural network, and (e) random forest, see Section \ref{sec_empirical1} and Appendix \ref{app_JC} for more details. 

The methods capable of sparsity yield a clear mixture pattern for the estimated distribution of $p_0(x)$. In particular, (c), (d), and (e) produce exact sample shares of numerical ones up to $45.16\%$. (a) imposes strong monotonicity and cannot produce exact ones in finite samples by construction. We suspect that (a) is most likely to be misspecified. (b) has relevant density around one, but also generates overly extreme values for the relative selection probabilities. The sparse methods agree on around 45\% -- 60\% of the close or equal to one classifications. This qualitative pattern can be observed along a long sequence of post-treatment periods. Over time there is an average distribution shift suggesting more positive employment effects. The sparsity pattern, however, is pervasive. 

For a more quantitative assessment, we also test the null of no point mass $P(\mathcal{X}^0) = 0$ using the diagnostic test developed in Section \ref{sec_Test-pointMass1}. For the learners that can produce sparsity (XGBoost, Neural Network and Random Forest), the null is rejected for the majority of the post-treatment periods, see Figure \ref{fig_Test_INTRO1}.
Overall, we take this as a signal that the mixture distributions are likely reflective of or a good approximation to an underlying structure and are not only spurious by-products of particular estimation algorithms, motivating Assumption \ref{ass_weakMon1} with $P(\mathcal{X}^0) > 0$. 

\begin{figure}[!h]
	\centering
	\caption{Distributions for $S$-, $R$-, and $DR$-Learners for the selection CATE at $t = 135$}
	\label{fig:hist_CATE_135} 
\begin{subfigure}{0.3\textwidth}
\caption{\footnotesize S Learner}
    \begin{flushright} \includegraphics[trim = 0 0 0 50,width=\textwidth, clip]{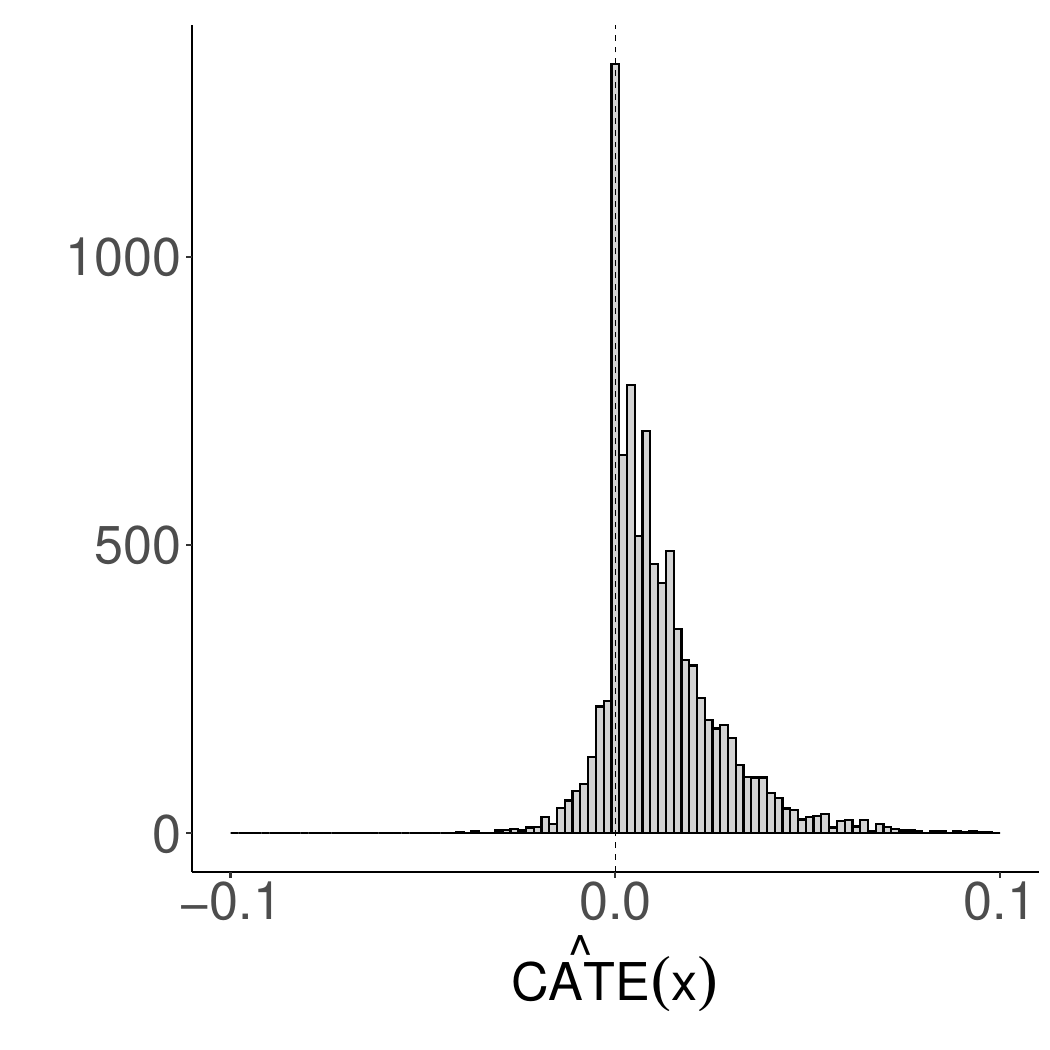} \end{flushright}
\end{subfigure}
\begin{subfigure}{0.3\textwidth}
\caption{\footnotesize R Learner}
    \begin{flushleft} \includegraphics[trim = 0 0 0 50,width=\textwidth, clip]{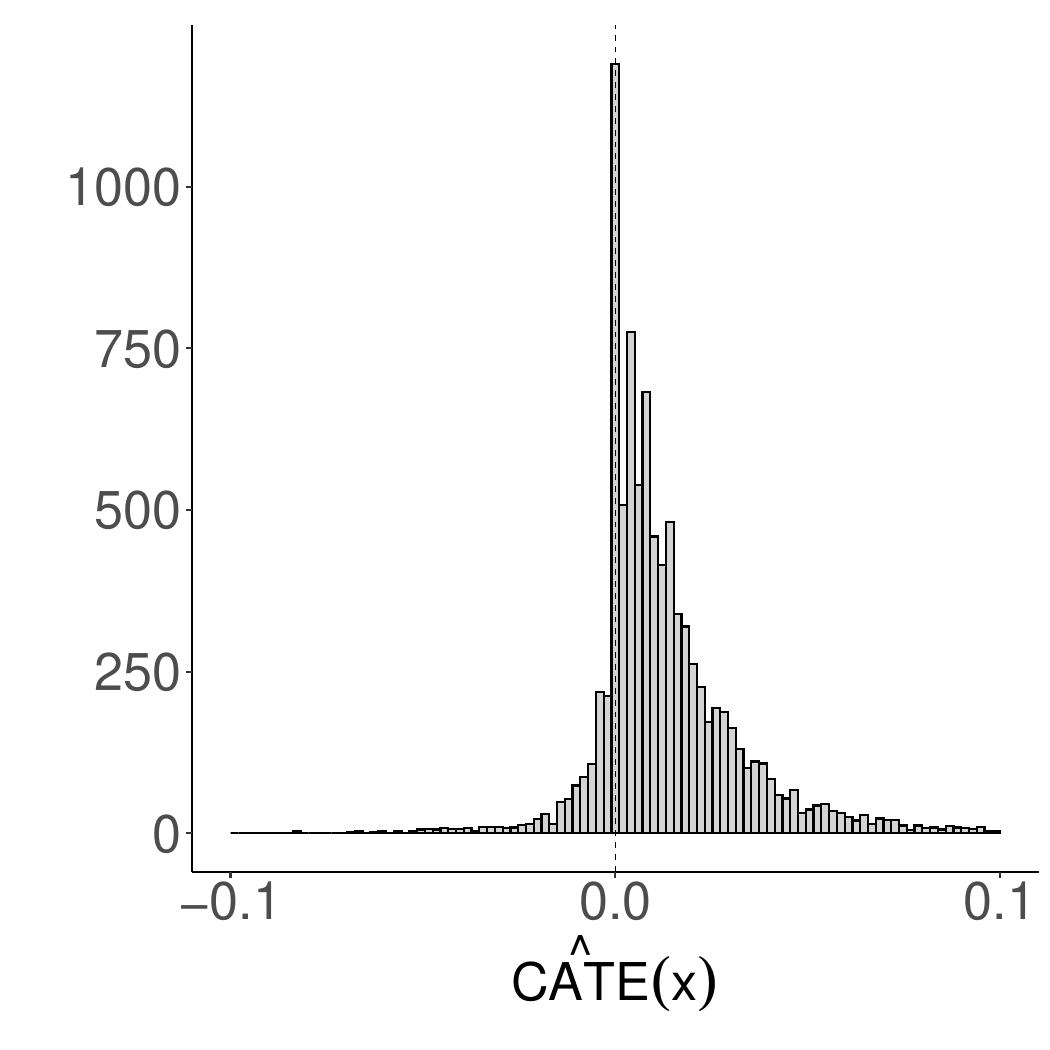} \end{flushleft}
\end{subfigure}
\begin{subfigure}{0.3\textwidth}
\caption{\footnotesize DR Learner}
    \begin{flushleft} \includegraphics[trim = 0 0 0 50,width=\textwidth, clip]{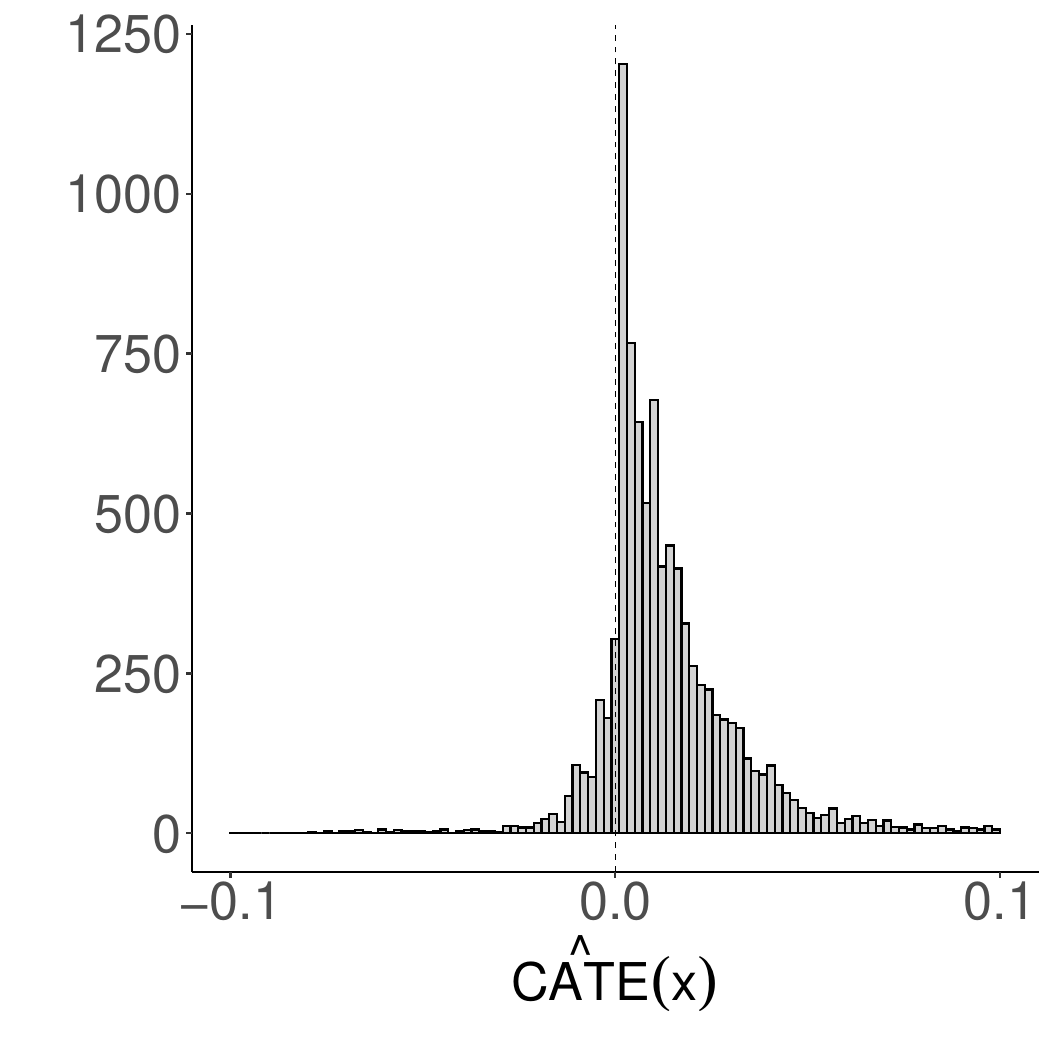} \end{flushleft}
\end{subfigure} \vspace{-12pt}
	\begin{justify} \footnotesize
		Histograms of estimated CATEs for treatment on selection using three different learners at week $t=135$. (a) $S$-Learner denotes the single learner that uses cross-fitted nuisances for selection probabilities directly \citep{kuenzel2019metalearners}. (b) $R$-Learner denotes a regularized flexible regression of residualized outcomes on residualized treatments with cross-fitting \citep{nie2020quasioracle} (c) $DR$-Learner denotes a regularized flexible regression with efficient score function outcomes \citep{kennedy2023towards}. For all learners, XGBoost has been used for all nuisances and regularized regression steps with separate tuning at each stage. 
	\end{justify}
\end{figure}

\paragraph{Remark 2 (Empirical Assessment of Point Mass):}
\textit{On a cautionary note, empirical distributions of individual predictions such as in Figure \ref{fig:hist_alL^208} should not be over-interpreted. On one hand, we can have clear theoretical reasons why true $p_0(x)$ ought to be one for many units, e.g.~units that did not intend to join the program are likely unaffected by a non-binding assignment to treatment. On the other hand, these are just finite-sample estimates from models that are all likely misspecified to a smaller or larger extent. Extreme values that suggest economically unreasonable unit selection effects are likely driven by a high-variance model. 
We can take a more agnostic perspective with regard to the evidence provided by the empirical distribution of $\hat{p}_0(x)$. In particular, in high dimensions or with complicated functional forms, convergence to true relative selection probabilities is hard to achieve for any model. In finite samples, producing a sparse representation by removing the impact of the treatment in parts of the covariate space could just be a method's solution to optimize fit. Methods for predicting conditional selection probabilities are not necessarily designed to best predict the implied relative $p_0(x)$. However, if a learner performs best among conventional performance metrics for probabilities, and does so by producing sparsity with respect to the treatment, we treat it as a potentially better approximation and need to deal with the point mass regardless of whether this is a true or approximate model. 
Ultimately, however, our assumptions are about the respective \textit{population} quantities. Thus, we can further investigate the robustness of the point mass phenomenon using alternative learning approaches that might be less subject to regularization bias. Figure \ref{fig:hist_CATE_135} contains an example for the Job Corps data at $t=135$ using $S$-, $R$-, and $DR$-learners \citep{kuenzel2019metalearners,nie2020quasioracle,kennedy2023towards}. Instead of reporting the estimated relative selection effects
, we report the more conventional CATE versions $E[S(1)|X=x] - E[S(0)|X=x]$. In this case, the point mass is expected to occur at zero, not at one. Additional results are available upon request. The $S$-learner here corresponds to the approach used for the previously presented $p_0(x)$ estimates. $R$- and $DR$-learners are less prone to regularization bias. For all approaches, the point mass phenomenon occurs and results seem overall comparable. This again indicates that results are not only spurious by-products of a particular learning algorithm motivating Assumption \ref{ass_weakMon1} with $P(\mathcal{X}^0) > 0$. 
Note again that, in practice and independently of the learning methods, we recommend complementing any inspection of a histogram with the diagnostic test of no point mass provided in Section \ref{sec_Test-pointMass1}.}

\section{Identification of Sharp and Smooth Bounds} \label{sec_identification}
\subsection{Sharp Bounds}
\subsubsection{The General Case}
 We denote $a = O(b)$ and $a=O_p(b)$ as $a\lesssim b$ and $a\lesssim_P b$, respectively. We first present identification of conditional causal effect bounds for any principal stratum, i.e.~parameters of the form \begin{align}
    \beta(x,s_0,s_1) := E[Y(1)-Y(0)|S(0)=s_0,S(1)=s_1,X=x].
\end{align}
This covers all relevant components for any of the principal strata and margin effects.\footnote{Trivially, for a principal stratum effect to be well defined, we require that the corresponding population share is non-zero. We assume this throughout.} 
Let $q_d(u,x)$ be the $u$-quantile of $Y$ conditional on $S=1, D=d, X=x$. For $d\in\{0,1\}$, we define \begin{align}
    \beta_{1,d}(x,u) := E[Y|S=1,D=d,X=x,Y\leq q_d(u,x)], \\
    \beta_{0,d}(x,u) := E[Y|S=1,D=d,X=x,Y\geq q_d(u,x)],
\end{align}
which yields
\begin{align}
 \beta_{1,d}(x,1) = \beta_{0,d}(x,0) =  E[Y|S=1,D=d,X=x].
\end{align}
In line with the literature on monotonicity bounds, we assume a continuous outcome.

\begin{ass}[Continuity]\label{ass_continuity} 
     For all $x\in \mathcal{X}$ and $d\in \{0,1\}$, the conditional outcome distribution $P(Y \leq y|S=1,D=d,X=x)$ is continuous.
\end{ass}
Moreover, there must be comparable units in terms of covariates in both selected treatment groups. \begin{ass}[Multiple Overlap]\label{ass_overlap} Let $m(x) := P(D=1|X=x)$ and $s(d,x) := P(S=1|D=d,X=x)$. For all $x\in \mathcal{X}$ and any $d \in \{0,1\}$, we have that $0 < m(x) < 1$ and $0 < s(d,x)< 1$.
\end{ass}

Assumptions \ref{ass_CIA} and \ref{ass_monotone} imply the specific bounds in terms of truncated means of observed conditional distributions. Assumptions \ref{ass_continuity} and \ref{ass_overlap} ensure that the corresponding conditioning sets are non-empty and the relevant truncation quantiles are unique. In particular, we obtain the following identification result for the sharp upper and lower bounds for all principal strata.

\begin{prop}[Identification] \label{prop_identification1}
For $x\in \mathcal{X}$ and $d\in \{0,1\}$, let $\mathcal{Y}_d(x)$ be the support of $Y(d)$ conditional on $X=x$ with $\underline{y}_d(x)$ and $\overline{y}_d(x)$ its respective infimum and supremum. Under Assumptions \ref{ass_CIA}, \ref{ass_monotone}, \ref{ass_continuity}, and \ref{ass_overlap}, the sharp lower bounds for the principal strata are given by \begin{align*}
        \beta_L(x,s_0,s_1) = \begin{cases}
            \beta_{1,1}(x,\min\{p_0(x),1\}) - \beta_{0,0}(x,1-\min\{1/p_0(x),1\}) &\text{ if } s_0 = 1,~s_1 = 1 \\
            \beta_{1,1}(x,1-\min\{p_0(x),1\}) - \beta_{0,0}(x,\min\{1/p_0(x),1\}) &\text{ if } s_0 \neq s_1 \\
            \underline{{y}}_1(x) - \overline{{y}}_0(x) &\text{ if } s_0 = 0,~s_1 = 0,
        \end{cases}
    \end{align*}
    and the sharp upper bounds are given by \begin{align*}
        \beta_U(x,s_0,s_1) = \begin{cases}
            \beta_{0,1}(x,1-\min\{p_0(x),1\}) - \beta_{1,0}(x,\min\{1/p_0(x),1\}) &\text{ if } s_0 = 1,~s_1 = 1 \\
            \beta_{0,1}(x,\min\{p_0(x),1\}) - \beta_{1,0}(x,1-\min\{1/p_0(x),1\}) &\text{ if } s_0 \neq s_1 \\
            \overline{{y}}_1(x) - \underline{{y}}_0(x) &\text{ if } s_0 = 0,~s_1 = 0.
        \end{cases}
    \end{align*}
\end{prop}
For never-takers, all lower and upper bounds are generally uninformative due to never observing them as part of a selected group. For defiers and compliers, the bounds are only informative when at least part of the conditional support is bounded, while for always-takers bounds are also informative without any support condition.

Before we present unconditional bounds, we collect in the next result identities about conditional covariate distributions.

\begin{prop} \label{prop_cond_covariate}
Under Assumptions \ref{ass_CIA} and \ref{ass_monotone}, the conditional covariate distributions for different strata are given by
 \begin{align*}
    dP(x|S(0)=s_0,S(1)=s_1) &= \begin{cases}
        \frac{\min\{s(0,x),s(1,x)\}}{E[\min\{s(0,X),s(1,X)\}]}dP(x) &\text{ if } s_0 = 1,~ s_1 = 1 \\
        \frac{\max\{0, s(1,x)-s(0,x)\}}{E[\max\{0, s(1,X)-s(0,X)\}]}dP(x) &\text{ if } s_0 = 0,~s_1 = 1 \\
        \frac{\max\{0, s(0,x)-s(1,x)\}}{E[\max\{0, s(0,X)-s(1,X)\}]}dP(x) &\text{ if } s_0 = 1,~s_1 = 0 \\
        \frac{1-\max\{s(0,x),s(1,x)\}}
        {E[1-\max\{s(0,X),s(1,X)\}]}dP(x) &\text{ if } s_0 = 0,~s_1 = 0.
    \end{cases}
\end{align*}
and the conditional covariate distribution for the extensive margin is given by \begin{align*}
    dP(x|S(0) \neq S(1))= \frac{|s(1,x) - s(0,x)| dP(x)}{E[|s(1,X) - s(0,X)|]}.
\end{align*}
\end{prop}

Unconditional sharp bounds for any principal stratum are obtained by integrating the conditional bounds with respect to the results provided in Proposition \ref{prop_cond_covariate} for $B\in\{L,U\}$:
\begin{align}\label{def_betaB_general}
    \beta_B(s_0,s_1) &:= \int\beta_B(x,s_0,s_1)dP(x|S(0)=s_0,S(1)=s_1).
\end{align}

\subsubsection{Example I: Intensive Margin Bounds}\label{ex_intensive}
The upper and lower conditional intensive margin or always-taker bounds correspond to the ones given by 
\cite{lee2009training} and \cite{semenova2025generalized}. Namely, we have that
\begin{equation}
\beta_L(x,1,1)=\beta_{L,1}(x,1,1)-\beta_{L,0}(x,1,1)
\end{equation} 
where
\begin{align*}
    \beta_{L,1}(x,1,1) &= \begin{cases}
        E[Y|S=1,D=1,X=x,Y\leq q_1(p_0(x),x)] &\quad~\quad~\text{ if } x \in \mathcal{X}^+ ,\\
        E[Y|S=1,D=1,X=x]  &\quad~\quad~\text{ if } x \in \mathcal{X}^- \cup \mathcal{X}^0,
    \end{cases} \\
%
    \beta_{L,0}(x,1,1) &= \begin{cases}
        E[Y|S=1,D=0,X=x] &\text{ if } x \in \mathcal{X}^+ \cup \mathcal{X}^0, \\
        E[Y|S=1,D=0,X=x,Y\geq q_0(1-1/p_0(x),x)] &\text{ if } x \in \mathcal{X}^-. \\
    \end{cases}
\end{align*}
Similarly, the conditional upper bounds are given by
\begin{equation}
\beta_U(x,1,1)=\beta_{U,1}(x,1,1)-\beta_{U,0}(x,1,1),
\end{equation}
where
 \vspace{-6pt}
\begin{align*}
    \beta_{U,1}(x,1,1) &= \begin{cases}
        E[Y|S=1,D=1,X=x,Y\geq q_1(1-p_0(x),x)] &\text{ if } x \in \mathcal{X}^+ \\
        E[Y|S=1,D=1,X=x] &\text{ if } x \in \mathcal{X}^- \cup \mathcal{X}^0, \\
    \end{cases} \\
    \beta_{U,0}(x,1,1) &= \begin{cases}
    E[Y|S=1,D=0,X=x] &\text{ if } x \in \mathcal{X}^+ \cup \mathcal{X}^0, \\
       E[Y|S=1,D=0,X=x,Y\leq q_0(1/p_0(x),x)] &\text{ if } x \in \mathcal{X}^-.
    \end{cases}
\end{align*}
In view of Proposition \ref{prop_cond_covariate} and \eqref{def_betaB_general},  the unconditional sharp bounds for the effect at the intensive margin can then be recovered for $B\in\{L,U\}$ as \begin{align}\label{def_betaB}
    \beta_B(1,1) = \frac{\int \beta_B(x,1,1)\min\{s(0,x),s(1,x)\}dP(x)}{\int\min\{s(0,x),s(1,x)\}dP(x)}.
\end{align}
\subsubsection{Example II: Extensive Margin Bounds}\label{ex_extensive}
The extensive margin is the combination of compliers and defiers where $s_0 \neq s_1$. Due to Proposition \ref{prop_identification1},  the conditional lower bound is given by $\beta_{L}(x,em)=\beta_{L,1}(x,em)-\beta_{L,0}(x,em),$ where
\begin{align*}
    \beta_{L,1}(x,em) = \begin{cases}
        E[Y|S=1,D=1,X=x,Y\leq q_1(1-p_0(x),x)] &\text{ if } x \in \mathcal{X}^+ ,\\
        \underline{y}_1(x)  &\text{ if } x \in \mathcal{X}^- \cup \mathcal{X}^0,
    \end{cases}
\end{align*}
and
\begin{align*}
    \beta_{L,0}(x,em) = \begin{cases}
        \overline{y}_0(x) &\text{ if } x \in \mathcal{X}^+ \cup \mathcal{X}^0, \\
        E[Y|S=1,D=0,X=x,Y\geq q_0(1/p_0(x),x)] &\text{ if } x \in \mathcal{X}^-. \\
    \end{cases}
\end{align*}
The conditional upper bound $\beta_U(x,em)$ is found analogously from Proposition  \ref{prop_identification1}.
As a result, the unconditional sharp bounds at the extensive margin for $B\in\{L,U\}$ follow from Proposition \ref{prop_cond_covariate} as
\begin{align}
\beta_B(em)&:=\int \beta_B(x,em)  dP(x|S(0) \neq S(1))\\
&=\frac{\int \beta_B(x,em) |s(1,x)-s(0,x)|dP(x)}{\int |s(1,x)-s(0,x)|dP(x)}
\end{align}

\subsection{Smooth Bounds}\label{sec_smooth}
\subsubsection{Identification}
The bounds in Proposition \ref{prop_identification1} are sharp. However, they, as well as the strata densities, contain a variety of non-differentiable components. This poses a challenge to statistical inference. We demonstrate the specific regularity problem and its consequences for inference in Section \ref{sec_regularity}. 
We now introduce the smooth bounds that provide an outer identification region to circumvent these problems. In particular, the identification region is constructed to eliminate the impact of misclassifying different monotonicity types while still remaining close to the sharp identified set. We focus on the intensive margin bounds for simplicity, see Appendix \ref{sec_smoothBounds.EM} for the extensive margin bounds.
More specifically, the form of the bounds in Proposition \ref{prop_identification1} reveal two sources where classification of the partition will enter, (i) conditional effects bounds $\beta_B(x,1,1)$, and (ii) conditional intensive margin share $\min\{s(0,x),s(1,x)\}$. We suggest to apply repeated, asymmetric smoothing to both components. For (i), consider first the conditional bound $\beta_L(x,1,1)$. Note that, for any $x \in \mathcal{X}$, the following parameter is a valid lower bound for $\beta_L(x,1,1)$
\begin{align}\label{smoot-bound-lower-term}
    {\beta}_{L,h}(x,1,1) &= E[Y|S=1,D=1,X=x,Y\leq q_1(g_{1,h}(p_0(x)),x )] \notag \\
    &\quad - E[Y|S=1,D=0,X=x,Y\geq q_0(1-g_{1,h}(1/p_0(x)),x)] \notag  \\
    &= \beta_{1,1}(x,g_{1,h}(p_0(x))) - \beta_{0,0}(x,1-g_{1,h}(1/p_0(x)))
\end{align}
where $g_{1,h}(z)$ is a smooth monotonic function indexed by a smoothing parameter $h \in \mathcal{H}_n \subset \mathbb{R}^+$ such that $0 \leq g_{1,h}(z) \leq \min\{z,1\}.$ 
This ensures that
${\beta}_{L,h}(x,1,1) \leq \beta_L(x,1,1).$

Next, consider that the smoothing of the conditional intensive margin is given by
$\min\{s(0,x),s(1,x)\} = s(1,x) \min\{p_0(x),1\}$.
We obtain lower and upper bounds using smooth functions 
$g_{1,h}(z)$ and $g_{3,h}(z)$ such that $g_{1,h}(z) \leq \min\{z,1\} \leq g_{3,h}(z).$
In which direction to smooth depends on the sign of ${\beta}_{L,h}(x,1,1)$. We decompose this into the difference of two non-negative terms via $z=\max\{z,0\}-\max\{-z,0\}.$
We approximate these two functions with a smoothing function $g_{4,h}(z)\leq \max\{z,0\},$ and the earlier $g_{2,h}(-z) \geq \max\{-z,0\}.$
The unconditional smooth lower effect bound is then given by
\begin{align}\label{def_smoothbeta}
\beta_{L,h}(1,1) &=
\frac{E\left[g_{4,h}({\beta}_{L,h}(X,1,1))g_{1,h}(p_0(X))s(1,X)\right]}{E[g_{3,h}(p_0(X))s(1,X)]} \notag \\
&\quad -\frac{E[g_{2,h}(-{\beta}_{L,h}(X,1,1)) g_{3,h}(p_0(X))s(1,X)]}
{E[g_{1,h}(p_0(X))s(1,X)]}.
\end{align}
The aforementioned relations together yield
$\beta_{L,h}(1,1) \leq \beta_L(1,1).$
Analogously, the conditional smooth upper bound is given by
\begin{align}\label{smoot-bound-upper-term}
\beta_{U,h}(x,1,1) 
&= \beta_{0,1}(x,1-g_{1,h}(p_0(x))) - \beta_{1,0}(x,g_{1,h}(1/p_0(x)))
\end{align}
which obeys $\beta_{U,h}(x,1,1)  \geq \beta_U(x,1,1)$. The unconditional smooth upper bound is 
\begin{align}\label{def_smoothbeta_upper}
\beta_{U,h}(1,1) &=
\frac{E\left[g_{2,h}({\beta}_{U,h}(x,1,1))g_{3,h}(p_0(X))s(1,X)\right]}{E[g_{1,h}(p_0(X))s(1,X)]} \notag \\
&\quad -\frac{E[g_{4,h}(-{\beta}_{U,h}(x,1,1)) g_{1,h}(p_0(X))s(1,X)]}
{E[g_{3,h}(p_0(X))s(1,X)]},
\end{align}
which analogously satisfies the reversed inequality
$\beta_{U,h}(1,1) \geq \beta_U(1,1).$

\subsubsection{Approximation Functions and Bias}
Next, we propose a class of functions that satisfy the imposed conditions and show that the distance between smooth and sharp unconditional bounds is negligible as $h \to 0$ with bounded support. 
\begin{ass}[Outcome Regularity I]\label{ass_bddmoment}  For $d\in\{0,1\}$, the outcome distribution $P(Y \leq y| S=1, D=d, X=x)$ has uniformly bounded support
, i.e. $\sup_{x\in \mathcal{X}} (|\underline{y}_d(x)|+|\overline{y}_d(x)|)<\infty.$
\end{ass}
We also impose the following assumption about approximation functions.
\begin{ass}[Approximation Functions]\label{ass_errors-of-g} For any $h>0, z \in \mathbb{R}$, functions $g_{i,h}$ $i=1, \ldots, 4$ satisfy
$g_{1,h}(z) \leq \min\{z,1\} \leq g_{3,h}(z)$ and $g_{4,h}(z)\leq \max\{z,0\} \leq g_{2,h}(z).$ 
Moreover, 
$\sup_{z \in \mathbb{R}} |g_{i,h}(z)-g_i(z)| 
\lesssim h$ 
where $g_1(z)=g_3(z)=\min\{z,1\}$ and $g_2(z)=g_4(z)=\max\{z,0\}.$ 
In addition, $g_{1,h}(z) \geq 0$ for all $z \in \mathbb{R}$ and $h \in \mathcal{H}_n.$\footnote{All quantile trimming thresholds are well defined as long as $\inf_{x \in \mathcal{X}} g_{1,h}(p_0(x)) \geq 0$ and $\inf_{x \in \mathcal{X}} g_{1,h}(1/p_0(x)) \geq 0.$ Given the strong multiple overlap assumption, one can always find an upper bound for $h$ such that this applies. These large-$h$ values are not relevant from a practical perspective and we consider $h$ to be below such a ceiling in what follows.} 
\end{ass}

\begin{figure}[!h]
    \centering
    \caption{Smooth Approximation: $g_{1,h}(z)$ Example}
    \label{fig_smoothapprox1}
    \includegraphics[trim = 0 130 0 110,  width = 0.6\textwidth]{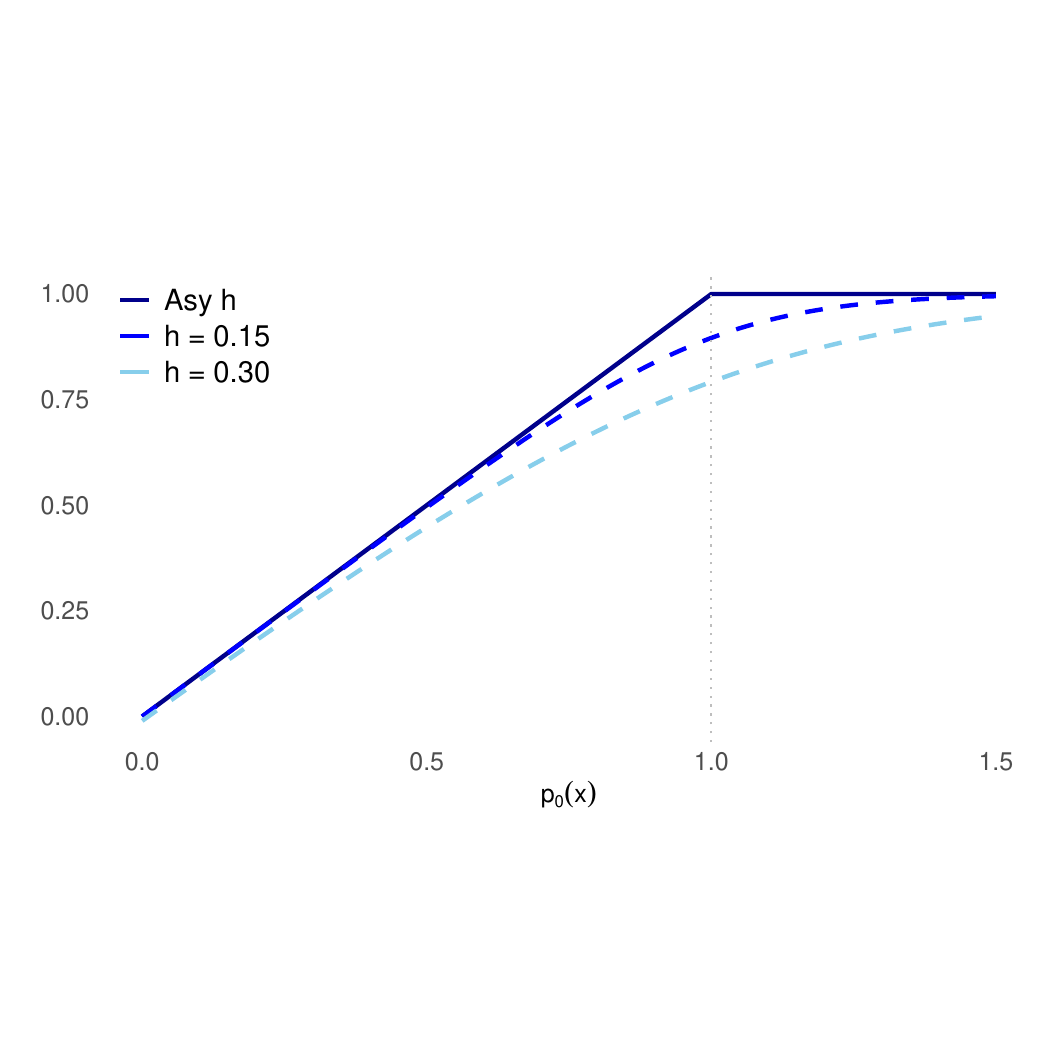}
   \footnotesize \begin{justify}
		The figure contains $\min\{p_0(x),1\}$ as well as two smooth approximations using $g_{1,h}(z) = 1 - h\log(1 + \exp(-(z - 1)/h))$ for $h=0.30$, $h=0.15$, and $h=0.0015$ (Asy). The size of the $h$-cover at $p_0(x) = 1$ determines how close the bounds are to the point-identified effect for $x \in \mathcal{X}^0$. The $h$-cover for units with $p_0(x) \neq 1$ introduces an additional distance compared to the sharp trimming bounds. 
	\end{justify}
\end{figure}

There exist many functions with these properties (LogSumExp, hyperbolic, \dots). Figure \ref{fig_smoothapprox1} depicts an example using a LogSumExp function for smoothing via $g_{1,h}(z)$. 
The specific function is secondary compared to smoothing parameter choice. We return to this in Section \ref{sec_smoothingALL}. We obtain the following theorem.
\begin{theorem}[Smoothing Bias]\label{thm:convergence}
Let $\mathcal{P}$ be the set of probability measures satisfying Assumptions \ref{ass_CIA}, \ref{ass_monotone}, \ref{ass_continuity}, \ref{ass_overlap}, \ref{ass_bddmoment}, and \ref{ass_errors-of-g}. Denote $\underline{g}_n = \inf_x g_{1,h}(p_0(x))$. Then, for $B \in \{L, U\},$ and $(s_0, s_1) \neq (0,0),$ we have that
\begin{equation*}
\sup_{P \in \mathcal{P}} |\beta_{B,h}(s_0,s_1) -\beta_B(s_0, s_1)|\lesssim \frac{h}{\underline{g}_n^2}.
\end{equation*}
\end{theorem}

Theorem \ref{thm:convergence} gives the approximation error/bias of the smooth bounds. Under a strong overlap condition, $\underline{g}_n$ is bounded away from zero and thus the difference decays linearly in $h$. Without strong overlap, there is an implicit restriction on the rate at which the smooth relative probabilities $g_{1,h}(p_0(x))$ can approach zero.

\section{Regularity and Semiparametric Efficiency Bounds} \label{sec_regularity}
In this section, we present our main results on regularity and semiparametric efficiency. Again, we focus on the lower bound for the intensive margin for simplicity. 
Table \ref{tab_EIFs1} provides the influence function for the intensive margin lower bound and its smoothed counterpart. All other influence functions are in Appendix \ref{sec_proofs}. We add the following assumptions.

\begin{ass}[Outcome Regularity II]\label{ass_density} For all $x\in \mathcal{X}$ and $d \in \{0,1\},$ the conditional outcome distribution $P(Y \leq y| S=1, D=d, X=x)$ has compact support $[\underline{y}_d, \overline{y}_d]$ and continuous density $f(y|X=x, D=d, S=1)$ uniformly bounded from above and away from zero. 
\end{ass}

\begin{ass}[Strong Multiple Overlap]\label{ass_strong_overlap} There exist constants $\underline{m},\underline{s} \in (0,1/2)$ such that \begin{align*}
			\underline{m} < \inf_{x\in\mathcal{X}} m(x) &\leq \sup_{x\in\mathcal{X}} m(x) < 1- \underline{m}, \\
			\underline{s} < \inf_{x\in\mathcal{X},d\in{\{0,1\}}} s(d,x) &\leq \sup_{x\in\mathcal{X},d\in{\{0,1\}}} s(d,x) < 1- \underline{s}. 
		\end{align*} \label{ass_STRONGoverlap}
\end{ass}
\vspace{-12pt}
 Assumptions \ref{ass_density} and \ref{ass_strong_overlap} are stronger versions of continuity and overlap in Assumptions \ref{ass_continuity} and \ref{ass_overlap}, respectively. 
  Without strong overlap, there is an additional irregularity problem for the population bounds analogous to irregular identification of average treatment effects under unconfoundedness with many extreme propensity scores \citep{khan2010irregular,HEILER2021valid}. We abstract from such issues in this paper to focus on the irregularity obtained from $P(\mathcal{X}^0) > 0$.  We obtain the following theorem.

\begin{theorem}\label{thm_differentiable}
 Suppose that Assumptions \ref{ass_CIA}, \ref{ass_monotone}, \ref{ass_density}, and \ref{ass_strong_overlap} hold. For $B \in \{L, U \},$ $\beta_B(1,1)$, $\beta_B(0,1)$, $\beta_B(1,0),$ $\beta_B(\mbox{em})$ are pathwise differentiable if and only if $P(\mathcal{X}^0)=0.$ 
 \end{theorem}

 \begin{corr}\label{corr_var_bounds}
 Let $1_{\mathcal{X}^{+}} = \mathbbm{1}(X \in \mathcal{X}^+)$ and $1_{\mathcal{X}^{-}} = \mathbbm{1}(X \in \mathcal{X}^-)$. Suppose that Assumptions \ref{ass_CIA}, \ref{ass_monotone}, \ref{ass_density} and \ref{ass_strong_overlap} hold, and $P(\mathcal{X}^0)=0.$ The semiparametric efficiency bound $Var_a(\beta_L(1,1))$ for $\beta_L(1,1)$  is defined by 
{\footnotesize\begin{align*}
&E[\min\{s(0,X), s(1,X)\}]^2Var_a(\beta_L(1,1)) = 
 E\left[\frac{s(1,X) \sigma_1^2 (X) }{m(X)}+\frac{s(0,X) \sigma_0^2 (X) }{1-m(X)} \right] \\
&+E\left[1_{\mathcal{X}^{+}}(\beta_L(X,1,1)-\beta_L(1,1))^2 \frac{s(0,X)(1-s(0,X) m(X))}{1-m(X)} \right] \\
&+E\left[1_{\mathcal{X}^{-}}(\beta_L(X,1,1)-\beta_L(1,1))^2 \frac{s(1,X)(1-s(1,X)+ s(1,X) m(X))}{m(X)} \right] \\
&+E\left[ 1_{\mathcal{X}^{+}} \frac{s(1,X) q_1(p_0(X),X)^2 p_0(X) (1-p_0(X))}{m(X)} \right] \\
&+E\left[ 1_{\mathcal{X}^{-}}  \frac{s(0,X) q_0(1-1/p_0(X),X)^2 p_0(X)^{-1} (1-p_0(X)^{-1})}{1-m(X)} \right]\\
&+E\left[1_{\mathcal{X}^{+}} (q_1(p_0(X),X)-\beta_{1,1}(X,p_0(X)))^2\left(\frac{ s(0,X)(1-s(0,X))}{1-m(X)}+
  \frac{p_0(X)^2 s(1,X)(1-s(1,X))}{m(X)} \right) \right]  \\
  &+E\bigg[1_{\mathcal{X}^{-}}(q_0(1-1/p_0(X),X)-\beta_{0,0}(X,1-1/p_0(X)))^2 \\
  &\quad \times \left(\frac{ p_0(X)^{-2} s(0,X)(1-s(0,X))}{1-m(X)}+
  \frac{s(1,X)(1-s(1,X))}{m(X)} \right) \bigg]\\
&- 2E\left[ 1_{\mathcal{X}^{+}}
  \frac{q_1(p_0(X),X) \beta_{1,1}(X,p_0(X)) s(1, X) p_0(X)(1-p_0(X))}{m(X)} \right]\\
&-2E\left[ 1_{\mathcal{X}^{-}}
  \frac{q_0(1-1/p_0(X),X) \beta_{0,0}(X,1-1/p_0(X)) s(0, X) p_0(X)^{-1}(1-p_0(X)^{-1})}{1-m(X)} \right]\\
 &+2E\left[ 1_{\mathcal{X}^{+}}
  \frac{(\beta_L(X,1,1)-\beta_L(1,1))(q_1(p_0(X),X)-\beta_{1,1}(X,p_0(X))) s(0, X) (1-s(0,X))}{1-m(X)} \right] \\
  &-2E\left[ 1_{\mathcal{X}^{-}}
  \frac{(\beta_L(X,1,1)-\beta_L(1,1))(q_0(1-1/p_0(X),X)-\beta_{0,0}(X,1-1/p_0(X))) s(1, X) (1-s(1,X))}{m(X)} \right],
\end{align*} }
 where
{\footnotesize \begin{align*}
    \sigma_{1}^2(x) &=
        Var[Y1_{\{ Y \leq q_1(\min\{p_0(x),1\},x)  \}}|S=1,D=1,X=x], \\
    \sigma_{0}^2(x) &=
        Var[Y  1_{\{ Y \geq q_0(\max\{1-1/p_0(x),0\},x)  \}}|S=1,D=0,X=x].
\end{align*} }
 \end{corr}

Theorem \ref{thm_differentiable} characterizes that the necessary and sufficient condition for the unconditional lower bound introduced in \eqref{def_betaB} to be pathwise differentiable is the absence of the point mass on the set $\mathcal{X}^0$, i.e.~$P(\mathcal{X}^0)=0$.\footnote{
When $P(\mathcal{X}^0)>0$, the result in Theorem \ref{thm_differentiable} that $\beta_L(1,1)$ is not pathwise differentiable is related but different from \cite{luedtke2016statistical}. Their target is essentially the denominator of $\beta_L(1,1)$ in \eqref{def_betaB}. However, divergence of the derivative of the denominator and/or divergence of the numerator separately does not necessarily imply divergence of the ratio. Moreover, the numerator includes the conditional treatment effect bounds $\beta_L(x,1,1)$ and consequently conditional quantiles $q_d(u,x)$ which change the analysis. } Corollary \ref{corr_var_bounds} presents the semiparametric efficiency bound for $\beta_L(1,1)$. It depends neither on the knowledge of the partitions $\mathcal{X}^+$ and $\mathcal{X}^-$ nor the treatment propensity score.

The setting with full selection, $S=1$ almost surely, renders any truncation redundant and thus $\beta_L(x,1,1) = \beta_U(x,1,1) = \beta(x,1,1)$ are equal to the conditional average treatment effect. Consequently, the efficiency bound in Corollary \ref{corr_var_bounds} becomes 
 \begin{equation}
 E\left[ \frac{\sigma_1^2 (X) }{m(X)}+\frac{\sigma_0^2 (X) }{1-m(X)}+(\beta(X,1,1)-\beta(1,1))^2  \right],
 \end{equation} with
 \begin{align*}
    \sigma_{d}^2(x) &=
        Var[Y|S=1,D=d,X=x], 
\end{align*}
 which is exactly the efficiency bound for the ATE obtained in Theorem 2 of \cite{hahn1998role}, i.e.~Corollary \ref{corr_var_bounds} nests this as a special case.\footnote{
The result also suggests that if there are covariates that are not required for Assumptions \ref{ass_CIA} and \ref{ass_weakMon1} but still help to predict outcomes/propensities, then their inclusion or exclusion can affect efficiency. For instance, as in standard ATE estimation when $S=1$, such covariates should be used when informative for estimation of the (truncated) conditional expectations but not when they only affect propensity scores. This can be seen e.g.~by the leading conditional variation terms $\sigma_1^2(x)$ and $\sigma_0^2(x)$ in the efficiency bound in Corollary \ref{corr_var_bounds}, which are reduced by inclusion of such covariates. For the purpose of this paper, we assume that $\mathcal{X}$ is selected such that Assumptions \ref{ass_CIA} and \ref{ass_weakMon1} are credible and leave a thorough discussion of efficiency, e.g.~along the lines of \cite{hahn2004functionalrestriction}, for future work. 
 }
The smooth bounds \eqref{def_smoothbeta} circumvent the irregularity problem by relaxing the width of the identified set using smooth approximators, i.e.~the $g$-functions. If the latter are sufficiently smooth, then the resulting outer bounds are pathwise differentiable. In particular, we obtain the following theorem:

\begin{theorem}\label{thm_differentiable_smooth}
 Suppose that Assumptions \ref{ass_CIA}, \ref{ass_monotone}, \ref{ass_errors-of-g}, \ref{ass_density}, \ref{ass_strong_overlap} hold. If $\sup_{z,i}|g_{i,h}'(z)| \lesssim 1$ for some $h>0$, then $\beta_{L,h}(1,1)$ is pathwise differentiable. 
 \end{theorem}
 The corresponding efficiency bound is in Appendix \ref{app:proof_differentiable_smooth}. The additional smoothness assumption applies to the LogSumExp example shown in Section \ref{sec_smooth}.

\section{Estimation and Inference} \label{sec_estimationinference}
\subsection{Estimators}
We propose to construct estimators by using empirical analogues to the efficient influence functions provided throughout the paper. For example, Tables \ref{tab_EIFs1}-\ref{tab_EIFs_smooth} provide the influence functions of $\beta_L(1,1)$ and $\beta_{L,h}(1,1).$ 

 Let $E_n[X] = n^{-1}\sum_i^nX_i$. For given nuisance estimates $\hat{\eta}$, any estimator $\hat{\beta}_B = \hat{\beta}_B(s_0,s_1)$ solves \begin{align}
    E_n[\psi_{\beta_B}(W,\hat{\eta},\hat{\beta}_B)] = 0.
\end{align}
Note that all principal strata and margin influence functions have the following linear structure
 \begin{align} \label{eq_IF_linear1}
     \psi_{\beta_B}(W,\eta,\beta_B) = \psi^{[B]}_{\beta_B}(W,\eta) - \psi^{[S]}_{\beta_B}(W,\eta)\beta_B.
 \end{align}
Thus, the influence function-based estimators have a ratio structure of the form \begin{align} \label{eq_esthat1}
    \hat{\beta}_{B} &= \frac{E_n[\psi^{[B]}_{\beta_B}(W,\hat{\eta})]}{E_n[\psi^{[S]}_{\beta_B}(W,\hat{\eta})] }.
\end{align}
For the smooth bounds, parameters are given by a sum 
\begin{align}
    \beta_{B,h} = \beta_{B,+,h} + \beta_{B,-,h},
\end{align}
whose components also have linear influence functions with structure
 \begin{align} \label{eq_IF_linear_smooth}
     \psi_{\beta_{B,+,h}}(W,\eta,\beta_B) &= \psi^{[B]}_{\beta_{B,+,h}}(W,\eta) - \psi^{[S]}_{\beta_{B,+,h}}(W,\eta)\beta_{B,+,h}, \\
     \psi_{\beta_{B,-,h}}(W,\eta,\beta_{B,-,h}) &= \psi^{[B]}_{\beta_{B,-,h}}(W,\eta) - \psi^{[S]}_{\beta_{B,-,h}}(W,\eta)\beta_{B,-,h}, 
 \end{align}
which combined with estimated nuisances yield moment estimators
\begin{align}
     \hat{\beta}_{B,h}
     &= \hat{\beta}_{B,+,h} + \hat{\beta}_{B,-,h} \notag \\
   &= \frac{E_n[\psi^{[B]}_{\beta_{B,+,h}}(W,\hat{\eta})]}{E_n[\psi^{[S]}_{\beta_{B,+,h}}(W,\hat{\eta})]}  + \frac{E_n[\psi^{[B]}_{\beta_{B,-,h}}(W,\hat{\eta})]}{E_n[\psi^{[S]}_{\beta_{B,-,h}}(W,\hat{\eta})]}.
\end{align}
For the intensive margin, we also consider the moment functions suggested by \cite{heiler2024heterogeneous} and \cite{semenova2025generalized} under known propensity scores that have influence function \begin{align}
   \tilde{\psi}_{\beta_{B}}(W,\eta,\beta_B) = \psi_{\beta_B}(W,\eta,\beta_B) - \delta(D,X,\eta), 
\end{align}
with mean-zero term $\delta(D,X,\eta)$ characterized by {\begin{align}
 &\qquad ~ \qquad E[\min\{s(0,X),s(1,X)\}]\delta(D,X,\eta) \notag \\ &= \mathbbm{1}_{(X \in \mathcal{X}^+)}s(0,X)\bigg[\beta_{1,1}(X,p_0(X))\bigg(1-\frac{D}{m(X)}\bigg)  \notag   -\beta_{0,0}(X,0)\bigg(1-\frac{1-D}{1-m(X)}\bigg)\bigg]  \notag \\
&+ \mathbbm{1}_{(X \in \mathcal{X}^-)}s(1,X)\bigg[\beta_{1,1}(X,1)\bigg(1-\frac{D}{m(X)}\bigg) \notag - \beta_{0,0}(X,1-1/p_0(X))\bigg(1-\frac{1-D}{1-m(X)}\bigg)\bigg]. 
\end{align}}
Solving for the parameter at the sample average then yields the alternative estimator  \begin{align} \label{eq_est_btilde}
     \tilde{\beta}_{B} &= \frac{E_n[\tilde{\psi}^{[B]}_{\beta_{B}}(W,\hat{\eta})]}{E_n[\psi^{[S]}_{\beta_{B}}(W,\hat{\eta})] }.
\end{align}

\begin{table}[!h] \caption{Influence Functions: Regular Bound $\beta = \beta_L(1,1)$} \label{tab_EIFs1}
    \centering
{\footnotesize    \begin{tabular}{lc|c} \hline \hline && \\[-1.5ex]
        Partition & Component &  {Non-centered Influence Function}  \\ \hline 
                    & &  \\ 
        $\mathcal{X}^+$& $\psi_{\beta}^{[L^+]}$  & $
 \frac{SD}{m(X)} Y 1\{Y \le q_1(p_0(X),X)\} 
-\frac{S(1-D)}{1-m(X)}
Y $  \\ &&
$- \frac{SD}{m(X)}q_1(p_0(X),X)
\left[1\{Y \le q_1(p_0(X),X)\}-p_0(X) \right]
$ \\ &&
$+q_1(p_0(X),X)\left[\frac{1-D}{1-m(X)}(S-s(0,X))- p_0(X)
\frac{D}{m(X)}(S-s(1, X)) \right]$  \\ &&
$+ s(0,X)\left[\beta_{1,1}(X,p_0(X))\left(1-\frac{D}{m(X)}\right) - \beta_{0,0}(X,0)\left(1-\frac{1-D}{1-m(X)}\right)\right]$
     \\  && \\
           ~ & $\psi_{\beta}^{[S^+]}$  & $ s(0,X) + \frac{(1-D)(S-s(0,X))}{1-m(X)}$ \\
        && \\
        $\mathcal{X}^-$& $\psi_{\beta}^{[L^-]}$  & $
 \frac{SD}{m(X)} Y 
-\frac{S(1-D)}{1-m(X)}Y1\{Y \geq q_0(1-1/p_0(X),X)\} 
 $  \\ &&
$- \frac{S(1-D)}{1-m(X)}q_0(1-1/p_0(X),X)
\left[1/p_0(X) - 1\{Y \geq q_0(1-1/p_0(X),X)\} \right]
$ \\ &&
$-q_0(1-1/p_0(X),X)\left[\frac{D}{m(X)}(S-s(1,X))- p_0(X)^{-1}
\frac{1-D}{1-m(X)}(S-s(0,X)) \right]$  \\ &&
$+ s(1,X)\left[\beta_{1,1}(X,1)\left(1-\frac{D}{m(X)}\right) - \beta_{0,0}(X,1-1/p_0(X))\left(1-\frac{1-D}{1-m(X)}\right)\right]$
     \\  && \\
           ~ & $\psi_{\beta}^{[S^-]}$  & $ s(1,X) + \frac{D(S-s(1,X))}{m(X)}$ \\   \hline \\[-0.5ex]
\multicolumn{3}{c}{$   \psi_{\beta}  = \big[\mathbbm{1}{(X \in \mathcal{X}^+)}\psi_{\beta}^{L^+} + \mathbbm{1}{(X \in \mathcal{X}^-)}\psi_{\beta}^{L^-}\big] - \big[\mathbbm{1}{(X \in \mathcal{X}^+)}\psi_{\beta}^{S^+} + \mathbbm{1}{(X \in \mathcal{X}^-)}\psi_{\beta}^{S^-}\big] \beta$ } \\[1ex] \hline && \\[-1.5ex] \multicolumn{1}{l}{Nuisance} && Definition \\ \hline && \\[-0.5ex]
 $m(x)$ && $P(D=1|X=x)$ \\
 $s(d,x)$ && $P(S=1|D=d,X=x)$ \\
 $q_d(u,x)$ && $\inf\{q \in \mathcal{Y}: u \leq P(Y\leq q|S=1,D=d,X=x)\}$ \\
 $\beta_{1,d}(x,u)$ && $E[Y|S=1,D=d,X=x,Y\leq q_d(u,x)]$ \\
 $\beta_{0,d}(x,u)$ && $E[Y|S=1,D=d,X=x,Y\geq q_d(u,x)]$ \\ && \\[-0.5ex]
 \hline \hline 
    \end{tabular} }
\end{table}

\begin{table}[!h] \caption{Influence Functions: Smooth Bound $\beta_{L,h}(1,1) = \beta_{L,+,h}(1,1)  + \beta_{L,-,h}(1,1)$} \label{tab_EIFs_smooth}
    \centering
{\footnotesize   \begin{tabular}{l|c} \hline \hline \\[-1.5ex] 
      Function  &    {Definition}  \\ \hline  & \\[-0.5ex] 
      $f^{[L]}(j,k)$ & $\frac{SD}{m(X)}\frac{g'_{k,h}(\beta_{L,h}(X))g_{j,h}(p_0(X))}{g_{1,h}(p_0(X))}\left[Y\mathbbm{1}(Y \leq q_1(g_{1,h}(p_0(X)),X)) - g_{1,h}(p_0(X))\beta_{1,1}(X,g_{1,h}(p_0(X))\right]   $ \\
      & $-\frac{S(1-D)}{1-m(X)}\frac{g'_{k,h}(\beta_{L,h}(X))g_{j,h}(p_0(X))}{p_0(X)g_{1,h}(1/p_0(X))}$\\&$\times \left[Y\mathbbm{1}(Y \geq q_0(1-g_{1,h}(1/p_0(X)),X)) - g_{1,h}(1/p_0(X))\beta_{0,0}(X,1-g_{1,h}(1/p_0(X)))\right]$ \\
      & $+ g_{k,h}(\beta_{L,h}(X))\left[g_{j,h}(p_0(X))s(1,X) + g'_{j,h}(p_0(X))\frac{1-D}{1-m(X)}(S-s(0,X))\right]$\\
      & $+ g_{k,h}(\beta_{L,h}(X))\left[g_{j,h}(p_0(X)) - p_0(X) g'_{j,h}(p_0(X))\frac{D}{m(X)}(S-s(1,X))\right]$ \\
      & $-\frac{SD}{m(X)}\frac{g'_{k,h}(\beta_{L,h}(X))g_{j,h}(p_0(X))}{g_{1,h}(p_0(X))}q_1(g_{1,h}(p_0(X)),X)\left[\mathbbm{1}(Y\leq q_1(g_{1,h}(p_0(X)),X)) -g_{1,h}(p_0(X))\right]$ \\
      & $+ \frac{S(1-D)}{1-m(X)}\frac{g'_{k,h}(\beta_{L,h}(X))g_{j,h}(p_0(X))}{p_0(X)g_{1,h}(1/p_0(X))}q_0(1-g_{1,h}(1/p_0(X)),X)$ \\ & $\times \left[\mathbbm{1}(Y \geq q_0(1-g_{1,h}(1/p_0(X)),X)) - g_{1,h}(1/p_0(X))\right]$ \\
      & $+ g'_{k,h}(\beta_{L,h}(X))\left[\frac{1-D}{1-m(X)}(S-s(0,X)) - p_0(X)\frac{D}{m(X)}(S-s(1,X)) \right]$\\
      & $\times \bigg[\frac{g'_{1,h}(p_0(X))g_{j,h}(p_0(X))}{g_{1,h}(p_0(X))}(q_1(g_{1,h}(p_0(X)),X) - \beta_{1,1}(X,g_{1,h}(p_0(X))) $ \\
      & $+ \frac{g'_{1,h}(1/p_0(X))g_{j,h}(p_0(X))}{p^2_0(X)g_{1,h}(1/p_0(X))}(q_0(1-g_{1,h}(1/p_0(X)),X) - \beta_{0,0}(X,1-g_{1,h}(1/p_0(X))))\bigg]$ \\
      & \\
       $f^{[S]}(j)$ & $\left[g_{j,h}(p_0(X))s(1,X) + g'_{j,h}(p_0(X))\frac{1-D}{1-m(X)}(S-s(0,X))\right]$ \\
        & + $\left[g_{j,h}(p_0(X)) - p_0(X)g'_{j,h}(p_0(X))\frac{D}{m(X)}(S-s(1,X))\right]$ \\
        & \\
        $\psi^{[L]}_{\beta_{L,+,h}}$ & $f^{[L]}(1,4)$ \\
        $\psi^{[S]}_{\beta_{L,+,h}}$ & $f^{[S]}(3)$ \\
        $\psi^{[L]}_{\beta_{L,-,h}}$ & $f^{[L]}(3,5)$ \\
        $\psi^{[S]}_{\beta_{L,-,h}}$ & $f^{[S]}(1)$ \\
        $\psi_{\beta_{L,+,h}}$ & $\psi^{[L]}_{\beta_{L,+,h}} - \psi^{[S]}_{\beta_{L,+,h}}\beta_{L,+,h}$\\
        $\psi_{\beta_{L,-,h}}$ & $\psi^{[L]}_{\beta_{L,-,h}} - \psi^{[S]}_{\beta_{L,-,h}}\beta_{L,-,h}$ \\[1ex] \hline \\[-1.5ex] Nuisance &Definition \\ \hline & \\[-0.5ex]
 $m(x)$ & $P(D=1|X=x)$ \\
 $s(d,x)$ & $P(S=1|D=d,X=x)$ \\
 $q_d(u,x)$ & $\inf\{q \in \mathcal{Y}: u \leq P(Y\leq q|S=1,D=d,X=x)\}$ \\
 $\beta_{L,h}(x)$ & $\beta_{1,1}(x,g_{1,h}(p_0(x))) - \beta_{0,0}(x,1-g_{1,h}(1/p_0(x)))$ \\
 $\beta_{1,d}(x,u)$ & $E[Y|S=1,D=d,X=x,Y\leq q_d(u,x)]$ \\
 $\beta_{0,d}(x,u)$ & $E[Y|S=1,D=d,X=x,Y\geq q_d(u,x)]$ \\[1ex] \hline \\[-1.5ex]  & \multicolumn{1}{c}{$g$-functions} \\ \hline & \\[-0.5ex]
&\multicolumn{1}{c}{$g_{1,h}(z) \leq \min\{z,1\} \leq g_{3,h}(z)$} \\
&\multicolumn{1}{c}{$g_{4,h}(z) \leq \max\{z,0\} \leq g_{2,h}(z)$} \\
&\multicolumn{1}{c}{$g_{5,h}(z) = -g_{2,h}(-z)$}  \\
 \hline \hline 
    \end{tabular} }
\end{table}

\subsection{Definitions and Assumptions}
In this section, we provide and discuss the assumptions and results for large sample inference for the regular and irregular case.
 Denote $||\cdot||_p$ as the $L^p$ norm. Let $\eta \in T$ where $T$ is a convex subset of some normed vector space. We assume all nuisance functions are cross-fitted as in \cite{chernozhukov2018double}, Definition 3.1 and use the LogSumExp function for smoothing, see Section \ref{sec_smooth}. Denote nuisance realization set $\mathcal{T}_n = (\mathcal{S}_{0,n}\times\mathcal{S}_{1,n}\times \mathcal{Q}_{0,n} \times \mathcal{Q}_{1,n} \times \mathcal{M}_n \times \mathcal{B}_{0,0,n}\times \mathcal{B}_{0,1,n}\times \mathcal{B}_{1,0,n}\times \mathcal{B}_{1,1,n}) \subset T$ as the set that with high probability contains estimators $\hat{\eta} = \{\hat{s}(0,X), \hat{s}(1,X),\hat{q}_0(u,X),\hat{q}_1(u,X),\hat{m}(X),\hat{\beta}_{0,0}(u,X),\hat{\beta}_{0,1}(u,X),$ ~ $\hat{\beta}_{1,0}(u,X),\hat{\beta}_{1,1}(u,X)\}$ for nuisance quantities  $\eta = \{s(0,X),s(1,X),q_0(u,X),q_1(u,X),$ $m(X),{\beta}_{0,0}(u,X), {\beta}_{0,1}(u,X),$ ${\beta}_{1,0}(u,X),{\beta}_{1,1}(u,X)\}$. Let their corresponding $L^p$ error rates be

\begin{align*} 
	\lambda_{s,n,p} &= \sup_{d\in\{0,1\}}\sup_{\hat{s}(d) \in \mathcal{S}_{d,n}} E[|\hat{s}(d,X) - s(d,X)|^p]^{1/p}, \\
	\lambda_{q,n,p} &= \sup_{d\in\{0,1\}}\sup_{u \in \tilde{U}}\sup_{\hat{q}_d(u) \in \mathcal{Q}_{d,n}} E[|\hat{q}_d(u,X) - q_d(u,X)|^p]^{1/p}, \\
    \lambda_{m,n,p} &= \sup_{\hat{m} \in \mathcal{M}_{n}} E[|\hat{m}(X) - m(X)|^p]^{1/p}, \\
	\lambda_{b,n,p} &= \sup_{j,d\in\{0,1\}}\sup_{u \in \tilde{U}}\sup_{\hat{\beta}_{j,d}(u) \in \mathcal{B}_{j,d,n}} E[|\hat{\beta}_{j,d}(u,X) - \beta_{j,d}(u,X)|^p]^{1/p}, 
\end{align*}
where $\tilde{U}$ is a compact subset of $(0,1)$ containing the union of the supports of the relevant quantile-trimming thresholds.\footnote{For the regular case, this will be $([\mbox{supp}(p_0(X)) \cup \mbox{supp}(1-p_0(X))] \cap \mathcal{X}^{+}) \cup ([\mbox{supp}(1/p_0(X)) \cup \mbox{supp}(1-1/p_0(X))] \cap \mathcal{X}^{-})$ while for the smooth bounds, the subset will depend on $h$ and is defined equivalently with all bounds replaced by their $g_{1,h}$-smoothed counterparts.}

\begin{ass}[Outcome Regularity III]
For all $x\in \mathcal{X}$ and $d\in \{0,1\}$, the outcome has 
      a continuously differentiable conditional density  $f(y|S=1,D=d,X=x)$ with finite support on  $[\underline{y}_d, \overline{y}_d]$ that is uniformly bounded from above and away from zero and with uniformly bounded first derivative.\label{ass_regularEst}
\end{ass}

Assumption \ref{ass_regularEst}  here strengthens the smoothness and moment assumptions. This is required to bound (higher order) terms in some of the expansions and a central limit theorem for transformations of the outcome with bounded weights.

\subsection{Regular Bounds}
We first consider the regular case that sets $P(\mathcal{X}^0) = 0$. For this, we require the density of $s(1,x) - s(0,x)$ in a neighborhood around zero to be well-behaved. This neighborhood is allowed to shrink at a rate proportional to the error over the realization set for the selection probabilities. In particular,

\begin{ass}[Margin]
     There exist a finite constant $C > 0$ and $\alpha > 0$ such that \begin{align*}
        P(|s(1,x) - s(0,x)| \leq \delta) \leq C\delta^{\alpha} \quad \text{for any }\ 0 \leq \delta \leq c_n
    \end{align*}
    and $c_n \lesssim \lambda_{s,n,2}^{1/(2+\alpha)}$.  \label{ass_margin1}
\end{ass}

Note that this implies that $P(\mathcal{X}^0) = 0$ which we know is necessary for the existence of a first-order uniformly unbiased estimator by Theorem \ref{thm_differentiable}. Any continuous bounded density for $s(1,x) - s(0,x)$ is sufficient yielding $\alpha = \infty$. However, a density may also diverge around zero at a controlled rate. 
We further require the following learning rates.
\begin{ass}[Machine Learning Bias I] Let $e_n = o(1)$. For all folds, the nuisance parameters obtained via cross-fitting belong to a shrinking neighborhood $\mathcal{T}_n$ around $\eta$ with probability of at least $1-e_n$, such that 
		\begin{align*}
\lambda_{s,n,1} + \lambda_{q,n,1}= o(1)
\end{align*}
    and \begin{align*}
  \sqrt{n}(\lambda_{s,n,2}^{\frac{2\alpha}{2 + \alpha}} + \lambda_{q,n,2}^2  + \lambda_{m,n,2}^2 + \lambda_{b,n,2}^2) = o(1).
\end{align*} \label{ass_MLbias1}
\end{ass}

One can see that we require $L^1$ consistency for quantiles and conditional selection probabilities. The $L^1$ requirement is due to the variance of the trimming indicator for the outcome. Moreover, to control the machine learning bias, the squared $L^2$ rates have to be sufficiently fast. For most nuisances, these are equivalent to the usual $o(n^{-1/4})$ root mean squared error (RMSE) requirement in the debiased machine learning literature \citep{chernozhukov2018double}. The more demanding rate requirement for the selection probabilities is due to difficulties of classifying positive and negative monotonicity types around the boundary. Given these rates, Assumption \ref{ass_margin1} is only really plausible for $\alpha > 2$ as, even in the parametric case where $\lambda_{s,n,2} \sim n^{-1/2}$, $\sqrt{n}\lambda_{s,n,2}^{2\alpha/(2+\alpha)} \nrightarrow 0$ when $\alpha \leq 2$. For more demanding selection probabilities in high dimensions or with complicated functional forms, the density needs to be increasingly well-behaved around the margin. As $\alpha \rightarrow \infty$, the requirement reduces to the typical $o(n^{-1/4})$ rate as in \cite{heiler2024heterogeneous}. 
We obtain the following theorem.
\begin{theorem} \label{thm_asyN_regular}
    Under Assumptions \ref{ass_CIA}, \ref{ass_weakMon1}, \ref{ass_STRONGoverlap}, \ref{ass_regularEst}, \ref{ass_margin1}, and \ref{ass_MLbias1}, the regular estimator is asymptotically normal and semiparametrically efficient, i.e.\begin{align*}
    \sqrt{n}(\hat{\beta}_L(1,1) - \beta_L(1,1)) \overset{d}{\rightarrow} \mathcal{N}\big(0,Var_a(\beta_L(1,1))\big),
\end{align*}
where $Var_a(\beta_L(1,1))$ is the semiparametric efficiency bound in Corollary \ref{thm_differentiable}. 
\end{theorem}

\subsection{Smooth Bounds}
We now consider the large-sample behavior of the smooth bounds. We impose the following learning requirements.
\begin{ass}[Machine Learning Bias II] Let $e_n = o(1)$. For all folds, the nuisance parameters obtained via cross-fitting belong to a shrinking neighborhood $\mathcal{T}_n$ around $\eta$ with probability of at least $1-e_n$, such that 
		\begin{align*}
\lambda_{s,n,1} + \lambda_{q,n,1} = o(1)
\end{align*}
    and \begin{align*}
  \sqrt{n}(\lambda_{s,n,2}^2 + \lambda_{q,n,2}^2 +  \lambda_{m,n,2}^2 + \lambda_{b,n,2}^2 ) = o(1).
\end{align*} \label{ass_MLbias2}
\end{ass}

We obtain the following theorem.
\begin{theorem} \label{thm_asyN_smooth}
    Under Assumptions \ref{ass_CIA}, \ref{ass_weakMon1}, \ref{ass_errors-of-g}, \ref{ass_STRONGoverlap}, \ref{ass_regularEst}, and \ref{ass_MLbias2}, and fixed $h > 0$ such that $\sup_{z,j}|g_{j,h}'(z)| \lesssim 1$ and $\sup_{z,j}|g_{j,h}''(z)| \lesssim 1$, the smooth bound estimator is asymptotically normal and semiparametrically efficient, i.e.\begin{align*}
    \sqrt{n}(\hat{\beta}_{L,h}(1,1) - \beta_{L,h}(1,1)) \overset{d}{\rightarrow} \mathcal{N}\big(0,Var_a(\beta_{L,h}(1,1))\big),
\end{align*}
where $Var_a(\beta_{L,h})$ is the semiparametric efficiency bound for $\beta_{L,h}$. 
\end{theorem}

We can see that, without restricting the distribution of $s(1,x) - s(0,x)$, i.e.~allowing for point mass $P(\mathcal{X}^0) > 0$ or arbitrary density around zero, the smooth bounds are asymptotically normal under weaker assumptions than the non-smooth bounds with restricted distributions. In particular, they only require root mean squared error rates for all nuisance quantities of order $o(n^{-1/4})$ as in standard debiased machine learning \citep{chernozhukov2018double}.
This is fundamentally due to the smoothing turning the target parameter into a pathwise differentiable object. 


\subsection{The Efficiency Gap and Known Propensity Scores} \label{sec_efficiency-gap1}
If propensity scores are known, the moment functions from \cite{semenova2025generalized} or \cite{heiler2024heterogeneous} without correction can be used as well. The corresponding estimator \eqref{eq_est_btilde} does not require estimation of $\beta_{j,d}(u,X)$ for any $j,d\in\{0,1\}$ and thus all $\lambda_{b,n,p}$ terms that enter Assumption \ref{ass_MLbias1} and \ref{ass_MLbias2} can be omitted. However, there is an information loss from ignoring the conditional variation in the truncated means. In particular, we obtain the following theorem.
\begin{theorem} \label{thm_efficiency_gap}
    Under Assumptions \ref{ass_CIA}, \ref{ass_weakMon1}, \ref{ass_STRONGoverlap}, \ref{ass_regularEst}, \ref{ass_margin1}, and \ref{ass_MLbias1} with $\lambda_{b,n,p} = 0$ known, $\tilde{\beta}_{L}(1,1)$ is asymptotically normal \begin{align*}
    \sqrt{n}(\tilde{\beta}_L(1,1) - \beta_L(1,1)) \overset{d}{\rightarrow} \mathcal{N}\big(0,Var_{b}(\beta_L(1,1))\big),
\end{align*}
with efficiency gap
\begin{align*}
    &E[\min\{s(0,X),s(1,X)\}]^2(Var_{b}(\beta_L(1,1)) - Var_a(\beta_L(1,1))) \\
    &={E\bigg[\mathbbm{1}_{\{X \in \mathcal{X}^+\}}s(0,X)^2 \bigg( \beta_{1,1}(X,p_0(X))\sqrt{\frac{1-m(X)}{m(X)}} - \beta_{0,0}(X,0)\sqrt{\frac{m(X)}{1-m(X)}}\bigg)^2 \bigg]} \\
    &\quad + {E\bigg[\mathbbm{1}_{\{X \in \mathcal{X}^-\}}s(1,X)^2 \bigg( \beta_{1,1}(X,1)\sqrt{\frac{1-m(X)}{m(X)}} - \beta_{0,0}(X,1-1/p_0(X))\sqrt{\frac{m(X)}{1-m(X)}}\bigg)^2 \bigg]}.
\end{align*}
\end{theorem}
Theorem \ref{thm_efficiency_gap} demonstrates that, under known propensity scores, there is a trade-off between efficiency and learning rates for the truncated conditional means $\beta_{j,d}(u,X)$. A lack of efficiency from not including information about the conditional truncated potential outcome means in the respective influence function arises as the variance of the truncated variables are bounded from below by their conditional analogues. This is equivalent to the comparison between the variance of a dependent variable and its residual variance in standard regression analysis. Theorem \ref{thm_efficiency_gap} also nests the case without sample selection. In particular, when $S=1$ almost surely, the ATE is point identified $\beta_L(1,1) = \beta_U(1,1)$ and all truncated means simplify to their untruncated counterparts $\beta_{d,1}(x,1) = \beta_{d,0}(x,0) = E[Y|D=d,X=x]$.
In this case all correction terms vanish as well and the alternative estimator collapses to \begin{align*}
    \tilde{\beta}_{L}(1,1) = {E_n\bigg[\frac{YD}{m(X)}\bigg]} - {E_n\bigg[\frac{Y(1-D)}{1-m(X)}\bigg]}.
\end{align*}
This is the standard Horvitz-Thompson-type IPW estimator for the ATE using true propensities which is known to not reach the semiparametric efficiency bound \citep{hahn1998role}. More precisely, Theorem \ref{thm_efficiency_gap} yields efficiency gap \begin{align*}
    Var_{b}(\beta_L(1,1)) - Var_a(\beta_L(1,1)) = E\bigg[ \bigg( \beta_{1,1}(X,1)\sqrt{\frac{1-m(X)}{m(X)}} - \beta_{0,0}(X,0)\sqrt{\frac{m(X)}{1-m(X)}}\bigg)^2 \bigg].
\end{align*}
This comparison to the point-identified case suggests that, in the regular regime, one could potentially obtain an efficient estimator for the intensive margin ATE bounds without modeling the conditional truncated quantiles analogously to \cite{hirano2003efficient} using nonparametrically estimated treatment propensities with sufficiently rich basis functions that asymptotically encode all information about the truncated conditional outcome means. We leave the development of such an alternative estimation approach for future research.

\section{Smoothing Parameter Selection} \label{sec_smoothingALL}
\subsection{Statistical Smoothing Choice} \label{sec_smoothingSTAT}
For every fixed $h>0$, smooth bounds are pathwise differentiable and admit influence-function representations. Their asymptotic variance remains bounded along sequences $h\downarrow 0$ provided $\sup_{z,j}|g'_{j,h}(z)|\lesssim 1$, as the influence-function weights remain square-integrable even when $P(\mathcal{X}^0)>0$. Hence, potential reintroduction of irregularity along sequences $h\downarrow 0$ does not arise from variance explosion, but from increased sensitivity to nuisance estimation through the curvature of the smoothing functions. 

In particular, for typical $g_{j,h}(\cdot)$ functions, the curvature $g''_{j,h}(\cdot)$ of the smoothing near the kink grows at rate $1/h$, see Appendix \ref{app:smoothing_rates} for all details. An expansion of the estimator therefore yields linearization 
\begin{align}
\sqrt{n}(\hat\beta_{B,h}(1,1)-\beta_{B,h}(1,1))
&=
\frac{1}{\sqrt{n}}\sum_{i=1}^n 
\psi_{\beta_{B,h}(1,1)}(W_i,\eta)
+
O_p\!\left(\frac{\sqrt{n}\lambda_n^2}{h}\right)
+
o_p(1),
\end{align}
where $\lambda_n^2$ denotes the aggregate nuisance remainder rate. Regardless of whether $P(\mathcal{X}^0)>0$ or not, the smoothing bias is linear in $h$, i.e.~$\beta_{B,h}(1,1)-\beta_B(1,1)=O(h)$, see Theorem \ref{thm:convergence}. 
Moreover, smooth bounds are always outer bounds: $\beta_{L,h}(1,1)\le\beta_L(1,1)$ and $\beta_{U,h}(1,1)\ge\beta_U(1,1)$. Hence the sign of the smoothing bias is known. As a result, root-$n$ inference centered at the smooth bounds remains conservative for the sharp bounds.
In particular, root-$n$ asymptotic linearity at the smooth bounds 
and, hence, conservative inference for the bounds only require $\sqrt{n}\lambda_n^2/h = o(1)$. 
If instead one required root-$n$ centering at the sharp bounds, the additional condition $\sqrt{n}h\to 0$ would be needed. 
Combining this with $\sqrt{n}\lambda_n^2/h\to 0$ would imply $\lambda_n^2=o(n^{-1})$, which is stronger than parametric rates and therefore implausible in practice.

Thus, when $P(\mathcal{X}^0)>0$, sufficiently slow smoothing sequences that do not reintroduce irregularity are key for valid statistical inference.
For example, in our empirical application, we consider values between $n^{-1/4}\log^{-1}n$ and $n^{-1/12}\log^{-1}n$.

\subsection{Economic Smoothing Choice} \label{sec_smoothingECON}

The previous subsection guides smoothing parameter selection but is purely asymptotic.  
We now provide a complementary approach that guides reasonable $h$ selection based on substantive/economic conditions, i.e.~margin sizes. First, note that, for any fixed $h > 0$, the smooth bounds can alternatively be viewed as valid bounds under a relaxed version of conditional monotonicity that permits a limited amount of simultaneous compliance and defiance within covariate cells depending on $h$.\footnote{More explicitly, by construction, any $h$ implies a possible outer identification region for the target population shares compared to Assumption \ref{ass_weakMon1}. The smooth bounds are consistent with using weights and truncations according to this outer region for the shares in numerator and denominator of the bounds at any margin. However, the resulting unconditional smooth bounds are not sharp in general as they are obtained by deliberately optimizing for the worst-case over numerator and denominator separately, see \eqref{def_smoothbeta} for intensive margin and Appendix \ref{sec_smoothBounds.EM} for the extensive margin. Joint optimization would again result in a non-differentiable functional defeating the purpose for smoothing in the first place.} 

To do so, we first express the implied relaxation in terms of the extensive margin
$\pi_{ex}(x)=P(S(1)\neq S(0)\mid X=x)$.
Let $\pi_{im}(x)=P(S(1)=S(0)=1\mid X=x)$ denote the intensive-margin share.
By Assumption  \ref{ass_CIA}, the two margins satisfy the accounting identity
\begin{align}
    \pi_{ex}(x)
    =
    s(1,x)+s(0,x)-2\pi_{im}(x).
    \label{eq_pi-ex-pi_im-relation1}
\end{align}
Equation \eqref{eq_pi-ex-pi_im-relation1} is useful as the smooth bounds modify the intensive-margin
benchmark under Assumption \ref{ass_weakMon1} that would set $\pi_{im}(x) = \min\{s(1,x),s(0,x)\}$. The economic size of any relaxation is
then naturally expressed as an allowed increase in the extensive margin and, most importantly, its respective joint complier and defier shares.
To interpret such a relaxation, let
\begin{align}
\pi_c(x) &=P(S(1)=1,S(0)=0\mid X=x), \\
\pi_d(x) &=P(S(1)=0,S(0)=1\mid X=x),
\end{align}
denote the conditional shares of compliers and defiers. Then
$\pi_{ex}(x)=\pi_c(x)+\pi_d(x)$ and
$s(1,x)-s(0,x)=\pi_c(x)-\pi_d(x)$, and therefore
\begin{align}
\pi_{ex}(x)-|s(1,x)-s(0,x)|
=
2\min\{\pi_c(x),\pi_d(x)\}. \label{eq:comp_def_identity}
\end{align}
Conditional monotonicity sets one of these two shares equal to zero, so the
extensive margin would reduce to the absolute selection first stage

\begin{align}
\pi_{ex}(x)=|s(1,x)-s(0,x)|.
\end{align}
Now consider smooth estimands such as $\beta_{B,h}(1,1)$ in
\eqref{def_smoothbeta} and \eqref{def_smoothbeta_upper}. Instead of the non-differentiable component $\min\{s(1,x),s(0,x)\}$, these estimands contain smooth envelopes such as $g_{1,h}(p_0(x))s(1,x)$ or $g_{3,h}(p_0(x))s(1,x)$. For simplicity, we focus attention on the latter in what follows. The resulting smoothing slack relative to the conditional monotonicity benchmark is then given by

\begin{align}
    \varphi_h(x)
    :=
    g_{3,h}(p_0(x))s(1,x)-\min\{s(1,x),s(0,x)\},
    \label{eq_smoothing_slack}
\end{align}
which is nonnegative by construction.
Thus, the smooth version can be interpreted as allowing the extensive margin
to satisfy

\begin{align}
    \pi_{ex}(x)
    \leq
    |s(1,x)-s(0,x)|+2\varphi_h(x),
    \label{eq_expost_relaxed}
\end{align}
so that the excess relative to the conditional-monotonicity benchmark is

\begin{align}
    2\varphi_h(x)
    =
    2\left[
    g_{3,h}(p_0(x))s(1,x)-\min\{s(1,x),s(0,x)\}
    \right].
    \label{eq_expost-diff}
\end{align}
For example, in the LogSumExp case, this becomes

\begin{align}
    2\varphi_h(x)
    =
    2h s(1,x)
    \log\left[
    1+\exp\left(
    -\frac{|s(1,x)-s(0,x)|}{h s(1,x)}
    \right)
    \right].
    \label{eq_expost-diff-lse}
\end{align}
Thus, the relaxation is locally adaptive.
In particular, smoothing
mainly affects covariate cells close to the irregular region
$s(1,x)=s(0,x)$. The permitted excess is bounded above by
$2h s(1,x)\log 2$ and decays rapidly as the selection first stage moves away from zero.

Using the complier-defier identity \eqref{eq:comp_def_identity}, the relaxed inequality
\eqref{eq_expost_relaxed} permits simultaneous presence of compliers and defiers
up to
\begin{align}
    \min\{\pi_c(x),\pi_d(x)\}
    \leq
    \varphi_h(x)
    = g_{3,h}(p_0(x))s(1,x)-\min\{s(1,x),s(0,x)\}.
    \label{eq_expost-C-Df}
\end{align}
We can thus report estimates for summary statistics over \eqref{eq_expost-diff} or \eqref{eq_expost-C-Df} such as 
$E[\varphi_h(X)]$ 
to assess the size of relaxation of conditional monotonicity in terms of the size of the margin(s) and/or the joint compliance and defiance probabilities. If these are implausibly large from a substantive perspective, smaller values of $h$ should be considered. We present some explicit calculations and additional discussion in the context of the empirical application in Section \ref{sec_empirical_EXTENSIVE1}.

\section{Testing for Point Mass} \label{sec_Test-pointMass1}
This section provides a diagnostic test for the null hypothesis of no point mass $P(\mathcal X^0) = 0$ that is necessary and sufficient for regularity of the sharp bounds. 
Under conditional monotonicity, $\mathcal X^0$ consists of covariate values for which treatment does not affect selection. Writing $\Delta(x)=s(1,x)-s(0,x)$, we thus have $x\in \mathcal X^0$ if and only if $\Delta(x)=0$. Consequently, 
the testing problem reduces to detecting a point mass at zero in the distribution of the conditional selection effect. We therefore consider hypothesis
\begin{align}
H_0: P(\Delta(X)=0)=0
\qquad\text{versus}\qquad
H_1: P(\Delta(X)=0)>0 . \label{eq_test_H0}
\end{align}
Let $\hat s(d,x)$ denote the cross-fitted estimator introduced in Section 6 and define the estimated selection effect $\hat\Delta_i=\hat s(1,X_i)-\hat s(0,X_i)$. For a threshold sequence $\kappa_n>0$, define the near-zero mass statistic
\begin{align}
T_n(\kappa_n)=\frac1n\sum_{i=1}^n1\{|\hat\Delta_i|\le\kappa_n\}.
\end{align}
Large values of $T_n(\kappa_n)$ indicate that the estimated treatment effect on selection is frequently close to zero. Under suitable choices for $\kappa_n$ and some regularity on the distribution of $\Delta(x)$ near the margin, this statistic can be used for assessing \eqref{eq_test_H0}. In particular, we obtain the following theorem: 

\begin{theorem}[Asymptotic Properties of $T_n(\kappa_n)$] \label{thm_TEST_properties1}
Suppose Assumptions 3.1 and 3.2 hold, estimators satisfy convergence rate $\lambda_{s,n,1}$ defined in Section 6 and there exists a constant $C>0$ such that
\begin{align*}
P(0<|\Delta(X)|\le\delta)\le C\delta
\end{align*}
for all sufficiently small $\delta>0$. Let $M_n \rightarrow \infty $ and let $\kappa_n$ be any sequence satisfying $\kappa_n\downarrow0$ and $\lambda_{s,n,1}/\kappa_n\to0$. Define the critical value
\begin{align*}
c_n=M_n\Big(\kappa_n+\lambda_{s,n,1}/\kappa_n+\sqrt{\log n/n}\Big)
\end{align*}
and consider the test that rejects $H_0$ whenever $T_n(\kappa_n)>c_n$. Let $c_n \rightarrow 0$. Then the test is
\begin{enumerate}
\item conservative, i.e.~$P(T_n(\kappa_n)>c_n)\to 0$  under $H_0$.
\item consistent, i.e.~$P(T_n(\kappa_n)>c_n)\to 1$ under $H_1$.  
\end{enumerate}
\end{theorem}

The test provides a simple empirical diagnostic for the condition $P(\mathcal X^0)=0$. We opt for a conservative test as exact size-control with generic machine learning nuisances is hard to achieve. Rejection of $H_0$ indicates evidence that $P(\mathcal X^0)>0$, implying irregularity and that the smooth bounds developed should be used for inference. 

The near-margin condition $P(0<|\Delta(X)|\le\delta)\le C\delta$ restricts the amount of probability mass that the distribution of $\Delta(X)$ may place near zero under the null. 
The test rejects when the empirical near-zero mass of $\hat\Delta_i$ exceeds what can be expected under this assumption while simultaneously accounting for the nuisance estimation error. $M_n$ may be any sequence diverging to infinity. We suggest $M_n=2\log\log n$ to match a constant in the Hoeffding bound, see Appendix \ref{app_sec_Theorem_test1} for the proof and more details. 
The theoretical rate $\lambda_{s,n,1}$ is not directly observable in practice. A data-driven proxy can be obtained via repeated cross-fitting. Let $\hat \Delta^{(f_1)}$ and $\hat \Delta^{(f_2)}$ denote two independent cross-fitted first-stage estimators of $\Delta$. Set $
\hat r_n = \tfrac{1}{n} \sum_{i=1}^n | \hat\Delta_i^{(f_1)} - \hat\Delta_i^{(f_2)} |$.
Thus one can set $\kappa_n = \hat r_n^{1/2}$ and replace $\lambda_{s,n,1}$ by $\hat r_n$ in the $c_n$.
A possibly more stable and simpler alternative that avoids data-dependency is $\kappa_n=n^{-1/4}\log^{-1}n$ matching the moment shifting rate in \cite{semenova2025generalized}. This deterministic choice satisfies the conditions of the theorem whenever $\lambda_{s,n,1}=o(n^{-1/4}(\log n \log \log n)^{-1})$ which is often achievable in practice. 

\section{Empirical Study: Labor Market Policy} \label{sec_empirical1} 
\subsection{Job Corps and Data}
Job Corps (JC) is a US Department of Labor program aimed at empowering young individuals of economically or otherwise disadvantaged background with important labor market skills and qualifications. The intense program combines general education, vocational training, extracurricular activities, placement services, and more. Most participants are in residential slots in centers of varying size all over the US. 
The National Job Corps Study, executed by Mathematica Policy Research in the mid-90s, analyzed the effects of the Job Corps program on labor market prospects. The experiment implemented stratified randomized assignment of applicants, incorporating over 15400 units between ages 16 to 24. Data on various outcomes such as income, job status, educational achievements, and criminal activities were gathered at different intervals.
The outcomes of the original Job Corps study were mixed. Initially, the program boosted the participants' educational achievements and income. However, most aggregate effects waned over time. There is evidence for increased earnings in the older participant population \citep{schochet2008does}. There is also well-documented heterogeneity in terms of vocational and academic training returns as well as gender, see e.g.~\cite{flores2012estimating} or \cite{heiler2026heterogeneity}. 

We employ the public use files which contain all survey data up to 208 weeks after the initial assignment \citep{burghardt1999national,schochet2008does}. Our main outcome measure is log hourly earnings as in \cite{lee2009training}. If workers are priced around their marginal productivity, this measure can be used as a proxy for increases in the latter caused by assignment to Job Corps.  
We extract an extended set of covariates containing detailed information regarding demographics, employment, criminal history, education, health, expectations, regional characteristics, and other JC-related information.\footnote{We collect all variables used in \cite{lee2009training} and additional information on individual background and behavior similar to \cite{flores2012estimating} and \cite{heiler2023effect}, as well as all variables that were used for stratification in the original experiment. Some of the latter have previously been overlooked but are necessary for correct specification of the propensity score, see Appendix \ref{app_JC_data1} for more details.}
All impact analyses are conditional on having a non-missing earnings entry, including zero, at weeks $t\in\{1,\dots,208\}$, as in \cite{lee2009training}, leaving a sample of 9415 units.\footnote{\cite{lee2009training} suggests that non-response at this stage is a second-order issue and initial treatment-control balance is preserved, see his Remark 2. We reassess this claim with the extended data using both formal tests for average differences as well as standardized differences and find little evidence for imbalance, see Appendix \ref{app_JC_data1}. The only exception is significantly larger worries to attend Job Corps in the treatment group. 
Our prior is that worries are more likely negatively related to any treatment effect. Thus, we expect any potential bias to be negative, i.e.~earnings and employment effects are likely to be at least as large.}

In the following, results for nuisance functions are presented for the research sample. For all impact analyses as well as general descriptive statistics, we use the nationally representative design weights as in \cite{schochet2008does}.
For the impact evaluation, we analyze the effect per eligible participant, i.e.~the causal effect of assignment to Job Corps. 
As participation or outside alternatives after assignment are not controlled, all Job Corps effects here are relative to the control state of not being assigned, including partaking in any alternative program and/or working.

\begin{figure}[!h]
	\centering \caption{Job Corps: Employment and Earnings} \label{fig_JC_desc1}
	\begin{subfigure}{0.32\textwidth}
		\includegraphics[width = \textwidth, trim = 0 80 0 50, clip]{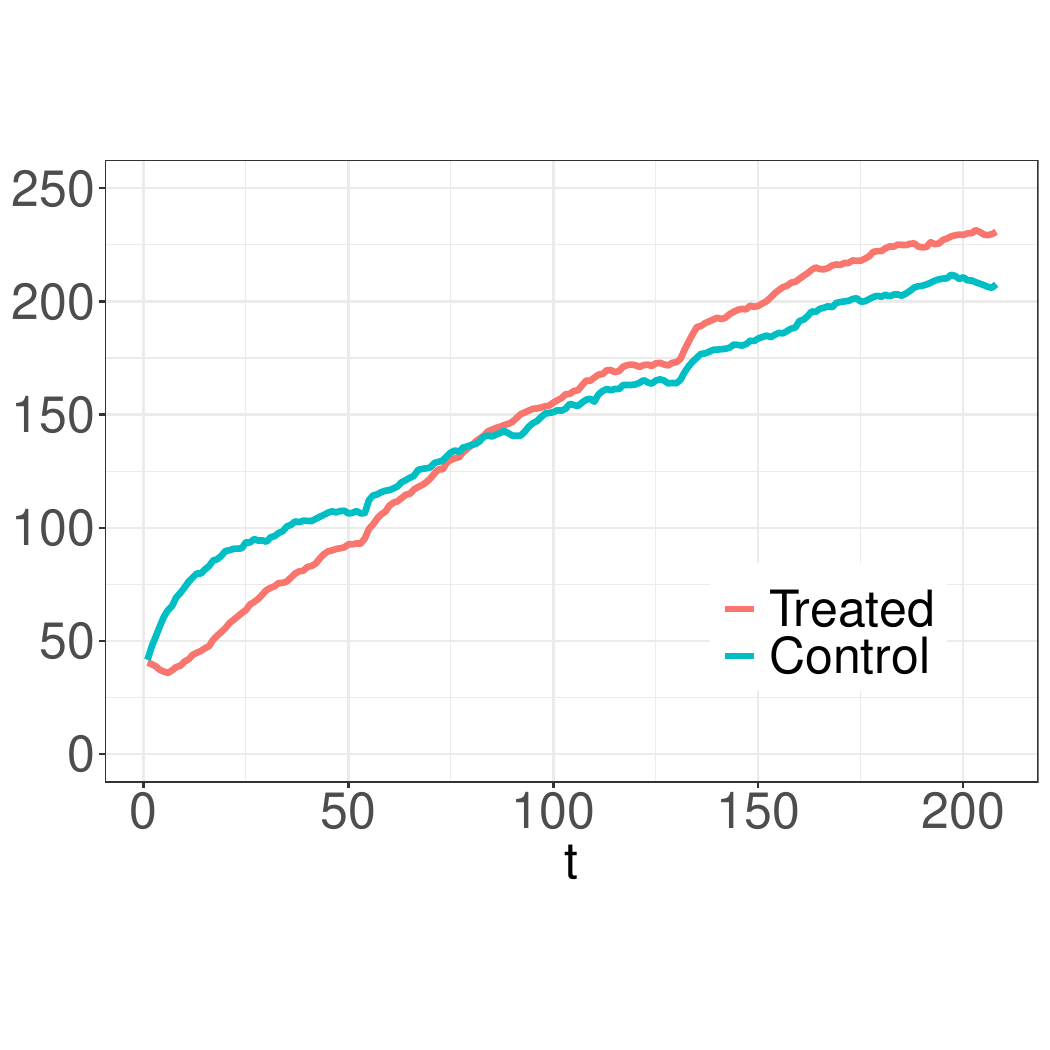}
		\caption{Weekly Earnings in USD} 
	\end{subfigure}
	\begin{subfigure}{0.32\textwidth}
		\includegraphics[width = \textwidth, trim = 0 80 0 50, clip]{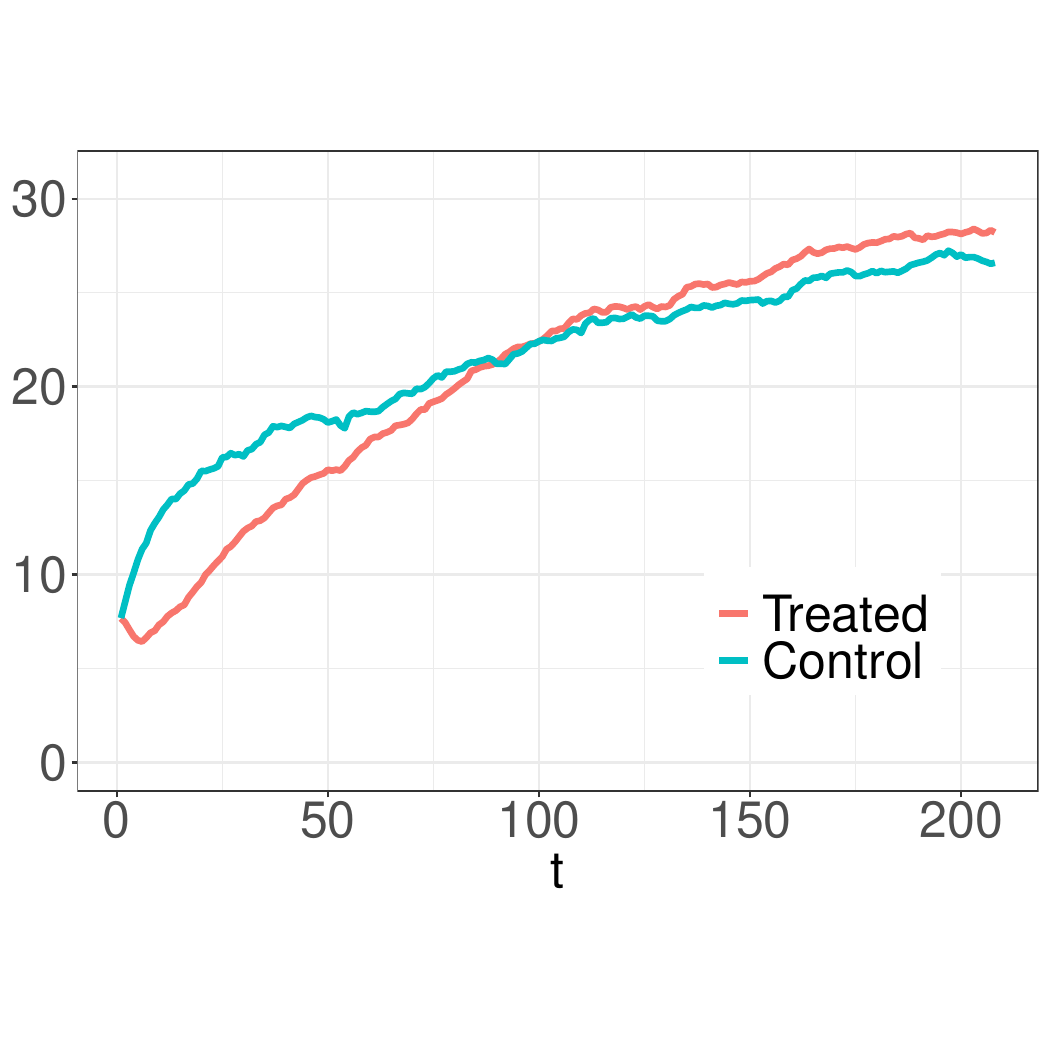}
		\caption{Weekly Hours Worked}
	\end{subfigure}	
	\begin{subfigure}{0.32\textwidth}
	\includegraphics[width = \textwidth, trim = 0 80 0 50, clip]{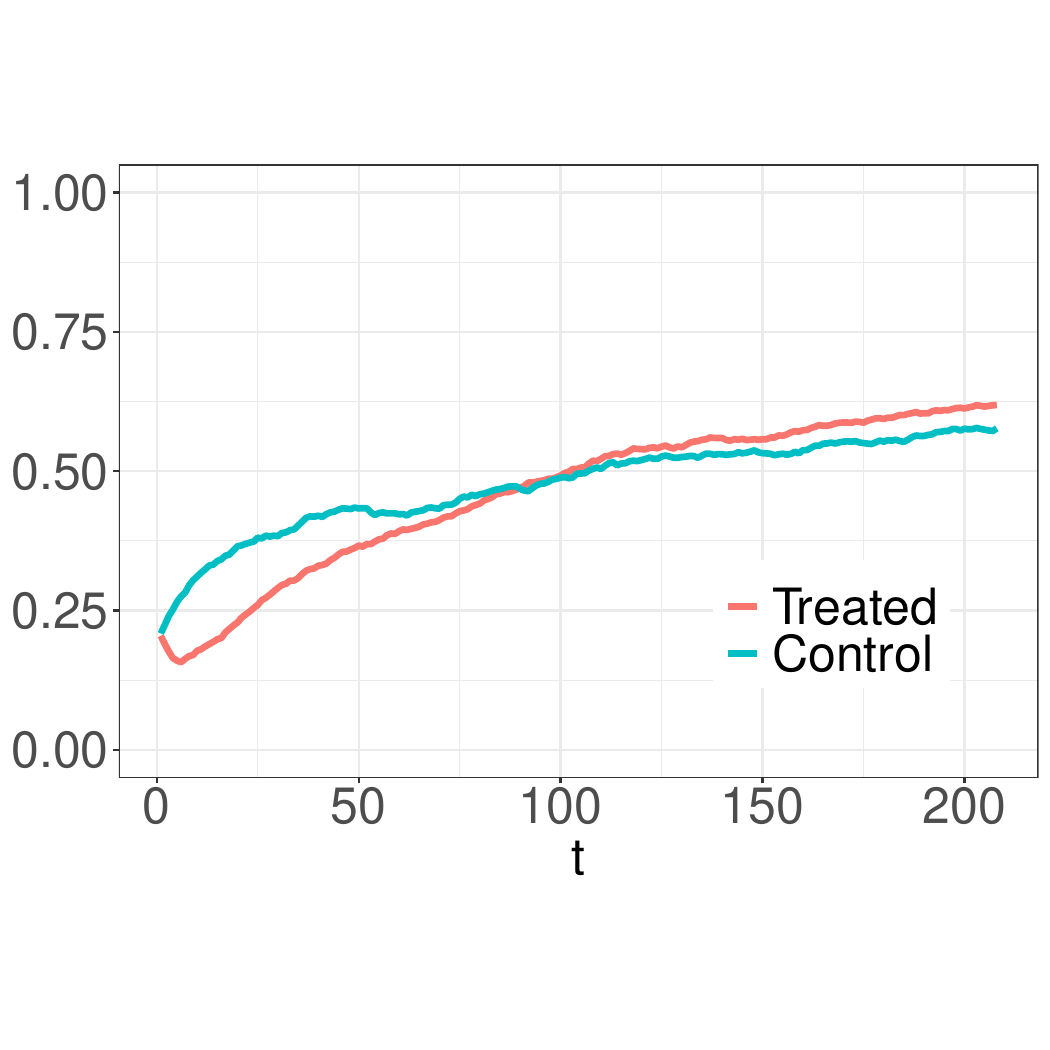}
	\caption{Weekly Employment Rate}
	\end{subfigure} \footnotesize
 \begin{justify}
     Weekly earnings, hours worked, and employment averages of control and treated units. The $n=9415$ sample is restricted to units that report hours and earnings for all periods $t\in\{1,\dots,208\}$. All calculations use nationally representative design weights. 
 \end{justify}
\end{figure}

Figure \ref{fig_JC_desc1} contains the weekly average earnings and employment rate for the 208 weeks after treatment assignment. There are systematic trends and differences between treatment and control groups in both earnings and employment over time. 
There is a clear upwards trend in earnings and employment over time. The control group tends to be ahead early after assignment but the treatment group catches up later in time exceeding the trajectory of the control group for both measures. 
However, any difference between treatment and control earnings effectively consists of a mixture of units (the working) that could be made up of varying proportions of always-takers, compliers, and defiers, effectively leading to selected comparisons. Raw differences in employment rates are also uninformative about heterogeneity and composition of the intensive and extensive margins. We take this into account when constructing the bounds in what follows. 

\subsection{Models and Estimation}
We give a brief overview of the estimation methods used. See Appendix \ref{app_JC_estimation} for more details. Treatment propensities are known by the design of the stratified experiment, see Appendix \ref{app_JC_data1}. All other nuisance parameters are estimated using 5-fold cross-fitting with all available observations and pre-treatment predictors at each time period reported. No cross-time information was used.

For the conditional selection probabilities we compare five different main specifications: (a) Logistic regression, (b) logistic regression with interacted treatment, (c) gradient-boosted trees (XGBoost), (d) artificial neural network, and (e) random forest. These methods differ with respect to their capabilities to (i) allow for weak monotonicity, (ii) produce sparsity, (iii) be robust to extreme observations, and (iv) fit and predict employment probabilities and status well. Table \ref{tab:models_selection} contains an aggregate comparison between the different methods with respect to these properties. By construction, all models except for the simple logit allow for weak monotonicity. XGBoost, neural network, and random forest can produce sparsity.\footnote{Both the interacted logit and neural network tend to produce many or some relatively extreme estimates for the relative selection probabilities, see also Figure \ref{app_JC_p0x_all} in Appendix \ref{app_JC_results}. It seems questionable whether e.g.~extreme relative selection probabilities that suggest e.g.~a quadrupling in relative employment chances due to Job Corps are credible or mostly a by-product of an overly high-variance model.} 
XGBoost has, by far, the best out-of-sample accuracy and loss, where the latter is measured by the negative log-likelihood. It outperforms all other methods in 202 and 206 of the 208 periods for accuracy and loss, respectively, see Appendix \ref{app_JC_estimation} for detailed results.
This is in line with many studies that demonstrate boosted trees as or among the best performing methods for tabular data \citep{shwartz2022tabular}.

\begin{table}[!h] \centering\caption{Selection Model Estimates: Comparison} \label{tab:models_selection}
\footnotesize
\begin{tabular}{lccccc} \hline \hline \\[-1.5ex]
 Method    & Weak Monotonicity & Sparsity & Extreme Predictions & Accuracy & Likelihood  \\[-0.5ex]
 &&&&(rank)&(rank) \\[0.5ex]
 Logit                  &  \xmark & \xmark & \xmark & 3 & 4 \\
 Logit Interacted       &  \cmark & \xmark & \cmark & 5 & 5 \\
 XGBoost                &  \cmark & \cmark & \xmark & 1 & 1 \\
 Neural Network         &  \cmark & \cmark & \cmark & 2 & 3 \\
 Random Forest          &  \cmark & \cmark & \xmark & 4 & 2 \\ \hline
\end{tabular}
\begin{justify}
    This table contains the different estimation methods, their properties as well as an overview over the resulting $p_0(x)$ estimates. Weak Monotonicity refers to Assumption \ref{ass_monotone}. Logit has that either $P(\mathcal{X}^+)$ or $P(\mathcal{X}^-)$ equal zero. Sparsity is achieved whenever $P(\mathcal{X}^0)$ can be larger than zero. Extreme Predictions refers to an average of at least 10 selection estimates outside of the $(0.25,4)$ interval. Accuracy and Likelihood contain the rank in the respective performance metric based on the cross-fitting error averaged over all periods $t\in\{1,\dots,208\}$. 
\end{justify}
\end{table}

For conditional means, conditional quantiles, and truncated means required for the bound estimators, we use off-the-shelf honest regression and quantile forests across all specifications \citep{athey2019generalized}.
This was chosen to isolate the contribution of differences in relative selection probabilities, which is the main issue addressed in this paper.\footnote{Quantile forests are also computationally attractive in high dimensions and have been used for sample selection bounds in previous research \citep{andersen2023guide, heiler2024heterogeneous}.}

Our main analysis centers around the intensive margin ATE. Here, we consider the bounds that (i) remove observations with $p_0(x) = 1$ and use the corresponding influence function (``Trim''), (ii) the moment shifting or selection method by \cite{semenova2025generalized} (``Shift''), (iii) smooth bounds (``Smooth''), and (iv) conventional Lee bounds using strong monotonicity. For the shifting method, we use the tuning parameter as suggested by \cite{semenova2025generalized}, $\rho_n = n^{-1/4}\log^{-1}n$. For both trimming and shifting, we report results with and without using the propensity score correction that affects efficiency. For the smooth bounds, we report the analysis for smoothing parameters $n^{-1/4}\log^{-1}n \approx 0.01$ and $n^{-1/12}\log^{-1}n \approx 0.05$ in line with the discussion in Section \ref{sec_smoothingSTAT}.

For the extensive margin effect, fewer methods are available. Thus, we consider (i) the same smooth bounds (``Smooth'') with equal parameterization as for the intensive margin as well as (ii) naive bounds that treat the problem as if there was no irregularity problem (``Regular'').  

\subsection{Causal Effects Bounds}
We now present the impact results for assignment to JC on log hourly earnings. We first present the results for the intensive margin main evaluation in $t=208$ using the new methods and contrast them with results based on the literature and replications with comparable sample and design weights. 
We then present the results using the best performing nuisance specification for methods that allow for conditional monotonicity for periods $t\in\{45,90,135,180,208\}$ as in \cite{lee2009training}. We follow-up with the impact evaluation on the extensive margin at $t = 45$. Robustness checks and results for additional methods, time periods, and smoothing parameters can be found in Appendix \ref{app_JC_results}.

\subsubsection{Intensive Margin: Main Effects at $t=208$} \label{sec_INTENSIVE-1}

\afterpage{
\begin{landscape}
\begin{table}[!h]
    \caption{Impact Results at $t=208$: Overview}
    \label{tab:JC_results_208_overview1}
\begin{threeparttable}
    \centering \scriptsize
    \begin{tabular}{llccccccc} \hline \hline 
        	&	Method	&	Impact Estimate	&	95\% CI	&	Sample 	&	Selection	&	Selection 	&	Covariates	&	Sample	\\
	&		&		&		&	Selection	&	Model	&	Key Assumptions	&		&		\\ \hline \\[-0.5ex]
Schochet et al.~(2008)	&	Mean Difference	&	0.059	&	$(0.031,~0.086)$	&	\xmark	&	--	&	Independence$^1$	&	0	&	10602	\\
	&		&		&		&		&		&		&		&		\\
	&		&		&		&		&		&		&		&		\\
Lee (2009)	&	Mean Difference	&	0.058	&	$(0.027,~0.088)$	&	\xmark	&	--	&	Independence$^1$ &	0	&	9415	\\
	&		&		&		&		&		&		&		&		\\
	&	Bounds	&	$[-0.019,~0.093]$	&	$(-0.049,~0.114)$	&\cmark	&	--	&	Monotonicity	&	0	&	9415	\\
	&	+ Covariates	&	$[-0.012,~0.089]$	&	$(-0.037,~0.112)$	&\cmark	&	Nonparametric	&	Monotonicity	&	$28/1/5^2$	&	9415	\\
	&		&		&		&		&		&		&		&		\\
	&	Heckman (1979)	&	0.015	&	$(-0.008,~0.038)$	&\cmark	&	Probit	&	Exclusion$^3$, Normality 	&	28	&	9415	\\
	&	Das et al.~(2003)	&	0.014	&	$(-0.010,~0.038)$	&\cmark	&	Linear$^4$	&	Exclusion$^3$, Single Index &	$28^4$ &	9415	\\
	&		&		&		&		&		&		&		&		\\
	&		&		&		&		&		&		&		&		\\
Shift$^{5}$	&	Known Propensity	&		$[-0.646,~1.514]$	&	$(-0.438,~1.650)$	&\cmark	&	Nonparametric	&	Weak Monotonicity	&	77	&	9415	\\
	&	Unknown Propensity	&		$[0.138,~0.162]$	&	$(-0.003,~0.303)$	&	\cmark	&	Nonparametric	&	Weak Monotonicity	&	77	&	9415	\\
	&		&		&		&		&		&		&		&		\\
	&		&		&		&		&		&		&		&		\\
Trim$^{5,6}$	&	Known Propensity &		$[-0.008,~0.054]$	&	$(-0.063,~0.111)$		&\cmark	&	Nonparametric	&	Weak Monotonicity	&	77	&	9415	\\
	&	Unknown Propensity  &		$[-0.036,~0.077]$	&	$(-0.201,~0.245)$	&	\cmark	&	Nonparametric	&	Weak Monotonicity	&	77	&	9415	\\
	&		&		&		&		&		&		&		&	\\ 
Smooth Bounds$^{5,7}$	&	$h = 5$ &		$[-0.057,~0.275]$	&	$(-0.095,~0.312)$		&\cmark	&	Nonparametric	&	Weak Monotonicity	&	77	&	9415	\\
	&	$h = 1$  &		$[-0.005,~0.113]$	&	$(-0.040,~0.148)$	&	\cmark	&	Nonparametric	&	Weak Monotonicity	&	77	&	9415	\\
	&		&		&		&		&		&		&		&	\\ \hline	
    \end{tabular}
   Impact estimates using different methods and samples with key assumptions for validity of the impact estimate. All estimates and intervals except for Lee (2009) are based on own calculations using design weights. Confidence intervals for point-identified methods use standard critical values. Confidence intervals for bounds use the \cite{imbens2004confidence} refinement. All methods use known propensity scores of the stratified experiment if necessary.  
$^1$Independence of potential selection from potential outcomes.
$^2$28 covariates are used to generate a single predictive score. Bounds are calculated within 5 strata of this score.
$^3$Months employed is used as excluded variable in the selection model.
$^4$Semiparametric model also includes transformations of the original variables, see Lee (2009).
$^5$Best performing nuisance parameter specifications, see Appendix \ref{app_JC_estimation} for details.
$^6$Trim are the bounds that remove all units for which $\hat{p}_0(x) = 1$ with unknown/efficient and known/inefficient propensity score and related moment function.
$^7$Smoothing parameters $h$ scaled by 100.
\end{threeparttable}
\end{table}
\end{landscape}
}

Table \ref{tab:JC_results_208_overview1} contains the impact analysis for various methods and specifications at period $t=208$. All entries except for Lee (2009) are based on our own calculations and with identical outcomes and design weights. \cite{schochet2008does} is a simple treatment-control difference for the sample of reported log hourly earnings. It is of similar magnitude when compared with the equivalent mean difference in the restricted Lee (2009) sample in line with the balancing analysis.

Shift, regular, and smooth bounds all admit flexible estimation of nuisance parameters. Using the best performing XGBoost model for selection, there are 4252 observations with exact $\hat{p}_0(x) = 1$. Thus, trimming bounds effectively have to use a much reduced sample. Nevertheless, estimates are in a similar range but with larger standard errors. The moment shifting approach by \cite{semenova2025generalized} with known propensity scores provides overly wide bounds. This seems to be due to a strong sensitivity when there are many estimates for $p_0(x)$ around one. With unknown scores, estimates are tighter, however with standard errors much larger compared to alternatives. 

Our preferred specification is smooth bounds with $h=1$. The impact estimate of $[-0.005,0.113]$ with 95\% confidence interval of $(-0.040,0.148)$ rules out large negative earnings effects and is surprisingly close and of similar width compared to results obtained under strong monotonicity as in \cite{lee2009training}. As strong monotonicity is rejected by the data \citep{semenova2025generalized}, this provides more evidence on the robustness of this final impact evaluation. 
We also include a larger smoothing specification for comparison; for more values see Appendix \ref{app:JC_effect_bounds_all1}. As expected, the identified set becomes larger. In this case, the standard errors are still comparable such that overall less smoothing seems preferable.

\subsubsection{Intensive Margin: Additional Periods and Robustness} \label{sec_INTENSIVE-2}
We now analyze the effect for multiple periods using all methods. We present the results for the two best nuisance models (XGBoost and neural network). For the shifting and the trimming bounds, we only consider the unknown propensity score estimates, as they are more reliable across specifications. For smooth bounds we present two smoothing parameter results as above. Results for the other methods and parameters are robust with the exception of the interacted logistic model, that failed to converge in some folds.\footnote{See Appendix \ref{app_JC_results} for histograms and descriptive statistics.}

\begin{table}[!h]
\begin{threeparttable}
    \centering
    \caption{Job Corps Impact Results, by week}
    \label{tab:JC_results_both1}
    \scriptsize
	\begin{tabular}{llccccc}
		\hline
		\hline ~ \\[-0.5ex]
  \multicolumn{7}{c}{Simple Monotonicity Bounds} \\[1ex]
		~ & ~ & $t = 45$ & $t = 90$ & $t = 135$ & $t = 180$ & $t = 208$\\[-0.5ex] ~ \\ 
		Lee (2009) & ~ & $[-0.074,~0.125]$ & $[0.042,~0.043]$ & $[-0.016,~0.076]$ & $[-0.033,~0.087]$ & $[-0.019,~0.093]$ \\ 
		~ & ~ & $(-0.097,~0.152)$ & $(-0.005,~0.092)$ & $(-0.051,~0.099)$ & $(-0.064,~0.108)$ & $(-0.049,~0.114)$ \\ 
		~ & ~ & ~ & ~ & ~ & ~ & ~ \\ 
  ~ \\[-0.5ex]
  \multicolumn{7}{c}{Results for XGBoost Selection Model} \\[1ex]
		~ & ~ & $t = 45$ & $t = 90$ & $t = 135$ & $t = 180$ & $t = 208$\\[-0.5ex] ~ \\ 
		Smooth & $h = 5$ & $[-0.109,~0.145]$ & $[-0.112,~0.212]$ & $[-0.093,~0.213]$ & $[-0.072,~0.215]$ & $[-0.057,~0.275]$ \\ 
		~ & ~ & $(-0.172,~0.199)$ & $(-0.161,~0.256)$ & $(-0.139,~0.250)$ & $(-0.107,~0.251)$ & $(-0.095,~0.312)$ \\ 
		~ & ~ & ~ & ~ & ~ & ~ & ~ \\ 
		~ & $h=1$ & $[-0.000,~0.161]$ & $[0.008,~0.096]$ & $[-0.028,~0.090]$ & $[-0.029,~0.093]$ & $[-0.005,~0.113]$ \\ 
		~ & ~ & $(-0.068,~0.209)$ & $(-0.036,~0.135)$ & $(-0.066,~0.125)$ & $(-0.064,~0.128)$ & $(-0.040,~0.148)$ \\ 
		~ & ~ & ~ & ~ & ~ & ~ & ~ \\ 
		Trim 
		~ & ~ & $[-0.257,~0.126]$ & $[-0.175,~0.004]$ & $[-0.021,~0.152]$ & $[-0.180,-0.011]$ & $[-0.036,~0.077]$ \\ 
		~ & ~ & $(-0.403,~0.269)$ & $(-0.745,~0.573)$ & $(-0.197,~0.327)$ & $(-0.328,0.134)$ & $(-0.201,~0.245)$ \\ 
		~ & ~ & ~ & ~ & ~ & ~ & ~ \\ 
		Shift 
		~ & ~ & $[-0.324,-0.076]$ & $[-0.021,-0.011]$ & $[0.022,~0.067]$ & $[0.075,~0.122]$ & $[0.138,~0.162]$ \\ 
		~ & ~ & $(-0.470,~0.067)$ & $(-0.194,~0.162)$ & $(-0.116,~0.204)$ & $(-0.055,~0.253)$ & $(-0.003,~0.303)$ \\ 
		~ & ~ & ~ & ~ & ~ & ~ & ~ \\ 
		$|\hat{p}_0(x)-1|$ & $\leq \frac{n^{-1/4}}{\log(n)}$ & 2592 & 8680 & 6350 & 4871 & 5688 \\ 
		~ & ~ & ~ & ~ & ~ & ~ & ~ \\ 
		\hline
	\end{tabular}
 \begin{tabular}{llccccc}
  ~ \\[-0.5ex]
  \multicolumn{7}{c}{Results for Neural Network Selection Model} \\[1ex]
		~ & ~ & $t = 45$ & $t = 90$ & $t = 135$ & $t = 180$ & $t = 208$\\[-0.5ex] ~ \\ 
		Smooth & $h = 5$ & $[-0.177,~0.089]$ & $[-0.137,~0.214]$ & $[-0.089,~0.187]$ & $[-0.092,~0.182]$ & $[-0.058,~0.228]$ \\ 
		~ & ~ & $(-0.234,~0.137)$ & $(-0.183,~0.255)$ & $(-0.132,~0.220)$ & $(-0.125,~0.215)$ & $(-0.094,~0.263)$ \\ 
		~ & ~ & ~ & ~ & ~ & ~ & ~ \\ 
		~ & $h=1$ & $[-0.469,~0.777]$ & $[-0.005,~0.105]$ & $[-0.023,~0.107]$ & $[-0.045,~0.112]$ & $[-0.022,~0.127]$ \\ 
		~ & ~ & $(-0.543,~0.855)$ & $(-0.049,~0.147)$ & $(-0.061,~0.144)$ & $(-0.080,~0.149)$ & $(-0.058,~0.165)$ \\ 
		~ & ~ & ~ & ~ & ~ & ~ & ~ \\ 
		Trim & ~ & $[-0.347,~0.192]$ & $[-0.031,~0.003]$ & $[-0.107,~0.046]$ & $[-0.141,~0.038]$ & $[-0.157,~0.072]$ \\ 
		~ & ~ & $(-0.481,~0.319)$ & $(-0.179,~0.149)$ & $(-0.223,~0.164)$ & $(-0.266,~0.159)$ & $(-0.284,~0.203)$ \\ 
		~ & ~ & ~ & ~ & ~ & ~ & ~ \\ 
		Shift &  ~ & $[-0.424,-0.122]$ & $[-0.054,~0.011]$ & $[0.049,~0.108]$ & $[-0.062,~0.076]$ & $[0.046,~0.121]$ \\ 
		~ & ~ & $(-0.571,~0.019)$ & $(-0.205,~0.162)$ & $(-0.078,~0.235)$ & $(-0.187,~0.199)$ & $(-0.084,~0.252)$ \\ 
		~ & ~ & ~ & ~ & ~ & ~ & ~ \\ 
		$|\hat{p}_0(x)-1|$ & $\leq \frac{n^{-1/4}}{\log(n)}$ & 737 & 4842 & 3220 & 2180 & 3669 \\ 
		~ & ~ & ~ & ~ & ~ & ~ & ~ \\ 
		\hline
	\end{tabular}
 \footnotesize 
 This table contains the estimated intensive margin bounds of assignment to JC on log hourly earnings using smoothed, trimmed, shifted, and regular (generalized) Lee bounds using $n = 9415$ for all weeks. Identified sets are in brackets and 95\% \cite{imbens2004confidence} confidence intervals in parentheses below. The threshold $\frac{n^{-1/4}}{\log(n)}$ is the shifting threshold suggested by \cite{semenova2025generalized} to choose between moments. Smoothing parameters are scaled by 100. All calculations use design weights.
\end{threeparttable}
\end{table}

Table \ref{tab:JC_results_both1} contains the estimates and confidence intervals for XGBoost and neural network selection scores. There are a few general observations. The choice of the selection model matters. In particular, the larger dispersion of predicted $p_0(x)$ for the neural net leads to wide intervals at $t=45$ for most methods. Trimming and shifting seem more sensitive in general, see also Appendix \ref{app:JC_effect_bounds_all1}. For all other periods, results are relatively consistent across selection models. Trimming and shifting bounds, however, tend to have relatively large standard errors compared to the smooth bounds. 

The intervals for the smooth bounds at $h=1$ are, again, surprisingly close to the original \cite{lee2009training} specification. Even a strictly positive identified set at $t=90$ is recovered with XGBoost, albeit with larger width which seems more credible given the estimates of the surrounding periods. The Lee unconditional monotonicity assumption can be rejected from the data \citep{semenova2025generalized}. However, it seems that the general finding for the later periods, with log hourly earnings effects in the range of around $-0.03$ to $0.11$, is robust to conditional monotonicity without parametric assumptions. 

\subsubsection{Extensive Margin: Main Effects at $t=45$} \label{sec_empirical_EXTENSIVE1}
We now present results for the effect bounds at the extensive margin. We focus on period
$t=45$, which exhibits the largest extensive-margin shares among the main evaluation
periods considered in \cite{lee2009training}, see Appendix \ref{sec_app_EXTENSIVE} for additional
time periods.\footnote{For periods after week 45, the estimated extensive-margin shares
are generally below 5\% of the sample and decline quickly to less than 2\% after week 57
as long-run employment effects fade. As a result, the corresponding bounds become very
wide and inference may be further complicated by weak identification.}

\begin{table}[!h]
\centering
\caption{Extensive Margin Effect Bounds and Sizes in Job Corps: Week 45}
\label{tab_JC.EM_week45}
\begin{threeparttable}
\footnotesize
\begin{tabular*}{\linewidth}{@{\extracolsep{\fill}} lcccc}
\hline \hline
Method & Smoothing & Target & \multicolumn{2}{c}{Nuisance Model} \\ 
&&& XGBoost & Neural Network \\ \cline{1-5} \\[1ex]
Smooth & $h=1$ & $[\beta_{L,h}(em),\beta_{U,h}(em)]$ &  [-1.359, 0.448]  & [-2.412, 0.527] \\
       &       &                   & (-1.417, 0.756) & (-2.572, 1.318) \\[1ex]
       &       & $\pi_{em,h}$      & 0.058 & 0.030 \\
       &       &                   & (0.056, 0.060) & (0.028, 0.032) \\[2ex]
       & $h=5$ & $[\beta_{L,h}(em),\beta_{U,h}(em)]$ & [-4.436, 0.950] & [-17.523, 5.316] \\
       &       &                   & (-5.299, 2.301) & (-22.139, 9.640) \\[1ex]
       &       & $\pi_{em,h}$      & 0.039  & 0.017 \\
       &       &                   & (0.037, 0.041) & (0.016, 0.018) \\[2ex]
Regular &       & $[\beta_{L}(em),\beta_{U}(em)]$  & [-0.862, 0.562] & [-3.828, -0.281] \\
       &       &                   & (-1.043, 0.859) & (-4.613, 0.341) \\[1ex]
       &       & $\pi_{em}$        & 0.067   & 0.048 \\
       &       &                   & (0.048, 0.087) & (0.028, 0.068) \\[1ex] 
\hline
\end{tabular*}
    This table reports the estimated extensive-margin share and the bounds on the effect of assignment to Job Corps on log hourly earnings in week 45 using $n=9415$
observations. ``Smooth'' refers to the bounds developed in this paper, while ``Regular''
treats the problem as if the target parameters $\beta_L(em)$ and $\beta_U(em)$ were differentiable. The associated confidence intervals for the regular bounds are therefore invalid. Identified sets are reported in brackets, with 95\% confidence intervals shown in parentheses below. Because the extensive-margin shares are small, we use misspecification-robust confidence intervals following \cite{stoye2020simple}. Smoothing parameters are scaled by 100. All calculations use design weights.
\end{threeparttable}
\end{table}

Table \ref{tab_JC.EM_week45} summarizes the results for the same nuisance-model and
smoothing specifications as in Section \ref{sec_INTENSIVE-2}, together with naive
(``Regular'') bounds that ignore the non-differentiability of the target parameters
$\beta_B(em)$. Overall, identification uncertainty dominates sampling uncertainty. Bounds are wide across all specifications and contain the intensive margin bounds from Section \ref{sec_INTENSIVE-2}. Under our preferred specification (XGBoost with $h=1$), the data are consistent with decreases of 74\% up to increases of 57\% in wages. This substantial uncertainty is driven by the small estimated extensive-margin share of about 6\%. By comparison, the intensive margin at $t=45$ is around 32\% which yields the considerably tighter bounds in Section \ref{sec_INTENSIVE-2}. This contrast between the intensive- and extensive-margin results highlights that the empirical relevance of the latter is limited by the small employment effect of JC.


The results are also informative about the implied relaxation associated with the
smoothing parameters, as discussed in Section 7.2, see \eqref{eq_expost_relaxed} and \eqref{eq_expost-C-Df}. In particular, additional calculations using \eqref{eq_expost-C-Df} show that the empirical 
maximum admitted estimated value of
$\min\{\pi_c(x),\pi_d(x)\}$ is around 2.6\% for $h=5$
and 0.5\% for $h=1$, a moderate deviation from weak monotonicity. Moreover, for $h=5$, the lower smoothed extensive-margin
share is 41.70\% below the unsmoothed benchmark share, falling from 6.7 to
3.9 percentage points. This seems overly conservative relative to $h=1$, for which
the implied reduction is only 13.43\% (from 6.7 to 5.8 percentage points).

\section{Comparison to Semenova (2025)} \label{sec_S25}
This paper builds on and extends \cite{semenova2025generalized}, henceforth S25. The two papers are closely connected, but differ in scope, methods, assumptions and results. Regarding scope, S25 only considers identification of the intensive margin ATE under conditional monotonicity as in \cite{lee2009training}. We characterize bounds for all principal strata and extensive margin effects under the same identification assumptions (Proposition \ref{prop_identification1}). We also provide extensions to heterogeneous effects and stochastic dominance in Appendix \ref{sec_extensions}.

A central theoretical contribution of this paper is the characterization of the boundary between regular and irregular regimes, which has not been addressed by S25. Theorem \ref{thm_differentiable} shows that absence of point mass, i.e.~$P(\mathcal{X}^0) = 0$, is necessary and sufficient for pathwise differentiability of strata and margin bounds in the nonparametric sample selection model under conditional monotonicity.

Regarding estimation and inference, both papers adopt debiased moment-based approaches that allow for high-dimensional machine learning nuisances. S25 verifies required convergence rates in the context of regularized high-dimensional models. We provide estimators for the extended set of parameters and inference under both regular regimes, $P(\mathcal{X}^0)=0$, and possibly irregular regimes, $P(\mathcal{X}^0)>0$.  In regular regimes, our influence function for the intensive-margin ATE coincides with the influence function implied by the moment estimator in S25 under unknown propensities. Consequently, our efficiency results imply that the estimator in S25 with unknown propensities is semiparametrically efficient. We also quantify the efficiency gap that arises when using moments based on known propensity scores, paralleling the results for ATE estimation \citep{hahn1998role}.

A main methodological difference emerges in irregular regimes. 
S25 suggests a moment selection procedure in the spirit of \cite{andrews2010inference} that switches between truncation-based and IPW contributions in the selected population, $SY(D-m(X))\big/m(X)(1-m(X))$, depending on whether an estimated selection CATE falls within a shrinking corridor $[-\rho_n, \rho_n].$ In contrast, we construct a smooth outer identification region by replacing the underlying nonsmooth objects with smooth approximations indexed by a tuning parameter $h$.
As a first consequence, our nuisance requirements are weaker than the ones in S25. In particular, we do not require uniform convergence rates for the selection probabilities and standard $L^2$ rates of order $o(n^{-1/4})$ suffice. 
Second, for any fixed $h>0$, the resulting bounds remain pathwise differentiable and admit well-defined efficient influence functions. Third, as $h \to 0$, the smooth bounds approach the sharp bounds, yielding a convergence corridor under which inference is asymptotically valid but conservative. 
Fourth, unlike the switching parameter $\rho_n$ in S25, $h$ can be equipped with an economic interpretation as an implied admitted share for the margins further guiding admissible choices. Fifth, while both approaches are asymptotically valid, they reflect different trade-offs between sharpness and stability in finite samples. In particular, depending on $\rho_n$, in designs with substantial mass of the estimated selection CATE
near zero (see Figure 3.2), a non-negligible share of observations may fall into the
shifting corridor and contribute through the IPW-type component that would be used
under point identification.
As a consequence, S25 bounds may be narrower in finite samples relative to
estimators that retain truncation-based contributions. While asymptotically valid, this
feature can lead to more liberal inference in finite samples compared to smoothing.
This distinction is also reflected in our empirical application, where the two approaches can yield noticeably different estimates and standard errors.

The scope of our empirical analysis and testing methodology extends beyond S25. We provide a more comprehensive empirical evaluation using the JC dataset across multiple margins, time periods, and estimation strategies, whereas the JC impact evaluation in S25 focuses exclusively on the intensive-margin application in week 90. Finally, S25 develops a test of monotonicity against conditional monotonicity based on \cite{chernozhukov2019inference}. We complement this by providing a test for the absence of point mass under various nuisance estimators.

\section{Concluding Remarks} \label{sec_conclusion}
This paper demonstrates the importance of heterogeneity in selection behavior with respect to estimation and inference on causal effects at the intensive and extensive margins in selected samples.
Allowing for different types of monotonicity crucially affects the (ir)regularity of sharp effect bounds and thus the ability to provide precise inference for the associated causal effects. This paper provides a solution in the form of 
outer identification regions that can be estimated efficiently from a semiparametric perspective. The approach can handle modern machine learning methods that are able to better approximate such important heterogeneity in sample selection. 
We discover an empirically relevant trade-off between tightness of identification and regularity of inference that is likely to apply to a much larger set of models and parameters beyond the ones considered in this paper.

\addcontentsline{toc}{section}{References}	
	 {\setstretch{1}
	\bibliography{HKV}
	}

\subsection*{Declaration of generative AI and AI-assisted technologies in the manuscript preparation process}	

During the preparation of this work, the authors used OpenAI's ChatGPT to assist with mathematical proofs, coding, formatting of tables and outputs, and spelling. The authors reviewed and edited the content as needed and take full responsibility for the content of the article. 

\newpage

\appendix

{\centering
\huge{Supplementary Appendix}
}

\section{Proofs and Supplementary Material} \label{sec_proofs}
\subsection{Proof of Proposition \ref{prop_identification1}}
We present the derivations for the lower bound.
First note that the only observed strata for which we have outcome information are the selected treated and control groups. Moreover, for any $x \in \mathcal{X}^+$ {{\footnotesize \begin{align*}
    E&[Y|DS=1,X=x] = E[Y(1)|S(1)=1,X=x] \\
    &=E[Y(1)|S(0)=1,S(1)=1,X=x]p_0(X) + E[Y(1)|S(0)=0,S(1)=1,X=x](1-p_0(X))
\end{align*}} }
and analogously, for any $x \in \mathcal{X}^-$, {{\footnotesize \begin{align*}
    E&[Y|(1-D)S=1,X=x] = E[Y(0)|S(0)=1,X=x] \\
    &=E[Y(0)|S(0)=1,S(1)=1,X=x] 1/p_0(X) + E[Y(0)|S(0)=1,S(1)=0,X=x](1-1/p_0(X)).
\end{align*}} }
Thus, by \cite{horowitz1995identification}, Corollary 4.1, we obtain the following sharp lower bounds
{{\footnotesize \begin{align*}
    E[Y(1)|S(0)=1,S(1)=1,X=x]& \geq E[Y|DS=1,Y\leq q_{1}(p_0(X),X),X=x] \text{ for any } x \in \mathcal{X}^+ \text{ and } \\ 
    -E[Y(0)|S(0)=1,S(1)=1,X=x]& \geq - E[Y|(1-D)S=1,Y\geq q_{0}(1-1/p_0(X),X),X=x]  \text{ for any } x \in \mathcal{X}^-. 
\end{align*}}}
Moreover, for the intensive margin, the remaining components are point identified as
{{\footnotesize \begin{align*}
    E[Y(0)|S(0)=1,S(1)=1,X=x] = E[Y|(1-D)S=1,X=x]  \text{ for any } x \in \mathcal{X}^+, \\
    E[Y(1)|S(0)=1,S(1)=1,X=x] = E[Y|DS=1,X=x]  \text{ for any } x \in \mathcal{X}^-. 
\end{align*}}}
On $\mathcal{X}^0$, the same applies to $E[Y(d)|X=x]$, as $D \ind S~|~X=x$ for all $x\in\mathcal{X}^0$. 
For compliers and defiers, we can use the same mixtures as above to obtain {{\footnotesize \begin{align*}
    E[Y(1)|S(0)=0,S(1)=1,X=x]  &\geq E[Y|DS=1,Y\leq q_{1}(1-p_0(X),X),X=x] \text{ for any } x \in \mathcal{X}^+, \\
    -E[Y(0)|S(0)=1,S(1)=0,X=x]  &\geq - E[Y|(1-D)S=1,Y\geq q_{0}(1/p_0(X),X),X=x]  \text{ for any } x \in \mathcal{X}^-. 
\end{align*}}}
For never-takers, no observed stratum is informative and thus, the lower bound is only informed by the support {{\footnotesize \begin{align*}
    E[Y(1)|S(0)=0,S(1)=0,X=x] &\geq \underline{y}_1(x), \\
    -E[Y(0)|S(0)=0,S(1)=0,X=x] &\geq -\overline{y}_0(x). 
\end{align*}} }
This completes all lower bounds. For the upper bounds, the corresponding expression can be obtained by using the complementary trimming thresholds and replacing the upper (lower) support bounds by their respective counterparts. 
\subsection{Proof of Proposition \ref{prop_identification_dominance1}}
Proposition \ref{prop_identification_dominance1} follows directly from the previous proof by plugging in Assumption \ref{ass_dominance} into the corresponding expressions.

\subsection{Proof of Proposition  \ref{prop_cond_covariate}}
We only prove always-takers and extensive-margin cases since other cases can be shown analogously. In the always-takers case, we obtain that
{\footnotesize\begin{align*}
dP(x|S(0)=1,S(1)=1)= \frac{P(S(0)=1, S(1)=1|X=x) dP(x)}{P(S(0)=1, S(1)=1)}.
\end{align*}}
Invoking Assumption \ref{ass_monotone}, we deduce that  $P(S(0)=1, S(1)=1|X=x)=P(S(0)=1|X=x)=s(0,x)$ for $x \in \mathcal{X}^{+}\cup \mathcal{X}^{0}$ and similarly $P(S(0)=1, S(1)=1|X=x)=s(1,x)$ for $x \in \mathcal{X}^{-}\cup \mathcal{X}^{0}.$
In compact form, this can be written as $\min\{s(0,X),s(1,X)\}.$

In the extensive margins case, we have that
{\footnotesize\begin{align*}
dP(x|S(0) \neq S(1))= \frac{P(S(0) \neq S(1)|X=x) dP(x)}{P(S(0) \neq S(1))}.
\end{align*}}
 Now, in view of Assumption \ref{ass_monotone}, for $x \in \mathcal{X}^{+}$ we have that 
{\footnotesize\begin{align*}
P(S(0) \neq S(1)|X=x)=&P(S(0)=0,S(1)=1|X=x)\\
=&P(S(1)=1|X=x)-P(S(0)=1, S(1)=1|X=x)\\
=&s(1,x) - s(0,x),
\end{align*}}
and similarly we deduce that $P(S(0) \neq S(1)|X=x)=s(0,x)-s(1,x)$ for $x \in \mathcal{X}^{-}$, and this term vanishes for $x \in \mathcal{X}^{0}.$ This can compactly be expressed as $|s(1,x)-s(0,x)|.$

\subsection{Example for Smooth Bounds}
Let $g(z)=h\log(1+e^{z/h})$ be a log-sum-exponential function, where $h>0$ is a smoothing parameter. For each $h>0$ and $z \in \mathbb{R},$ we have
{\footnotesize \begin{equation}\label{LogSumExpIneq}
\max\{z,0\}<g(z) \leq \max\{z,0\}+h \log(2),
 \end{equation}}
where the equality is attained if and only if $z=0.$ Based on this LogSumExp function $g$, we introduce the following four smoothing functions.
{\footnotesize \begin{align*}
g_{1,h}(z)=&1-g(1-z), \\
g_{2,h}(z)=&g(z)    , \\
g_{3,h}(z)=&1-g(1-z)+ h\log(2), \\
g_{4,h}(z)=&g(z)-h\log(2).
\end{align*}}

Relying on inequalities in \eqref{LogSumExpIneq}, it is straightforward to show that 
{\footnotesize \begin{align*}
g_{1,h}(z) & < \min\{z,1\} \leq g_{3,h}(z),\\
g_{4,h}(z)& \leq \max\{z,0\} < g_{2,h}(z).
\end{align*}}
Further, for each $i\in\{1, \ldots, 4\},$ and $h>0$, we also have that 
{\footnotesize \begin{equation}
\sup_{z \in \mathbb{R}} |g_{i,h}(z)-g_i(z)| \leq h \log(2), 
\end{equation}}
where $g_1(z)=g_3(z)=\min\{z,1\}$ and $g_2(z)=g_4(z)=\max\{z,0\}.$


\subsection{Proof of Theorem \ref{thm:convergence}}
We only prove this result for $\beta_{L,h}(1,1)$ since the other cases are analogous.
We introduce the following intermediate Lee lower bound.
{\footnotesize \begin{equation}
\tilde{\beta}_{L,h}= \frac{E[\beta_{L,h}(X)\min\{s(0,X),s(1,X)\}]}{E[\min\{s(0,X),s(1,X)\}]}.
\end{equation}}
It suffices to prove that 
{\footnotesize \begin{equation}\label{convh-1}
\sup_{P \in \mathcal{P}}|\tilde{\beta}_{L,h}-\beta_L(1,1)|\lesssim h,
\end{equation}}
and
{\footnotesize \begin{equation}\label{convh-2}
\sup_{P \in \mathcal{P}}|\beta_{L,h}(1,1)-\tilde{\beta}_{L,h}|\lesssim h.
\end{equation}}

First, we deal with the convergence in \eqref{convh-1} and write this term as
{\footnotesize \begin{equation}\label{tilde-beta-eq}
\tilde{\beta}_{L,h}-\beta_L(1,1)=
\frac{E\left[ \big(\beta_{L,h}(X,1,1)-\beta_L(X,1,1) \big)\min \{s(0,X),s(1,X)\} \right] }
{E[\min\{s(0,X),s(1,X)\}]}.
\end{equation}}
Next, we note that
{\footnotesize \begin{equation}
\beta_{L,h}(X,1,1)-\beta_L(X,1,1)= R_1^h(x)-R_2^h(x),
\end{equation}}
where

{\footnotesize \begin{align*}
R_1^h(x)=& \frac{1}{g_{1,h}(p_0(x))}\int_{\underline{y}_1(x)}^{q_1(g_{1,h}(p_0(x)),x)} y  f_{1}(y|x) dy-\frac{1}{\min\{p_0(x),1\}}\int_{\underline{y}_1(x)}^{q_1(\min\{p_0(x),1\},x)} y  f_{1}(y|x) dy,\\
R_2^h(x)=& \frac{1}{g_{1,h}(1/p_0(x))}\int_{q_0(1-g_{1,h}(1/p_0(x)),x)}^{\overline{y}_0(x)} y  f_{0}(y|x) dy\\
&-\frac{1}{\min\{1/p_0(x),1\}}\int_{q_0(1-\min\{1/p_0(x),1\},x)}^{\overline{y}_0(x)} y  f_{0}(y|x) dy.
\end{align*}}

We further decompose the $R_1^h(x)$ term as
{\footnotesize \begin{align*}
R_1^h(x)=&\left(\frac{1}{g_{1,h}(p_0(x))}-\frac{1}{\min\{p_0(x),1\}} \right)\int_{\underline{y}_1(x)}^{q_1(g_{1,h}(p_0(x)),x)} y  f_{1}(y|x) dy \\
&-\frac{1}{\min\{p_0(x),1\}} \int_{q_1(g_{1,h}(p_0(x)),x)}^{q_1(\min\{p_0(x),1\},x)} y  f_{1}(y|x) dy.
\end{align*}}

Relying on Assumption \ref{ass_bddmoment} and the inequality $g_{1,h}(z) \leq \min\{z,1\}$, we obtain that
{\footnotesize \begin{equation}
|R_1^h(x) | \lesssim \frac{1}{\underline{g}_n^2}\sup_{x \in\mathcal{X}} \left| g_{1,h}(p_0(x))-\min\{p_0(x),1\}\right|
\end{equation}}

Similarly, we obtain that
{\footnotesize \begin{equation}
|R_2^h(x)| \ \lesssim \frac{1}{\underline{g}_n^2} \sup_{x\in\mathcal{X}}  \left|
g_{1,h}(1/p_0(x))-\min\{1/p_0(x),1\} \right|.
\end{equation}}

Now, using Assumption \ref{ass_errors-of-g} and recalling the expression in \eqref{tilde-beta-eq}, we deduce that
{\footnotesize \begin{equation}
\sup_{P \in \mathcal{P}}|\tilde{\beta}_{L,h}-\beta_L(1,1)|\lesssim \frac{h}{\underline{g}_n^2}.
\end{equation}}

Next, we turn to the term $\beta_{L,h}(1,1)-\tilde \beta_{L,h}$ and observe that
{\footnotesize \begin{align*}
\beta_{L,h}(1,1)-\tilde \beta_{L,h}=\left(\frac{A_h}{C_h}+\frac{B_h}{D_h}\right)-\left(\frac{\tilde A_h+\tilde B_h}{D}    \right),
\end{align*}}
where
{\footnotesize \begin{align*}
A_h&=E\left[g_{4,h}(\beta_{L,h}(X,1,1))g_{1,h}(p_0(X))s(1,X)\right],
\quad C_h=E[g_{3,h}(p_0(X))s(1,X)] , \\
B_h&=E[g_{4,h}(\beta_{L,h}(X,1,1)) g_{3,h}(p_0(X))s(1,X)]    , \quad
D_h =E[g_{1,h}(p_0(X))s(1,X)], \\
\tilde{A}_h&= E\left[\max\{\beta_{L,h}(X,1,1), 0\} \min\{p_0(X), 1 \} s(1,X)\right]   , \quad C=E[\min\{s(0,X), s(1,X) \}]. \\
\tilde{B}_h&=E\left[\min\{\beta_{L,h}(X,1,1), 0\} \min\{p_0(X), 1 \} s(1,X)\right].
\end{align*}}

In view of Assumptions \ref{ass_bddmoment} and \ref{ass_errors-of-g}, we note that 
{\footnotesize \begin{equation}
|A_h-\tilde A_h|+|C_h-C|+|B_h-\tilde B_h|+|D_h-C| \lesssim
\sum_{i=1}^5\sup_{z \in \mathbb{R}}\left(|g_{i,h}(z)-g_i(z)| \right).
\end{equation}}
Further, from $C>0$ and the assumption $|g_{1,h}(z)-\min\{z, 1\}| \lesssim h,$ we deduce that $D_h$ is bounded away from zero for all small $h$, which implies the same for $C_h.$
Consequently, we obtain that
{\footnotesize \begin{equation}
|\beta_{L,h}(1,1)-\tilde \beta_{L,h}| \lesssim |A_h-\tilde A_h|+|C_h-C|+|B_h-\tilde B_h|+|D_h-C|\lesssim h.
\end{equation}}

We finish the proof by adding both bounds.

\subsection{Proof of Theorem \ref{thm_differentiable}}
We only prove this theorem for always-takers/intensive margin (lower and upper bounds), extensive margin (lower bound), and compliers (lower bound). The proofs of other cases are analogous.

\subsubsection{Notations and Auxiliary Results}\label{Sec_Notation}
In this part, we introduce some notations and preliminary results necessary for the proof of Theorem \ref{thm_differentiable}. 
We are interested in the estimands $\beta_L(1,1)$, $\beta_U(1,1),$ $\beta_L(0,1)$ and $\beta_L(em)$. We note that all these estimands depend on $f_d(y|x)$, $s_d(x)$ for $d\in\{0,1\}$ and $f(x)$, where $f_d(y|x)$ is the conditional density of $Y$ given $S=1, D=d, X=x.$  
In semiparametric efficiency proofs, we consider a submodel. Letting $\varepsilon=\theta-\theta_0,$ we consider a parametric submodel for $d\in\{0,1\}.$

 {\footnotesize \begin{align*}
 f_d(y|x, \theta)&=(1+\varepsilon \tau_d(y|x, \theta_0)) f_d(y|x), \quad
 s(d,x, \theta)=s(d,x)+\varepsilon s'(d,x, \theta_0), \\ 
 f(x, \theta)&=(1+\varepsilon t(x)) f(x),
\end{align*}}
which are equal to $f_d(y|x)$, $s_d(x)$ and $f(x)$ when $\theta=\theta_0,$ and where
{\footnotesize \begin{align*}
\tau_{d}(y|x, \theta) &= \frac{\partial \log f_{d}(y|x, \theta)  }{\partial \theta}  ,  \quad
s'(d, x,\theta)=\frac{\partial s(d,x, \theta)}{\partial \theta},\\
t(x,\theta) &= \frac{\partial \log f(x, \theta)  }{\partial \theta},
\end{align*}}
such that for $d\in\{0,1\}$
{\footnotesize \begin{align*}
 &\int \tau_{d}(y|x, \theta_0) f_{d}(y|x) dy=0 \mbox{ for all } x, \quad |\tau_d(y|x, \theta_0)|<\infty \mbox{ and } |s'(d,x, \theta_0)|<\infty \mbox{ for all } x, y,\\
 &\int t(x, \theta_0) f(x) dx=0, \quad |t(x, \theta_0)|<\infty \mbox{ for all } x.
\end{align*}}

In the following result, we use the decompositions $\beta_L(x,1,1) = \beta_{L,1}(x,1,1) - \beta_{L,0}(x,1,1)$, 
$\beta_U(x,1,1) = \beta_{U,1}(x,1,1) - \beta_{U,0}(x,1,1)$, $\beta_L(x,0,1) = \beta_{L,1}(x,0,1) - \beta_{L,0}(x,0,1)$, which are derived from the components in Proposition \ref{prop_identification1}. In addition, we have from Example \ref{ex_intensive} that $\beta_L(x,em) = \beta_{L,1}(x,0,1).$

\begin{lem}\label{lemma-beta(x)-diff}
Suppose that the assumptions of Theorem \ref{thm_differentiable} hold and that $P(\mathcal{X}^0)=0.$ Then, for each $x \in \mathcal{X},$
$\beta_L(x,1,1),$ $\beta_U(x,1,1),$ $\beta_L(x,0,1),$ and $\beta_L(x,em)$ are pathwise differentiable. In particular, we have that
{\footnotesize \begin{align*}
\frac{\partial \beta_{L,1}(x,1,1, \theta_0)}{\partial \theta}=
&\frac{1}{\min\{p_0(x),1\}} \int_{\underline{y}_1}^{q_1(\min\{p_0(x),1\},x)} y \tau_{1}(y|x, \theta_0) f_{1}(y|x)  dy\\
&-\mathbbm{1}(x\in \mathcal{X}^+)\frac{q_1(p_0(x),x)}{p_0(x)} \int_{\underline{y}_1}^{q_1(p_0(x),x)} \tau_{1}(y|x, \theta_0) f_{1}(y|x)  dy\\ 
&+\mathbbm{1}(x\in \mathcal{X}^+) \frac{(q_1(p_0(x),x)- \beta_{L,1}(x,1,1)) }{p_0(x)} \frac{\partial p_0(x, \theta_0)}{\partial \theta},\\
\frac{\partial \beta_{L,0}(x,1,1,\theta_0)}{\partial \theta}=&
\frac{1}{\min\{1/p_0(x),1\}} \int_{q_0(1-\min\{1/p_0(x),1\}, x)}^{\overline{y}_0} y \tau_{0}(y|x, \theta_0) f_{0}(y|x)  dy \\
&-\mathbbm{1}(x\in \mathcal{X}^-)p_0(x) q_0(1-1/p_0(x),x) \int_{q_0(1-1/p_0(x),x)}^{\overline{y}_0} \tau_{0}(y|x, \theta_0) f_{0}(y|x)  dy\\
&-\mathbbm{1}(x\in \mathcal{X}^-)\frac{(q_0(1-1/p_0(x),x)- \beta_{L,0}(x,1,1)) }{p_0(x)} \frac{\partial p_0(x, \theta_0)}{\partial \theta},\\
\frac{\partial \beta_{U,1}(x,1,1, \theta_0)}{\partial \theta}=&\frac{1}{\min\{p_0(x),1\} } \int_{q_1(1-\min\{p_0(x),1\}, x)}^{\overline{y}_1} y \tau_{1}(y|x, \theta_0) f_{1}(y|x)  dy\\
&-\mathbbm{1}(x\in \mathcal{X}^+)\frac{q_1(1-p_0(x),x)}{p_0(x)} 
\int_{q_1(1-p_0(x),x)}^{\overline{y}_1} \tau_{1}(y|x, \theta_0) f_{1}(y|x)  dy\\ 
&+\mathbbm{1}(x\in \mathcal{X}^+) \frac{(q_1(1-p_0(x),x)- \beta_{U,1}(x,1,1)) }{p_0(x)} \frac{\partial p_0(x, \theta_0)}{\partial \theta},\\
\frac{\partial \beta_{U,0}(x,1,1, \theta_0)}{\partial \theta}=&
\frac{1}{\min\{1/p_0(x),1\}} \int_{\underline{y}_0}^{q_0(\min\{1/p_0(x),1\},x)} y \tau_{0}(y|x, \theta_0) f_{0}(y|x)  dy \\
&-\mathbbm{1}(x\in \mathcal{X}^-)p_0(x) q_0(1/p_0(x),x)
\int_{\underline{y}_0}^{q_0(1/p_0(x),x)} \tau_{0}(y|x, \theta_0) f_{0}(y|x)  dy\\
&-\mathbbm{1}(x\in \mathcal{X}^-)\frac{(q_0(1/p_0(x),x)- \beta_{U,0}(x,1,1)) }{p_0(x)} \frac{\partial p_0(x, \theta_0)}{\partial \theta},
\end{align*}}
{\footnotesize \begin{align*}
\frac{\partial \beta_{L,1}(x, 0,1, \theta_0)}{\partial \theta}=&
\mathbbm{1}(x\in \mathcal{X}^+)\frac{1}{1-p_0(x)} \int_{\underline{y}_1}^{q_1(1-p_0(x),x)} y \tau_{1}(y|x, \theta_0) f_{1}(y|x)  dy\\
&-\mathbbm{1}(x\in \mathcal{X}^+)\frac{q_1(1-p_0(x),x)}{1-p_0(x)} \int_{\underline{y}_1}^{q_1(1-p_0(x),x)} \tau_{1}(y|x, \theta_0) f_{1}(y|x)  dy\\
&- \mathbbm{1}(x\in \mathcal{X}^+) \frac{(q_1(1-p_0(x),x)- \beta_{L,1}(x, 0,1)) }{1-p_0(x)} \frac{\partial p_0(x, \theta_0)}{\partial \theta}, \\
\frac{\partial \beta_{L,0}(x,0,1, \theta_0)}{\partial \theta}=&
\mathbbm{1}(x\in \mathcal{X}^-) \frac{1}{1-1/p_0(x)} \int_{q_0(1/p_0(x), x)}^{\bar{y}_0(x)} y \tau_{0}(y|x, \theta_0) f_{0}(y|x)  dy \\
&- \mathbbm{1}(x\in \mathcal{X}^-) \frac{q_0(1/p_0(x),x)}{1-1/p_0(x)} \int_{q_0(1/p_0(x),x)}^{\bar{y}_0(x)} \tau_{0}(y|x, \theta_0) f_{0}(y|x)  dy\\
&-\mathbbm{1}(x\in \mathcal{X}^-) \frac{(q_0(1/p_0(x),x)- \beta_{L,0}(x,0,1)) }{p_0(x) (1-p_0(x))} \frac{\partial p_0(x, \theta_0) }{\partial \theta}.
\end{align*}}
\end{lem}
 
We observe that $\beta_L(x,1,1)$, $\beta_U(x,1,1)$ and $\beta_L(x,0,1)$ have components of form $\beta_{1,1}(x, h(x))$, $\beta_{0,0}(x, h(x))$, $\beta_{0,1}(x, h(x))$ and  $\beta_{1,0}(x, h(x))$ for different choices of $h(x).$ 

Below, we spell out the details for the $\beta_L(x,1,1)$ case which is based on $\beta_{1,1}(x, h(x) )$ and $\beta_{0,0}(x, h(x))$ with $h(x)=\min\{p_0(x),1\}$ and $h(x)=1-\min\{1/p_0(x),1\},$ respectively. 

Let $\beta(x)=\beta_L(x,1,1).$ In the decomposition $\beta(x, \theta)=\beta_1(x, \theta)-\beta_0(x, \theta)$, we note that 
{\footnotesize \begin{align*}
\beta_1(x, \theta)=\frac{1}{\min\{p(x,\theta),1\}} \int_{\underline{y}_1}^{q_1(\min\{p(x,\theta),1\},x)} y f_{1}(y|x, \theta)  dy.
\end{align*}}
This term hints to the use of product and Leibniz integral rules for differentiation. Since it contains minimum function, we cannot take derivatives directly. Instead, we carefully mimic these rules via decompositions. Letting $m1(x,\theta)=\min\{p_0(x,\theta),1\}$, we have that

{\footnotesize \begin{align*}
&\beta_1(x, \theta)=\beta_1(x)+\frac{\varepsilon}{m1(x)} \int_{\underline{y}_1}^{q_1(m1(x,\theta),x)} y \tau_{1}(y|x, \theta_0) f_{1}(y|x)  dy \\
&+\left(\frac{1}{m1(x,\theta)}-\frac{1}{m1(x)}\right)
 \int_{\underline{y}_1}^{q_1(m1(x,\theta),x)} y f_{1}(y|x, \theta)  dy+\frac{1}{m1(x)}
 \int_{q_1(m1(x),x)}^{q_1(m1(x,\theta),,x)} y f_{1}(y|x))  dy.
\end{align*}}
Regarding the last term, we further write this as
{\footnotesize \begin{align*}
\int_{q_1(m1(x),x)}^{q_1(m1(x,\theta),x)} y f_{1}(y|x)  dy =& q_1(m1(x),x) \int_{\underline{y}_1}^{q_1(m1(x, \theta),x)} (f_{1}(y|x)-f_{1}(y|x, \theta))  dy\\
&+q_1(m1(x),x)) \underbrace{\int_{\underline{y}_1}^{q_1(m1(x, \theta),x)} f_{1}(y|x, \theta)  dy}_{m1(x,\theta)}\\
&-q_1(m1(x),x))\underbrace{ \int_{\underline{y}_1}^{q_1(m1(x),x)} f_{1}(y|x))  dy}_{m1(x)}\\
&+\int_{q_1(m1(x),x)}^{q_1(m1(x,\theta),x)} (y-q_1(m1(x),x)) f_{1}(y|x))  dy.
\end{align*}}
Denoting $\Delta(x, \theta)=s(1,x, \theta)-s(0,x, \theta),$ we also observe that
{\footnotesize \begin{align*}
m1(x,\theta)-m1(x)=&(p_0(x, \theta)-p_0(x))\mathbbm{1}(\Delta(x)>0)\\
&+(1-p_0(x, \theta)) (\mathbbm{1}(\Delta(x, \theta)<0)-\mathbbm{1}(\Delta(x)<0)).
\end{align*}}
Combining these identities, we obtain that $\beta_1(x, \theta)=\hat{\beta}_1(x, \theta)+\bar{\beta}_1(x, \theta),$ where
{\footnotesize \begin{align*}
\hat{\beta}_1(x, \theta)=& \beta_1(x)+\frac{\varepsilon}{m1(x)} \int_{\underline{y}_1}^{q_1(m1(x,\theta),x)} y \tau_{1}(y|x, \theta_0) f_{1}(y|x))  dy \\
&-\frac{\varepsilon q_1(m1(x),x)}{m1(x)} \int_{\underline{y}_1}^{q_1(m1(x, \theta),x)} \tau_{1}(y|x, \theta_0) f_{1}(y|x)  dy\\
&+\mathbbm{1}(\Delta(x)>0) \frac{q_1(m1(x),x)-\beta_1(x, \theta)}{p_0(x)}\left[
 p_0(x, \theta)-p_0(x) \right]
, \\
\bar{\beta}_1(x, \theta)=&  (\mathbbm{1}(\Delta(x, \theta)<0)-\mathbbm{1}(\Delta(x)<0)) \frac{q_1(m1(x),x)-\beta_1(x, \theta)}{m1(x)}\left[
 1-p_0(x, \theta) \right]\\
 &+  \frac{1}{m1(x)}
 \int_{q_1(m1(x),x)}^{q_1(m1(x,\theta),x)} (y-q_1(m1(x),x)) f_{1}(y|x))  dy.
\end{align*}}
Here $\hat{\beta}_1(x, \theta)$ is the main term and now it is easy to derive that $\varepsilon(\hat{\beta}_1(x, \theta)-\beta_1(x))$ converges to the claimed limiting term by noting that
{\footnotesize \begin{equation}
\int_{\underline{y}_1}^{q_1(m1(x),x)} \tau_{1}(y|x, \theta_0) f_{1}(y|x)  dy=0 \mbox{ if } x \in \mathcal{X}^{-}.
\end{equation}}
Further, $\bar{\beta}_1(x, \theta)$ is the remainder term. If
$x \in \mathcal{X}^{-}$, we have that $\varepsilon^{-1} \bar{\beta}_1(x, \theta)=0$ for all small $|\varepsilon|.$ The reason is that $\Delta(x, \theta)<0$ and $p(x, \theta)>1$.

If $x \in \mathcal{X}^{+}$, we have that $ |\varepsilon^{-1} \bar{\beta}_1(x, )| \leq C |\varepsilon|$ for all small $|\varepsilon|$ and a constant $C>0$ in view of Assumption \ref{ass_density}. The reason is that $$|q_1(m1(x,\theta),x)-q_1(m1(x),x)|\leq c_1^{-1} |p_0(x, \theta)-p_0(x) \leq C_1 |\varepsilon|$$ for positive constants $c_1, c_2.$

On the other hand, we can do a similar decomposition for the term $\beta_0(x).$ To be precise, we have that
$\beta_0(x, \theta)=\hat{\beta}_0(x, \theta)+\bar{\beta}_0(x, \theta),$ where $m0(x, \theta)=\min\{1/p_0(x,\theta),1\}$ and
{\footnotesize \begin{align*}
\hat{\beta}_0(x, \theta)=& \beta_0(x)+\frac{\varepsilon}{m0(x)} \int^{\underline{y}_0}_{q_0(1-m0(x,\theta),x)} y \tau_{0}(y|x, \theta_0) f_{0}(y|x))  dy \\
&-\frac{\varepsilon q_0(1-m0(x),x)}{m0(x)} \int^{\underline{y}_0}_{q_0(1-m0(x, \theta),x)} \tau_{0}(y|x, \theta_0) f_{0}(y|x)  dy\\
&-\mathbbm{1}(\Delta(x)<0) \frac{q_0(1-m0(x),x)-\beta_0(x, \theta)}{p_0(x, \theta)}\left[
 p_0(x, \theta)-p_0(x) \right]
, \\
\bar{\beta}_0(x, \theta)=&  (\mathbbm{1}(\Delta(x, \theta)>0)-\mathbbm{1}(\Delta(x)>0)) \frac{q_0(1-m0(x),x)-\beta_0(x, \theta)}{m1(x)}\left[
 1-1/p_0(x, \theta) \right]\\
 &-\frac{1}{m0(x)}
 \int_{q_0(1-m0(x),x)}^{q_0(1-m0(x,\theta),x)} (y-q_0(1-m0(x),x)) f_{0}(y|x))  dy.
\end{align*}}
Similarly, we obtain that $\varepsilon^{-1}\hat{\beta}_0(x, \theta)$ converges to the claim limiting term and $\varepsilon^{-1}\bar{\beta}_0(x, \theta)$ turns out to be negligible.
The results for other cases are analogous and therefore omitted.

\subsubsection{Proof of Theorem \ref{thm_differentiable} - pathwise differentiability part}
\noindent \textbf{Step 0: Preliminaries and Decomposition}\\
Under the assumption $P(\mathcal{X}^0)=0,$ we prove that $\beta_B(s_0, s_1)$ is pathwise differentiable. For brevity, we will only derive this for $\beta_L(1,1),$ $\beta_U(1,1),$ $\beta_L(0,1),$ and $\beta_L(em)$, since proofs of other cases are analogous.

We consider a parametric submodel described in Section \ref{Sec_Notation}.
 We can represent each $\beta \in \{ \beta_L(1,1), \beta_U(1,1), \beta_L(0,1),\beta_L(em)\}$ as $\beta(\theta)=N(\theta)/D(\theta)$, where its numerator and denominator are parameterized as
\begin{align}
N(\theta) &= \int_{\mathcal{X}} \beta(x, \theta) g_{\beta}(x, \theta) f(x, \theta) dx, \quad
D(\theta)=\int_{\mathcal{X}} g_{\beta}(x, \theta) f(x, \theta) dx .
\end{align}
The functions $\beta(x, \theta)$ and $g_{\beta}(x, \theta)$ for each case are found in Proposition \ref{prop_identification1}, \eqref{def_betaB_general} and Example \ref{ex_intensive}.
We note that
{\footnotesize \begin{equation}\label{expansion-ratio1}
\frac{\beta(\theta)-\beta(\theta_0)}{\varepsilon}=\frac{1}{D(\theta)} \left[\frac{N(\theta)-N(\theta_0)}{\varepsilon}-\beta\frac{D(\theta)-D(\theta_0)}{\varepsilon}\right].
\end{equation}}
Since $\lim_{\varepsilon \to 0} D(\theta)=D(\theta_0),$ it suffices to prove the convergence of the term in the brackets. \\[1ex]
\textbf{Step 1: Proof of the Convergence in \eqref{expansion-ratio1}}\\
First, we prove this convergence for the case $\beta=\beta_L(1,1)$ with $g_{\beta}(x)=\min\{s(0,x), s(1,x)\}.$ Denoting $\Delta(x)=s(1,x)-s(0,x)$ and $\Delta(x, \theta)=s(1,x, \theta)-s(0,x, \theta)$, we note that 

{\footnotesize \begin{align*}
\min\{s(0,x, \theta), s(1,x, \theta)\}=&s(\mathbbm{1}( \Delta(x, \theta)<0),x, \theta)\\
=&s(0,x, \theta)+\Delta(x, \theta) \mathbbm{1}( \Delta(x, \theta)<0)\\
=&s(\mathbbm{1}( \Delta(x)<0),x, \theta)+\Delta(x, \theta) (\mathbbm{1}(\Delta(x, \theta)<0)-\mathbbm{1}(\Delta(x)<0)).
\end{align*}}
Recalling the decomposition from the proof of Lemma \ref{lemma-beta(x)-diff}, we also obtain for $\beta(x, \theta)$ that

{\footnotesize \begin{align*}
\beta(x, \theta)=\hat{\beta}(x, \theta)+\bar{\beta}(x, \theta),
\end{align*}}
where
$\hat{\beta}(x, \theta)=$ and $\bar{\beta}(x, \theta)=\bar{\beta}_1(x, \theta)-\bar{\beta}_0(x, \theta)$.
Then, we have that
{\footnotesize \begin{align*}
N(\theta)-N(\theta_0)=&\left(E_{\theta}[ \hat{\beta}(X, \theta) s(\mathbbm{1}(\Delta(X)<0),X, \theta)]
-E[ \beta(X) s(\mathbbm{1}(\Delta(X)<0),X)]\right)\\
&+E_{\theta}[ \beta(X, \theta) \Delta(X, \theta) (\mathbbm{1}(\Delta(X, \theta)<0)-\mathbbm{1}(\Delta(X)<0))] \\
&+E_{\theta}[ \bar{\beta}(X, \theta) s(\mathbbm{1}(\Delta(X)<0),X, \theta)]\\
\equiv & \hat{N}(\theta)+\tilde{N}(\theta)+\bar{N}(\theta),
\end{align*}}Similarly, we decompose $D(\theta)-D(\theta_0)$ into $\hat{D}(\theta)+\tilde{D}(\theta)$ by setting $\beta(x,\theta)$ and $\beta(x)$ to one.
It turns out that $\hat{N}(\theta)$ and $\hat{D}(\theta)$ are main terms, which include the component $D=\mathbbm{1}(\Delta(X)<0)$ that does not depend on $\theta$ and hence is easier to handle. On the other hand,  the terms $\tilde{N}(\theta)-\beta \tilde{D}(\theta)$ and $\bar{N}(\theta)-\beta \bar{D}(\theta)$ include non-smooth components that depend on $\theta$ and turn out to be negligible.
In particular, we show that

{\footnotesize \begin{align*}
\lim_{\varepsilon \to 0}  \frac{\hat{N}(\theta)-\beta \hat{D}(\theta)}{\varepsilon}=& 
\int_{\mathcal{X}} (\beta(x)-\beta) \left(s'(\mathbbm{1}(\Delta(x)<0),x)+s(\mathbbm{1}(\Delta(x)<0),x) t(x, \theta_0) \right) f(x) dx, \\
&+\int_{\mathcal{X}} \frac{\partial \beta (x,\theta_0)}{\partial \theta}
s(\mathbbm{1}(\Delta(x)<0),x)  f(x) dx,\\
\lim_{\varepsilon \to 0}  \frac{\tilde{N}(\theta)-\beta \tilde{D}(\theta)}{\varepsilon}=&0, \quad 
\lim_{\varepsilon \to 0}  \frac{\bar{N}(\theta)-\beta \bar{D}(\theta)}{\varepsilon}=0.
\end{align*}}
The limiting result on $(\hat{N}-\beta \hat{D})(\theta)$ quickly follows by writing
{\footnotesize \begin{align*}
(\hat{N}-\beta \hat{D})(\theta)=&\int_{\mathcal{X}}(\hat{\beta}(x, \theta)-\beta) s(\mathbbm{1}(\Delta(x)<0),x, \theta) f(x, \theta) dx\\
&-\int_{\mathcal{X}}(\beta(x)-\beta) s(\mathbbm{1}(\Delta(x)<0),x) f(x) dx,
\end{align*}}
and exploiting Lemma \ref{lemma-beta(x)-diff} with
{\footnotesize \begin{align*}
&\lim_{\varepsilon \to 0}  \frac{\hat{\beta}(x,\theta)-\beta(x)}{\varepsilon}=\frac{\partial \beta(x, \theta_0)}{\partial \theta}, \quad \lim_{\varepsilon \to 0}  \frac{f(x,\theta)-f(x)}{\varepsilon}=t(x, \theta_0) f(x), \\
&\lim_{\varepsilon \to 0}  \frac{s(1_{\{\Delta(x)<0\}},x, \theta)-s(1_{\{\Delta(x)<0\}},x)}{\varepsilon}=s'(1_{\{\Delta(x)<0\}},x, \theta_0),
\end{align*}}
along with product rule on differentiation and the boundedness of $\beta(x),$ $s(d, x)$ and $f(x).$

Next, we deal with the term decompose $\tilde{N}(\theta)-\beta \tilde{D}(\theta)$. From Section \ref{Sec_Notation}, we first recall that 
$s(d, x, \theta)=s(d,x)+\varepsilon s'(d,x, \theta_0),$ which yields

{\footnotesize \begin{equation}\label{delta-difference}
\Delta(x, \theta)-\Delta(x)=\varepsilon \left[s'(1,x, \theta_0)-s'(0,x, \theta_0) \right],
\end{equation}}
and hence $|\Delta(x, \theta)-\Delta(x)|\leq C_1 |\varepsilon|$ for some constant $C_1>0$ since $s'(d,x, \theta_0)$ is bounded. We also obtain that
{\footnotesize\begin{align}
|\mathbbm{1}(\Delta(x, \theta)<0)-\mathbbm{1}(\Delta(x)<0)|\leq \mathbbm{1}(|\Delta(x)|\leq |\Delta(x, \theta)-\Delta(x)|), \label{eq_deltaclass}
\end{align}}
which exploits that the left-hand side is zero when both terms have same signs. 
Consequently, using the identity $\mathcal{X}^{+}\cup \mathcal{X}^{-}=\{ x \in \mathcal{X}: |\Delta(x)|>0 \},$ we deduce that 

{\footnotesize \begin{align*}
|\tilde{N}(\theta)-\beta \tilde{D}(\theta)| &\leq E_{\theta}[ |\beta(X,\theta)-\beta| |\Delta(X,\theta)| \mathbbm{1}(|\Delta(X)|\leq |\Delta(X, \theta)-\Delta(X)|) \mathbbm{1}( |\Delta(X)|>0 )] \\
& \leq 2 C_1 C_2 |\varepsilon | E_{\theta}[ \mathbbm{1}(|\Delta(X)|\leq 2 C_1  |\varepsilon|) \mathbbm{1}( |\Delta(X)|>0 )] \\
& \leq 2 C_1 C_2 C_3 |\varepsilon| E[ \mathbbm{1}(0<|\Delta(X)|\leq 2 C_1  |\varepsilon|) ]=o(\varepsilon),
\end{align*}}where we used $|\beta(x, \theta)-\beta|\leq C_2$ and  $|\Delta(x, \theta)| \leq |\Delta(x)|+|\Delta(x, \theta)-\Delta(x)|\leq 2 C_1 |\varepsilon|$ under the  indicator function $\mathbbm{1}(|\Delta(X)|\leq |\Delta(X, \theta)-\Delta(X)|)$ and $|1+\varepsilon t(x)| \leq C_3,$ and $\lim_{\varepsilon \to 0 }P(0<|\Delta(X)|\leq 2 C_1|\varepsilon|)=0.$

We can deal with the term $\bar{N}(\theta)-\beta \bar{D}(\theta)$ similarly. In particular, the terms involving the difference of two indicator functions are treated exactly as above. The other terms involving conditional quantiles are dealt with exploiting Assumption \ref{ass_density} and turn out to be of order $O(\varepsilon^2).$

As a result, we have shown the convergence in \eqref{expansion-ratio1} for the cases $\beta=\beta_L(1,1)$ and the obtained limiting integral can be written as

{\footnotesize \begin{align*}
\frac{\partial \beta(\theta_0)}{\partial \theta}&=
D(\theta_0)^{-1}\int_{\mathcal{X}^+} \left[\left(\beta(x)-\beta) (s'(0,x, \theta_0)+s(0,x)t(x, \theta_0) \right)+\frac{\partial \beta (x,\theta_0)}{\partial \theta} s(0,x) \right]  f(x) dx \\
&+
D(\theta_0)^{-1} \int_{\mathcal{X}^-} \left[(\beta(x)-\beta)\left(s'(1,x, \theta_0)+
s(1,x) t(x, \theta_0) \right)+\frac{\partial \beta (x,\theta_0)}{\partial \theta} s(1,x)\right] f(x) dx.
\end{align*}}

The case $\beta=\beta_U(1,1)$ has the same $g_{\beta}(x, \theta)=\min\{s(0,x, \theta),s(1,x, \theta)\}$ but based on $\beta_U(1,1,x).$ Hence, its proof is analogous.

For the cases $\beta \in \{ \beta_L(0,1), \beta_L(em)\},$ we have $g_{\beta}(x, \theta)=\max\{s(1,x, \theta)-s(0,x, \theta),0\}$ and  
$g_{\beta}(x, \theta)=|s(1,x, \theta)-s(0,x, \theta)|,$ respectively. We observe that

{\footnotesize \begin{align*}
\max\{s(1,x, \theta)-s(0,x, \theta),0\}&=s(1,x, \theta)-\min\{s(0,x, \theta), s(1,x, \theta)\} , \\
|s(1,x, \theta)-s(0,x, \theta)|&=s(1,x, \theta)+s(0,x, \theta)-2\min\{s(0,x, \theta), s(1,x, \theta)\},
\end{align*}}and hence the convergence in \eqref{expansion-ratio1} for these two cases is handled analogously to the earlier $\min\{s(0,x, \theta), s(1,x, \theta)\}$ function case. \\[1ex]
\textbf{Step 2: Calculating the Tangent Space and the Score}\\
The density of the observed data $(SY, S, D, X)$ is given by

{\footnotesize \begin{align*}
L(y, s, d, x)=&\left[ f_{1}(y|x)^d f_{0}(y|x)^{1-d}   s(1,x)^d s(0,x)^{1-d} \right]^s  \\
&\times \left[ (1-s(1,x))^{d} (1-s(0,x))^{1-d} \right]^{1-s} m(x)^d (1-m(x))^{1-d} f(x)   
\end{align*}}
Now, we consider a regular parametric submodel indexed by $\theta$ with density

{\footnotesize \begin{align*}
L(y, s, d, x|\theta)=&\left[ f_{1}(y|x, \theta)^d  f_{0}(y|x, \theta)^{1-d}  s(1,x,\theta)^d s(0,x|\theta)^{1-d} \right]^s  \\
&\times \left[ (1- s(1,x, \theta))^{d} (1-s(0,x, \theta))^{1-d} \right]^{1-s} m(x, \theta)^d (1-m(x, \theta))^{1-d} f(x, \theta),
\end{align*}}
which equals $L(y,s,d,x)$ when $\theta=\theta_0.$
If $m(x, \theta)=m(x)$ is known (does not depend on $\theta$),
its score is given by

{\footnotesize \begin{align*}
Q_1(y, s, d, x|\theta) =& s d  \tau_{1}(y|x, \theta)+s (1-d) \tau_{0}(y|x, \theta)+ t(x,\theta)\\
&+d \frac{(s-s(1,x,\theta)) s'(1,x,\theta)) }{s(1,x,\theta) (1-s(1,x, \theta)}+ (1-d) \frac{(s-s(0,x,\theta)) s'(0,x,\theta)) }{s(0,x,\theta) (1-s(0,x, \theta)}.
\end{align*}}
In view of this, the tangent space of this model is
{\footnotesize \begin{align*}
\mathcal{S}_1=\{  s d \tau_{1}(y,x)+ s (1-d) \tau_{ 0}(y,x)+\tau_x(x)+a(x) d (s-s(1,x)) + b(x) (1-d) (s-s(0,x)) \},
\end{align*}}where $a(x)$, $b(x)$ are any bounded functions, and $\tau_{d}(x),$ $d=0,1,$ $t(x)$ are bounded functions that satisfy
{\footnotesize \begin{align*}\int \tau_{d}(y,x) f_{d}(y|x) dy=0, \forall x, \mbox { and } \int t(x) f(x) dx=0.
\end{align*}}

If $m(x, \theta)$ is unknown (may depend on $\theta$),
its score is given by

{\footnotesize \begin{align*}
Q_2(y, s, d, x|\theta) =Q_1(y, s, d, x|\theta)+\frac{d-m(x, \theta) }{m (x, \theta) (1-m(x, \theta))} \frac{\partial m(x, \theta)}{\partial \theta}.
\end{align*}}
Hence, the tangent space of this model is

{\footnotesize \begin{align*}
\mathcal{S}_2=&\{  s d \tau_{1}(y,x)+ s (1-d) \tau_{ 0}(y,x)+\tau_x(x)+a(x) d (s-s(1,x)) \\
&~~+ b(x) (1-d) (s-s(0,x))+c(x) (d-m(x)) \},
\end{align*}}where $c(x)$ is any bounded function, and other functions were introduced earlier.\\[1ex]
\textbf{Step 3: Detailed Calculation of the Pathwise Derivative}\\
For each $\beta \in \{ \beta_L(1,1), \beta_U(1,1), \beta_L(0,1),\beta_L(em)\},$ our remaining task is to come up with the efficient influence function $\psi_{\beta}(W)$ such that for $i\in\{1, 2\}$

{\footnotesize \begin{equation}\label{pwderivative-score-equal}
\frac{\partial \beta(\theta_0)}{\partial \theta}=E[\psi_{\beta}(W) Q_i(W|\theta_0)].
\end{equation}}
For both sides of this equality, we proceed on a case-by-case basis. In this step, based on Lemma \ref{lemma-beta(x)-diff} and Step 1, we decompose the left-hand side of this as

{\footnotesize \begin{equation}\label{pwderivation-5T-decomp}
\frac{\partial \beta(\theta_0)}{\partial \theta}=(E[g_{\beta}(X)])^{-1}(T_1+T_2+T_3+T_4+T_5).
\end{equation}}
First, we let $\beta=\beta_L(1,1),$ where
$g_{\beta}(x)=\min\{s(0,x), s(1,x)\}.$
Then, we have that

{\footnotesize \begin{align*}
T_1 =&\int_{\mathcal{X} } \int_{\underline{y}}^{q_1(\min\{p_0(x),1\},x)} y \tau_{1}(y|x, \theta_0) f_{1}(y|x)  s(1,x) f(x) dy dx,\\
T_2 =&-\int_{\mathcal{X}} \int^{\bar y}_{q_0(\max\{1-1/p_0(x),0 \},x)} y \tau_{0}(y|x, \theta_0) f_{0}(y|x)  s(0,x) f(x) dy dx, \\
T_3 =&\int_{\mathcal{X}} (\beta(x)-\beta) \left(s'(1_{\{\Delta(x)<0\}},x)+s(1_{\{\Delta(x)<0\}},x) t(x, \theta_0) \right) f(x) dx,\\
T_4 =&-\int_{\mathcal{X}^+} \int_{\underline{y}}^{q_1(p_0(x),x)} q_1(p_0(x),x) \tau_{1}(y|x, \theta_0) f_{1}(y|x)  s(1,x) f(x) dy dx,\\
&+\int_{\mathcal{X}^-} \int^{\bar y}_{q_0(1-1/p_0(x),x)} q_0(1-1/p_0(x),x) \tau_{0}(y|x, \theta_0) f_{0}(y|x)  s(0,x) f(x) dy dx, \\
T_5 =&\int_{\mathcal{X}^+} [q_1(p_0(x),x)- \beta_1(x)]  \frac{\partial p_0(x, \theta_0)}{\partial \theta}  s(1,x) f(x) dx \\
&+\int_{\mathcal{X}^-} \frac{[q_0(1-1/p_0(x),x)- \beta_0(x)]}{p_0(x)} \frac{\partial p_0(x, \theta_0)}{\partial \theta}  s(1,x) f(x) dx.
\end{align*}}
Second, we let $\beta=\beta_U(1,1),$ where
$g_{\beta}(x)=\min\{s(0,x), s(1,x)\}.$
Then, we have that

{\footnotesize \begin{align*}
T_1 =&\int_{\mathcal{X} } \int_{q_1(\max\{1-p_0(x),0\},x)}^{\bar{y}} y \tau_{1}(y|x, \theta_0) f_{1}(y|x)  s(1,x) f(x) dy dx,\\
T_2 =&-\int_{\mathcal{X}}\int_{\underline{y}}^{q_0(\min\{1/p_0(x),1\},x)} y \tau_{0}(y|x, \theta_0) f_{0}(y|x)  s(0,x) f(x) dy dx, \\
T_3 =&\int_{\mathcal{X}} (\beta(x)-\beta) \left(s'(1_{\{\Delta(x)<0\}},x)+s(1_{\{\Delta(x)<0\}},x) t(x, \theta_0) \right) f(x) dx,\\
T_4 =&-\int_{\mathcal{X}^+}  
\int_{q_1(1-p_0(x),x)}^{\bar{y}} q_1(1-p_0(x),x) \tau_{1}(y|x, \theta_0) f_{1}(y|x)  s(1,x) f(x) dy dx,\\
&+\int_{\mathcal{X}^-} \int_{\underline{y}}^{q_0(1/p_0(x),x)} q_0(1/p_0(x),x) \tau_{0}(y|x, \theta_0) f_{0}(y|x)  s(0,x) f(x) dy dx, \\
T_5 =&\int_{\mathcal{X}^+} [q_1(1-p_0(x),x)- \beta_1(x)]\frac{\partial p_0(x, \theta_0)}{\partial \theta}  s(1,x) f(x) dx \\
&+\int_{\mathcal{X}^-} \frac{[q_0(1/p_0(x),x)- \beta_0(x)]}{p_0(x)} \frac{\partial p_0(x, \theta_0)}{\partial \theta}  s(1,x) f(x) dx.
\end{align*}}
Third, we let $\beta=\beta_L(0,1),$ where
$g_{\beta}(x)=\max\{s(1,x)-s(0,x),0\}.$ Then, we obtain that

{\footnotesize \begin{align*}
T_1 =&\int_{\mathcal{X}^{+} }  \int_{\underline{y}_1}^{q_1(1-p_0(x),x)} y \tau_{1}(y|x, \theta_0) f_{1}(y|x) s(1,x) f(x) dy dx,\\
T_2 =&-\int_{\mathcal{X}^{-}} \int^{\bar y_0(x)}_{q_0(1/p_0(x),x)} y \tau_{0}(y|x, \theta_0) f_{0}(y|x)  s(0,x) f(x) dy dx, \\
T_3 =&\int_{\mathcal{X}^{+}} (\beta(x)-\beta) \left(s'(1,x, \theta_0)-s'(0,x, \theta_0)+(s(1,x)-s(0,x)) t(x, \theta_0) \right) f(x) dx,\\
T_4 =&-\int_{\mathcal{X}^+} \int_{\underline{y}_1}^{q_1(1-p_0(x),x)} q_1(1-p_0(x),x) \tau_{1}(y|x, \theta_0) f_{1}(y|x)  s(1,x) f(x) dy dx,\\
&+\int_{\mathcal{X}^-} \int^{\bar y}_{q_0(1/p_0(x),x)} q_0(1/p_0(x),x) \tau_{0}(y|x, \theta_0) f_{0}(y|x)  s(0,x) f(x) dy dx, \\
T_5 =&-\int_{\mathcal{X}^+} (q_1(1-p_0(x),x)- \beta_1(x)) \frac{\partial p_0(x, \theta_0)}{\partial \theta}  s(1,x) f(x) dx \\
&+\int_{\mathcal{X}^-} \frac{[q_0(1/p_0(x),x)- \beta_0(x)]}{p_0(x)} \frac{\partial p_0(x, \theta_0)}{\partial \theta}  s(1,x) f(x) dx.
\end{align*}}

Finally, we let $\beta=\beta_L(em),$ where
$g_{\beta}(x)=|s(1,x)-s(0,x)|.$ Then, we obtain that the same $T_1, T_2, T_4$ and $T_5$ terms as in the previous case due to $\beta_L(x,em)=\beta_L(x, 0,1)$. The only new term here is
{\footnotesize \begin{align*}
T_3 =&\int_{\mathcal{X}} (\beta(x)-\beta) \left(s'(1,x, \theta_0)-s'(0,x, \theta_0)+(s(1,x)-s(0,x)) t(x, \theta_0) \right) f(x) dx,\\
&+\int_{\mathcal{X}^-} (\beta(x)-\beta) \left(s'(0,x, \theta_0)-s'(1,x, \theta_0)+(s(0,x)-s(1,x)) t(x, \theta_0) \right) f(x) dx.
\end{align*}} \\[1ex]
\textbf{Step 4: Proposing the Efficient Influence Function}\\
In this step, for each $\beta$, we propose $\psi_{\beta}(W)$ on the right-hand side of \eqref{pwderivative-score-equal} in the form:

{\footnotesize \begin{equation}
\psi_{\beta}(W)=(E[g_{\beta}(X)])^{-1}(M_1+M_2+M_3+M_4+M_5).
\end{equation}}
We obtained influence functions by following the heuristic approach of \cite{hines2022demystifying}. However, not to make the proof even longer, we refrain from showing these details and instead simply write these in the style of \cite{hahn1998role}.

First, let $\beta=\beta_L(1,1),$ where
$g_{\beta}(x)=\min\{s(0,x), s(1,x)\}.$
Then, we have that

{\scriptsize \begin{align*}
M_1 =& \frac{SD}{m(X)} \left[Y 1\{Y \le q_1(\min\{p_0(X),1\},X)\}-\min\{p_0(X),1\}\beta_{1}(X)\right], \\
M_2=&-\frac{S(1-D)}{1-m(X)}
\left[Y 1\{Y \geq q_0(1-\min\{1/p_0(X),1\},X)\}-\min\{1/p_0(X),1\} \beta_{0}(X)\right]   , \\
M_3 =& \mathbbm{1}(X \in \mathcal{X}^{+}) (\beta(X)-\beta) \left[s(0,X)+\frac{1-D}{1-m(X)}(S-s(0,X))  \right]  \\
&+\mathbbm{1}(X \in \mathcal{X}^{-}) (\beta(X)-\beta) \left[s(1,X)+
\frac{D}{m(X)}(S-s(1,X))  \right], \\
M_4=&- \mathbbm{1}(X \in \mathcal{X}^{+}) \frac{SD}{m(X)}q_1(p_0(X),X)
\left[\mathbbm{1}(Y \le q_1(p_0(X),X))-p_0(X) \right]\\
&+\mathbbm{1}(X \in \mathcal{X}^{-})\frac{S(1-D)}{1-m(X)} q_0(1-p_0(X)^{-1},X)
\left[\mathbbm{1}(Y \geq q_0(1-p_0(X)^{-1},X))-p_0(X)^{-1}\right]  , \\
M_5=&\mathbbm{1}(X \in \mathcal{X}^{+}) \left[q_1(p_0(X),X)- \beta_{1}(X) \right]
\left[\frac{1-D}{1-m(X)}(S-s(0,X))- p_0(X)
\frac{D}{m(X)}(S-s(1, X)) \right] \\  
&+\mathbbm{1}(X \in \mathcal{X}^{-}) \left[q_0(1-p_0(X)^{-1},X)- \beta_{0}(X) \right]
\left[p_0(X)^{-1}\frac{1-D}{1-m(X)}(S-s(0,X))-\frac{D}{m(X)}(S-s(1, X)) \right].
\end{align*}}

Second, let $\beta=\beta_U(1,1),$ where
$g_{\beta}(x)=\min\{s(0,x), s(1,x)\}.$
Then, we have that

{\scriptsize \begin{align*}
M_1 =& \frac{SD}{m(X)} \left[Y \mathbbm{1}(Y \geq  q_1(1-\min\{p_0(X),1\},X))-\min\{p_0(X),1\}\beta_{1}(X)\right], \\
M_2=&-\frac{S(1-D)}{1-m(X)}
\left[Y \mathbbm{1}(Y \leq q_0(\min\{1/p_0(X),1\},X))-\min\{1/p_0(X),1\}\beta_{0}(X)\right]   , \\
M_3 =& \mathbbm{1}( X \in \mathcal{X}^{+}) (\beta(X)-\beta) \left[s(0,X)+\frac{1-D}{1-m(X)}(S-s(0,X))  \right]  \\
&+\mathbbm{1}(X \in \mathcal{X}^{-}) (\beta(X)-\beta) \left[s(1,X)+
\frac{D}{m(X)}(S-s(1,X))  \right], \\
M_4=&- \mathbbm{1}(X \in \mathcal{X}^{+}) \frac{SD}{m(X)}q_1(1-p_0(X),X)
\left[\mathbbm{1}(Y \geq q_1(1-p_0(X),X))-p_0(X) \right]\\
&+\mathbbm{1}(X \in \mathcal{X}^{-})\frac{S(1-D)}{1-m(X)} q_0(1/p_0(X),X)
\left[\mathbbm{1}(Y \leq q_0(p_0(X)^{-1},X))-p_0(X)^{-1}\right]  , \\
M_5=&\mathbbm{1}(X \in \mathcal{X}^{+}) \left[q_1(1-p_0(X),X)- \beta_{1}(X) \right]
\left[\frac{1-D}{1-m(X)}(S-s(0,X))- p_0(X)
\frac{D}{m(X)}(S-s(1, X)) \right] \\  
&+\mathbbm{1}(X \in \mathcal{X}^{-}) \left[q_0(p_0(X)^{-1},X)- \beta_{0}(X) \right]
\left[p_0(X)^{-1}\frac{1-D}{1-m(X)}(S-s(0,X))-\frac{D}{m(X)}(S-s(1, X)) \right].
\end{align*}}
Third, let $\beta=\beta_L(0,1),$ where
$g_{\beta}(x)=\max\{s(1,x)-s(0,x),0\}.$ Then, we obtain that

{\scriptsize \begin{align*}
M_1 =& \mathbbm{1}(X \in \mathcal{X}^{+})\frac{SD}{m(X)} 
\left[Y \mathbbm{1}(Y \le q_1(1-p_0(X),X))-(1-p_0(X))\beta_{1}(X)\right], \\
M_2=&-\mathbbm{1}(X \in \mathcal{X}^{-}) \frac{S(1-D)}{1-m(X)}
\left[Y \mathbbm{1}(Y \geq q_0(1/p_0(X),X))-(1-1/p_0(X))\beta_{0}(X)\right]   , \\
M_3 =& \mathbbm{1}(X \in \mathcal{X}^{+}) (\beta(X)-\beta) \left[s(1,X)-s(0,X)+\frac{D}{m(X)}(S-s(1,X))-\frac{1-D}{1-m(X)}(S-s(0,X))  \right]  \\
&+\mathbbm{1}(X \in \mathcal{X}^{-}) (\beta(X)-\beta) \left[s(0,X)-s(1,X)+
\frac{1-D}{1-m(X)}(S-s(0,X))-\frac{D}{m(X)}(S-s(1,X))  \right], \\
M_4=&- \mathbbm{1}( X \in \mathcal{X}^{+}) \frac{SD}{m(X)}q_1(1-p_0(X),X)
\left[\mathbbm{1}(Y \le q_1(1-p_0(X),X))-(1-p_0(X)) \right]\\
&+\mathbbm{1}(X \in \mathcal{X}^{-})\frac{S(1-D)}{1-m(X)} q_0(p_0(X)^{-1},X)
\left[\mathbbm{1}(Y \geq q_0(p_0(X)^{-1},X))-(1-p_0(X)^{-1})\right]  , \\
M_5=&-\mathbbm{1}( X \in \mathcal{X}^{+}) \left[q_1(1-p_0(X),X)- \beta_{1}(X) \right]
\left[\frac{1-D}{1-m(X)}(S-s(0,X))- p_0(X)
\frac{D}{m(X)}(S-s(1, X)) \right] \\  
&+\mathbbm{1}(X \in \mathcal{X}^{-}) \left[q_0(p_0(X)^{-1},X)- \beta_{0}(X) \right]
\left[p_0(X)^{-1}\frac{1-D}{1-m(X)}(S-s(0,X))-\frac{D}{m(X)}(S-s(1, X)) \right]
\end{align*}}

Finally, let $\beta=\beta_L(em),$ where
$g_{\beta}(x)=|s(1,x)-s(0,x)|.$ Again, we obtain the same $M_1, M_2, M_4$ and $M_5$ terms as in the previous case due to $\beta_L(x,em)=\beta_L(x, 0,1)$. The only new term is

{\footnotesize \begin{align*}
M_3 =& \mathbbm{1}(X \in \mathcal{X}^{+}) (\beta(X)-\beta) \left[s(1,X)-s(0,X)+\frac{D}{m(X)}(S-s(1,X))-\frac{1-D}{1-m(X)}(S-s(0,X))  \right].
\end{align*}}

\textbf{Step 5: Checking the Equality in \eqref{pwderivative-score-equal}}\\
We note that $M \in \mathcal{S}_1$ and $M \in \mathcal{S}_2.$ In addition, for each $i=1, \ldots, 5,$, we have that 

{\footnotesize \begin{align*}
T_i&=E[M_i Q_1(Y, S, D, X|\theta_0)], \\
T_i&=E[M_i Q_2(Y, S, D, X|\theta_0)], 
\end{align*}}
which shows that $\beta$ is pathwise differentiable irrespective of $m(x,\theta)$ is known or unknown.

\subsubsection{Pathwise differentiability in proof of Theorem \ref{thm_differentiable}}

In this section, we show that 
$\beta \in \{ \beta_L(1,1), \beta_L(em)\}$ is not pathwise differentiable if 
$P(\mathcal{X}^0)>0$ holds. Other cases are similar and therefore skipped. Our counterexample proposes $\tau_d(y|x, \theta_0)$, $s'(d, x, \theta_0)$, $t(x, \theta_0)$ (see Section \ref{Sec_Notation}) such that 

{\footnotesize \begin{equation}
\lim_{\varepsilon \downarrow 0} \frac{\beta(\theta)-\beta(\theta_0)}{\varepsilon} \neq \lim_{\varepsilon \uparrow 0} \frac{\beta(\theta)-\beta(\theta_0)}{\varepsilon}.
\end{equation}}
We borrow the proof idea from \cite{luedtke2016statistical}. However, our case is more delicate because the divergence of the derivative of denominator $D(\theta)$ and numerator $N(\theta)$ does not necessarily imply the divergence of the ratio $\beta.$ Recalling \eqref{expansion-ratio1} and since $\lim_{\varepsilon \to 0} D(\theta)=D(\theta_0),$ it suffices to show that $L^1 \neq L^2,$ where

{\footnotesize \begin{align*}
L^1&\equiv\lim_{\varepsilon \downarrow 0} \frac{N(\theta)-N(\theta_0)-\beta(D(\theta)-D(\theta_0))}{\varepsilon}, \\ 
L^2 & \equiv\lim_{\varepsilon \uparrow 0} \frac{N(\theta)-N(\theta_0)-\beta(D(\theta)-D(\theta_0))}{\varepsilon}.
\end{align*}}
First, we let $\beta=\beta_L(1,1).$ In our decomposition introduced after \eqref{expansion-ratio1}, we now include extra $\mathcal{X}^0$ terms in 

{\footnotesize \begin{align*}
\check{N}(\theta)-\beta\check{D} (\theta)=&E_{\theta}[ 1_{\{ X \in \mathcal{X}^0 \}}(\beta(X, \theta)-\beta) E_{\theta}[S|D=1_{\{ \Delta(X)<0  \}}, X]] \\
&-E[1_{\{ X \in \mathcal{X}^0 \}}(\beta(X)-\beta) E[S|D=1_{\{ \Delta(X)<0  \}}, X]],\\
\ddot{N}(\theta)-\beta\ddot{D} (\theta)=&E_{\theta}[ 1_{\{ X \in \mathcal{X}^0 \}} (\beta(X, \theta)-\beta) \Delta(X, \theta) (1_{\{ \Delta(X, \theta)<0  \}}-1_{\{ \Delta(X)<0  \}})],
\end{align*}}
while retaining $\mathcal{X}^{+} \cup \mathcal{X}^{-}$ terms in
$\hat{N}(\theta)-\beta \hat{D}(\theta)$, $\tilde{N}(\theta)-\beta \tilde{D}(\theta)$ and 
$\bar{N}(\theta)-\beta \bar{D}(\theta).$

We let $\tau_d(y|x, \theta_0)=0,$ $t(x, \theta_0)=0,$ $s'(0,x, \theta_0)=0$ and $s'(1,x, \theta_0)=1$ for all $d, y$ and $x \in \mathcal{X}.$ Then, our results obtained in Step 1 of Theorem \ref{thm_differentiable}  imply

{\footnotesize \begin{align*}
\lim_{\varepsilon \to 0} \frac{\hat{N}(\theta)-\beta \hat{D}(\theta)}{\varepsilon}=&\int_{\mathcal{X}^-} (\beta(x)-\beta) f(x) dx+\int_{\mathcal{X}^+ \cup \mathcal{X}^-} \frac{\partial \beta (x,\theta_0)}{\partial \theta} s(1_{\{ \Delta(x)<0 \}},x)f(x) dx,\\
\lim_{\varepsilon \to 0} \frac{\tilde{N}(\theta)-\beta \tilde{D}(\theta)}{\varepsilon}=&0=\lim_{\varepsilon \to 0} \frac{\bar{N}(\theta)-\beta \bar{D}(\theta)}{\varepsilon},
\end{align*}}
where some simplifications are due to $t(x, \theta_0)=s'(0,x, \theta_0)=0$ and $s'(1,x, \theta_0)=1.$

Before we proceed to the other two terms, for $x \in \mathcal{X}^0$, we will show that $\beta(x, \theta)$ is not pathwise differentiable. In particular, we will prove that 

{\footnotesize \begin{align*}
&\lim_{\varepsilon \downarrow 0}\frac{\beta(x, \theta)-\beta(x, \theta_0)}{\varepsilon}=L(x)+ (\beta_1(x)-\overline{y}_1) \frac{s'(1,x, \theta_0)- s'(0, x, \theta_0)}{s(1,x)}, \\
&\lim_{\varepsilon \uparrow 0}\frac{\beta(x, \theta)-\beta(x, \theta_0)}{\varepsilon}=L(x)+(\beta_0(x)-\underline{y}_0) \frac{s'(1,x, \theta_0)- s'(0, x, \theta_0)}{s(1,x)},
\end{align*}}
where

{\footnotesize \begin{align*}
  L(x)= \int y \tau_1(y|x, \theta_0) f_1(y|x) dy-\int y \tau_0(y|x, \theta_0) f_0(y|x) dy.
   \end{align*}}

We have $p_0(x)=1$ for $x \in \mathcal{X}^0$. 
Since  $s'(1, x, \theta)=1>0=s'(0,x, \theta_0),$ we have that $p_0(x, \theta)<1$ for $\varepsilon>0,$ and $p_0(x, \theta)>1$ for $\varepsilon<0.$
Then, it suffices to show that

{\footnotesize \begin{align*}
&\lim_{\varepsilon \downarrow 0}\frac{\beta_{1,1}(x, p_0(x, \theta),\theta)-\beta_{1,1}(x, 1, \theta)}{\varepsilon}=(\beta_1(x)-\overline{y}_1) \frac{s'(1,x, \theta_0)- s'(0, x, \theta_0)}{s(1,x)}, \\
&\lim_{\varepsilon \uparrow 0}\frac{\beta_{0,0}(x, 1-1/p_0(x, \theta),\theta)-\beta_{0,0}(x, 0, \theta)}{\varepsilon}=-(\beta_0(x)-\underline{y}_0) \frac{s'(1,x, \theta_0)- s'(0, x, \theta_0)}{s(1,x)}.
\end{align*}}
To prove the first convergence, we first write this as

{\footnotesize \begin{align*}
\frac{\beta_{1,1}(x, p_0(x, \theta),\theta)-\beta_{1,1}(x, 1, \theta)}{\varepsilon}=&\frac{1-p_0(x, \theta)}{p_0(x, \theta)} \int_{\underline{y}_1}^{q_1(p_0(x, \theta),x)} y f_{1}(y|x, \theta)  dy\\
&-\int_{q_1(p_0(x, \theta),x)}^{\overline{y}_1} y f_{1}(y|x, \theta)  dy,
\end{align*}}and hence the result follows from $\lim_{\varepsilon \downarrow 0} \varepsilon^{-1}(1-p_0(x, \theta))=(s'(1,x, \theta_0)- s'(0, x, \theta_0))/s(1,x),$
and due to $\varepsilon^{-1}$ times the second integral converging to $\overline{y}_1 (s'(1,x, \theta_0)- s'(0, x, \theta_0))/s(1,x).$

The second convergence follows similarly by using the decomposition

{\footnotesize \begin{align*}
\beta_{0,0}(x, 1-1/p_0(x, \theta),\theta)-\beta_{0,0}(x, 0, \theta)=&(p_0(x, \theta)-1) \int^{\overline{y}_0}_{q_0(1-p_0^{-1}(x, \theta),x)} y f_{0}(y|x, \theta)  dy\\
&-\int_{\underline{y}_0}^{q_0(1-p_0^{-1}(x, \theta),x)} y f_{0}(y|x, \theta)  dy.
\end{align*}}

 Having established this result, for $x \in \mathcal{X}^0$, we deduce that 
 
{\footnotesize \begin{align*}
\lim_{\varepsilon \downarrow 0} \frac{\check{N}(\theta)-\beta \check{D}(\theta)}{\varepsilon}=&\int_{\mathcal{X}^0}  (\beta_1(x)-\overline{y}_1) f(x) dx, \\
\lim_{\varepsilon \uparrow 0} \frac{\check{N}(\theta)-\beta \check{D}(\theta)}{\varepsilon}=&\int_{\mathcal{X}^0} (\beta_0(x)-\underline{y}_0) f(x) dx.
\end{align*}}

Finally, we deal with the term $\ddot{N}(\theta)-\beta\ddot{D} (\theta).$ For $x \in \mathcal{X}^0$ we note that 
$\Delta(x)=0$ and $\Delta(x, \theta)>0$ if $\varepsilon>0.$
This quickly implies that

{\footnotesize \begin{align*}
\lim_{\varepsilon \downarrow 0} \frac{\ddot{N}(\theta)-\beta \ddot{D}(\theta)}{\varepsilon}=0.
\end{align*}}
On the other hand, we have $\Delta(x, \theta)<0$ if $\varepsilon<0$, which leads to

{\footnotesize \begin{align*}
\lim_{\varepsilon \uparrow 0} \frac{\ddot{N}(\theta)-\beta \ddot{D}(\theta)}{\varepsilon}=\int_{\mathcal{X}^0} (\beta(x)-\beta) f(x) dx.
\end{align*}}
where we used $\beta(x, \theta) \to \beta(x)$ and $f(x, \theta)=f(x)$ and $\Delta(x)=0$ for $x \in \mathcal{X}^0.$

Combining all these limits and denoting  we obtain that

{\footnotesize \begin{align*}
 L^1-L^2&=\int_{\mathcal{X}^0} (\beta-(\overline{y}_1-\underline{y}_0) f(x) dx=P(\mathcal{X}^{0})(\beta-(\overline{y}_1-\underline{y}_0))<0,
\end{align*}} 
where we used that $P(\mathcal{X}^{0})>0$ and $\beta<(\overline{y}_1-\underline{y}_0).$ The last inequality follows from $f_d(y|x)$ being continuous density with support on $[\underline{y}_d, \overline{y}_d]$ and hence $\beta_{1,1}(x, u)<\overline{y}_1$ and $\beta_{0,0}(x, u)>\underline{y}_0.$

Next, we consider the extensive margin case by letting $\beta=\beta_L(em)$. We again set $\tau_d(y|x, \theta_0)=0,$ $t(x, \theta_0)=0,$ 
$s'(0,x, \theta_0)=0$ and $s'(1,x, \theta_0)=1.$

We will only look at the difference between the terms evaluated on the set $\mathcal{X}^0$ since the terms converge on $\mathcal{X}^{+} \cup \mathcal{X}^-$ due to the results obtained in Step 1 of Theorem \ref{thm_differentiable}. We note that $|s(1,x)-s(0,x)|=0$ for $x \in \mathcal{X}^0,$ which implies that

{\footnotesize \begin{align*}
N(\theta_0)-\beta D(\theta_0)=\int_{\mathcal{X}^0} (\beta(x)-\beta) |s(1,x)-s(0,x)| f(x) dx=0.
\end{align*}}
On the other hand, using $f(x, \theta)=f(x)$ due to $t(x, \theta_0)=0,$ we have that

{\footnotesize \begin{align*}
N(\theta)-\beta D(\theta) &= \int_{\mathcal{X}^0} (\beta(x, \theta)-\beta) |s(1,x, \theta)-s(0,x,\theta)| f(x) dx.
\end{align*}}
Since $|s(1,x, \theta)-s(0,x,\theta)|=|\varepsilon|$ for $x \in \mathcal{X}^0,$ we obtain that

{\footnotesize \begin{align*}
\varepsilon^{-1}\left[N(\theta)-\beta D(\theta)-(N(\theta_0)-\beta D(\theta_0)) \right]&=\mbox{sign}(\varepsilon)\int_{\mathcal{X}^0} (\beta(x, \theta)-\beta) f(x) dx.
\end{align*}}
Due to $\lim_{\varepsilon \to 0} \beta(x, \theta)=\beta(x)=\underline{y}_1-\overline{y}_0,$ we obtain that

{\footnotesize \begin{align*}
&\lim_{\varepsilon \downarrow 0}\varepsilon^{-1}\left[N(\theta)-\beta D(\theta)-(N(\theta_0)-\beta D(\theta_0)) \right]=\int_{\mathcal{X}^0} (\beta(x)-\beta) f(x) dx<0 \\
&\lim_{\varepsilon \uparrow 0} \varepsilon^{-1}\left[N(\theta)-\beta D(\theta)-(N(\theta_0)-\beta D(\theta_0)) \right]=-\int_{\mathcal{X}^0} (\beta(x)-\beta) f(x) dx >0,
\end{align*}}where we used that $\beta(x)=\underline{y}_1-\overline{y}_0$ for $x \in \mathcal{X}^0$ and $\beta>\underline{y}_1-\overline{y}_0.$ The last inequality follows from $f_d(y|x)$ being continuous density with support on $[\underline{y}_d, \overline{y}_d]$ and hence $\beta_{1,1}(x, u)>\underline{y}_1$ for $u>0$ and $\beta_{0,0}(x, u)<\overline{y}_0$ for $u<1$ and $P(\mathcal{X}^{+} \cup \mathcal{X}^-)>0.$

\subsection{Proof of Corollary \ref{corr_var_bounds}}
The efficiency lower bound is equal to $E[M^2 ],$ where the contribution terms to the total variance are
{\scriptsize \begin{align*}
E[M_1^2]+&E[M_2^2]= E\left[\frac{s(1,X) \sigma_1^2 (X) }{m(X)}+\frac{s(0,X) \sigma_0^2 (X) }{1-m(X)} \right], \\
E[M_3^2] =& E\left[1_{\mathcal{X}^{+}}(X)(\beta_L(X)-\beta)^2 \frac{s(0,X)(1-s(0,X) m(X))}{1-m(X)} \right] \\
&+E\left[1_{\mathcal{X}^{-}}(X)(\beta_L(X)-\beta)^2 \frac{s(1,X)(1-s(1,X)+ s(1,X) m(X))}{m(X)} \right] \\
E[M_4^2] =&E\left[ 1_{\mathcal{X}^{+}} \frac{s(1,X) q_1(p_0(X),X)^2 p_0(X) (1-p_0(X))}{m(X)} \right] \\
&+E\left[ 1_{\mathcal{X}^{-}}  \frac{s(0,X) q_0(1-1/p_0(X),X)^2 p_0(X)^{-1} (1-p_0(X)^{-1})}{1-m(X)} \right]\\
E[M_5^2] =&E\left[1_{\mathcal{X}^{+}} (q_1(p_0(X),X)-\beta_1(X))^2\left(\frac{ s(0,X)(1-s(0,X))}{1-m(X)}+
  \frac{p_0(X)^2 s(1,X)(1-s(1,X))}{m(X)} \right) \right]  \\
  &+E\left[1_{\mathcal{X}^{-}}(q_0(1-1/p_0(X),X)-\beta_0(X))^2\left(\frac{ p_0(X)^{-2} s(0,X)(1-s(0,X))}{1-m(X)}+
  \frac{s(1,X)(1-s(1,X))}{m(X)} \right) \right]\\
E[M_1 M_4] =&- E\left[ 1_{\mathcal{X}^{+}}
  \frac{q_1(p_0(X),X) \beta_1(X) s(1, X) p_0(X)(1-p_0(X))}{m(X)} \right]\\
E[M_2 M_4]  =&-E\left[ 1_{\mathcal{X}^{-}}
  \frac{q_0(1-1/p_0(X),X) \beta_0(X) s(0, X) p_0(X)^{-1}(1-p_0(X)^{-1})}{1-m(X)} \right]\\
 E[M_3 M_5] =&E\left[ 1_{\mathcal{X}^{+}}
  \frac{(\beta(X)-\beta)(q_1(p_0(X),X)-\beta_1(X)) s(0, X) (1-s(0,X))}{1-m(X)} \right] \\
  &-E\left[ 1_{\mathcal{X}^{-}}
  \frac{(\beta(X)-\beta)(q_0(1-1/p_0(X),X)-\beta_0(X)) s(1, X) (1-s(1,X))}{m(X)} \right]
\end{align*}}

\subsection{Proof of Theorem \ref{thm_differentiable_smooth}}
\label{app:proof_differentiable_smooth}

First, we provide an auxiliary result similar to Lemma \ref{lemma-beta(x)-diff}. For $B \in \{L, U\},$ we use the decomposition $\beta_{B,h}(x,1,1)=\beta_{B,1,h}(x,1,1)-\beta_{B,0,h}(x,1,1).$ 

\begin{lem}
Suppose that the assumptions of Theorem \ref{thm_differentiable_smooth} hold. 
Then, for each $x \in \mathcal{X},$
$\beta_{L,h}(x,1,1)$ and $\beta_{U,h}(x,1,1)$  are pathwise differentiable. In particular, we have that

{\footnotesize \begin{align*}
\frac{\partial \beta_{L,1,h}(x,1,1,\theta_0)}{\partial \theta}=
&\frac{1}{g_{1,h}(p_0(x))} \int_{\underline{y}_1}^{q_1(g_{1,h}(p_0(x)),x)} y \tau_{1}(y|x, \theta_0) f_{1}(y|x)  dy\\
&-\frac{q_1(g_{1,h}(p_0(x)),x)}{g_{1,h}(p_0(x))} \int_{\underline{y}_1}^{q_1(g_{1,h}(p_0(x)),x)} \tau_{1}(y|x, \theta_0) f_{1}(y|x)  dy \\
&+(q_1(g_{1,h}(p_0(x)),x)- \beta_{L1,h}(x,1,1)) \frac{g'_{1,h}(p_0(x))}{g_{1,h}(p_0(x))} \frac{\partial p_0(x, \theta_0)}{\partial \theta}, \\
\frac{\partial \beta_{L,0,h}(x,1,1,\theta_0)}{\partial \theta}=
&\frac{1}{g_{1,h}(1/p_0(x))} \int_{q_0(1-g_{1,h}(1/p_0(x)),x)}^{\overline{y}_0} y \tau_{0}(y|x, \theta_0) f_{0}(y|x)  dy\\
&-\frac{q_0(1-g_{1,h}(1/p_0(x)),x)}{g_{1,h}(1/p_0(x))} \int_{q_0(1-g_{1,h}(1/p_0(x)),x)}^{\overline{y}_0} \tau_{0}(y|x, \theta_0) f_{0}(y|x)  dy \\
&-(q_0(1-g_{1,h}(1/p_0(x)),x)- \beta_{L0,h}(x,1,1,\theta_0)) \frac{g_{1,h}'(1/p_0(x))}{g_{1,h}(1/p_0(x))) p_0^2(x)} \frac{\partial p_0(x, \theta_0)}{\partial \theta},
\end{align*}}
and

{\footnotesize \begin{align*}
\frac{\partial \beta_{U,1,h}(x,1,1,\theta_0) }{\partial \theta}=&\frac{1}{ g_{1,h}(p_0(x))} \int_{q_1(1-g_{1,h}(p_0(x)), x)}^{\overline{y}_1} y \tau_{1}(y|x, \theta_0) f_{1}(y|x)  dy\\
&-\frac{q_1(1-g_{1,h}(p_0(x)),x)}{g_{1,h}(p_0(x))} 
\int_{q_1(1-g_{1,h}(p_0(x)), x)}^{\overline{y}_1} \tau_{1}(y|x, \theta_0) f_{1}(y|x)  dy\\ 
&+ (q_1(1-g_{1,h}(p_0(x)),x)- \beta_{U1,h}(x,1,1))  \frac{g_{1,h}'(p_0(x))}{g_{1,h}(p_0(x))} \frac{\partial p_0(x, \theta_0)}{\partial \theta},\\
\frac{\partial \beta_{U,0,h}(x,1,1,\theta_0)}{\partial \theta}=&
\frac{1}{g_{1,h}(1/p_0(x))} \int_{\underline{y}_0}^{q_0(g_{1,h}(1/p_0(x)),x)} y \tau_{0}(y|x, \theta_0) f_{0}(y|x)  dy \\
&-\frac{q_0(g_{1,h}(1/p_0(x)),x)}{g_{1,h}(1/p_0(x))}
\int_{\underline{y}_0}^{q_0(g_{1,h}(1/p_0(x)),x)} \tau_{0}(y|x, \theta_0) f_{0}(y|x)  dy\\
&-(q_0(g_{1,h}(1/p_0(x)),x)- \beta_{U0,h}(x,1,1)) \frac{g_{1,h}'(1/p_0(x))}{g_{1,h}(1/p_0(x)) p_0^2(x)} \frac{\partial p_0(x, \theta_0)}{\partial \theta}.
\end{align*}}
\end{lem}

Its proof is a simpler version of the proof of Lemma \ref{lemma-beta(x)-diff} since the irregularity issues are not present here. Details are omitted for brevity.

 The rest of the proof is analogous and we only present the most important parts. First, we deal with the lower bound case. We write $\beta_{L,h}$ instead of $\beta_{L,h}(1,1),$
and split it as

{\footnotesize \begin{equation}
\beta_{L,h}=\beta_{L,+,h}+\beta_{L,-,h}
 \end{equation}}
  with
  
{\footnotesize \begin{align*}
\beta_{L,+,h}=\frac{E[g_{4,h}(\beta_{L,h}(X)) g_{1,h}(p_0(X))s(1,X)]}{E[g_{3,h}(p_0(X))s(1,X)]}, \quad
\beta_{L,-,h}=\frac{E[g_{5,h}(\beta_{L,h}(X)) g_{3,h}(p_0(X))s(1,X)]}
{E[g_{1,h}(p_0(X))s(1,X)]}.
\end{align*}}
where we use the notation $g_{5,h}(z)=-g_{2,h}(-z).$ 
This leads to

{\footnotesize \begin{align*}
\frac{\partial \beta_{L,h}(\theta_0)}{\partial \theta}=&(T_1^{+}+T_2^{+}+T_3^{+}+T_4^{+}+T_5^{+})/E[g_{3,h}(p_0(X))s(1,X)]\\
&+(T_1^{-}+T_2^{-}+T_3^{-}+T_4^{-}+T_5^{-})/E[g_{1,h}(p_0(X))s(1,X)],
\end{align*}}
where
{\scriptsize \begin{align*}
T_1^{+}=&  \int_{\mathcal{X}} \int_{\underline{y}_1}^{q_1(g_{1,h}(p_0(x)),x)} g_{4,h}'(\beta_{L,h}(x)) y \tau_{1}(y|x, \theta_0) f_{1}(y|x)  s(1,x) f(x) dy dx, \\
T_1^{-}=&  \int_{\mathcal{X}} \int_{\underline{y}_1}^{q_1(g_{1,h}(p_0(x)),x)} g_{5,h}'(\beta_{L,h}(x)) y \tau_{1}(y|x, \theta_0) f_{1}(y|x) \frac{g_{3,h}(p_0(x))}{g_{1,h}(p_0(x))}s(1,x) f(x) dy dx, \\
T_2^{+}=&-\int_{\mathcal{X}} \int_{q_0(1-g_{1,h}(1/p_0(x)),x)}^{\overline{y}_0} \frac{g_{4,h}'(\beta_{L,h}(x))}{g_{1,h}(1/p_0(x))} y \tau_{0}(y|x, \theta_0) f_{0}(y|x)  g_{1,h}(p_0(x))s(1,x) f(x)  dy dx, \\
T_2^{-}=&-\int_{\mathcal{X}} \int_{q_0(1-g_{1,h}(1/p_0(x)),x)}^{\overline{y}_0} \frac{g_{5,h}'(\beta_{L,h}(x))}{g_{1,h}(1/p_0(x))} y \tau_{0}(y|x, \theta_0) f_{0}(y|x)  g_{3,h}(p_0(x))s(1,x) f(x)  dy dx, \\
T_3^{+}=&  \int_{\mathcal{X}}  g_{4,h}(\beta_{L,h}(x)) ((g_{1,h}(p_0(x))s(1,x))'+g_{1,h}(p_0(x))s(1,x) t(x)) f(x) dx \\
& -\int_{\mathcal{X}}  \beta_{L,+,h} ((g_{1,h}(p_0(x))s(1,x))'+g_{1,h}(p_0(x))s(1,x) t(x)) f(x) dx,\\
T_3^{-}=&  \int_{\mathcal{X}}  g_{5,h}(\beta_{L,h}(x)) ((g_{3,h}(p_0(x))s(1,x))'+g_{3,h}(p_0(x))s(1,x) t(x)) f(x) dx \\
& -\int_{\mathcal{X}}  \beta_{L,-,h} ((g_{3,h}(p_0(x))s(1,x))'+g_{3,h}(p_0(x))s(1,x) t(x)) f(x) dx,\\
T_4^{+}=&  -\int_{\mathcal{X}} \int_{\underline{y}_1}^{q_1(g_{1,h}(p_0(x)),x)} q_1(g_{1,h}(p_0(x)),x) g_{4,h}'(\beta_{L,h}(x))  \tau_{1}(y|x, \theta_0) f_{1}(y|x)  s(1,x) f(x) dy dx \\
&+\int_{\mathcal{X}} \int_{q_0(1-g_{1,h}(1/p_0(x)),x)}^{\overline{y}_0} q_0(1-g_{1,h}(1/p_0(x)),x) \frac{g_{4,h}'(\beta_{L,h}(x))}{g_{1,h}(1/p_0(x))}  \tau_{0}(y|x, \theta_0) f_{0}(y|x)  g_{1,h}(p_0(x))s(1,x) f(x)  dy dx,\\
T_4^{-}=&  -\int_{\mathcal{X}} \int_{\underline{y}_1}^{q_1(g_{1,h}(p_0(x)),x)} q_1(g_{1,h}(p_0(x)),x) g_{5,h}'(\beta_{L,h}(x)) \tau_{1}(y|x, \theta_0) f_{1}(y|x) \frac{g_{3,h}(p_0(x))}{g_{1,h}(p_0(x))}s(1,x) f(x) dy dx\\
&+\int_{\mathcal{X}} \int_{q_0(1-g_{1,h}(1/p_0(x)),x)}^{\overline{y}_0} q_0(1-g_{1,h}(1/p_0(x)),x) \frac{g_{5,h}'(\beta_{L,h}(x))}{g_{1,h}(1/p_0(x))} \tau_{0}(y|x, \theta_0) f_{0}(y|x)  g_{3,h}(p_0(x))s(1,x) f(x)  dy dx \\
T_5^{+}=& \int_{\mathcal{X}} (q_1(g_{1,h}(p_0(x)),x)- \beta_{L,1,h}(x)) g_{4,h}'(\beta_{L,h}(x)) g'_{1,h}(p_0(x)) \frac{\partial p_0(x, \theta_0)}{\partial \theta} s(1,x) f(x) dx \\
&+\int_{\mathcal{X}} (q_0(1-g_{1,h}(1/p_0(x)),x)- \beta_{L,0,h}(x)) \frac{g_{4,h}'(\beta_{L,h}(x)) g'_{1,h}(1/p_0(x)) }{g_{1,h}(1/p_0(x)) p_0^2(x)} \frac{\partial p_0(x, \theta_0)}{\partial \theta} g_{1,h}(p_0(x)) s(1,x) f(x) dx, \\
T_5^{-}=& \int_{\mathcal{X}} (q_1(g_{1,h}(p_0(x)),x)- \beta_{L,1,h}(x)) g_{5,h}'(\beta_{L,h}(x)) g'_{1,h}(p_0(x)) \frac{\partial p_0(x, \theta_0)}{\partial \theta} \frac{g_{3,h}(p_0(x))}{g_{1,h}(p_0(x))} s(1,x) f(x) dx \\
&+\int_{\mathcal{X}} (q_0(1-g_{1,h}(1/p_0(x)),x)- \beta_{L,0,h}(x)) \frac{g_{5,h}'(\beta_{L,h}(x))  g'_{1,h}(1/p_0(x)) }{g_{1,h}(1/p_0(x)) p_0^2(x)} \frac{\partial p_0(x, \theta_0)}{\partial \theta} g_{3,h}(p_0(x)) s(1,x) f(x) dx.
\end{align*}}

Next, we find the corresponding influence functions for
$\beta_{L,+,h}$ and $\beta_{L,-,h},$ separately.

For $\beta_{L,+,h}$, we consider the following candidate influence function 

{\footnotesize \begin{equation}
M_{L,+,h}=(M_1^{+}+M_2^{+}+M_3^{+}+M_4^{+}+M_5^{+})/E[g_{3,h}(p_0(X))s(1,X)],
\end{equation}}
where

{\footnotesize \begin{align*}
M_1^{+} =& \frac{SD g_{4,h}'(\beta_{L,h}(X)) }{m(X)} \left[Y \mathbbm{1}(Y \le q_1(g_{1,h}(p_0(X)),X))-g_{1,h}(p_0(X))\beta_{L,1,h}(X)\right], \\
M_2^{+}=&-\frac{S(1-D)}{1-m(X)} \frac{g_{4,h}'(\beta_{L,h}(X)) g_{1,h}(p_0(X))}{p_0(X) g_{1,h}(1/p_0(X))} \\
& \times
\left[Y \mathbbm{1}(Y \geq q_0(1-g_{1,h}(1/p_0(X)),X))-g_{1,h}(1/p_0(X)) \beta_{L,0,h}(X)\right]   , \\
M_3^{+} =& g_{4,h}(\beta_{L,h}(X)) \left[ g_{1,h}(p_0(X))s(1,X)+g_{1,h}'(p_0(X)) \frac{1-D}{1-m(X)}(S-s(0,X)) \right] \\
& +g_{4,h}(\beta_{L,h}(X)) \left[(g_{1,h}(p_0(X))-p_0(X) g_{1,h}'(p_0(X))) \frac{D}{m(X)}(S-s(1,X)) \right] \\
& -\beta_{L,+,h} \left[ g_{3,h}(p_0(X))s(1,X)+g_{3,h}'(p_0(X)) \frac{1-D}{1-m(X)}(S-s(0,X)) \right] \\
& -\beta_{L,+,h}\left[(g_{3,h}(p_0(X))-p_0(X) g_{3,h}'(p_0(X))) \frac{D}{m(X)}(S-s(1,X)) \right], \\
M_4^{+}=&- \frac{SD g_{4,h}'(\beta_{L,h}(X))} {m(X)} q_1(g_{1,h}(p_0(X)),X)
\left[\mathbbm{1}(Y \le q_1(g_{1,h}(p_0(X)),X))-g_{1,h}(p_0(X)) \right]\\
&+\frac{S(1-D)}{1-m(X)} \frac{g_{4,h}'(\beta_{L,h}(X)) g_{1,h}(p_0(X))}{p_0(X) g_{1,h}(1/p_0(X))} q_0(1-g_{1,h}(1/p_0(X)),X) \\
&\times
\left[\mathbbm{1}(Y \geq q_0(1-g_{1,h}(1/p_0(X)),X))-g_{1,h}(1/p_0(X))\right]  , \\
M_5^{+}=& g_{4,h}'(\beta_{L,h}(X)) \bigg[(q_1(g_{1,h}(p_0(X)),X)- \beta_{L,1,h}(X)) g'_{1,h}(p_0(X))\\
&+(q_0(1-g_{1,h}(1/p_0(X)),X)-\beta_{L,0,h}(X)) \frac{ g'_{1,h}(1/p_0(X)) g_{1,h}(p_0(X)) }{g_{1,h}(1/p_0(X)) p_0^2(X)} \bigg]\\
& \times \left[\frac{1-D}{1-m(X)}(S-s(0,X))- p_0(X)
\frac{D}{m(X)}(S-s(1, X)) \right].
\end{align*}}

Similarly, for $\beta_{L,-,h}$, we have the following candidate influence function

{\footnotesize \begin{equation}
M_{L,-,h}=(M_1^{-}+M_2^{-}+M_3^{-}+M_4^{-}+M_5^{-})/E[g_{1,h}(p_0(X))s(1,X)],
\end{equation}}
where

{\footnotesize \begin{align*}
M_1^{-} =& \frac{SD g_{5,h}'(\beta_{L,h}(X)) }{m(X)} \frac{g_{3,h}(p_0(X))}{g_{1,h}(p_0(X))} \left[Y \mathbbm{1}
(Y \le q_1(g_{1,h}(p_0(X)),X))-g_{1,h}(p_0(X))\beta_{L,1,h}(X)\right], \\
M_2^{-}=&-\frac{S(1-D)}{1-m(X)} \frac{g_{5,h}'(\beta_{L,h}(X)) g_{3,h}(p_0(X))}{p_0(X) g_{1,h}(1/p_0(X))} \\
& \times \left[Y \mathbbm{1}(Y \geq q_0(1-g_{1,h}(1/p_0(X)),X))-g_{1,h}(1/p_0(X)) \beta_{L,0,h}(X)\right]   , \\
M_3^{-} =& g_{5,h}(\beta_{L,h}(X)) \left[ g_{3,h}(p_0(X))s(1,X)+g_{3,h}'(p_0(X)) \frac{1-D}{1-m(X)}(S-s(0,X)) \right] \\
& +g_{5,h}(\beta_{L,h}(X)) \left[(g_{3,h}(p_0(X))-p_0(X) g_{3,h}'(p_0(X))) \frac{D}{m(X)}(S-s(1,X)) \right] \\
& -\beta_{L,-,h} \left[ g_{1,h}(p_0(X))s(1,X)+g_{1,h}'(p_0(X)) \frac{1-D}{1-m(X)}(S-s(0,X)) \right] \\
& -\beta_{L,-,h}\left[(g_{1,h}(p_0(X))-p_0(X) g_{1,h}'(p_0(X))) \frac{D}{m(X)}(S-s(1,X)) \right], \\
M_4^{-}=&- \frac{SD g_{5,h}'(\beta_{L,h}(X))} {m(X)} \frac{g_{3,h}(p_0(X))}{g_{1,h}(p_0(X))} q_1(g_{1,h}(p_0(X)),X) \\
&\times \left[\mathbbm{1}(Y \le q_1(g_{1,h}(p_0(X)),X))-g_{1,h}(p_0(X)) \right]\\
&+\frac{S(1-D)}{1-m(X)} \frac{g_{5,h}'(\beta_{L,h}(X)) g_{3,h}(p_0(X))}{p_0(X) g_{1,h}(1/p_0(X))} q_0(1-g_{1,h}(1/p_0(X)),X) \\
&\times
\left[\mathbbm{1}(Y \geq q_0(1-g_{1,h}(1/p_0(X)),X))-g_{1,h}(1/p_0(X))\right]  , \\
M_5^{-}=& g_{5,h}'(\beta_{L,h}(X)) \bigg[(q_1(g_{1,h}(p_0(X)),X)- \beta_{L,1,h}(X)) \frac{g'_{1,h}(p_0(X)) g_{3,h}(p_0(X))}{g_{1,h}(p_0(X))} \\
&+(q_0(1-g_{1,h}(1/p_0(X)),X)-\beta_{L,0,h}(X)) \frac{ g'_{1,h}(1/p_0(X)) g_{3,h}(p_0(X)) }{g_{1,h}(1/p_0(X)) p_0^2(X)} \bigg]\\
& \times \left[\frac{1-D}{1-m(X)}(S-s(0,X))- p_0(X)
\frac{D}{m(X)}(S-s(1, X)) \right].
\end{align*}}

For each $i\in\{1, \ldots, 5\}$, we have that

{\footnotesize \begin{align*}
T_i^{+}=E[M_{i}^{+}  Q(Y, S, D, X|\theta_0)], \quad
T_i^{-}=E[M_{i}^{-}  Q(Y, S, D, X|\theta_0)].
\end{align*}}
As a result, $\beta_{L,+,h}$ and $\beta_{L,-,h}$ are pathwise differentiable and their efficient influence functions are given by $M_{L,+,h}$ and $M_{L,-,h},$ respectively. Due to the linearity property of the influence function, $\beta_{L,h}$ is pathwise differentiable and its efficient influence function is  $M_{L,+,h}+M_{L,-,h}$. 
In addition, $M_{L,+,h} \in \mathcal{S}_1$ and $M_{L,-,h} \in \mathcal{S}_1,$ where $\mathcal{S}_1$ denotes the score of the observed data $(S Y, S, D, X)$ Hence, the efficiency bound for $\beta_{L,h}$ is equal to $E[(M_{+}^h+M_{-}^h)^2].$ 
Results for the smooth upper bound $\beta_{U,h}$ are similar and therefore omitted. Below, we only write its influence function. Again, we use the decompositions $\beta_{U,h}=\beta_{U,+,h}+\beta_{U,-,h}$ and
$M_{U,h}=M_{U,+,h}+M_{U,-,h}.$
First, we have that

{\footnotesize \begin{equation}
M_{U,+,h}=(M_1^{+}+M_2^{+}+M_3^{+}+M_4^{+}+M_5^{+})/E[g_{1,h}(p_0(X))s(1,X)],
\end{equation}}
where 

{\footnotesize \begin{align*}
M_1^{+} =& \frac{SD g_{2,h}'(\beta_{U,h}(X)) }{m(X)} \frac{g_{3,h}(p_0(X))}{g_{1,h}(p_0(X))} \left[Y \mathbbm{1}(Y \geq q_1(1-g_{1,h}(p_0(X)),X))-g_{1,h}(p_0(X))\beta_{U,1,h}(X)\right], \\
M_2^{+}=&-\frac{S(1-D)}{1-m(X)} \frac{g_{2,h}'(\beta_{U,h}(X)) g_{3,h}(p_0(X))}{p_0(X) g_{1,h}(1/p_0(X))} \\
& \times \left[Y \mathbbm{1} (Y \leq q_0(g_{1,h}(1/p_0(X)),X))-g_{1,h}(1/p_0(X)) \beta_{U,0,h}(X)\right]   , \\
M_3^{+} =& g_{2,h}(\beta_{U,h}(X)) \left[ g_{3,h}(p_0(X))s(1,X)+g_{3,h}'(p_0(X)) \frac{1-D}{1-m(X)}(S-s(0,X)) \right] \\
& +g_{2,h}(\beta_{U,h}(X)) \left[(g_{3,h}(p_0(X))-p_0(X) g_{3,h}'(p_0(X))) \frac{D}{m(X)}(S-s(1,X)) \right] \\
& -\beta_{U,+,h} \left[ g_{1,h}(p_0(X))s(1,X)+g_{1,h}'(p_0(X)) \frac{1-D}{1-m(X)}(S-s(0,X)) \right] \\
& -\beta_{U,+,h}\left[(g_{1,h}(p_0(X))-p_0(X) g_{1,h}'(p_0(X))) \frac{D}{m(X)}(S-s(1,X)) \right], \\
M_4^{+}=&- \frac{SD g_{2,h}'(\beta_{U,h}(X))} {m(X)}  \frac{g_{3,h}(p_0(X))}{g_{1,h}(p_0(X))} q_1(1-g_{1,h}(p_0(X)),X) \\
& \times \left[\mathbbm{1}(Y \geq q_1(1-g_{1,h}(p_0(X)),X))-g_{1,h}(p_0(X)) \right]\\
&+\frac{S(1-D)}{1-m(X)} \frac{g_{2,h}'(\beta_{U,h}(X)) g_{3,h}(p_0(X))}{p_0(X) g_{1,h}(1/p_0(X))} q_0(g_{1,h}(1/p_0(X)),X) \\
&\times
\left[\mathbbm{1}(Y \leq q_0(g_{1,h}(1/p_0(X)),X))-g_{1,h}(1/p_0(X))\right]  , \\
M_5^{+}=& g_{2,h}'(\beta_{U,h}(X)) \bigg[(q_1(1-g_{1,h}(p_0(X)),x)- \beta_{U,1,h}(X)) g'_{1,h}(p_0(X)) \frac{g_{3,h}(p_0(X))}{g_{1,h}(p_0(X))}\\
&+(q_0(g_{1,h}(1/p_0(X)),X)- \beta_{U,0,h}(X)) \frac{ g'_{1,h}(1/p_0(X)) g_{3,h}(p_0(X)) }{g_{1,h}(1/p_0(X)) p_0^2(X)} \bigg]\\
& \times \left[\frac{1-D}{1-m(X)}(S-s(0,X))- p_0(X)
\frac{D}{m(X)}(S-s(1, X)) \right].
\end{align*}}
Further, we have that
{\footnotesize \begin{equation}
M_{U,-,h}=(M_1^{-}+M_2^{-}+M_3^{-}+M_4^{-}+M_5^{-})/E[g_{3,h}(p_0(X))s(1,X)],
\end{equation}}
where we use the notation $g_{6,h}(z)=-g_{4,h}(-z)$ and

{\footnotesize \begin{align*}
M_1^{-} =& \frac{SD g_{6,h}'(\beta_{U,h}(X)) }{m(X)} 
\left[Y \mathbbm{1}(Y \geq q_1(1-g_{1,h}(p_0(X)),X))-g_{1,h}(p_0(X))\beta_{U,1,h}(X)\right], \\
M_2^{-}=&-\frac{S(1-D)}{1-m(X)} \frac{g_{6,h}'(\beta_{U,h}(X)) g_{1,h}(p_0(X))}{p_0(X) g_{1,h}(1/p_0(X))}\\
& \times
\left[Y \mathbbm{1}(Y \leq q_0(g_{1,h}(1/p_0(X)),X))-g_{1,h}(1/p_0(X)) \beta_{U,0,h}(X)\right]   , \\
M_3^{-} =& g_{6,h}(\beta_{U,h}(X)) \left[ g_{1,h}(p_0(X))s(1,X)+g_{1,h}'(p_0(X)) \frac{1-D}{1-m(X)}(S-s(0,X)) \right] \\
& +g_{6,h}(\beta_{U,h}(X)) \left[(g_{1,h}(p_0(X))-p_0(X) g_{1,h}'(p_0(X))) \frac{D}{m(X)}(S-s(1,X)) \right] \\
& -\beta_{U,-,h} \left[ g_{3,h}(p_0(X))s(1,X)+g_{3,h}'(p_0(X)) \frac{1-D}{1-m(X)}(S-s(0,X)) \right] \\
& -\beta_{U,-,h}\left[(g_{3,h}(p_0(X))-p_0(X) g_{3,h}'(p_0(X))) \frac{D}{m(X)}(S-s(1,X)) \right], \\
M_4^{-}=&- \frac{SD g_{6,h}'(\beta_{U,h}(X))} {m(X)} q_1(1-g_{1,h}(p_0(X)),X)
\left[\mathbbm{1}(Y \geq q_1(1-g_{1,h}(p_0(X)),X))-g_{1,h}(p_0(X)) \right]\\
&+\frac{S(1-D)}{1-m(X)} \frac{g_{6,h}'(\bar{\beta}(X)) g_{1,h}(p_0(X))}{p_0(X) g_{1,h}(1/p_0(X))} q_0(g_{1,h}(1/p_0(X)),X) \\
&\times
\left[\mathbbm{1}(Y \leq q_0(g_{1,h}(1/p_0(X)),X))-g_{1,h}(1/p_0(X)))\right]  , \\
M_5^{-}=& g_{6,h}'(\beta_{U,h}(X)) \bigg[(q_1(1-g_{1,h}(p_0(X)),X)- \beta_{U,1,h}(X)) g'_{1,h}(p_0(X)) \\
&+(q_0(g_{1,h}(1/p_0(X)),X)- \beta_{U,0,h}(X)) \frac{ g'_{1,h}(1/p_0(X)) g_{1,h}(p_0(X)) }{g_{1,h}(1/p_0(X)) p_0^2(X)} \bigg]\\
& \times \left[\frac{1-D}{1-m(X)}(S-s(0,X))- p_0(X)
\frac{D}{m(X)}(S-s(1, X)) \right].
\end{align*}}

\subsection{Proof of Theorem \ref{thm_asyN_regular}}
 We suppress the parameter $\beta_L = \beta_L(1,1)$ in what follows whenever it does not cause confusion. We denote the uncentered influence functions for the corresponding numerator and denominator as $\psi^{[L]}(W,\eta) = \psi^{[L]}(\eta)$ and $\psi^{[S]}(W,\eta) = \psi^{[S]}(\eta)$. Recall that the sample analogues using estimated nuisances $\hat{\eta} = \hat{\eta}(X)$ are then given by

{\footnotesize \begin{align*}
    \hat{\beta}_L = \frac{E_n[\psi^{[L]}(W,\hat{\eta})]}{E_n[\psi^{[S]}(W,\hat{\eta})]}.
\end{align*}}

Thus, we obtain the following linearization 

{\footnotesize \begin{align*}
    \sqrt{n}(\hat{\beta}_L - \beta_L) &= \sqrt{n}E_n[M(\hat{\eta})]\bigg(1 + O\bigg(\big|E_n[\psi^{[S]}(\hat{\eta})] - E[\psi^{[S]}(\eta)]\big|\bigg)\bigg),
\end{align*}}
where $M(\hat \eta) = \psi_{\beta_L}(W,\hat\eta)/E[\psi^{[S]}(W,\eta)]$ is the influence function for $\beta_L$, and $O(\cdots)$ term is negligible by Assumption \ref{ass_MLbias1}. 

Let $G_n[X] = n^{-1/2}\sum_i^n(X_i - E[X_i])$. Further decomposing the leading term yields

{\footnotesize \begin{align*}
    \sqrt{n}E_n[M(\hat{\eta})] &= G_n[M({\eta})] + (G_n[M(\hat{\eta})]  - G_n[M({\eta})]) + \sqrt{n}E[M(\hat{\eta})-M({\eta})].
\end{align*}}
The CLT implies that the first term above yields the asymptotic limit. To prove that the other two terms are negligible, we need some preparation. 
For any values of the nuisances parameters, note that the moments can be decomposed as the (finite) sum over its (finite) product components. In particular, we can write 

{\footnotesize \begin{align*}
    M(\eta) &= \sum_r m_r(\eta) \\
    &=\sum_r \prod_{j\in \mathcal{J}(r)}m_{r,j}(\eta),
\end{align*}} 
where $\mathcal{J}(r)$ is the set of factors for multiplication in the $r$-th summand of $M(\eta)$. 
Decomposing the difference in moments evaluated at the different nuisance parameters then yields

{\footnotesize 
\begin{align*}
    M(\hat{\eta}) - M(\eta)
    &= \sum_r\left(\prod_{j\in \mathcal{J}(r)} m_{r,j}(\hat{\eta})
      - \prod_{j\in \mathcal{J}(r)} m_{r,j}(\eta)\right) \\
    &= \sum_r\left(
        \sum_{j\in \mathcal{J}(r)}
        \big(m_{r,j}(\hat{\eta}) - m_{r,j}(\eta)\big)
        \prod_{k<j} m_{r,k}(\hat{\eta})
        \prod_{k>j} m_{r,k}(\eta)
      \right) \\
    &= \sum_r\Bigg(
        \sum_{j\in \mathcal{J}(r)}
        \big(m_{r,j}(\hat{\eta}) - m_{r,j}(\eta)\big)
        \prod_{k\neq j} m_{r,k}(\eta) \\
    &\qquad\quad
        + \sum_{j\in \mathcal{J}(r)}\sum_{k<j}
        \big(m_{r,j}(\hat{\eta}) - m_{r,j}(\eta)\big)
        \big(m_{r,k}(\hat{\eta}) - m_{r,k}(\eta)\big)
        \prod_{l\neq j,k} m_{r,l}(\eta) \\
    &\qquad\quad
        + \cdots
        + \prod_{j\in \mathcal{J}(r)}
        \big(m_{r,j}(\hat{\eta}) - m_{r,j}(\eta)\big)
      \Bigg).
\end{align*}

}

This decomposition has the following properties: For any $j$, all 
   $\prod_{k\neq j}m_{r,k}(\eta)$ are either bounded or have at least two moments by Assumption \ref{ass_STRONGoverlap} and \ref{ass_regularEst}. Moreover, by the pathwise differentiability as shown in Proof of Theorem \ref{thm_differentiable}, we have that 
   
   {\footnotesize \begin{align*}
       \sup_{\tilde{\eta}\in\mathcal{E}_n}\sum_r\sum_{j\in \mathcal{J}(r)} E\Big[(m_{r,j}(\tilde{\eta}) - m_{r,j}({\eta}))\prod_{k\neq j}m_{r,k}(\eta) \Big] = 0.
   \end{align*}}

To make the final bounds transparent, we isolate three representative moment types.

\noindent
\textbf{(i) Smooth nuisance moments.}
For any score component $m_{r,j}(\eta)$ involving only smooth nuisance transformations such as $m(X), 1-m(X), s(1,X), p_0(X), q_1(u,x), \beta_{0,0}(X),$
and their inverses on trimmed sets, standard mean-square expansions yield
\begin{align*}
E\big[(m_{r,j}(\hat\eta)-m_{r,j}(\eta))^2\big]
\lesssim
\lambda_{q,n,2}+\lambda_{s,n,2}^2+\lambda_{m,n,2}^2+\lambda_{b,n,2}^2.
\label{eq:smooth_moment_bound}
\end{align*}
\noindent
\textbf{(ii) Quantile composition moments.}
Consider next a conditional quantile term $m_{r,j}(\eta)=\mathbbm{1}\{Y \leq q_1(p_0(X),X)\}.$
Under Assumption \ref{ass_regularEst}, the conditional density is uniformly bounded away from zero on its support, so 
$|q_1(u_1,x)-q_1(u_2,x)|\lesssim |u_1-u_2|$ for all $u_1,u_2\in(0,1).$
Hence, we deduce that
{\footnotesize
\begin{align*}
E[(m_{r,j}(\hat \eta)-m_{r,j}(\eta))^2] & =  E[|m_{r,j}(\hat \eta)-m_{r,j}(\eta)|] \\
& \lesssim E\big[\big| \hat{q}_1(\hat p_0(X),X)-q_1(p_0(X),X)\big|\big] \\
& \le E\big[\big|\hat q_1(\hat p_0(X),X)-q_1(\hat p_0(X),X)\big|+\big|q_1(\hat p_0(X),X)-q_1(p_0(X),X)\big|\big]\\
&\lesssim \lambda_{q,n,1}+\lambda_{s,n,1}.
\end{align*}
}

\noindent
\textbf{(iii) Classification moments.}
Finally, we study $m_{r,j}=\mathbbm{1}(X\in\mathcal X^+)=\mathbbm{1}(\Delta(X)>0),$ which is a classification indicator.
We use the same decomposition as in \eqref{eq_deltaclass} to obtain 

{\footnotesize \begin{align*}
    E[|\mathbbm{1}(\hat{\Delta}(X) > 0) - \mathbbm{1}({\Delta}(X) > 0)|]
    &\leq P(|\Delta(X)| \leq |\hat{\Delta}(X) - \Delta(X)|) \\
    &\leq P(|\hat{\Delta}(X) - \Delta(X)| > \delta) + P(|\Delta(X)| \leq \delta) \\
    &\leq \frac{E[|\hat{\Delta}(X) - \Delta(X)|^2]}{\delta^2} + P(|\Delta(X)| \leq \delta) \\
    &\leq \lambda_{s,n,2}^2 \delta^{-2} + C\delta^{\alpha},
    \end{align*}}
    for any $\delta > 0$ by Assumption \ref{ass_margin1}. Optimizing rates by setting $\delta^{2+\alpha} \sim \lambda_{s,n,2}^2$ then yields 
    
    {\footnotesize \begin{align*}
     E[|\mathbbm{1}(\hat{\Delta}(X) > 0) - \mathbbm{1}({\Delta}(X) > 0)|]
    \lesssim \lambda_{s,n,2}^{\frac{2\alpha}{2+\alpha}},
\end{align*}}
where $2\alpha/(2+\alpha) \in (1,2)$ for any $\alpha > 2$.

Proceeding as in Semenova (2024), Version 3, Lemma A.7, for the correctly classified remainder terms, and using the classification bound above, we obtain

{\footnotesize \begin{align*}
  \sqrt{n}\sup_{\tilde{\eta} \in \mathcal{E}_n}|E[M(\tilde{\eta})-M({\eta})]|
  &\lesssim \sqrt{n}(\lambda_{q,n,2}^2 + \lambda_{s,n,2}^2 + \lambda_{m,n,2}^2 +\lambda_{b,n,2}^2 + \lambda_{s,n,2}^{\frac{2\alpha}{2 + \alpha}}).
\end{align*}} 
Moreover, as the nuisances are cross-fitted, we also obtain as in \cite{chernozhukov2018double} or \cite{kennedy2024semiparametric}, Lemma 1 that 

{\footnotesize \begin{align*}
 \sup_{\tilde{\eta} \in \mathcal{E}_n}E[(M(\tilde{\eta})-M({\eta}))^2]  \lesssim \lambda_{q,n,1} + \lambda_{s,n,1}+\lambda_{q,n,2}^2 + \lambda_{s,n,2}^2 + \lambda_{m,n,2}^2 + \lambda_{b,n,2}^2 + \lambda_{s,n,2}^{\frac{2\alpha}{2 + \alpha}}.
\end{align*}}
Hence, by Assumption \ref{ass_MLbias1} and Slutsky's Theorem, we obtain that

{\footnotesize \begin{align*}
    \sqrt{n}(\hat{\beta}_L - \beta_L) \overset{d}{\rightarrow} \mathcal{N}(0,Var_a(\beta_L)),
\end{align*}}
where $Var_a(\beta_L)$ is the semiparametric efficiency bound from Theorem \ref{thm_differentiable}.

\subsection{Proof of Theorem \ref{thm_asyN_smooth}}
First, we provide some definitions and auxiliary convergence results and then apply an analogous decomposition as in the proof of Theorem \ref{thm_asyN_regular} with these new, modified terms.  Recall that

{\footnotesize \begin{align*}
    \beta_{L,h} = \beta_{L,+,h} + \beta_{L,-,h}
\end{align*}} and sample analogue using estimated nuisances $\hat{\eta} = \hat{\eta}(X)$ given by 

{\footnotesize \begin{align*}
     \hat{\beta}_{L,h}
     &= \hat{\beta}_{L,+,h} + \hat{\beta}_{L,-,h} \\
   &= \frac{E_n[\psi^{[L]}_{\beta_{L,+,h}}(W,\hat{\eta})]}{E_n[\psi^{[S]}_{\beta_{L,+,h}}(W,\hat{\eta})]}  + \frac{E_n[\psi^{[L]}_{\beta_{L,-,h}}(W,\hat{\eta})]}{E_n[\psi^{[S]}_{\beta_{L,-,h}}(W,\hat{\eta})]}.
\end{align*}}
Based on Table \ref{tab_EIFs_smooth}, we denote
\[
M_h(\tilde \eta)
=
\sum_{j\in\{+,-\}}
\frac{
\psi^{[L]}_{\beta_{L,j,h}}(W,\tilde\eta)
-
\beta_{L,j,h}\psi^{[S]}_{\beta_{L,j,h}}(W,\tilde\eta)
}
{E[\psi^{[S]}_{\beta_{L,j,h}}(W,\eta)]} .
\]
Applying the same denominator expansion as in the proof of Theorem \ref{thm_asyN_regular} to each ratio component yields

{\footnotesize \begin{align*}
    \sqrt{n}(\hat{\beta}_{L,h} - \beta_{L,h}) 
    &= \sqrt{n} E_n[M_h(\hat{\eta})]+o(1),
\end{align*}}
by Assumption \ref{ass_MLbias2}. Further decomposing the leading term yields

{\footnotesize \begin{align*}
    \sqrt{n}E_n[M_h(\hat{\eta})] &= G_n[M_h({\eta})] + (G_n[M_h(\hat{\eta})]  - G_n[M_h({\eta})])+\sqrt{n}E[M_h(\hat{\eta})-M_h({\eta})].
\end{align*}}
Again the CLT yields the asymptotic limit for the first term and we next prove that the other two terms are negligible. To this end, we decompose the difference in moments evaluated at different nuisance parameters yields

{\footnotesize \begin{align*}
    M_h(\hat{\eta}) - M_h(\eta)
    &=\sum_r\bigg(\prod_{j\in \mathcal{J}(r)}m_{h,r,j}(\hat{\eta}) - \prod_{j\in \mathcal{J}(r)}m_{h,r,j}(\eta)\bigg) \\
    &= \sum_r\Bigg(
        \sum_{j\in \mathcal{J}(r)}
        \big(m_{h,r,j}(\hat{\eta}) - m_{h,r,j}(\eta)\big)
        \prod_{k\neq j} m_{h,r,k}(\eta) \\
    &\qquad\quad
        + \sum_{j\in \mathcal{J}(r)}\sum_{k<j}
        \big(m_{h,r,j}(\hat{\eta}) - m_{h,r,j}(\eta)\big)
        \big(m_{h,r,k}(\hat{\eta}) - m_{h, r,k}(\eta)\big)
        \prod_{l\neq j,k} m_{h, r,l}(\eta) \\
    &\qquad\quad
        + \cdots
        + \prod_{j\in \mathcal{J}(r)}
        \big(m_{h,r,j}(\hat{\eta}) - m_{h,r,j}(\eta)\big)
      \Bigg).
\end{align*}}

This decomposition has again the following properties. For any $j$, all 
   $\prod_{k\neq j}m_{h,r,k}(\eta)$ are either bounded or have at least two moments by Assumption \ref{ass_STRONGoverlap} and \ref{ass_regularEst}. Moreover, by pathwise differentiability as in Theorem \ref{thm_differentiable_smooth}, we have that 
   
   {\footnotesize \begin{align*}
       \sup_{\tilde{\eta}\in\mathcal{E}_n}\sum_r\sum_j E[(m_{h,r,j}(\tilde{\eta}) - m_{h,r,j}({\eta}))\prod_{k\neq j}m_{h,r,k}(\eta)] = 0.
   \end{align*}} 

Inspection of the structure of $M_h$ shows that there are two types of nuisance dependent factors: (i) unsmoothed terms as in regular bounds, e.g. $m(X)^{-1}$, $s(d,x)$, $q_d(u, X)$, $\beta_{1,1}(u,X),$ (ii) smooth transforms of such terms, e.g. $g_{j,h}(p_0(X),$ $g'_{j,h}(p_0(X),$ $g_{j,h}(\beta_{L,h})$.

We can deal with the first type term as in the proof of Theorem \ref{thm_asyN_regular}. We note that the second type terms take the form $g_{j,h}(z(\eta))$ or $g'_{j,h}(z(\eta))$, where $z(\eta)$ is a scalar nuisance term such as $p_0(X).$ An application of the mean value theorem yields

{\footnotesize \begin{align*}
    g_{j,h}(z(\hat{\eta}))-g_{j,h}(z(\eta))&=g'_{j,h}(\tilde z)(z(\hat{\eta})-z(\eta)),\\
    g'_{j,h}(z(\hat{\eta}))-g'_{j,h}(z(\eta))&=g''_{j,h}(\bar z)(z(\hat{\eta})-z(\eta)),
\end{align*}}
for some intermediate values $\tilde z$ and $\bar z.$ For fixed $h>0,$ in view of the assumption $\sup_{z, j} |g'_{j,h}(z)| \lesssim 1$ and $\sup_{z, j} |g''_{j,h}(z)| \lesssim 1$, we obtain that

{\footnotesize 
\begin{align*}
    |g_{j,h}(z(\hat{\eta}))-g_{j,h}(z(\eta))|+| g'_{j,h}(z(\hat{\eta}))-g'_{j,h}(z(\eta))| \lesssim  |(z(\hat{\eta})-z(\eta))|.
\end{align*}}
Hence all such terms satisfy the same bounds as in Theorem \ref{thm_asyN_regular}.

The only remaining type (ii) component are \(g_{j,h}(\beta_h(X))\) and \(g'_{j,h}(\beta_h(X))\), $j=4,5$  which are composite instances of \(g_{j,h}(z(\eta))\) since \(\beta_h(X)\) itself depends on multiple nuisance functions. These replace the classification error from regular bounds. Since \(h\) is fixed away from zero, the smoothed trimming indices remain in a compact subset of \((0,1)\), both at the truth $g_{1,h}(p_0(x))$ and $1-g_{1,h}(1/p_0(x))$ and on the realization set using $\hat{p}_0(x)$ instead. Therefore, by Lipschitz continuity of $g_{j,h},$ we obtain that

{\footnotesize
\begin{align*}
E[(g_{j,h}(\hat\beta_h(X)) - g_{j,h}(\beta_h(X)))^2]
\lesssim E[(\hat\beta_h(X)-\beta_h(X))^2]
\lesssim \lambda_{b,n,2}^2.
\end{align*}
}

Proceeding analogously as in the proof of Theorem \ref{thm_asyN_regular}, we deduce that
{\footnotesize \begin{align*}
  \sqrt{n}\sup_{\tilde{\eta} \in \mathcal{E}_n}|E[M_h(\tilde{\eta})-M_h({\eta})]|
  &\lesssim \sqrt{n}(\lambda_{q,n,2}^2 + \lambda_{s,n,2}^2 + \lambda_{m,n,2}^2 +\lambda_{b,n,2}^2), \\
  \sup_{\tilde\eta\in\mathcal E_n}
E[(M_h(\tilde\eta)-M_h(\eta))^2]
& \lesssim
\lambda_{q,n,1}+\lambda_{s,n,1}+\lambda_{q,n,2}^2+\lambda_{s,n,2}^2+\lambda_{m,n,2}^2+\lambda_{b,n,2}^2.
\end{align*}
}
Combining these results we prove asymptotic normality and efficiency.

\subsection{Proof of Theorem \ref{thm_efficiency_gap}}
We first show the proof under the positive monotonicity assumption. We suppress the nuisance functions, parameter, and principal strata in any moment function in what follows. First note that the estimator $\tilde{\beta}_L$ of \cite{semenova2025generalized} is based on proposing separate moments for numerator and denominator. The corresponding variance bound is established in Theorem 1 of \cite{semenova2025generalized} by an application of the delta method. First, we consider the moment function suggested for the known propensity score case, namely, denoting $W=(YS,S,D,X)$ the uncentered moments for numerator and denominator are given by 

{\footnotesize \begin{align*}
\tilde{\psi}^{[L]}_{\beta}(W)
&= \frac{SD}{m(X)} Y 1\{Y \le q_1(p_0(X),X)\}-\frac{S(1-D)}{(1-m(X)}
Y\\
&- \frac{SD}{m(X)}q_1(p_0(X),X)
\left[1\{Y \le q_1(p_0(X),X)\}-p_0(X) \right]\\
&+ q_1(p_0(X),X)
\left[\frac{1-D}{1-m(X)}(S-s(0,X))- p_0(X)
\frac{D}{m(X)}(S-s(1, X)) \right] \\
\tilde{\psi}^{[S]}_{\beta} &= s(0,X)+\frac{1-D}{1-m(X)}(S-s(0,X)).
\end{align*}}

The centered versions of these moments are given by
{\footnotesize \begin{align*}
 \tilde{\psi}^{[L]}_{\beta}&(W)-E[\tilde{\psi}^{[L]}_{\beta}(W)] \\
&= \frac{SD}{m(X)} Y 1\{Y \le q_1(p_0(X),X)\} 
-\frac{S(1-D)}{(1-m(X))}
Y - \beta E[s(0,X)]\\
&- \frac{SD}{m(X)}q_1(p_0(X),X)
\left[1\{Y \le q_1(p_0(X),X)\}-p_0(X) \right]\\
&+q_1(p_0(X),X)
\left[\frac{1-D}{1-m(X)}(S-s(0,X))- p_0(X)
\frac{D}{m(X)}(S-s(1, X)) \right],
\end{align*}} 

{\footnotesize \begin{align*}
\tilde{\psi}^{[S]}_{\beta}(W)-E[\tilde{\psi}^{[S]}_{\beta}(W)] = s(0,X)-E[s(0,X)]+\frac{1-D}{1-m(X)}(S-s(0,X)).
\end{align*}}
Define the centered ratio moment as 

{\footnotesize \begin{align*}
  \tilde{\psi}_{\beta}(W) =  E[s(0,X)]^{-1}(\tilde{\psi}^{[L]}_{\beta}(W)-\beta_L \tilde{\psi}^{[S]}_{\beta}(W)-E[\tilde{\psi}^{[L]}_{\beta}(W)-\beta_L \tilde{\psi}^{[S]}_{\beta}(W)])
\end{align*}}
Thus, we can write the difference between our influence function and the linearized ratio moment as

{\footnotesize \begin{align*}
\delta^+(D,X) &= {\psi}^{[L]}_{\beta}(W) - \tilde{\psi}^{[L]}_{\beta}(W) \\
&= E[s(0,X)]^{-1}s(0,X)\bigg[\beta_{1,1}(X,p_0(X))\bigg(1-\frac{D}{m(X)}\bigg) - \beta_{0,0}(X,0)\bigg(1-\frac{1-D}{1-m(X)}\bigg)\bigg]  \\  
\end{align*}}

It turns out that this is the difference to the moment function under unknown propensity scores as suggested by  \cite{semenova2025generalized}, Section 6.6, i.e.~the linearized version of the latter is equivalent to $\psi^{[L]}_{\beta}(W)$. 
Now note that, as $\psi^{[L]}_{\beta}(W)$ is efficient when $P(\mathcal{X}_0) = 0$, we must have that $Cov(\psi^{[L]}_{\beta}(W),\tilde{\psi}^{[L]}_{\beta}(W)) = E[\psi^{[L]}_{\beta}(W)^2]$ and thus the variance gap in the regular case is given by 

{\footnotesize \begin{align*}
    E&[\delta^+(D,X)^2] \\
    &= -E[s(0,X)]^{-2}E\bigg[s(0,X)^2 \bigg( \beta_{1,1}(X,p_0(X))\sqrt{\frac{1-m(X)}{m(X)}} - \beta_{0,0}(X,0)\sqrt{\frac{m(X)}{1-m(X)}}\bigg)^2 \bigg].
\end{align*}}

For conditional monotonicity, the proof is analogous within partitions where now {{\footnotesize \begin{align*}
    E&[\min\{s(0,X),s(1,X)\}]\delta(D,X) \\
    &=
    \mathbbm{1}_{X\in\mathcal{X}^+}s(0,X)\bigg[\beta_{1,1}(X,p_0(X))\bigg(1-\frac{D}{m(X)}\bigg) - \beta_{0,0}(X,0)\bigg(1-\frac{1-D}{1-m(X)}\bigg)\bigg] \\
    &+ \mathbbm{1}_{X\in\mathcal{X}^-}s(1,X)\bigg[\beta_{1,1}(X,1)\bigg(1-\frac{D}{m(X)}\bigg) - \beta_{0,0}(X,1-1/p_0(X))\bigg(1-\frac{1-D}{1-m(X)}\bigg)\bigg], 
\end{align*}} }
which yields variance difference
{\footnotesize \begin{align*}
    &E[\min\{s(0,X),s(1,X)\}]^{^2}(E[\psi^{[L]}_{\beta}(W)^2] - E[\tilde{\psi}^{[L]}_{\beta}(W)^2]) \\ 
    &= -{E\bigg[\mathbbm{1}_{\{X \in \mathcal{X}^+\}}s(0,X)^2 \bigg( \beta_{1,1}(X,p_0(X))\sqrt{\frac{1-m(X)}{m(X)}} - \beta_{0,0}(X,0)\sqrt{\frac{m(X)}{1-m(X)}}\bigg)^2 \bigg]} \\
    &\quad - {E\bigg[\mathbbm{1}_{\{X \in \mathcal{X}^-\}}s(1,X)^2 \bigg( \beta_{1,1}(X,1)\sqrt{\frac{1-m(X)}{m(X)}} - \beta_{0,0}(X,1-1/p_0(X))\sqrt{\frac{m(X)}{1-m(X)}}\bigg)^2 \bigg]}.
\end{align*}} 

\subsection{Proof of Theorem \ref{thm_TEST_properties1}} \label{app_sec_Theorem_test1}
Let the sample be partitioned into $K$ folds $I_1,\dots,I_K$, where $K\ge 2$ is fixed and $|I_k|=n_k$ with $n_k\asymp n$. 
The test statistic can be written as

{ \footnotesize \begin{align*}
T_n(\kappa_n)
=
\sum_{k=1}^K \frac{n_k}{n} T_{n,k}(\kappa_n), \quad \mbox{ with } \quad T_{n,k}(\kappa_n)
=
\frac1{n_k}\sum_{i\in I_k}1\{|\hat\Delta(X_i)|\le \kappa_n\}.
\end{align*} }

For any $\kappa>0$ and $i \neq 1,$ the following indicator inequalities hold:

{ \footnotesize  \begin{align}
1\{|\hat\Delta(X_i)|\le\kappa\}
&\le
1\{|\Delta(X_i)|\le2\kappa\}
+
1\{|\hat\Delta(X_i)-\Delta(X_i)|>\kappa\}, \label{eq:ineq1}\\
1\{\Delta(X_i)=0\}
&\le
1\{|\hat\Delta(X_i)|\le\kappa\}
+
1\{|\hat\Delta(X_i)-\Delta(X_i)|>\kappa\}. \label{eq:ineq2}
\end{align} }
We first prove the part on $H_0.$ Fix a fold $k$. Conditional on the training sample $\mathcal{D}_k = \{I_f : I_f \cap I_k = \emptyset \},$ the observations $\{X_i:i\in I_k\}$ are i.i.d. Averaging \eqref{eq:ineq1} over $i\in I_k$ and taking conditional expectation yields

{ \footnotesize \begin{align*}
E\!\left[T_{n,k}(\kappa_n)\mid \mathcal{D}_k \right]
&\le
P\big(|\Delta(X)|\le 2\kappa_n\big)
+
P\big(|\hat\Delta(X)-\Delta(X)|>\kappa_n \mid \mathcal{D}_k \big),
\end{align*} }
The first-stage rate implies
$
E[|\hat\Delta(X)-\Delta(X)| \mid \mathcal{D}_k ]\lesssim\lambda_{s,n,1},$
and Markov's inequality therefore yields
$P(|\hat\Delta-\Delta|>\kappa_n \mid \mathcal{D}_k)
\lesssim
\lambda_{s,n,1}/\kappa_n$. The near-margin condition implies
$P(|\Delta(X)|\le2\kappa)\lesssim\kappa,$
so
{ \footnotesize \begin{align*}
E[T_n(\kappa_n)\mid \mathcal{D}_k]
\lesssim
\kappa_n+\lambda_{s,n,1}/\kappa_n.
\end{align*} }
Next, conditional on $\mathcal{D}_k,$ the summands in $T_{n,k}(\kappa_n)$ are i.i.d. Bernoulli and bounded in $[0,1].$ Hence, Hoeffding's inequality yields, for any $t>0$,

{ \footnotesize \begin{align*}
P\!\left(|T_{n,k}(\kappa_n)-E[T_{n,k}(\kappa_n)\mid \mathcal{D}_k]|>t \mid \mathcal{D}_k\right)\le 2 e^{-2 n_k t^2}.
\end{align*} }
Since $K$ is fixed and $n_k\asymp n$, a union bound implies

{ \footnotesize \begin{align*}
P\!\left(
\max_{1\le k\le K}
\left|T_{n,k}(\kappa_n)-E[T_{n,k}(\kappa_n)\mid \mathcal{D}_k]\right|>t
\right)
\le
2K e^{-c n t^2}    
\end{align*} }
for some constant $c>0$. Taking $t=\sqrt{\log n/n},$ 
we obtain,

{ \footnotesize \begin{align*}
\max_{1\le k\le K}
\left|T_{n,k}(\kappa_n)-E[T_{n,k}(\kappa_n) \mid \mathcal{D}_k]\right|
=
O_p\!\left(\sqrt{\log n/n}\right),    
\end{align*} }
and therefore, under $H_0$,

{ \footnotesize \begin{align*}
T_n(\kappa_n)
=
O_p\!\left(
\kappa_n+\lambda_{s,n,1}/\kappa_n+\sqrt{\log n/n}
\right).    
\end{align*} }
Since $c_n=
M_n \left(\kappa_n+\lambda_{s,n,1}/\kappa_n+\sqrt{\log n/n}
\right),$ and $M_n\to\infty,$
it follows that $P\big(T_n(\kappa_n)>c_n\big)\to 0$. 
Next we prove the consistency part under $H_1$. Let $\pi_0=P(\mathcal X^0)>0$. Fix a fold $k$. Averaging \eqref{eq:ineq2} over $i\in I_k$ and taking conditional expectation given $\mathcal{D}_k$ yields

{ \footnotesize \begin{align*}
E[T_{n,k}(\kappa_n)\mid \mathcal{D}_k]\ge
\pi_0-P\big(|\hat\Delta(X)-\Delta(X)|>\kappa_n \mid \mathcal{D}_k \big).
\end{align*} }
By applying Markov's inequality again and noting $\lambda_{s,n,1}/\kappa_n\to 0,$ we obtain

{ \footnotesize \begin{align*}
E[T_{n,k}(\kappa_n)\mid \mathcal{D}_k]
\ge
\pi_0-o_p(1),
\end{align*} }
uniformly over $k=1,\dots,K$. Since $K$ is fixed, it follows that

{ \footnotesize \begin{align*}
\sum_{k=1}^K \frac{n_k}{n} E[T_{n,k}(\kappa_n)\mid \mathcal{D}_k]
\ge
\pi_0-o_p(1).
\end{align*} }
Combining this with the concentration bound established above, we obtain

{ \footnotesize \begin{align*}
T_n(\kappa_n)
=
\sum_{k=1}^K \frac{n_k}{n} E[T_{n,k}(\kappa_n)\mid \mathcal{D}_k]
+
O_p\!\left(\sqrt{\log n/n}\right),
\end{align*} }
and therefore

{ \footnotesize \begin{align*}
T_n(\kappa_n)
\ge
\pi_0-o_p(1)-O_p\!\left(\sqrt{\log n/n}\right).
\end{align*} }

Since $\sqrt{\log n/n}\to0$, we conclude that
{ \footnotesize \begin{align*}
T_n(\kappa_n)\ge \pi_0-o_p(1).
\end{align*} }
In particular, for any $\varepsilon\in(0,\pi_0)$,
$P\big(T_n(\kappa_n)\ge \pi_0-\varepsilon\big)\to 1.$ Because $c_n\to0$, we have 
$P\big(T_n(\kappa_n)\le c_n\big)
\to0,$
and therefore $P\big(T_n(\kappa_n)>c_n\big)\to1.$
This establishes consistency.

\section{Supplementary Material for Section \ref{sec_smoothingALL}}
\label{app:smoothing_rates}

This appendix shows how smooth bounds can remain regular along sequences $h\to0$ and derives the rate restriction on $h$ used in Section \ref{sec_smoothingSTAT}.
For fixed $h>0$, Theorem 5.2 and Theorem 6.2 imply that the smooth bounds $\beta_{B,h}(1,1)$ are pathwise differentiable and admit linear representations with influence functions
$\psi_{\beta_{B,h}}(W,\eta)$. Inspection of the proof of Theorem 5.2 shows that smoothing enters the influence function only through $g_{j,h}$ and $g'_{j,h}$. Since $\sup_{z,j}|g'_{j,h}(z)|\lesssim 1$ for the smooth approximations considered in Section 4.2, the first-order variance remains bounded along sequences $h\to0$. In particular, the irregularity when $P(\mathcal X^0)>0$ does not arise from variance explosion.

Let $M_h(\eta)=\psi_{\beta_{B,h}}(W,\eta)$. As shown in Appendix A.9, $M_h(\eta)$ can be written as a finite sum of finite products of nuisance-dependent factors. Relative to the non-smooth case, the only modification is that some factors take the form $g_{j,h}(z(\eta))$ or $g'_{j,h}(z(\eta))$, where $z(\eta)$ denotes a scalar nuisance index such as $p_0(X)$. For factors of the form $g_{j,h}(z(\hat\eta))$, the mean value theorem gives

{\footnotesize
\begin{align*}
|g_{j,h}(z(\hat\eta))-g_{j,h}(z(\eta))|
& \leq
\sup_{z}|g'_{j,h}(z)| \, |z(\hat\eta)-z(\eta)|, \\
|g'_{j,h}(z(\hat\eta))-g'_{j,h}(z(\eta))|
&\leq
\sup_z|g''_{j,h}(z)|\,|z(\hat\eta)-z(\eta)|.
\end{align*}}

For the smooth approximations considered in Section 4.2, as $h \to 0$,

{\footnotesize
\begin{align*}
\sup_{z,j}|g'_{j,h}(z)|\lesssim 1 \quad \mbox{ and } \quad \sup_{z,j}|g''_{j,h}(z)|\lesssim h^{-1}.
\end{align*}}
Applying the telescoping-product decomposition used in Appendix A.9 shows that
{\footnotesize \begin{align*}
  \sqrt{n}\sup_{\tilde{\eta} \in \mathcal{E}_n}|E[M_h(\tilde{\eta})-M_h({\eta})]|
  &\lesssim \frac{\sqrt{n}}{h}(\lambda_{q,n,2}^2 + \lambda_{s,n,2}^2 + \lambda_{m,n,2}^2 +\lambda_{b,n,2}^2), \\
  \sup_{\tilde\eta\in\mathcal E_n}
E[(M_h(\tilde\eta)-M_h(\eta))^2]
& \lesssim \frac{1}{h^2} (\lambda_{q,n,1}+\lambda_{s,n,1}+\lambda_{q,n,2}^2+\lambda_{s,n,2}^2+\lambda_{m,n,2}^2+\lambda_{b,n,2}^2).
\end{align*}
} 
Let $\lambda_n^2=:\lambda_{q,n,2}^2+\lambda_{s,n,2}^2+\lambda_{m,n,2}^2+\lambda_{b,n,2}^2+n^{-1/2}(\lambda_{q,n,1}^{1/2}+\lambda_{s,n,1}^{1/2})$ be the aggregated nuisance remainder rate. In view of the usual $\lambda_{a, n,2}=o_p(n^{-1/4})$ for 
$a \in \{q,s, m,b \}$, influence-function representation yields

{\footnotesize\begin{align*}
\sqrt{n}(\hat\beta_{B,h}(1,1)-\beta_{B,h}(1,1))
&=
\frac{1}{\sqrt{n}}\sum_{i=1}^n
\psi_{\beta_{B,h}(1,1)}(W_i,\eta)
+
O_p\!\left(\frac{\sqrt{n}\lambda_n^2}{h}\right)
+
o_p(1).
\end{align*}}
Therefore, root-$n$ asymptotic linearity at the smooth bounds requires $\sqrt n\,\lambda_n^2/h \to 0$. This also imposes $(\lambda_{q,n,1}^{1/2}+\lambda_{s,n,1}^{1/2})/h \to 0$ through the empirical-process remainder, so $L^1$-rates may be binding for $h$.
For example, if $\lambda_{a,n,1}\asymp \lambda_{a,n,2}\asymp n^{-1/4+\varepsilon}$ for $a\in\{q,s,m,b\}$, then the drift condition is weak, whereas the empirical-process term requires $h \gg n^{-1/8+\varepsilon/2}$, so the admissible smoothing rate is governed by the $L^1$-behavior.

\section{Appendix: Extensive Margin}
\subsection{Smooth Bounds} \label{sec_smoothBounds.EM}
The sharp bounds are in Proposition \ref{prop_identification1}. We now introduce the smooth bounds. For simplicity, we make the assumption that the support bounds do not depend on treatment status,
{ \footnotesize \begin{align*}
\underline y_1(x) &= \underline y_0(x) = \underline y(x), \\
\overline y_1(x) &= \overline y_0(x) = \overline y(x).
 \end{align*}}
This avoids the need to classify the sign of the conditional bounds as then, for all $x$,
{ \footnotesize \begin{align*}
\beta_L(x,em) \le 0 \le \beta_U(x,em).
 \end{align*}} 
Let $g_{1,h},g_{2,h},g_{3,h},g_{4,h}$ be the same smoothing functions from Section 4.2.1 satisfying
{ \footnotesize \begin{align*}
g_{1,h}(z) &\le \min\{z,1\} \le g_{3,h}(z) \leq 1, \\
0\leq g_{4,h}(z) &\le \max\{z,0\} \le g_{2,h}(z).
 \end{align*}}
We now introduce a new smoothing function $g_h$ that satisfies { \footnotesize \begin{align*}
    0 \le g_h(z) \leq \max\{z,0\} \mbox { and } |g_h'(z)| \leq C 
 \end{align*}}
This would be satisfied, for example, by { \footnotesize \begin{align*}
    g_h(z)=
\begin{cases}
0, & z\le 0,\\[4pt]
z\exp\!\left(-\dfrac{h^2}{z^2}\right), & z>0 .
\end{cases}
 \end{align*}}
where $C = 2 e^{-1/2}$. We also define the new $g_{3,h}(z) = 1- g_h(z)$ and $g_{4,h}(z) = 1 - g_h(1-z)$.  
We can now define smooth lower and upper envelopes for the extensive-margin share
{ \footnotesize \begin{align*}
\omega_h^L(x) &= g_{4,h}(\Delta(x)) + g_{4,h}(-\Delta(x)), \\
\omega_h^U(x) &= g_{2,h}(\Delta(x)) + g_{2,h}(-\Delta(x)),
 \end{align*}}
which assures that, for all $x$,
{ \footnotesize \begin{align*}
\omega_h^L(x) \le |\Delta(x)| \le \omega_h^U(x).
 \end{align*}}
For $B\in\{L,U\}$, we can now define the conditional outer bounds
{ \footnotesize \begin{align*}
\beta_{B,h}(x,em) = \beta_{B,h,1}(x,em) - \beta_{B,h,0}(x,em),
 \end{align*}}
where { \footnotesize \begin{align*}
\beta_{L,h,1}(x,em) &=
E[Y \mid S=1,D=1,X=x,\, Y\le q_1(1-g_{3,h}(p_0(x)),x)],
\\
\beta_{L,h,0}(x,em) &=
E[Y \mid S=1,D=0,X=x,\, Y\ge q_0(g_{3,h}(1/p_0(x)),x)],
\\
\beta_{U,h,1}(x,em) &=
E[Y \mid S=1,D=1,X=x,\, Y\ge q_1(g_{3,h}(p_0(x)),x)],
\\
\beta_{U,h,0}(x,em) &=
E[Y \mid S=1,D=0,X=x,\, Y\le q_0(1-g_{3,h}(1/p_0(x)),x)].
 \end{align*}}

We can now define the unconditional smooth bounds for the extensive margin as
{ \footnotesize \begin{align*}
\beta_{B,h}(em)
=
\frac{
E\!\left[\beta_{B,h}(X,em)\,\omega_h^U(X)\right]
}{
E\!\left[\omega_h^L(X)\right]
},
\qquad B\in\{L,U\}.
 \end{align*}}
These bounds are conservative for every fixed $h>0$:
{ \footnotesize \begin{align*}
\beta_{L,h}(em) &\le \beta_L(em), \\
\beta_{U,h}(em) &\ge \beta_U(em).
 \end{align*}}
Thus, $\beta_{L,h}(x,em)\le \beta_L(x,em)\le 0$ and 
$\beta_{U,h}(x,em)\ge \beta_U(x,em)\ge 0$ point-wise, while 
$\omega_h^U$ and $\omega_h^L$ are the respective upper and lower envelopes for $|\Delta(x)|$. Consequently, the ratio construction yields valid outer bounds.

\subsection{Influence Functions}
\subsubsection{Regular Bounds} We now present the influence function for $\beta_B(em)$ for the regular regime $P(\mathcal{X}^0) = 0$. Throughout, the support bounds are assumed to be known. Let $\Delta(x) = s(1,x) - s(0,x)$. The extensive margin bounds are given by

{ \footnotesize \begin{align*}
\beta_B(em)
=
\frac{E\!\left[\beta_B(X,em)|\Delta(X)|\right]}
{E[|\Delta(X)|]},
 \end{align*}}
for $B\in\{L,U\}$. The efficient influence function for $\beta_B(em)$ is then

{ \footnotesize \begin{align*}
\psi_{\beta_B(em)}(W)
=
\frac{M_{1B}+M_{2B}+M_{3B}+M_{4B}+M_{5B}}
{D_{em}} .
 \end{align*}}
where $D_{em}=E[|\Delta(X)|]$. The remaining are defined below. 
\subsubsection*{Auxiliary Influence Functions}
The influence function of $|\Delta(X)|$ is

{ \footnotesize \begin{align*}
\psi_{|\Delta(x)|}(W)
&=
1_{\mathcal X^+}(X)
\left[
\Delta(X)
+
\frac{D}{m(X)}(S-s(1,X))
-
\frac{1-D}{1-m(X)}(S-s(0,X))
\right] \\
&\quad
+
1_{\mathcal X^-}(X)
\left[
-\Delta(X)
+
\frac{1-D}{1-m(X)}(S-s(0,X))
-
\frac{D}{m(X)}(S-s(1,X))
\right].
 \end{align*}}
The influence function of $p_0(X)=s(0,X)/s(1,X)$ is

{ \footnotesize \begin{align*}
\psi_{p_0(x)}(W)
=
\frac{1}{s(1,X)}
\frac{1-D}{1-m(X)}(S-s(0,X))
-
\frac{p_0(X)}{s(1,X)}
\frac{D}{m(X)}(S-s(1,X)).
 \end{align*}}


\subsubsection*{Lower Extensive-Margin Bound}

The conditional lower bound is

{ \footnotesize \begin{align*}
\beta_L(X,em)
&=
1_{\mathcal X^+}(X)
\Big[
\beta_{1,1}(X,1-p_0(X))
-
\overline y_0(X)
\Big] 
+
1_{\mathcal X^-}(X)
\Big[
\underline y_1(X)
-
\beta_{0,0}(X,p_0(X)^{-1})
\Big].
 \end{align*}}
The EIF components are

{ \footnotesize \begin{align*}
M_{1L}
=&
1_{\mathcal X^+}(X)
\frac{SD}{m(X)}
\Big[
Y\,1\{Y\le q_1(1-p_0(X),X)\}
-
(1-p_0(X))\beta_{1,1}(X,1-p_0(X))
\Big],\\
M_{2L}
=&-1_{\mathcal X^-}(X)
\frac{S(1-D)}{1-m(X)}
\Big[
Y\,1\{Y\ge q_0(p_0(X)^{-1},X)\}
-
(1-p_0(X)^{-1})\beta_{0,0}(X,p_0(X)^{-1})
\Big],\\
M_{3L}
=&
(\beta_L(X,em)-\beta_L(em))\,\psi_{|\Delta(x)|}(W),\\
M_{4L}
=&
-
1_{\mathcal X^+}(X)
\frac{SD}{m(X)}
q_1(1-p_0(X),X)
\Big[
1\{Y\le q_1(1-p_0(X),X)\}
-
(1-p_0(X))
\Big] \\
&\quad
+
1_{\mathcal X^-}(X)
\frac{S(1-D)}{1-m(X)}
q_0(p_0(X)^{-1},X)
\Big[
1\{Y\ge q_0(p_0(X)^{-1},X)\}
-
(1-p_0(X)^{-1})
\Big],\\
M_{5L}
=&-
1_{\mathcal X^+}(X)
\Big[
q_1(1-p_0(X),X)-\beta_{1,1}(X,1-p_0(X))
\Big]
s(1,X)\psi_{p_0(x)}(W) \\
&\quad
+1_{\mathcal X^-}(X)
p_0(X)^{-1}
\Big[
q_0(p_0(X)^{-1},X)
-
\beta_{0,0}(X,p_0(X)^{-1})
\Big]
s(1,X)\psi_{p_0(x)}(W).
 \end{align*}}

\subsubsection*{Upper Extensive-Margin Bound}

The conditional upper bound is

{ \footnotesize \begin{align*}
\beta_U(X,em)
=
1_{\mathcal X^+}(X)
\Big[
\beta_{0,1}(X,p_0(X))
-
\underline y_0(X)
\Big] 
+
1_{\mathcal X^-}(X)
\Big[
\overline y_1(X)
-
\beta_{1,0}(X,1-p_0(X)^{-1})
\Big].
 \end{align*}}
The EIF components are

{ \footnotesize \begin{align*}
M_{1U}
=&
1_{\mathcal X^+}(X)
\frac{SD}{m(X)}
\Big[
Y\,1\{Y\ge q_1(p_0(X),X)\}
-
(1-p_0(X))\beta_{0,1}(X,p_0(X))
\Big],\\
M_{2U}
=&
-
1_{\mathcal X^-}(X)
\frac{S(1-D)}{1-m(X)}
\Big[
Y\,1\{Y\le q_0(1-p_0(X)^{-1},X)\}
-
(1-p_0(X)^{-1})\beta_{1,0}(X,1-p_0(X)^{-1})
\Big],\\
M_{3U}
=&
(\beta_U(X,em)-\beta_U(em))\,\psi_{|\Delta(x)|}(W), \\
M_{4U}
=&
-
1_{\mathcal X^+}(X)
\frac{SD}{m(X)}
q_1(p_0(X),X)
\Big[
1\{Y\ge q_1(p_0(X),X)\}
-
(1-p_0(X))
\Big] \\
&\quad
-1_{\mathcal X^-}(X)
\frac{S(1-D)}{1-m(X)}
q_0(1-p_0(X)^{-1},X)
\Big[
1\{Y\le q_0(1-p_0(X)^{-1},X)\}
-
(1-p_0(X)^{-1})
\Big],\\
M_{5U}
=&
-
1_{\mathcal X^+}(X)
\Big[
q_1(p_0(X),X)-\beta_{0,1}(X,p_0(X))
\Big]
s(1,X)\psi_{p_0(x)}(W) \\
&\quad
-
1_{\mathcal X^-}(X)
p_0(X)^{-1}
\Big[
q_0(1-p_0(X)^{-1},X)
-
\beta_{1,0}(X,1-p_0(X)^{-1})
\Big]
s(1,X)\psi_{p_0(x)}(W).
 \end{align*}}

\subsubsection{Smooth Bounds}

The smooth unconditional bounds are defined as

{ \footnotesize \begin{align*}
\beta_{B,h}(em)
=
\frac{E[\beta_{B,h}(X,em)\,\omega_h^U(X)]}
{E[\omega_h^L(X)]},
 \end{align*}} 
for $B \in \{L,U\}$. The efficient influence function of $\beta_{B,h}(em)$ is

{ \footnotesize \begin{align*}
\psi_{\beta_{B,h}(em)}(W)
=
\frac{M_{1B,h}+M_{2B,h}+M_{3B,h}+M_{4B,h}+M_{5B,h}
}{D_h},
 \end{align*}} with $D_h = E[\omega_h^L(X)]$. The remaining terms are provided below.

\subsubsection*{Auxiliary Influence Functions}
Define the influence function for the difference in selection probabilities

{ \footnotesize \begin{align*}
\psi_{\Delta(x)}
=
\frac{D}{m(X)}(S-s(1,X))
-
\frac{1-D}{1-m(X)}(S-s(0,X)).
 \end{align*}}
The influence function for $p_0(X)$ is

{ \footnotesize \begin{align*}
\psi_{p_0(x)}(W)
=
\frac{1}{s(1,X)}
\frac{1-D}{1-m(X)}(S-s(0,X))
-
\frac{p_0(X)}{s(1,X)}
\frac{D}{m(X)}(S-s(1,X)).
 \end{align*}}
The derivatives of the smooth extensive-margin weights are

{ \footnotesize \begin{align*}
\kappa_h^U(X)
&=
g'_{2,h}(\Delta(X))-g'_{2,h}(-\Delta(X)) \text{ and } 
\kappa_h^L(X)
=
g'_{4,h}(\Delta(X))-g'_{4,h}(-\Delta(X)).
 \end{align*}}

\subsubsection*{Lower Smooth Bound}
The conditional smooth lower bound is

{ \footnotesize \begin{align*}
\beta_{L,h}(X,em)
=
1_{\mathcal X^+}(X)
\Big[
\beta_{1,1}(X,1-g_{3,h}(p_0(X)))
-
\overline y_0(X)
\Big]
+
1_{\mathcal X^-}(X)
\Big[
\underline y_1(X)
-
\beta_{0,0}(X,g_{3,h}(p_0(X)^{-1}))
\Big].
 \end{align*}}
 
Thus

{ \footnotesize \begin{align*}
M_{1L,h}
=&1_{\mathcal X^+}(X) \frac{\omega_h^U(X)}{(1-g_{3,h}(p_0(X))) s(1,X)}
\frac{SD}{m(X)}
 \times \\ &\qquad \Big[
Y1\{Y\le q_{1}(1-g_{3,h}(p_0(X)),X)\}-
(1-g_{3,h}(p_0(X)))\beta_{1,1}(X, 1-g_{3,h}(p_0(X)))
\Big],\\
M_{2L,h}
=&-1_{\mathcal X^-}(X) \frac{\omega_h^U(X)}{(1-g_{3,h}(p_0(X)^{-1})) s(0,X)}
\frac{S(1-D)}{1-m(X)}
\times \\ &\qquad \Big[
Y1\{Y\ge q_{0}(g_{3,h}(p_0(X)^{-1}), X)\}
-
(1-g_{3,h}(p_0(X)^{-1}))\beta_{0,0}(X, g_{3,h}(p_0(X)^{-1}))
\Big],\\
M_{3L,h}=&\beta_{L,h}(X,em)\omega_h^U(X)
-
\beta_{L,h}(em)\omega_h^L(X)
+
(\beta_{L,h}(X,em)\kappa_h^U(X)
-
\beta_{L,h}(em)\kappa_h^L(X))\psi_{\Delta(x)}, \\
M_{4L,h}
=&
-1_{\mathcal X^+}(X) \frac{\omega_h^U(X)}{(1-g_{3,h}(p_0(X))) s(1,X)}
\frac{SD}{m(X)}
q_{1}(1-g_{3,h}(p_0(X)), X)
\times \\ &\qquad \Big[
1\{Y\le q_{1}(1-g_{3,h}(p_0(X)), X)\}-(1-g_{3,h}(p_0(X)))
\Big]\\
&
+ 1_{\mathcal X^-}(X)
\frac{\omega_h^U(X)}{(1-g_{3,h}(p_0(X)^{-1})) s(0,X)}
\frac{S(1-D)}{1-m(X)}
q_{0}(g_{3,h}(p_0(X)^{-1}), X)
\times \\ &\qquad \Big[
1\{Y\ge q_{0}(g_{3,h}(p_0(X)^{-1}), X)\}-(1-g_{3,h}(p_0(X)^{-1}))
\Big],\\
M_{5L,h}
=&-1_{\mathcal X^+}(X)\omega_h^U(X)
\left[\frac{(q_{1}((1-g_{3,h}(p_0(X))), X)-\beta_{1,1}(X,(1-g_{3,h}(p_0(X))))}{(1-g_{3,h}(p_0(X)))}
g'_{3,h}(p_0(X)) \right]
\psi_{p_0(x)}(W) \\
&+1_{\mathcal X^-}(X)\omega_h^U(X)\left[
\frac{(q_{0}(g_{3,h}(p_0(X)^{-1}), X)-\beta_{0,0}(X, g_{3,h}(p_0(X)^{-1}))}{(1-g_{3,h}(p_0(X)^{-1}))}
\frac{g'_{3,h}(p_0(X)^{-1})}{p_0(X)^2} \right]
\psi_{p_0(x)}(W).
 \end{align*}}

\subsubsection*{Upper Smooth Bound}

The conditional smooth upper bound is
{ \footnotesize \begin{align*}
\beta_{U,h}(X,em)
=
1_{\mathcal X^+}(X)
\Big[
\beta_{0,1}(X,g_{3,h}(p_0(X)))
-
\underline y_0(X)
\Big]
+
1_{\mathcal X^-}(X)
\Big[
\overline y_1(X)
-
\beta_{1,0}(X,1-g_{3,h}(p_0(X)^{-1}))
\Big].
 \end{align*}}

Thus
{ \footnotesize \begin{align*}
M_{1U,h}
=& 1_{\mathcal X^+}(X) \frac{\omega_h^U(X)}{(1-g_{3,h}(p_0(X))) s(1,X)}
\frac{SD}{m(X)}
\times \\ &\qquad \Big[
Y1\{Y\ge q_{1}(g_{3,h}(p_0(X)), X)\}
-
(1-g_{3,h}(p_0(X)))\beta_{0,1}(X,g_{3,h}(p_0(X)))
\Big],\\
M_{2U,h}
=&
-1_{\mathcal X^-}(X) \frac{\omega_h^U(X)}{(1-g_{3,h}(p_0(X)^{-1})) s(0,X)}
\frac{S(1-D)}{1-m(X)}
\times \\ &\qquad \Big[
Y1\{Y\le q_{0}(1-g_{3,h}(p_0(X)^{-1}), X)\}
-
(1-g_{3,h}(p_0(X)^{-1}))\beta_{1,0}(X,1-g_{3,h}(p_0(X)^{-1}))
\Big],\\
M_{3U,h}=&\beta_{U,h}(X,em)\omega_h^U(X)
-
\beta_{U,h}(em)\omega_h^L(X)
+
(\beta_{U,h}(X,em)\kappa_h^U(X)
-
\beta_{U,h}(em)\kappa_h^L(X))\psi_{\Delta(x)}, \\
M_{4U,h}
=&-1_{\mathcal X^+}(X) \frac{\omega_h^U(X)}{(1-g_{3,h}(p_0(X))) s(1,X)}
\frac{SD}{m(X)}
q_{1}(g_{3,h}(p_0(X)), X)
\times \\ &\qquad \Big[
1\{Y\ge q_{1}(g_{3,h}(p_0(X)), X)\}
-
(1-g_{3,h}(p_0(X)))
\Big]\\
&\quad
-1_{\mathcal X^-}(X) \frac{\omega_h^U(X)}{(1-g_{3,h}(p_0(X)^{-1})) s(0,X)}
\frac{S(1-D)}{1-m(X)}
q_{0}(1-g_{3,h}(p_0(X)^{-1}), X)
\times \\ &\qquad \Big[
1\{Y\le q_{0}(1-g_{3,h}(p_0(X)^{-1}), X)\}
-(1-g_{3,h}(p_0(X)^{-1}))
\Big],\\
M_{5U,h}
=&
-1_{\mathcal X^+}(X) \omega_h^U(X)
\frac{(q_{1}(g_{3,h}(p_0(X)), X)-\beta_{0,1}(X,g_{3,h}(p_0(X))))}{(1-g_{3,h}(p_0(X)))}
g'_{3,h}(p_0(X)) \psi_{p_0(x)}(W)\\
-
&1_{\mathcal X^-}(X) \omega_h^U(X)
\frac{(q_{0}(1-g_{3,h}(p_0(X)^{-1}), X)-\beta_{1,0}(X,1-g_{3,h}(p_0(X)^{-1})))}{(1-g_{3,h}(p_0(X)^{-1}))}
\frac{g'_{3,h}(p_0(X)^{-1})}{p_0(X)^2}
\psi_{p_0(x)}(W).
 \end{align*}}


\section{Supplementary Material for Section \ref{sec_empirical1}}
\label{app_JC}
\subsection{Data} \label{app_JC_data1}
We use an extended set of covariates containing detailed information regarding demographics, employment, criminal history, education, health, expectations, regional characteristics, and other JC related information similar to \cite{lee2009training}, \cite{flores2012estimating}, and \cite{heiler2023effect}. In particular, they contain all the pre-treatment covariates from  \cite{lee2009training} and \cite{heiler2023effect}, including all randomization variables. We removed all observations with a missing earnings or working hours entry at any week $t\in\{1,\dots,208\}$, i.e.~the analysis is conditional on that. All missing covariate entries were imputed by their mean in the final research sample as in \cite{lee2009training}. The missingness rates for covariates are relatively low for most variables other than earnings.

The Job Corps has varying treatment and control probabilities even conditional on the research sample based on gender ($FEMALE$), week of randomization ($RAND\_WK$), area concentration of nonresidential female students ($IN57$), and prediction regarding the residential status ($NONRES$), see \cite{burghardt1999national}. Based on this, we constructed treatment propensities for the research sample. We also constructed design weights to account for differences between research and national JC population for the impact evaluation. Table \ref{tab:research_probs1} contains the research sample propensities using the variables $RAND\_WK$, $FEMALE$, $IN57$, and $NONRES$ from the public use files. 

\begin{table}[h!] \footnotesize
	\centering \caption{Research Sample Treatment Propensities} \label{tab:research_probs1}
	\begin{tabular}{lc|c|c|c|c|c|c|c} \hline \hline  \\[-1ex]
		$RAND\_WK$ & \multicolumn{8}{c}{$< 41$}  \\
		$FEMALE$ & \multicolumn{4}{c|}{$1$} & \multicolumn{4}{c}{$0$}\\
		$IN57$ & \multicolumn{2}{c|}{$1$} &  \multicolumn{2}{c|}{$0$} & \multicolumn{2}{c|}{$1$} &  \multicolumn{2}{c}{$0$}   \\
		$NONRES$ & 1 & 0 & 1 & 0 & 1 & 0 & 1 & 0  \\ \hline 
		$P(D=1|X=x)$		&
		 $\frac{15.4}{8 + 15.4}$& $\frac{10.7}{8 + 10.7}$& $\frac{15.4}{5 + 15.4}$& $\frac{10.7}{5 + 10.7}$& $\frac{15.4}{8 + 15.4}$& $\frac{10.7}{8 + 10.7}$& $\frac{15.4}{8 + 15.4}$& $\frac{10.7}{8 + 10.7}$ \\ \hline \\[-1ex]
		 $RAND\_WK$ &  \multicolumn{8}{c}{$\geq 41$} \\
		 $FEMALE$ & \multicolumn{4}{c|}{$1$} & \multicolumn{4}{c}{$0$}  \\
		 $IN57$ & \multicolumn{2}{c|}{$1$} &  \multicolumn{2}{c|}{$0$} & \multicolumn{2}{c|}{$1$} &  \multicolumn{2}{c}{$0$}   \\
		 $NONRES$ & 1 & 0 & 1 & 0 & 1 & 0 & 1 & 0  \\ \hline 
		 $P(D=1|X=x)$		&
		 $\frac{17}{9 + 17}$& $\frac{11.1}{9 + 11.1 }$& $\frac{17}{5 + 17}$& $\frac{11.1}{5 + 11.1}$& $\frac{17}{9 + 17}$& $\frac{11.1}{9 + 11.1 }$& $\frac{17}{8 + 17}$& $\frac{11.1}{8 + 11.1 }$\\ \hline \hline 
	\end{tabular}
\end{table}

We make use of 77 pre-treatment covariates on top of the treatment assignment variable. Table \ref{tab:covariates_balance1} contains the average covariates conditional on observing the earnings and hours at all periods for treated and control groups. We also report the standardized mean differences as well as $p$-values for the two-sided hypothesis of equal means between treated and control. The sample is still well-balanced with the exception of worries to attend Job Corps (HADWORRY) that are higher in the treatment group.  

\begin{table}[!ht]
	\centering
	\begin{threeparttable}
		\caption{Job Corps Data: Covariates Balance} \label{tab:covariates_balance1} \tiny
	\begin{tabular}{rrrrrr}
		\hline
		& Treated & Control & std.diff & p-marginal & p-multiple \\ 
		\hline
		FEMALE & 0.4609 & 0.4662 & -0.0095 & 0.6630 & 1.0000 \\ 
		AGE & 19.2696 & 19.1506 & 0.0288 & 0.1958 & 1.0000 \\ 
		MARRIED & 0.0202 & 0.0234 & -0.0217 & 0.3185 & 1.0000 \\ 
		TOGETHER & 0.0394 & 0.0410 & -0.0081 & 0.7088 & 1.0000 \\ 
		SEPARATED & 0.0247 & 0.0216 & 0.0201 & 0.3479 & 1.0000 \\ 
		HASCHLD & 0.1932 & 0.1960 & -0.0071 & 0.7408 & 1.0000 \\ 
		NCHLD & 0.2731 & 0.2697 & 0.0053 & 0.8072 & 1.0000 \\ 
		HGC & 10.3163 & 10.2935 & 0.0091 & 0.6806 & 1.0000 \\ 
		HGC\_MOTH & 11.7133 & 11.6768 & 0.0113 & 0.6040 & 1.0000 \\ 
		HGC\_FATH & 11.6424 & 11.7148 & -0.0227 & 0.2987 & 1.0000 \\ 
		EVARRST & 0.2452 & 0.2448 & 0.0010 & 0.9636 & 1.0000 \\ 
		HH\_INC2 & 0.2115 & 0.2124 & -0.0028 & 0.8974 & 1.0000 \\ 
		HH\_INC3 & 0.1188 & 0.1165 & 0.0086 & 0.6877 & 1.0000 \\ 
		HH\_INC4 & 0.2489 & 0.2490 & -0.0001 & 0.9962 & 1.0000 \\ 
		HH\_INC5 & 0.1822 & 0.1838 & -0.0050 & 0.8170 & 1.0000 \\ 
		PERS\_INC2 & 0.1303 & 0.1334 & -0.0094 & 0.6613 & 1.0000 \\ 
		PERS\_INC3 & 0.0535 & 0.0473 & 0.0289 & 0.1748 & 1.0000 \\ 
		PERS\_INC4 & 0.0321 & 0.0346 & -0.0141 & 0.5126 & 1.0000 \\ 
		CURRJOB.y & 0.2015 & 0.1956 & 0.0143 & 0.5045 & 1.0000 \\ 
		MOSINJOB & 3.6672 & 3.5971 & 0.0157 & 0.4637 & 1.0000 \\ 
		YR\_WORK & 0.6484 & 0.6392 & 0.0180 & 0.4036 & 1.0000 \\ 
		EARN\_YR & 2962.0485 & 2868.9746 & 0.0168 & 0.4137 & 1.0000 \\ 
		HRSWK\_JR & 22.2540 & 21.3020 & 0.0433 & 0.0428 & 1.0000 \\ 
		WKEARNR & 113.2187 & 104.8441 & 0.0307 & 0.1151 & 1.0000 \\ 
		RACE\_W & 0.2717 & 0.2680 & 0.0080 & 0.7081 & 1.0000 \\ 
		RACE\_B & 0.5029 & 0.5006 & 0.0044 & 0.8383 & 1.0000 \\ 
		RACE\_H & 0.1727 & 0.1761 & -0.0087 & 0.6846 & 1.0000 \\ 
		RACE\_O & 0.0727 & 0.0738 & -0.0041 & 0.8502 & 1.0000 \\ 
		EDUC\_GR & 1.4706 & 1.4720 & -0.0016 & 0.9406 & 1.0000 \\ 
		LIVESPOU & 0.0605 & 0.0645 & -0.0164 & 0.4475 & 1.0000 \\ 
		EVERWORK & 0.8079 & 0.7935 & 0.0322 & 0.1354 & 1.0000 \\ 
		JOB0\_3 & 0.1912 & 0.1902 & 0.0026 & 0.9029 & 1.0000 \\ 
		JOB3\_9 & 0.2817 & 0.2741 & 0.0168 & 0.4322 & 1.0000 \\ 
		JOB9\_12 & 0.1756 & 0.1749 & 0.0017 & 0.9364 & 1.0000 \\ 
		MOSTWELF & 0.2065 & 0.2040 & 0.0060 & 0.7785 & 1.0000 \\ 
		GOT\_AFDC & 0.3199 & 0.3200 & -0.0002 & 0.9915 & 1.0000 \\ 
		GOT\_FS & 0.4564 & 0.4685 & -0.0234 & 0.2787 & 1.0000 \\ 
		ED0\_6 & 0.2782 & 0.2566 & 0.0474 & 0.0269 & 1.0000 \\ 
		ED6\_12 & 0.3603 & 0.3866 & -0.0528 & 0.0139 & 0.9599 \\ 
		PUBLICH & 0.2137 & 0.2018 & 0.0294 & 0.1693 & 1.0000 \\ 
		BADHLTH & 0.1315 & 0.1399 & -0.0239 & 0.2670 & 1.0000 \\ 
		HARDUSE & 0.0655 & 0.0625 & 0.0118 & 0.5811 & 1.0000 \\ 
		POTUSE & 0.2513 & 0.2411 & 0.0233 & 0.2760 & 1.0000 \\ 
		PMSA & 0.3282 & 0.3187 & 0.0201 & 0.3483 & 1.0000 \\ 
		MSA & 0.4655 & 0.4595 & 0.0116 & 0.5879 & 1.0000 \\ 
		HS\_D & 0.2000 & 0.1984 & 0.0040 & 0.8540 & 1.0000 \\ 
		GED\_D & 0.0486 & 0.0572 & -0.0375 & 0.0835 & 1.0000 \\ 
		VOC\_D & 0.0218 & 0.0197 & 0.0145 & 0.4970 & 1.0000 \\ 
		ANY\_ED1 & 0.6821 & 0.6815 & 0.0012 & 0.9540 & 1.0000 \\ 
		NTV\_LANG & 1.2063 & 1.2016 & 0.0085 & 0.6939 & 1.0000 \\ 
		R\_HEAD & 0.1156 & 0.1256 & -0.0309 & 0.1530 & 1.0000 \\ 
		HHMEMB & 3.0350 & 2.9214 & 0.0622 & 0.0039 & 0.2717 \\ 
		HEALTH & 1.7064 & 1.7137 & -0.0088 & 0.6838 & 1.0000 \\ 
		PY\_CIG & 0.5283 & 0.5072 & 0.0402 & 0.0601 & 1.0000 \\ 
		PY\_ALCHL & 0.5547 & 0.5314 & 0.0444 & 0.0382 & 1.0000 \\ 
		PY\_POT & 0.3075 & 0.2966 & 0.0231 & 0.2802 & 1.0000 \\ 
		HADWORRY & 0.3726 & 0.3339 & 0.0768 & 0.0003 & 0.0239 \\ 
		HEAR\_JC & 3.3126 & 3.2811 & 0.0179 & 0.4058 & 1.0000 \\ 
		KNEWCNTR & 0.5184 & 0.5334 & -0.0286 & 0.1833 & 1.0000 \\ 
		E\_MATH & 0.7051 & 0.6911 & 0.0282 & 0.1904 & 1.0000 \\ 
		E\_READ & 0.5454 & 0.5444 & 0.0020 & 0.9269 & 1.0000 \\ 
		E\_ALONG & 0.6160 & 0.6111 & 0.0094 & 0.6616 & 1.0000 \\ 
		E\_CONTRL & 0.5887 & 0.6040 & -0.0293 & 0.1720 & 1.0000 \\ 
		E\_ESTEEM & 0.5926 & 0.5922 & 0.0008 & 0.9706 & 1.0000 \\ 
		E\_SPCJOB & 0.9696 & 0.9743 & -0.0167 & 0.4437 & 1.0000 \\ 
		E\_FRIEND & 0.7220 & 0.7070 & 0.0305 & 0.1567 & 1.0000 \\ 
		KNEW\_JC & 0.6858 & 0.6884 & -0.0053 & 0.8064 & 1.0000 \\ 
		NONRES & 0.1514 & 0.1581 & -0.0206 & 0.3451 & 1.0000 \\ 
		PRARRI & 0.7502 & 0.7443 & 0.0263 & 0.2244 & 1.0000 \\ 
		RAND\_WK & 34.4925 & 34.9669 & -0.0285 & 0.1859 & 1.0000 \\ 
		IN57 & 0.3239 & 0.3298 & -0.0133 & 0.5311 & 1.0000 \\ 
		\hline
	\end{tabular} \footnotesize
    Columns 1 and 2 contain the covariate means in the treatment and control groups, respectively. Standardized differences are reported in Column 3. Unadjusted and multiple-testing-adjusted $p$-values for differences in means are reported in Columns 4 and 5, respectively
	The sample of $n = 9415$ is obtained by requiring non-missing weekly hours worked and weakly earnings for all weeks $t\in\{1,\dots,208\}$. Calculations use research sample and treatment propensity weights. 
\end{threeparttable}
\end{table}

\subsection{Estimation} \label{app_JC_estimation}
This section contains details on the estimation methods in Table \ref{tab:models_selection}. Logit is a logistic regression that contains all confounders. Logit interacted is a logistic regression that contains all confounders interacted with the treatment variable. XGBoost uses gradient-boosted trees with the binary logistic objective function, a maximum depth of $5$, a learning rate $\eta = 0.1$,  and cross-validated boosting steps. Neural Network is a two-hidden-layer, fully connected feedforward artificial neural network with rectified linear activation functions and a final sigmoid layer. Every layer has 0.2 dropout regularization. It is trained using binary cross-entropy with batch size 128 and 50 epochs. The latter were selected for the best cross-validation error in period $t = 208$. Random Forest uses honest probability forests using 1000 trees and default parameters in \verb|grf|.

\begin{figure}
	\centering \caption{Relative Performance Metrics for all Selection Models}
	\begin{subfigure}{0.48\textwidth}
		\includegraphics[width = \textwidth, trim = 0 100 0 100, clip]{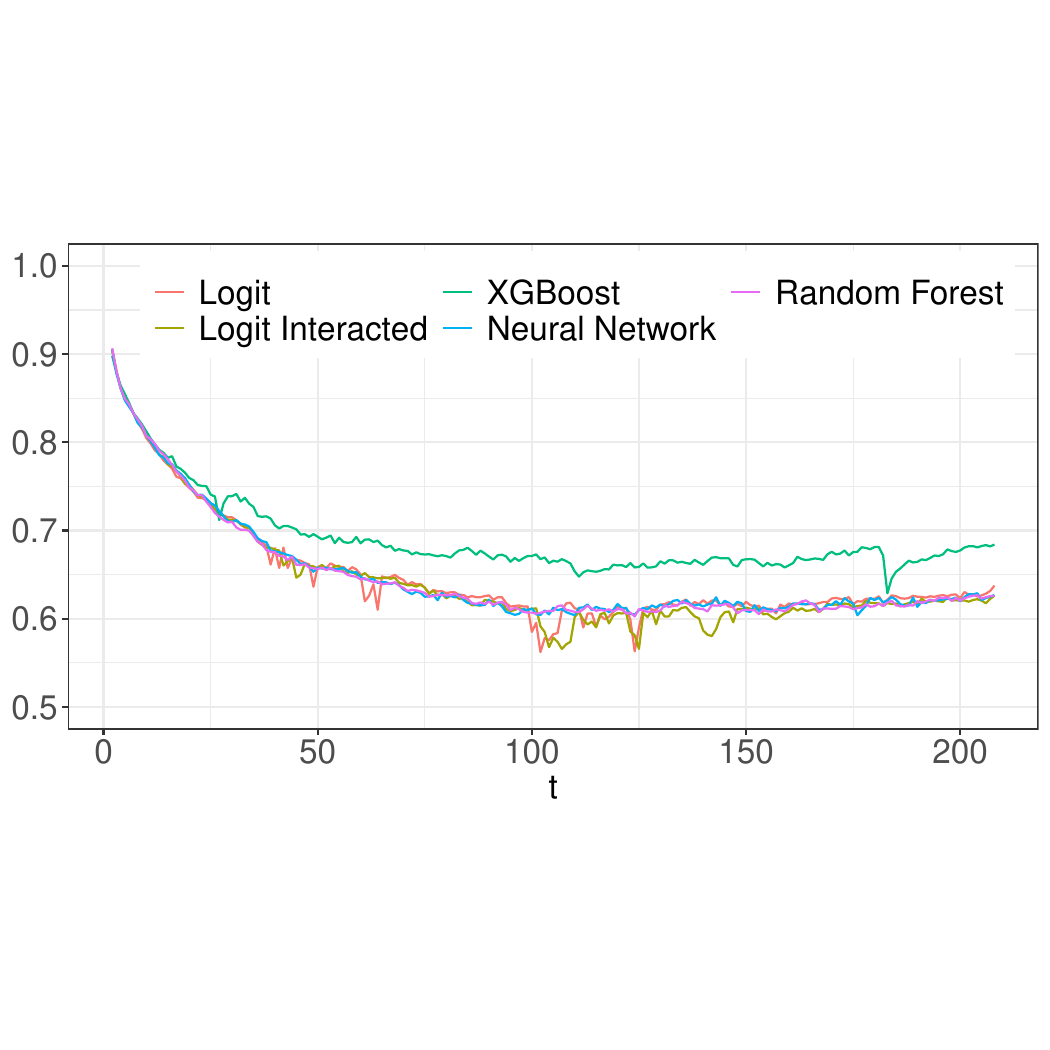}
		\caption{Accuracy} \label{fig_acc_all}
	\end{subfigure}
	\begin{subfigure}{0.48\textwidth}
		\includegraphics[width = \textwidth, trim = 0 100 0 100, clip]{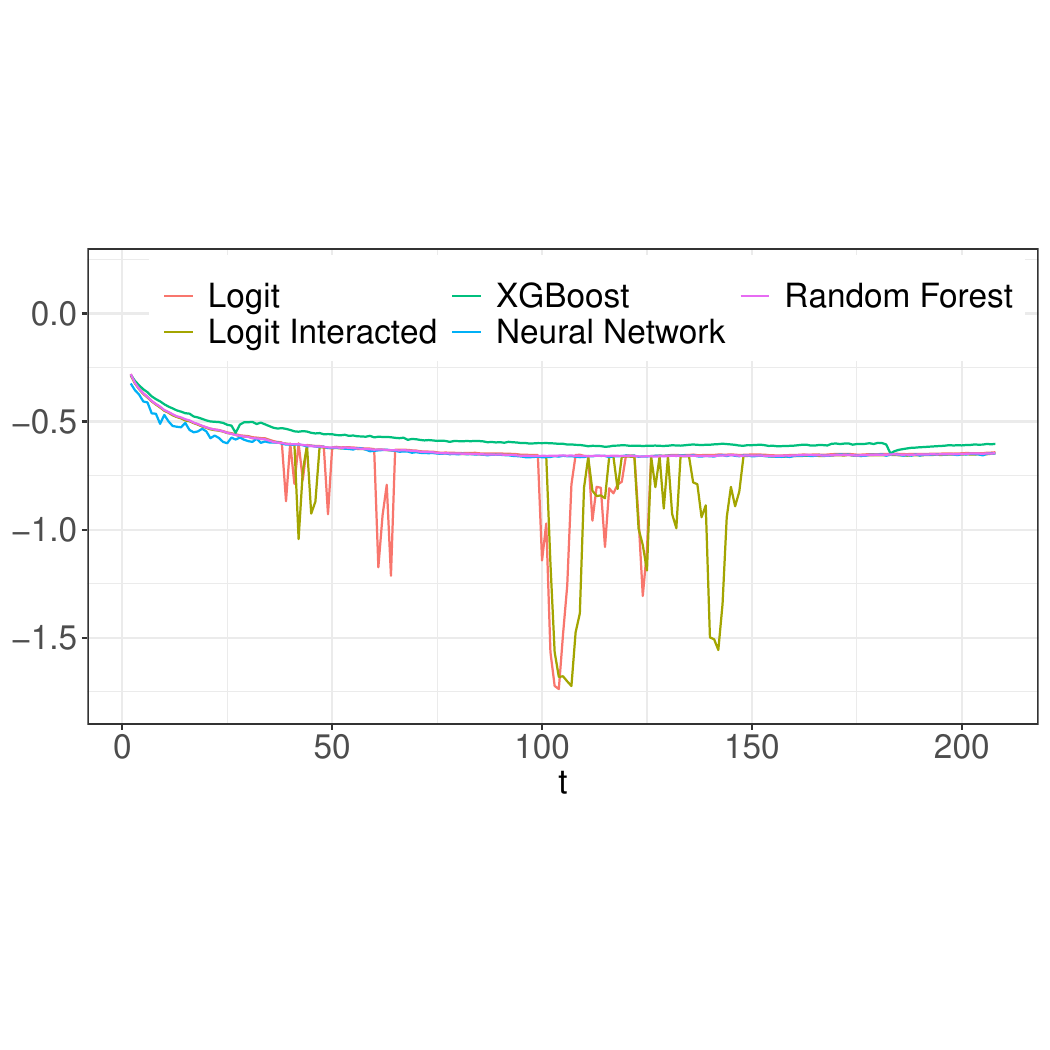}
		\caption{Log-likelihood} \label{fig_entropy_all}
	\end{subfigure}	
\end{figure}

Figures \ref{fig_acc_all} and \ref{fig_entropy_all} contain the accuracy and log-likelihoods for all periods $t\in\{1,\dots,208\}$. Higher accuracy and log-likelihood are better. One can see that XGBoost is the dominant model except for the very first period. The summary ranks over all periods are in Table \ref{tab:models_selection}.

For the conditional quantile models, we use honest quantile forests with 1000 trees and default parameters from \verb|grf|. The model was obtained using the fully selected sample and including the treatment indicator as well as all confounders.

For the truncated conditional expectations note that {\footnotesize \begin{align*}
    E\bigg[\frac{Y\mathbbm{1}(Y \leq q_{1}(p_0(X),X))}{p_0(X)}\bigg|SD=1,X=x\bigg] = E[Y|SD=1,Y\leq q_{1}(p_0(X),X)]
\end{align*}}
Thus, the truncated conditional expectations can be obtained from an auxiliary regression that uses the dependent variable $Y\mathbbm{1}(Y \leq q_1(p_0(X),X))/p_0(X)$ and regressors $X$ on the $S=D=1$ sample. The same applies equivalently to the other three truncated expectations. The specific models were estimated with a fully tuned regression forest. The unknown relative selection probabilities and quantiles were first estimated on the same training set on which the auxiliary model is estimated. This ensures that the models for the truncated conditional expectations preserve the same independence structure as conventional cross-fitting.

\subsection{Additional Results}\label{app_JC_results}

\subsubsection{Employment Effects}
\begin{figure}[!h]
	\centering \caption{Job Corps: Average Employment Effect Estimates} \label{fig_JC_impact_employment1}
 	\includegraphics[width = 0.6\textwidth, trim = 0 120 0 100, clip]{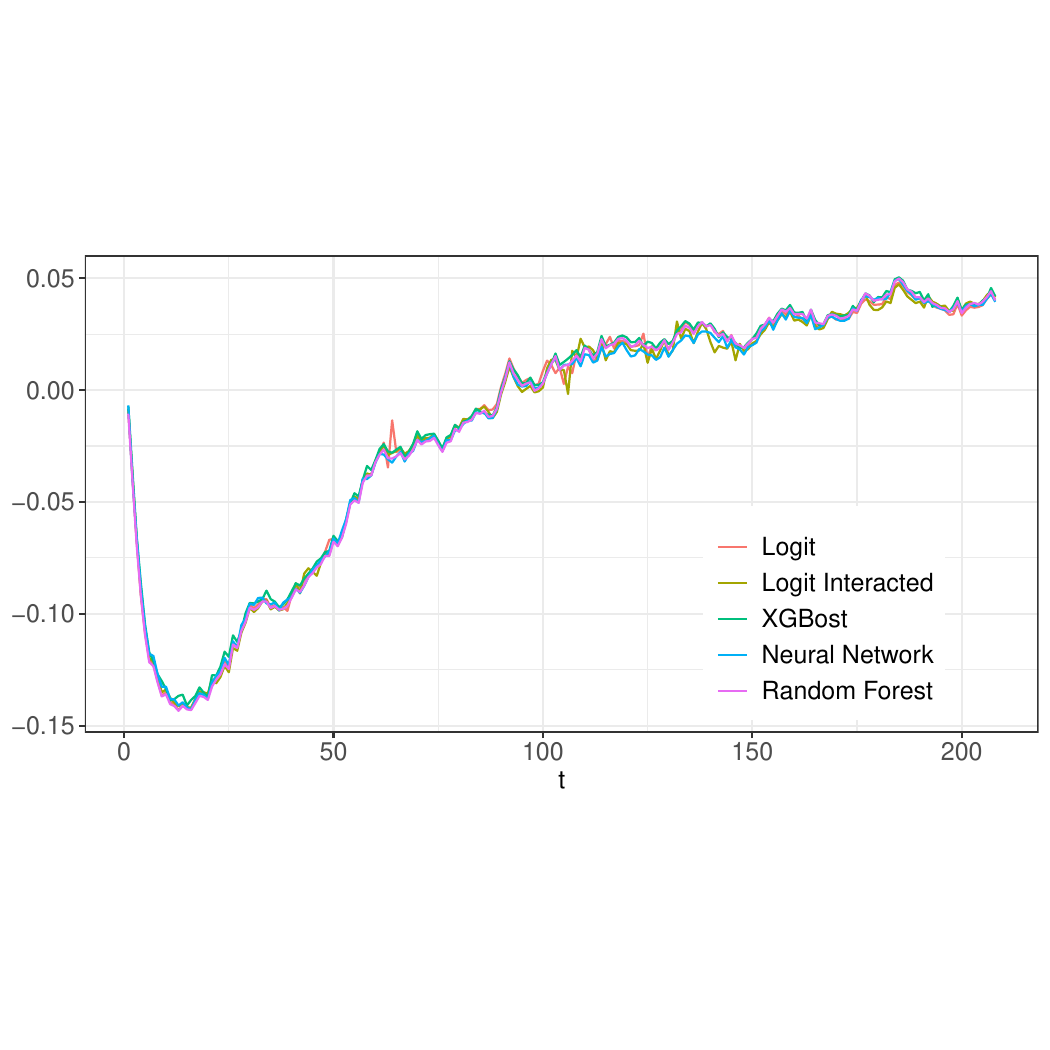} \footnotesize
  \begin{justify}
      Estimated average treatment effect of assignment to Job Corps on employment $E[S(1) - S(0)]$ using the efficient influence function with different selection probability models for $t\in\{1,\dots,208\}$. All calculations use nationally representative design weights. 
  \end{justify}
\end{figure}

Figure \ref{fig_JC_impact_employment1} contains the estimated average assignment effect on employment using the efficient estimator for all probability models. Estimates are almost identical. We replicate the finding that, in the short run, assignment reduces the likelihood of employment which only bounces back after about 1.5 years (the average duration of JC participation was around 8 months at the time). After about 2.25 years, positive employment effects tend to stabilize. In particular, most estimates are significantly negative for weeks 2 to 76 and significantly positive for weeks 116 to 208 ($p < 0.05$). 

\subsubsection{Relative Selection Probabilities} 

Figure \ref{fig:whole} contains the histograms of the estimated relative selection probabilities for all five post-intervention periods and all five estimation methods. 

\begin{figure}[!h]
	\centering
	\caption{Example Distributions for $\hat{p}_0(x)$}
	\label{fig:whole}
	\foreach \t in {45,90,135,180,208} { 
		\foreach \m in {1,2,3,4,5} {
			\begin{subfigure}{0.18\textwidth}
				\pgfkeysgetvalue{/captions/\m}{\captiontext}
				\caption{\captiontext, t=\t}
				\label{fig:sub\t}
				\includegraphics[width=\textwidth, clip]{v3_plot_p0x_m_\m_t_\t}
			\end{subfigure}
			\ifnum\m=5 
			\\ 
			\else
			\hfill
			\fi
		} 
	}
	\begin{justify} \footnotesize
		The figures contain histograms for the estimated $p_0(x)$ using five different estimation methods on the research sample at weeks $t \in \{ 45,90,135,180,208\}$. glm1 are logistic regressions with linear additive index. glm2 are fully treatment interacted logistic models. xgb are gradient-boosted trees with cross-validated boosting steps using \verb|xgboost|. nnet are batch-trained, fully connected artificial feed-forward neural network with three ReLU hidden layers, drop-out regularization, and final sigmoid layers using \verb|keras3|. grf are honest probability forests using \verb|grf| with default parameters. The spike at $p_0(x) = 1$ for glm2 at $t=45$ is due to non-convergence within some folds.
	\end{justify}  \label{app_JC_p0x_all}
\end{figure}

\subsubsection{Intensive Margin Effect Bounds} \label{app:JC_effect_bounds_all1}

Tables \ref{tab.IM:app1} -- \ref{tab.IM:app5} contain the intensive margin bound estimates for all periods and estimation methods.

{
\begin{table}[ht]
\centering
\caption{Job Corps Results: Bounds with Selection Model Logit} \label{tab.IM:app1}
\begingroup\scriptsize
\begin{tabular}{llccccc}
  \hline
  \hline
~ & ~ & $t = 45$ & $t = 90$ & $t = 135$ & $t = 180$ & $t = 208$\\[-0.5ex] ~ \\ 
  Smooth & $h = 5$ & $[-0.182,~0.223]$ & $[-0.133,~0.206]$ & $[-0.086,~0.149]$ & $[-0.096,~0.151]$ & $[-0.085,~0.171]$ \\ 
  ~ & ~ & $(-0.247,~0.263)$ & $(-0.182,~0.247)$ & $(-0.117,~0.181)$ & $(-0.123,~0.181)$ & $(-0.113,~0.201)$ \\ 
  ~ & ~ & ~ & ~ & ~ & ~ & ~ \\ 
  ~ & $h=1$ & $[-0.164,~0.214]$ & $[0.018,~0.081]$ & $[-0.015,~0.106]$ & $[-0.041,~0.111]$ & $[-0.030,~0.131]$ \\ 
  ~ & ~ & $(-0.232,~0.257)$ & $(-0.025,~0.123)$ & $(-0.054,~0.144)$ & $(-0.077,~0.148)$ & $(-0.067,~0.168)$ \\ 
  ~ & ~ & ~ & ~ & ~ & ~ & ~ \\ 
  ~ & $h=0.5$ & $[-3.601,~0.213]$ & $[-3.210,~0.064]$ & $[-0.043,~0.102]$ & $[-0.064,~0.110]$ & $[-0.062,~0.130]$ \\ 
  ~ & ~ & $(-3.660,~0.256)$ & $(-3.267,~0.108)$ & $(-0.089,~0.139)$ & $(-0.108,~0.147)$ & $(-0.106,~0.167)$ \\ 
  ~ & ~ & ~ & ~ & ~ & ~ & ~ \\ 
  Trim & Known & $[-0.135,~0.195]$ & $[0.052,~0.101]$ & $[-0.007,~0.121]$ & $[-0.034,~0.119]$ & $[-0.024,~0.142]$ \\ 
  ~ & ~ & $(-0.172,~0.237)$ & $(-0.021,~0.188)$ & $(-0.046,~0.155)$ & $(-0.071,~0.152)$ & $(-0.061,~0.176)$ \\ 
  ~ & ~ & ~ & ~ & ~ & ~ & ~ \\ 
  ~ & Unknown & $[-0.222,~0.112]$ & $[-0.052,-0.004]$ & $[-0.105,~0.041]$ & $[-0.112,~0.076]$ & $[-0.124,~0.068]$ \\ 
  ~ & ~ & $(-0.342,~0.227)$ & $(-0.187,0.140)$ & $(-0.214,~0.149)$ & $(-0.224,~0.188)$ & $(-0.239,~0.183)$ \\ 
  ~ & ~ & ~ & ~ & ~ & ~ & ~ \\ 
  Shift & Known & $[-0.582,-0.242]$ & $[0.237,~1.895]$ & $[0.464,~3.011]$ & $[0.541,~3.136]$ & $[0.566,~3.257]$ \\ 
  ~ & ~ & $(-0.828,-0.018)$ & $(-0.008,~2.084)$ & $(0.267,~3.136)$ & $(0.352,~3.255)$ & $(0.380,~3.373)$ \\ 
  ~ & ~ & ~ & ~ & ~ & ~ & ~ \\ 
  ~ & Unknown & $[-0.221,~0.113]$ & $[-0.023,-0.023]$ & $[-0.105,~0.042]$ & $[-0.111,~0.077]$ & $[-0.121,~0.070]$ \\ 
  ~ & ~ & $(-0.341,~0.229)$ & $(-0.202,~0.155)$ & $(-0.214,~0.150)$ & $(-0.223,~0.189)$ & $(-0.236,~0.185)$ \\ 
  ~ & ~ & ~ & ~ & ~ & ~ & ~ \\ 
  $|\hat{p}_0(x)-1|$ & $\leq n^{-1/4}/\log(n)$ & 4 & 5320 & 4 & 3 & 5 \\ 
  ~ & ~ & ~ & ~ & ~ & ~ & ~ \\ 
   \hline
\end{tabular}
\endgroup
\end{table}

\begin{table}[ht]
\centering
\caption{Job Corps Results: Bounds with Selection Model Logit Interacted} 
\begingroup\scriptsize
\begin{tabular}{llccccc}
  \hline
  \hline
~ & ~ & $t = 45$ & $t = 90$ & $t = 135$ & $t = 180$ & $t = 208$\\[-0.5ex] ~ \\ 
  Smooth & $h = 5$ & $[-12.082,~93.473]$ & $[-0.351,~0.204]$ & $[-4.914,~42.535]$ & $[-2.766,~1.120]$ & $[-2.146,~0.529]$ \\ 
  ~ & ~ & $(-13.379,~99.367)$ & $(-1.610,~1.484)$ & $(-6.244,~44.915)$ & $(-3.312,~1.581)$ & $(-2.314,~0.634)$ \\ 
  ~ & ~ & ~ & ~ & ~ & ~ & ~ \\ 
  ~ & $h=1$ & $[-13.300,~86.263]$ & $[-0.394,~0.276]$ & $[-6.885,~54.294]$ & $[-2.963,~1.066]$ & $[-2.105,~0.459]$ \\ 
  ~ & ~ & $(-14.801,~91.929)$ & $(-1.651,~1.554)$ & $(-8.329,~57.852)$ & $(-3.513,~1.536)$ & $(-2.273,~0.565)$ \\ 
  ~ & ~ & ~ & ~ & ~ & ~ & ~ \\ 
  ~ & $h=0.5$ & $[-13.750,~100.822]$ & $[-2.167,~1.668]$ & $[-6.267,~56.598]$ & $[-3.668,~3.270]$ & $[-2.940,~0.449]$ \\ 
  ~ & ~ & $(-15.404,~107.023)$ & $(-4.717,~4.178)$ & $(-7.643,~60.231)$ & $(-4.228,~3.732)$ & $(-3.111,~0.556)$ \\ 
  ~ & ~ & ~ & ~ & ~ & ~ & ~ \\ 
  Trim & Known & $[-0.222,~0.296]$ & $[0.006,~0.109]$ & $[-0.211,~0.316]$ & $[0.025,~0.076]$ & $[0.000,~0.101]$ \\ 
  ~ & ~ & $(-0.271,~0.348)$ & $(-0.035,~0.152)$ & $(-0.254,~0.363)$ & $(-0.011,~0.114)$ & $(-0.036,~0.140)$ \\ 
  ~ & ~ & ~ & ~ & ~ & ~ & ~ \\ 
  ~ & Unknown & $[-0.380,~0.228]$ & $[-0.084,~0.039]$ & $[-0.370,~0.372]$ & $[-0.045,~0.021]$ & $[-0.092,~0.037]$ \\ 
  ~ & ~ & $(-0.498,~0.344)$ & $(-0.190,~0.146)$ & $(-0.483,~0.486)$ & $(-0.150,~0.128)$ & $(-0.200,~0.146)$ \\ 
  ~ & ~ & ~ & ~ & ~ & ~ & ~ \\ 
  Shift & Known & $[-0.558,~0.416]$ & $[0.196,~1.677]$ & $[1.676,~2.059]$ & $[0.475,~2.117]$ & $[0.473,~2.293]$ \\ 
  ~ & ~ & $(-1.061,~0.720)$ & $(-0.019,~1.850)$ & $(0.096,~2.748)$ & $(0.299,~2.248)$ & $(0.298,~2.423)$ \\ 
  ~ & ~ & ~ & ~ & ~ & ~ & ~ \\ 
  ~ & Unknown & $[-0.400,~0.141]$ & $[-0.115,-0.010]$ & $[-0.177,~0.335]$ & $[-0.050,~0.063]$ & $[-0.057,~0.047]$ \\ 
  ~ & ~ & $(-0.524,~0.261)$ & $(-0.225,0.101)$ & $(-0.285,~0.442)$ & $(-0.155,~0.169)$ & $(-0.165,~0.157)$ \\ 
  ~ & ~ & ~ & ~ & ~ & ~ & ~ \\ 
  $|\hat{p}_0(x)-1|$ & $\leq n^{-1/4}/\log(n)$ & 1255 & 422 & 2749 & 467 & 510 \\ 
  ~ & ~ & ~ & ~ & ~ & ~ & ~ \\ 
   \hline
\end{tabular}
\endgroup
\end{table}
\begin{table}[ht]
\centering
\caption{Job Corps Results: Bounds with Selection Model XGBoost} 
\begingroup\scriptsize
\begin{tabular}{llccccc}
  \hline
  \hline
~ & ~ & $t = 45$ & $t = 90$ & $t = 135$ & $t = 180$ & $t = 208$\\[-0.5ex] ~ \\ 
  Smooth & $h = 5$ & $[-0.109,~0.145]$ & $[-0.112,~0.212]$ & $[-0.093,~0.213]$ & $[-0.072,~0.215]$ & $[-0.057,~0.275]$ \\ 
  ~ & ~ & $(-0.172,~0.199)$ & $(-0.161,~0.256)$ & $(-0.139,~0.250)$ & $(-0.107,~0.251)$ & $(-0.095,~0.312)$ \\ 
  ~ & ~ & ~ & ~ & ~ & ~ & ~ \\ 
  ~ & $h=1$ & $[-0.000,~0.161]$ & $[0.008,~0.096]$ & $[-0.028,~0.090]$ & $[-0.029,~0.093]$ & $[-0.005,~0.113]$ \\ 
  ~ & ~ & $(-0.068,~0.209)$ & $(-0.036,~0.135)$ & $(-0.066,~0.125)$ & $(-0.064,~0.128)$ & $(-0.040,~0.148)$ \\ 
  ~ & ~ & ~ & ~ & ~ & ~ & ~ \\ 
  ~ & $h=0.5$ & $[-3.641,~3.045]$ & $[-3.224,~0.099]$ & $[-0.078,~0.058]$ & $[-0.093,~0.069]$ & $[-0.091,~0.064]$ \\ 
  ~ & ~ & $(-3.705,~3.101)$ & $(-3.275,~0.141)$ & $(-0.126,~0.096)$ & $(-0.141,~0.105)$ & $(-0.139,~0.100)$ \\ 
  ~ & ~ & ~ & ~ & ~ & ~ & ~ \\ 
  Trim & Known & $[-0.138,~0.160]$ & $[0.005,~0.014]$ & $[-0.016,~0.054]$ & $[-0.022,~0.087]$ & $[-0.008,~0.054]$ \\ 
  ~ & ~ & $(-0.194,~0.217)$ & $(-0.051,~0.081)$ & $(-0.072,~0.106)$ & $(-0.074,~0.140)$ & $(-0.063,~0.111)$ \\ 
  ~ & ~ & ~ & ~ & ~ & ~ & ~ \\ 
  ~ & Unknown & $[-0.257,~0.126]$ & $[-0.175,~0.004]$ & $[-0.021,~0.152]$ & $[-0.180,-0.011]$ & $[-0.036,~0.077]$ \\ 
  ~ & ~ & $(-0.403,~0.269)$ & $(-0.745,~0.573)$ & $(-0.197,~0.327)$ & $(-0.328,0.134)$ & $(-0.201,~0.245)$ \\ 
  ~ & ~ & ~ & ~ & ~ & ~ & ~ \\ 
  Shift & Known & $[-0.756,~0.091]$ & $[0.238,~0.392]$ & $[0.549,~1.266]$ & $[0.559,~1.911]$ & $[0.646,~1.514]$ \\ 
  ~ & ~ & $(-1.085,~0.303)$ & $(-0.093,~0.542)$ & $(0.291,~1.396)$ & $(0.368,~2.052)$ & $(0.438,~1.650)$ \\ 
  ~ & ~ & ~ & ~ & ~ & ~ & ~ \\ 
  ~ & Unknown & $[-0.324,-0.076]$ & $[-0.021,-0.011]$ & $[0.022,~0.067]$ & $[0.075,~0.122]$ & $[0.138,~0.162]$ \\ 
  ~ & ~ & $(-0.470,~0.067)$ & $(-0.194,0.162)$ & $(-0.116,~0.204)$ & $(-0.055,~0.253)$ & $(-0.003,~0.303)$ \\ 
  ~ & ~ & ~ & ~ & ~ & ~ & ~ \\ 
  $|\hat{p}_0(x)-1|$ & $\leq n^{-1/4}/\log(n)$ & 2592 & 8680 & 6350 & 4871 & 5688 \\ 
  ~ & ~ & ~ & ~ & ~ & ~ & ~ \\ 
   \hline
\end{tabular}
\endgroup
\end{table}
\begin{table}[ht]
\centering
\caption{Job Corps Results: Bounds with Selection Model Neural Network} 
\begingroup\scriptsize
\begin{tabular}{llccccc}
  \hline
  \hline
~ & ~ & $t = 45$ & $t = 90$ & $t = 135$ & $t = 180$ & $t = 208$\\[-0.5ex] ~ \\ 
  Smooth & $h = 5$ & $[-0.177,~0.089]$ & $[-0.137,~0.214]$ & $[-0.089,~0.187]$ & $[-0.092,~0.182]$ & $[-0.058,~0.228]$ \\ 
  ~ & ~ & $(-0.234,~0.137)$ & $(-0.183,~0.255)$ & $(-0.132,~0.220)$ & $(-0.125,~0.215)$ & $(-0.094,~0.263)$ \\ 
  ~ & ~ & ~ & ~ & ~ & ~ & ~ \\ 
  ~ & $h=1$ & $[-0.469,~0.777]$ & $[-0.005,~0.105]$ & $[-0.023,~0.107]$ & $[-0.045,~0.112]$ & $[-0.022,~0.127]$ \\ 
  ~ & ~ & $(-0.543,~0.855)$ & $(-0.049,~0.147)$ & $(-0.061,~0.144)$ & $(-0.080,~0.149)$ & $(-0.058,~0.165)$ \\ 
  ~ & ~ & ~ & ~ & ~ & ~ & ~ \\ 
  ~ & $h=0.5$ & $[-4.014,~3.815]$ & $[-3.201,~0.103]$ & $[-0.186,~0.182]$ & $[-0.213,~0.210]$ & $[-0.091,~0.096]$ \\ 
  ~ & ~ & $(-4.089,~3.906)$ & $(-3.256,~0.145)$ & $(-0.250,~0.223)$ & $(-0.303,~0.282)$ & $(-0.140,~0.132)$ \\ 
  ~ & ~ & ~ & ~ & ~ & ~ & ~ \\ 
  Trim & Known & $[-0.252,~0.287]$ & $[0.050,~0.082]$ & $[-0.014,~0.132]$ & $[-0.036,~0.133]$ & $[-0.035,~0.187]$ \\ 
  ~ & ~ & $(-0.315,~0.351)$ & $(-0.038,~0.169)$ & $(-0.076,~0.192)$ & $(-0.090,~0.191)$ & $(-0.091,~0.258)$ \\ 
  ~ & ~ & ~ & ~ & ~ & ~ & ~ \\ 
  ~ & Unknown & $[-0.347,~0.192]$ & $[-0.031,~0.003]$ & $[-0.107,~0.046]$ & $[-0.141,~0.038]$ & $[-0.157,~0.072]$ \\ 
  ~ & ~ & $(-0.481,~0.319)$ & $(-0.179,~0.149)$ & $(-0.223,~0.164)$ & $(-0.266,~0.159)$ & $(-0.284,~0.203)$ \\ 
  ~ & ~ & ~ & ~ & ~ & ~ & ~ \\ 
  Shift & Known & $[-0.689,-0.033]$ & $[0.262,~2.277]$ & $[0.476,~2.819]$ & $[0.569,~2.947]$ & $[0.585,~3.080]$ \\ 
  ~ & ~ & $(-0.958,0.189)$ & $(0.023,~2.462)$ & $(0.283,~2.948)$ & $(0.385,~3.070)$ & $(0.402,~3.200)$ \\ 
  ~ & ~ & ~ & ~ & ~ & ~ & ~ \\ 
  ~ & Unknown & $[-0.424,-0.122]$ & $[-0.054,~0.011]$ & $[0.049,~0.108]$ & $[-0.062,~0.076]$ & $[0.046,~0.121]$ \\ 
  ~ & ~ & $(-0.571,~0.019)$ & $(-0.205,~0.162)$ & $(-0.078,~0.235)$ & $(-0.187,~0.199)$ & $(-0.084,~0.252)$ \\ 
  ~ & ~ & ~ & ~ & ~ & ~ & ~ \\ 
  $|\hat{p}_0(x)-1|$ & $\leq n^{-1/4}/\log(n)$ & 737 & 4842 & 3220 & 2180 & 3669 \\ 
  ~ & ~ & ~ & ~ & ~ & ~ & ~ \\ 
   \hline
\end{tabular}
\endgroup
\end{table}
\begin{table}[ht]
\centering 
\caption{Job Corps Results: Bounds with Selection Model Random Forest} \label{tab.IM:app5}
\begingroup\scriptsize
\begin{tabular}{llccccc}
  \hline
  \hline
~ & ~ & $t = 45$ & $t = 90$ & $t = 135$ & $t = 180$ & $t = 208$\\[-0.5ex] ~ \\ 
  Smooth & $h = 5$ & $[-0.126,~0.112]$ & $[-0.141,~0.214]$ & $[-0.081,~0.198]$ & $[-0.076,~0.222]$ & $[-0.074,~0.264]$ \\ 
  ~ & ~ & $(-0.187,~0.160)$ & $(-0.190,~0.256)$ & $(-0.129,~0.234)$ & $(-0.112,~0.259)$ & $(-0.113,~0.303)$ \\ 
  ~ & ~ & ~ & ~ & ~ & ~ & ~ \\ 
  ~ & $h=1$ & $[-0.086,~0.185]$ & $[0.030,~0.075]$ & $[-0.022,~0.096]$ & $[-0.011,~0.083]$ & $[0.016,~0.098]$ \\ 
  ~ & ~ & $(-0.156,~0.237)$ & $(-0.015,~0.114)$ & $(-0.062,~0.132)$ & $(-0.047,~0.118)$ & $(-0.019,~0.135)$ \\ 
  ~ & ~ & ~ & ~ & ~ & ~ & ~ \\ 
  ~ & $h=0.5$ & $[-3.666,~3.805]$ & $[-3.182,~3.047]$ & $[-0.071,~0.076]$ & $[-0.082,~0.050]$ & $[-0.061,~0.064]$ \\ 
  ~ & ~ & $(-3.731,~3.866)$ & $(-3.234,~3.091)$ & $(-0.122,~0.116)$ & $(-0.132,~0.086)$ & $(-0.109,~0.101)$ \\ 
  ~ & ~ & ~ & ~ & ~ & ~ & ~ \\ 
  Trim & Known & $[-0.162,~0.202]$ & $[0.043,~0.100]$ & $[-0.024,~0.138]$ & $[0.013,~0.121]$ & $[0.036,~0.120]$ \\ 
  ~ & ~ & $(-0.233,~0.273)$ & $(-0.037,~0.189)$ & $(-0.105,~0.209)$ & $(-0.055,~0.192)$ & $(-0.033,~0.196)$ \\ 
  ~ & ~ & ~ & ~ & ~ & ~ & ~ \\ 
  ~ & Unknown & $[-0.248,~0.116]$ & $[-0.038,~0.019]$ & $[-0.114,~0.054]$ & $[-0.074,~0.034]$ & $[-0.051,~0.044]$ \\ 
  ~ & ~ & $(-0.383,~0.244)$ & $(-0.173,~0.159)$ & $(-0.246,~0.178)$ & $(-0.202,~0.160)$ & $(-0.181,~0.177)$ \\ 
  ~ & ~ & ~ & ~ & ~ & ~ & ~ \\ 
  Shift & Known & $[-0.679,-0.047]$ & $[0.213,~1.646]$ & $[0.446,~2.178]$ & $[0.553,~2.429]$ & $[0.542,~2.401]$ \\ 
  ~ & ~ & $(-0.958,0.177)$ & $(-0.028,~1.829)$ & $(0.247,~2.325)$ & $(0.367,~2.571)$ & $(0.357,~2.540)$ \\ 
  ~ & ~ & ~ & ~ & ~ & ~ & ~ \\ 
  ~ & Unknown & $[-0.374,-0.110]$ & $[-0.029,-0.028]$ & $[0.088,~0.090]$ & $[0.139,~0.145]$ & $[0.156,~0.165]$ \\ 
  ~ & ~ & $(-0.524,~0.033)$ & $(-0.207,~0.150)$ & $(-0.066,~0.244)$ & $(-0.012,~0.295)$ & $(0.005,~0.315)$ \\ 
  ~ & ~ & ~ & ~ & ~ & ~ & ~ \\ 
  $|\hat{p}_0(x)-1|$ & $\leq n^{-1/4}/\log(n)$ & 1052 & 6848 & 6420 & 6211 & 6342 \\ 
  ~ & ~ & ~ & ~ & ~ & ~ & ~ \\ 
   \hline
\end{tabular}
\endgroup
\end{table}
}

\clearpage

\subsubsection{Extensive Margin: Effect Bounds} \label{sec_app_EXTENSIVE}
This section reports the estimated extensive margin share and bounds on the effect of assignment to Job Corps on log hourly earnings in weeks 90, 135, 180 and 208 using $n=9415$ observations and design weights. ``Smooth'' refers to the bounds developed in this paper, while ``Regular'' treats the problem as if the target parameters $\beta_L(em)$ and $beta_U(em)$ were differentiable. The associated confidence intervals for the regular bounds are therefore invalid. Identified sets are reported in brackets, with 95\% confidence \cite{stoye2020simple} intervals in parentheses below. Smoothing parameters are scaled by 100. 


\begin{table}[!h] 
\centering
\caption{Extensive Margin in Job Corps: Week 90}
\label{tab_JC.EM_week90}
\begin{threeparttable}
\footnotesize
\begin{tabular*}{\linewidth}{@{\extracolsep{\fill}} lcccc}
\hline \hline
Method & Smoothing & Parameter & \multicolumn{2}{c}{Nuisance Model} \\ 
&&& XGBoost & Neural Network \\ 
\cline{4-5}
\\[-1ex]
Smooth & $h=1$ & $\beta_{B,h}(em)$ & [-0.679, -0.583] & [-1.672, 1.605] \\
       &       &                   & (-3.684, 6.316) & (-1.973, 3.574) \\[1ex]
       &       & $\pi_{em,h}$      & 0.001 & 0.024 \\
       &       &                   & (0.000, 0.001) & (0.023, 0.026) \\[2ex]
       & $h=5$ & $\beta_{B,h}(em)$ & [-4.418, 5.184] & [-19.237, 4.317] \\
       &       &                   & (-24.235, 148.985) & (-22.917, 8.656) \\[1ex]
       &       & $\pi_{em,h}$      & 0.000 & 0.014 \\
       &       &                   & (0.000, 0.000) & (0.013, 0.015) \\[2ex]
Regular &       & $\beta_{B}(em)$  & [-6.449, -1.065] & [-2.698, 1.019] \\
       &       &                   & (-24.782, 53.994) & (-6.055, 2.342) \\[1ex]
       &       & $\pi_{em}$        & 0.001 & 0.016 \\
       &       &                   & (-0.004, 0.005) & (-0.005, 0.037) \\ 
\hline
\end{tabular*}
\end{threeparttable}
\end{table}

\begin{table}[!h] 
\centering
\caption{Extensive Margin in Job Corps: Week 135}
\label{tab_JC.EM_week135}
\begin{threeparttable}
\footnotesize
\begin{tabular*}{\linewidth}{@{\extracolsep{\fill}} lcccc}
\hline \hline
Method & Smoothing & Parameter & \multicolumn{2}{c}{Nuisance Model} \\ 
&&& XGBoost & Neural Network \\ 
\cline{4-5}
\\[-1ex]
Smooth & $h=1$ & $\beta_{B,h}(em)$ & [-1.725, 5.783] & [-0.846, 2.540] \\
       &       &                   & (-2.792, 16.296) & (-1.380, 8.343) \\[1ex]
       &       & $\pi_{em,h}$      & 0.006 & 0.016 \\
       &       &                   & (0.006, 0.007) & (0.015, 0.017) \\[2ex]
       & $h=5$ & $\beta_{B,h}(em)$ & [-55.823, -17.345] & [-20.028, 18.471] \\
       &       &                   & (-83.435, 362.058) & (-22.477, 47.454) \\[1ex]
       &       & $\pi_{em,h}$      & 0.001 & 0.008 \\
       &       &                   & (0.001, 0.001) & (0.008, 0.009) \\[2ex]
Regular &       & $\beta_{B}(em)$  & [-4.960, 5.194] & [-2.566, 1.676] \\
       &       &                   & (-18.389, 14.376) & (-3.601, 2.392) \\[1ex]
       &       & $\pi_{em}$        & 0.005 & 0.025 \\
       &       &                   & (-0.012, 0.021) & (0.003, 0.046) \\ 
\hline
\end{tabular*}
\end{threeparttable}
\end{table}

\begin{table}[!h] 
\centering
\caption{Extensive Margin in Job Corps: Week 180}
\label{tab_JC.EM_week180}
\begin{threeparttable}
\footnotesize
\begin{tabular*}{\linewidth}{@{\extracolsep{\fill}} lcccc}
\hline \hline
Method & Smoothing & Parameter & \multicolumn{2}{c}{Nuisance Model} \\ 
&&& XGBoost & Neural Network \\ 
\cline{4-5}
\\[-1ex]
Smooth & $h=1$ & $\beta_{B,h}(em)$ & [-1.035, 4.490] & [-0.501, 2.546] \\
       &       &                   & (-1.851, 9.584) & (-0.871, 4.000) \\[1ex]
       &       & $\pi_{em,h}$      & 0.009 & 0.033 \\
       &       &                   & (0.008, 0.009) & (0.032, 0.035) \\[2ex]
       & $h=5$ & $\beta_{B,h}(em)$ & [-18.563, 0.966] & [-5.366, -1.831] \\
       &       &                   & (-29.527, 66.590) & (-7.159, 23.304) \\[1ex]
       &       & $\pi_{em,h}$      & 0.002 & 0.019 \\
       &       &                   & (0.002, 0.003) & (0.017, 0.020) \\[2ex]
Regular &       & $\beta_{B}(em)$  & [-2.340, 4.414] & [-1.293, 1.245] \\
       &       &                   & (-4.569, 6.988) & (-1.639, 1.438) \\[1ex]
       &       & $\pi_{em}$        & 0.013 & 0.050 \\
       &       &                   & (-0.002, 0.028) & (0.029, 0.071) \\ 
\hline
\end{tabular*}
\end{threeparttable}
\end{table}

\begin{table}[!h] 
\centering
\caption{Extensive Margin in Job Corps: Week 208}
\label{tab_JC.EM_week208}
\begin{threeparttable}
\footnotesize
\begin{tabular*}{\linewidth}{@{\extracolsep{\fill}} lcccc}
\hline \hline
Method & Smoothing & Parameter & \multicolumn{2}{c}{Nuisance Model} \\ 
&&& XGBoost & Neural Network \\ 
\cline{4-5}
\\[-1ex]
Smooth & $h=1$ & $\beta_{B,h}(em)$ & [-0.269, 3.952] & [-1.273, 4.004] \\
       &       &                   & (-1.012, 7.699) & (-1.831, 7.530) \\[1ex]
       &       & $\pi_{em,h}$      & 0.013 & 0.020 \\
       &       &                   & (0.013, 0.014) & (0.018, 0.021) \\[2ex]
       & $h=5$ & $\beta_{B,h}(em)$ & [-11.431, 2.157] & [-5.560, -4.275] \\
       &       &                   & (-10.099, 124.198) & (-7.712, 44.263) \\[1ex]
       &       & $\pi_{em,h}$      & 0.004 & 0.011 \\
       &       &                   & (0.003, 0.004) & (0.010, 0.012) \\[2ex]
Regular &       & $\beta_{B}(em)$  & [-2.533, 3.335] & [-1.311, 1.181] \\
       &       &                   & (-3.421, 6.639) & (-1.704, 1.427) \\[1ex]
       &       & $\pi_{em}$        & -0.008 & 0.046 \\
       &       &                   & (-0.026, 0.010) & (0.026, 0.067) \\ 
\hline
\end{tabular*}
\end{threeparttable}
\end{table}

\section{Monte Carlo Study} \label{sec_montecarlo}
In this section, we analyze properties of methods for inference under conditional monotonicity in finite samples. In particular, across a set of principal strata shares we consider the bias, root mean-squared error, size, and power of trimming bounds, shifting bounds, and smooth bounds. Further, we analyze properties of smooth bounds along different values for the smoothing parameter $h$. We consider the following data-generating process

{\footnotesize
\begin{align*}
    & S(0) = \mathbbm{1}\{X_2 \ge V\}, \quad S(1) = \mathbbm{1}(X_1 + X_2 \ge V), \quad 
     Y(0) = 0, \\ &  Y(1) = \left[U \mathbbm{1}(S(1)=1, S(0)=1) + (U + \gamma) \mathbbm{1}(S(1)=1, S(0)=0)\right] \mathbbm{1}(X_1 = 1), \\
    & D \sim \text{Bernoulli}(0.5),\quad U \sim \text{Uniform}(0,1),\quad
    V \sim \mathcal{N}(0,1), \\
    & X_1 \sim \text{Categorical}\big(P(\mathcal{X}^+), P(\mathcal{X}^0), P(\mathcal{X}^-)\big) \text{ with support } X_1 \in \{1,0,-1\}, \\
    & X_2 \sim \text{truncated}~\mathcal{N}_{[-4,4]}(0,1). 
\end{align*}
}

This design implies that the target parameter, i.e.~the treatment effect on always-takers, equals the true lower bound of the identified set. This makes one-sided tests on the lower bound equivalent to testing inclusion of the target parameter in the identified set.
The target equals the lower bound because the always-takers and compliers are separated by a positive constant $\gamma$. This implies that units satisfying positive monotonicity are ordered such that always-takers never exhibit larger outcomes under treatment than compliers. This homogeneous design is set up in favor of the shifting method by \cite{semenova2025generalized} as the small mistakes when shifting moments around the margin based on selection probabilities do not introduce systematic bias. It also favors low bias over low variance smoothing as the conditional variance is constant around the truncation threshold.   We employ true nuisance functions throughout. 
All smooth approximations are based on the LogSumExp function.

We consider sample sizes $N = 400$ and $N = 2000$ across different strata shares $(P(\mathcal{X}^+), P(\mathcal{X}^0), P(\mathcal{X}^-))$. The first design $(1/2, 0, 1/2)$ is regular, the second design $(1/3, 1/3, 1/3)$ is irregular, and the third design $(5/100, 95/100, 0)$ is highly irregular. Table \ref{mctab} and Figure \ref{power} report results based on 2000 replications.

\subsection{Results}
\begin{figure}[p]
    \centering
    \caption{Power curves}

    \vspace{0.1em}
    \begin{subfigure}[b]{\textwidth}
        \centering
        \caption{$(P(\mathcal{X}^+), P(\mathcal{X}^0), P(\mathcal{X}^-)) = (1/2, 0, 1/2)$}
        \includegraphics[width=1\textwidth]{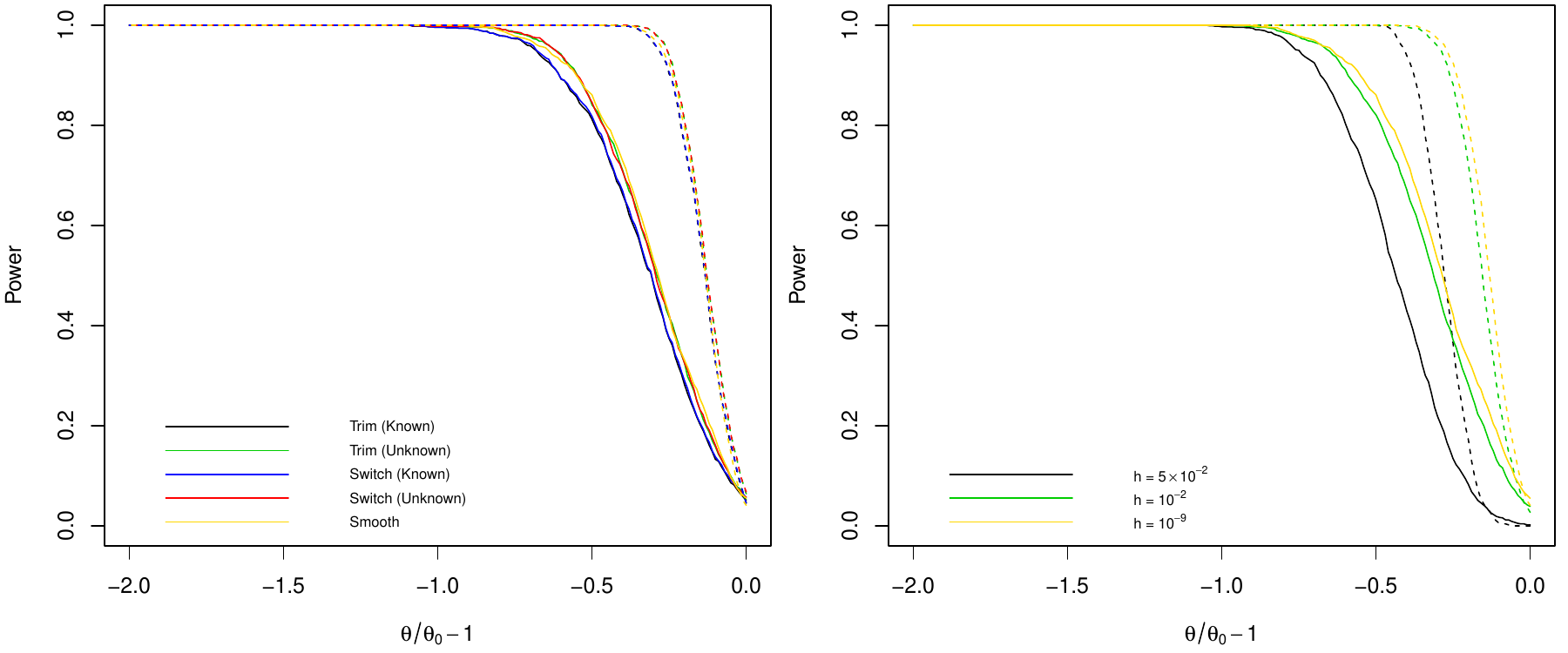}
        \label{pow1}
    \end{subfigure}

    \vspace{0.1em}
    
    \begin{subfigure}[b]{\textwidth}
        \centering
        \caption{$(P(\mathcal{X}^+), P(\mathcal{X}^0), P(\mathcal{X}^-)) = (1/3, 1/3, 1/3)$}
        \includegraphics[width=1\textwidth]{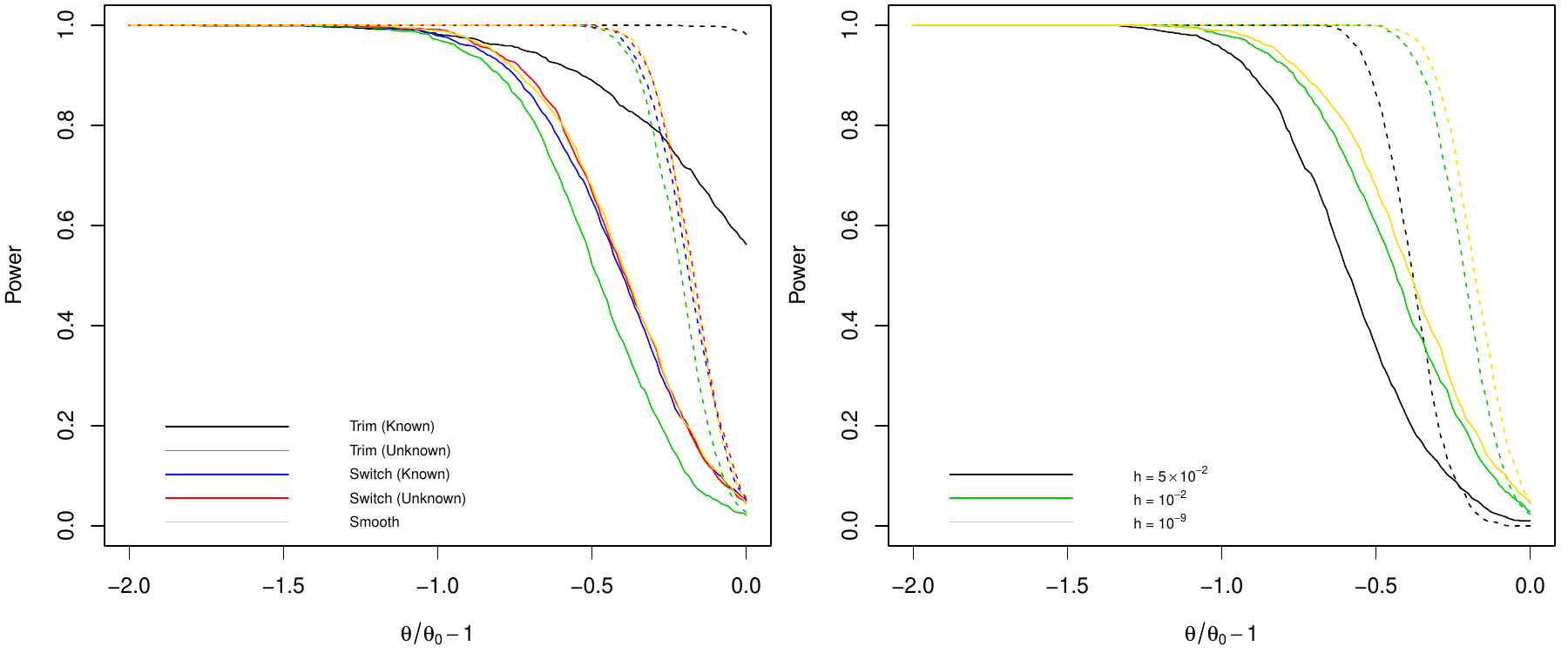}
        \label{pow2}
    \end{subfigure}

    \vspace{0.1em}

    \begin{subfigure}[b]{\textwidth}
        \centering
        \caption{$(P(\mathcal{X}^+), P(\mathcal{X}^0), P(\mathcal{X}^-)) = (5/100, 95/100, 0)$}
        \includegraphics[width=1\textwidth]{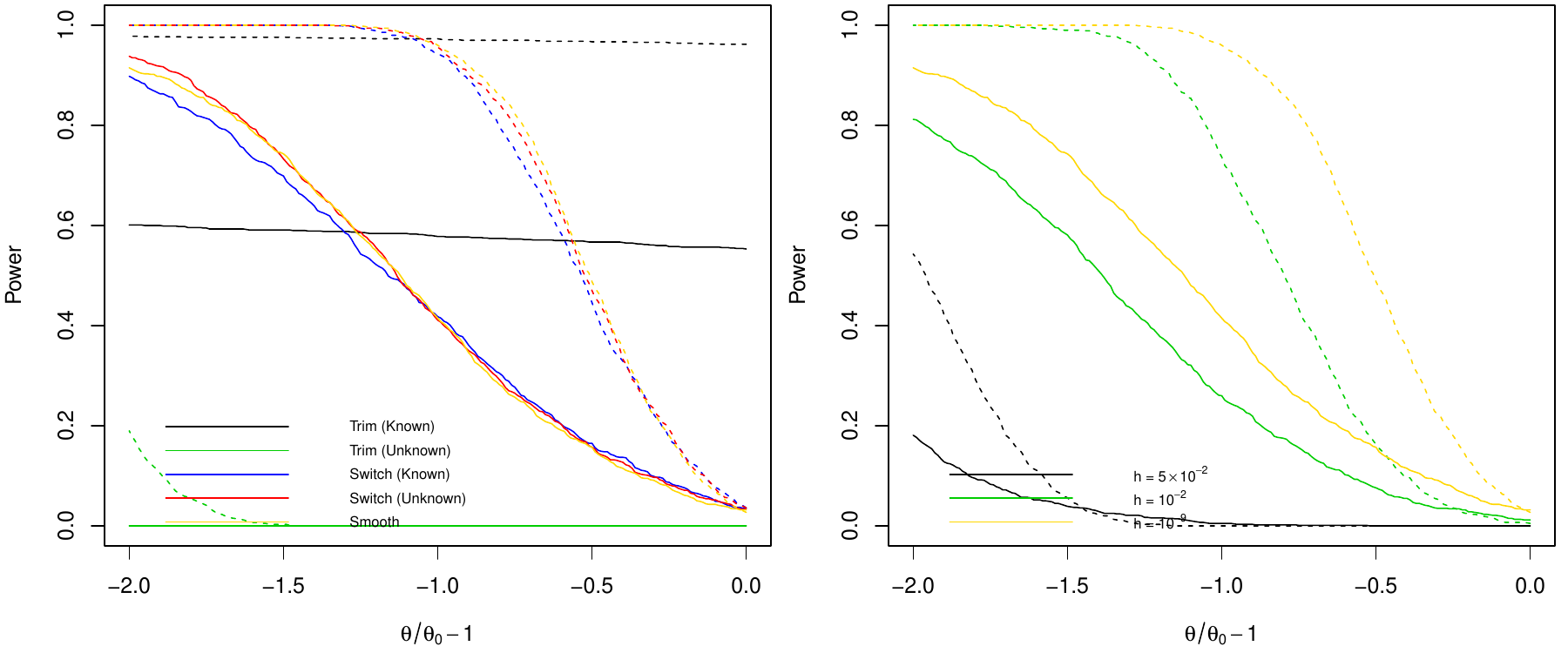}
        \label{pow3}
    \end{subfigure}
    
    \label{power}
    \vspace{-1.5em}
    \begin{justify} \footnotesize
    \centering
        Solid lines are $N=400$ and dotted lines are $N=2000$.
    \end{justify}
\end{figure}

Under the regular design reported in Table \ref{panelA}, all methods are unbiased. 
Bias and RMSE decrease with sample size. All methods are approximately correctly sized. Table \ref{panelA} and Figure \ref{pow1} show that methods with unknown propensity score display lower RMSE and higher power as predicted by Theorem \ref{thm_efficiency_gap}.

Under the irregular design reported in Table \ref{panelB}, all methods are unbiased except the trim method with known propensity score. Methods with unknown propensity score again display lower RMSE. The shift and the smooth ($h = 10^{-9}$) methods are approximately correctly sized. The trim method with unknown propensity score is incorrectly sized as expected. Figure \ref{pow2} shows that the shift method with unknown propensity score and the smooth method display higher power than the shift method with known propensity score. The power of trim methods is not well behaved.

Under the highly irregular design reported in Table \ref{panelC}, all methods are again unbiased except the trim method with known propensity score. RMSE is again smaller for methods with unknown propensity score. The shift and the smooth methods are approximately correctly sized, and the trim method with unknown propensity score is incorrectly sized, as in the irregular design. Figure \ref{pow3} shows again that the shift method with unknown propensity score and the smooth method display higher power than the shift method with known propensity score. Again, the power of trim methods is not well behaved.

In summary, in this data-generating process, the shift and smooth methods are unbiased and approximately correctly sized across designs. RMSE is lower and power is higher for methods with unknown propensity score. As expected, the trim methods are only well behaved in regular designs. Other results might arise under alternative data-generating processes, for example the shift methods are theoretically biased, while the smooth method remains unbiased, under alternative distributions of $p_0(x)$.

The bias decreases in $h$ across designs. RMSE also generally decreases except in the irregular design from $h = 10^{-2}$ to $h = 10^{-9}$. The size tends to one and power tends to zero as $h$ increases. 
Future designs can evaluate potential bias-variance trade-offs and size distortion of shifting methods when effects are heterogeneous.

\begin{table}[!p]
    \footnotesize
    \centering
    \caption{\normalsize Simulation results}
    \renewcommand{\arraystretch}{1.5}

    \newcolumntype{C}{>{\centering\arraybackslash}X} 

    \begin{subtable}{\textwidth}
        \centering
        \vspace{0.5em}
        \caption{$(P(\mathcal{X}^+), P(\mathcal{X}^0), P(\mathcal{X}^-)) = (1/2, 0, 1/2)$}
        \vspace{0.5em}
        \begin{tabularx}{0.7\textwidth}{l C C C}
            Method & Bias & RMSE & Size \\
            \hline
            Shift (Known) & $-0.001;0.000$ & $0.065;0.029$ & $0.056;0.052$ \\
            Shift (Unknown) & $-0.001;0.000$ & $0.060;0.027$ & $0.058;0.055$ \\
            Trim (Known) & $-0.002;-0.001$ & $0.065;0.029$ & $0.053;0.051$ \\
            Trim (Unknown) & $-0.002;0.000$ & $0.059;0.027$ & $0.052;0.054$ \\
            \hline
            Smooth &  &  &  \\                        
            $h = 5\times 10^{-2}$ & $-0.053;-0.051$ & $0.078;0.057$ & $0.005;0.000$ \\
            $h = 10^{-2}$ & $-0.010;-0.011$ & $0.061;0.028$ & $0.042;0.017$ \\
            $h = 10^{-9}$ & $-0.002;0.000$ & $0.060;0.027$ & $0.052;0.054$ \\
            \hline
            \label{panelA}
        \end{tabularx}
    \end{subtable}
    
    \vspace{1em}
    
    \begin{subtable}{\textwidth}
        \centering
        \vspace{0.5em}
        \caption{$(P(\mathcal{X}^+), P(\mathcal{X}^0), P(\mathcal{X}^-)) = (1/3, 1/3, 1/3)$}
        \vspace{0.5em}
        \begin{tabularx}{0.7\textwidth}{l C C C}
            Method & Bias & RMSE & Size \\
            \hline
            Shift (Known) & $0.002;0.000$ & $0.052;0.022$ & $0.052;0.042$ \\
            Shift (Unknown) & $0.002;0.000$ & $0.049;0.021$ & $0.054;0.048$ \\
            Trim (Known) & $0.137;0.136$ & $0.158;0.140$ & $0.547;0.982$ \\
            Trim (Unknown) & $0.001;0.000$ & $0.048;0.021$ & $0.020;0.020$ \\
            \hline
            Smooth &  &  &  \\                        
            $h = 5\times 10^{-2}$ & $-0.043;-0.042$ & $0.064;0.047$ & $0.006;0.000$ \\
            $h = 10^{-2}$ & $-0.008;-0.008$ & $0.047;0.023$ & $0.034;0.021$ \\
            $h = 10^{-9}$ & $0.001;0.000$ & $0.048;0.022$ & $0.054;0.046$ \\
            \hline
            \label{panelB}
        \end{tabularx}
    \end{subtable}
    
    \vspace{1em}
    
    \begin{subtable}{\textwidth}
        \centering
        \vspace{0.5em}
        \caption{$(P(\mathcal{X}^+), P(\mathcal{X}^0), P(\mathcal{X}^-)) = (5/100, 95/100, 0)$}
        \vspace{0.5em}
        \begin{tabularx}{0.7\textwidth}{l C C C}
            Method & Bias & RMSE & Size \\
            \hline
            Shift (Known) & $0.000;0.000$ & $0.018;0.008$ & $0.035;0.044$ \\
            Shift (Unknown) & $0.000;0.000$ & $0.017;0.008$ & $0.030;0.042$ \\
            Trim (Known) & $0.426;0.464$ & $0.624;0.478$ & $0.587;0.952$ \\
            Trim (Unknown) & $0.000;0.000$ & $0.017;0.008$ & $0.000;0.000$ \\
            \hline
            Smooth &  &  &  \\                        
            $h = 5\times 10^{-2}$ & $-0.037;-0.036$ & $0.040;0.037$ & $0.000;0.000$ \\
            $h = 10^{-2}$ & $-0.007;-0.007$ & $0.018;0.011$ & $0.011;0.003$ \\
            $h = 10^{-9}$ & $0.000;0.000$ & $0.017;0.008$ & $0.034;0.044$ \\
            \hline
            \label{panelC}
        \end{tabularx}
    \end{subtable}
    
    \vspace{0.8em}
    \begin{justify} \footnotesize
    \centering
        Result before semicolon is $N=400$ and after is $N=2000$.
    \end{justify}
    \label{mctab}
\end{table}

\clearpage

\section{Extensions} \label{sec_extensions}
\subsection{Mean Dominance}
The bounds in Proposition \ref{prop_identification1} can further be refined using mean-dominance assumptions. Mean dominance to bound causal effects in sample selection models has been previously suggested by \cite{zhang2003estimation} for always-takers and by \cite{huber2015sharp} for compliers under strong monotonicity without covariates.

\begin{ass}[Mean Dominance] \label{ass_dominance}
    The conditional potential outcomes at the intensive margin are larger than at the extensive margin, i.e. \begin{align}
        E[Y(d)|S(0)=1,S(1)=1,X=x] \geq E[Y(d)|S(0)\neq S(1), X=x]
    \end{align}
    for any $d=0,1$.
\end{ass}
This yields the following modified sharp bounds.\footnote{
If only either complier or defier bounds are of interest, Assumption \ref{ass_dominance} can be relaxed to hold for the respective group only.}
\begin{prop}[Identification with Mean Dominance] \label{prop_identification_dominance1}
Let $\mathcal{Y}_d(x)$ be the support of $Y(d)$ conditional on $X=x$ and $\overline{y}_d(x)$ and $\underline{y}_d(x)$ its respective infimum and supremum for $d\in\{0,1\}$. Under Assumptions \ref{ass_CIA}, \ref{ass_monotone}, \ref{ass_continuity}, \ref{ass_overlap}, and \ref{ass_dominance}, the lower bounds for the principal strata are given by \begin{align*}
        \beta_L(x,s_0,s_1) = \begin{cases}
            \beta_{1,1}(x,1) - \beta_{0,0}(x,1-\min\{1/p_0(x),1\}) &\text{ if } s_0 = 1,~s_1 = 1 \\
            \beta_{1,1}(x,1-\min\{p_0(x),1\}) - \beta_{0,0}(x,0) &\text{ if } s_0 \neq s_1 \\
            \underline{{y}}_1(x) - \overline{{y}}_0(x) &\text{ if } s_0 = 0,~s_1 = 0,
        \end{cases}
    \end{align*}
    and the upper bounds by \begin{align*}
        \beta_U(x,s_0,s_1) = \begin{cases}
            \beta_{0,1}(x,1-\min\{p_0(x),1\}) - \beta_{1,0}(x,1) &\text{ if } s_0 = 1,~s_1 = 1 \\
            \beta_{0,1}(x,0) - \beta_{1,0}(x,1-\min\{1/p_0(x),1\}) &\text{ if } s_0 \neq s_1 \\
            \overline{{y}}_1(x) - \underline{{y}}_0(x) &\text{ if } s_0 = 0,~s_1 = 0.
        \end{cases}
    \end{align*}
\end{prop}
 As the shares of the principal strata are already identified via conditional monotonicity, aggregation to unconditional bounds uses the same conditional probability weights as in the case without mean dominance in \eqref{def_betaB_general} and thus the same moment function for the denominator or density part in \eqref{eq_esthat1}. For the numerator, all truncated quantities can be estimated as in the pure monotonicity case with the components in the influence function corresponding to the now untruncated means replaced by their standard untruncated augmented IPW counterparts for the conditional outcome means \citep{robins1994estimation}. 

\subsection{Heterogeneous Bounds}
In many evaluation problems, objects of interest are heterogeneous effects, e.g.~the effect at a particular margin for an observable subgroup. \cite{heiler2024heterogeneous} argues that, in the nonparametric sample selection model and more broadly, there are two types of heterogeneity, 1) heterogeneous effects and 2) heterogeneity in the severity of the identification problem, i.e.~the width of the identified set. Exploiting these in combination can yield more precise inference and significant effect bounds even when unconditional aggregate bounds do not reject a null-effect. Our influence functions for any principal stratum or margin can readily be used for the estimation of such heterogeneous effect bounds following the method of \cite{heiler2024heterogeneous}. In particular, let $f:\mathcal{X}\rightarrow \mathcal{Z}$ and define $Z = f(X)$ as a low-dimensional subgroup indicator or mapping from the covariate space. Recall that all presented influence functions have the following linear structure
 \begin{align}
     \psi_{\beta_B}(W,\eta,\beta_B) = \psi_{\beta_B}^{[B]}(W,\eta) - \psi^{[S]}_{\beta_B}(W,\eta)\beta_B,
 \end{align}
where \begin{align}
    E[\psi^{[B]}_{\beta_B}(W,\eta)|X=x] &= \beta_L(x,s_0,s_1)P(S(0)=s_0,S(1)=s_1|X=x), \notag \\
    E[\psi^{[S]}_{\beta_B}(W,\eta)|X=x] &= P(S(0)=s_0,S(1)=s_1|X=x). 
\end{align}
  Thus, for any $z \in \mathcal{Z}$, \begin{align}
     E[&Y(1)-Y(0)|S(0)=s_0,S(1)=s_1,Z=z] \notag\\
     &\geq E[\beta_L(X,s_0,s_1)|S(0)=s_0,S(1)=s_1,Z=z] \notag\\
     &= \frac{E[\beta_L(X,s_0,s_1)P(S(0)=s_0,S(1)=s_1|X)|Z=z]}{E[P(S(0)=s_0,S(1)=s_1|X)|Z=z]} \notag\\
     &= \frac{E[\psi^{[B]}_{\beta_B}(W,\eta)|Z=z]}{E[\psi^{[S]}_{\beta_B}(W,\eta)|Z=z]},
 \end{align}
 by the law of iterated expectations. Hence we can obtain conditional effect bounds by aggregating the components of the presented influence functions within covariate partitions or by (nonparametric) projection onto (basis transformations of) $Z$. Regularity and/or the necessity for smoothing can be obtained under analogous conditions to the ones presented in this paper. For more details consider \cite{heiler2024heterogeneous}.

\end{document}